\documentclass[12pt]{iopart}

\usepackage{graphicx}
\usepackage{iopams}
\usepackage{setstack}
\usepackage{tabstackengine}
\usepackage{multirow}
\usepackage{appendix}
 \usepackage{comment}
  \usepackage{hyperref}
  \hypersetup{
    colorlinks=true,
    linkcolor=blue,
    filecolor=magenta,      
    urlcolor=cyan,
}
\newcommand{\bd}{\begin{displaymath} }
\newcommand{\ed}{\end{displaymath} }

\newcommand{\numu}[0]{\nu_{\mu}}

\newcommand{\nub}{\overline{\nu}}

\newcommand{\ubar}{\overline{u}}
\newcommand{\dbar}{\overline{d}}
\newcommand{\cbar}{\overline{c}}
\newcommand{\sbar}{\overline{s}}

\newcommand{\minerva}{\mbox{\hbox{MINER}$\nu$\hbox{A}}}

\begin{document}
%[Review article]
\title{Neutrino(Antineutrino)-Nucleus  ~Interactions  ~in  ~the  ~Shallow- ~and  ~Deep-Inelastic  ~Scattering  ~Regions}

\author{M. Sajjad Athar}
\address{Department of Physics, Aligarh Muslim University, Aligarh - 202 002, India}
\ead{sajathar@gmail.com}
\author{Jorge G. Morf\'{i}n}
\address{Fermi National Accelerator Laboratory, Batavia, Illinois 60510, USA}
\ead{morfin@fnal.gov}
\vspace{10pt}
%\date{\today}
\begin{center}
%\today
\end{center}
% \begin{indented}
% \item[]August 2017
% \end{indented}
%\submitto{\jpg}

\begin{abstract}

In $\nu/\nub$-Nucleon/Nucleus interactions Shallow Inelastic Scattering (SIS) is technically defined in terms of the four-momentum transfer to the hadronic system as non-resonant meson production with $Q^2 \lessapprox 1~GeV^2$.  This non-resonant meson production intermixes with resonant meson production in a regime of similar effective hadronic mass W of the interaction.    As $Q^2$ grows and surpasses this $\approx 1~GeV^2$ limit, non-resonant interactions begin to take place with quarks within the nucleon indicating the start of Deep Inelastic Scattering (DIS). To essentially separate this resonant plus non-resonant meson production from DIS quark-fragmented meson production, a cut of 2 GeV in 
%effective hadronic mass 
W of the interactions is generally introduced. However, since experimentally mesons from resonance decay cannot be separated from non-resonant produced mesons,  SIS for all practical purposes in this review has been defined as inclusive meson production that includes non-resonant plus resonant meson production and the interference between them.  Experimentally then for W $\lessapprox$ 2 GeV inclusive meson production with W $\gtrapprox (M_N + M_\pi)$ and all $Q^2$ is here defined as SIS, while for W $\gtrapprox$ 2 GeV,  
%the region with $Q^2 \lessapprox 1~GeV^2$ is a continuation of SIS 
the kinematic region with $Q^2 \gtrapprox 1~GeV^2$ is defined as DIS. 
%However, experimentally, the SIS region is defined to also includes resonant meson production, which cannot be separated from non-resonant meson production, and the interference between the non-resonant a\end{comment}nd resonant processes in this W $\lessapprox$ 2 GeV region.  
%Note then that  SIS events are present with W $\gtrapprox$ 2~GeV as long as $Q^2 \lessapprox 1~GeV^2$.
The so defined SIS and DIS regions have received varying degrees of attention from the community.  While the theoretical / phenomenological study of $\nu$-nucleon and $\nu$-nucleus DIS scattering is advanced, such studies of a large portion of the SIS region, particularly the SIS to DIS transition region, have hardly begun.  Experimentally, the SIS and the DIS regions for $\nu$-nucleon scattering have minimal results and only in the experimental study of the $\nu$-nucleus DIS region are there  significant results for some nuclei.
%mainly low-statistics bubble chamber data from the 1970's and 1980's.  
%Therefore, both theoretically and experimentally a large portion of the SIS region for $\nu$-nucleus scattering, especially the SIS to DIS transition region, is in need of considerable study.  
%Consequently this review 
%It is only in the experimental study of the $\nu$-nucleus DIS region that there are statistically significant results from several experiments. 
Since current and future neutrino oscillation experiments have contributions from both higher W SIS and DIS kinematic regions and these regions are in need of both considerable theoretical and experimental study, this review will concentrate on these SIS to DIS transition and DIS kinematic regions surveying our knowledge and the current challenges.

%have been quite rigorously investigated experimentally and theoretically / phenomenologically for e/$\mu$-N scattering and somewhat less thoroughly for e/$\mu$-A scattering.  In the study of the kinematic region above quasi-elastic scattering in $\nu$-A scattering
\end{abstract}
\maketitle
\tableofcontents
\markboth{}{}

\section{Introduction}
 %The present neutrino and antineutrino ($\nu_l/\bar\nu_l$) oscillation experiments using reactor, solar, accelerator and atmospheric $\nu_l$ and $\bar\nu_l$ have measured the oscillation parameters $\theta_{ij}(i<j=1-3)$  and mass squared differences $\Delta m_{ij}^2(i,j=1-3; i \ne j)$ with varying degree of accuracy. The sign of $\Delta m_{32}^2$ that fixes normal vs inverted mass hierarchy has not yet been determined  although there are some hints from the recent analyses that favor normal hierarchy~\cite{Simpson:2017qvj}.The next generation of oscillation experiments with $\nu$ and $\bar \\end{comment}nu$ have the goal of determining both this mass hierarchy as well as measure $\delta$, which quantifies CP violation, in the Pontecorvo-Maki-Nakagawa-Sakata (PMNS matrix) mixing matrix.The knowledge of $\delta$ and mass hierarchy will completely describe the physics of three flavor $\nu_l$ oscillation phenomenology in the lepton sector of weak interactions. 
 
 The study of neutrino and antineutrino ($\nu_l/\bar\nu_l$) interactions with nuclei covers an extended range of energies from the coherent elastic scattering off nuclei studied by experiments like CE$\nu$NS~\cite{Scholberg:2005qs,Akimov:2017ade} to the ultra high energy cosmological (multi-messenger) neutrinos studied by experiments like IceCube~\cite{Halzen:2013dva}.
 In the energy range of accelerator-based and atmospheric neutrinos, the experimental study of neutrino physics is currently focused on understanding the three flavor $\nu_l$ oscillation phenomenology in the lepton sector of weak interactions. In particular an accurate measurement of any CP violation as well as determining the mass hierarchy of the three neutrino mass states is the goal of current and future neutrino oscillation experiments. 
 
 The experimental determination of these important properties depend on accurate knowledge of the energy ($E_\nu$) of the interacting $\nu_l$  and the produced particles at the interaction point.
 %The experimental determination of $\theta_{ij}$ and $\Delta m_{ij}^2$ depends on accurate knowledge of the energy ($E_\nu$) of the interacting $\nu_l$  and the produced particles at the interaction point.
 However, due to the weak nature of these interactions, to obtain necessary statistics $\nu_l$ oscillation experiments using accelerator and atmospheric  $\nu_l/\bar\nu_l$ have been using moderate to heavy nuclear targets like $^{12}C$, $^{16}O$, $^{40}Ar$ and $^{56}Fe$.  %$^{208}Pb$. 
 This complicates the precision measurement of  these properties since to obtain the initial energy and produced topology of the interacting neutrino, as opposed to the energy and topology measured in the detectors, model-dependent nuclear corrections, referred to as the "nuclear model", must be applied to the interpretation of the data. This nuclear model contains the current knowledge of the initial $\nu_l/\bar\nu_l$ - nucleon cross sections, the initial state nuclear medium effects and the final state interactions of the produced hadrons within the nucleus. The introduction of this nuclear model to the interpretation of experimental data is performed by Monte Carlo simulation programs (neutrino event generators) that apply these nuclear effects to the free nucleon interaction cross sections. Note that in this procedure, even before introducing  uncertainties associated with the nuclear model ~\cite{Katori:2016yel}-\cite{Buss:2011mx}, uncertainties are
 %\end{comment} 
 already introduced into the analysis due to the lack of precise knowledge of the $\nu_l$ \emph{nucleon} interaction cross sections. 
 
%  together, yield large uncertainties in the true energy of neutrinos.  Combined with this is uncertainty in the $\nu_l/\bar\nu_l$ flux and detector-associated uncertainties.
 
%Presently, the neutrino-nucleus scattering cross section is known within a precision of 20-25$\%$. 
In the energy region of $\approx$ 1-10 GeV, covering present and future oscillation experiments,
the final states are dominated by quasielastic(QE) scattering, resonant and non-resonant (mainly) $\pi$ production
 %shallow inelastic scattering (SIS) that includes resonant and non-resonant meson production 
 and deep inelastic (DIS) scattering processes. These scattering processes are possible via charged(CC) as well as neutral(NC) current channels for which the main basic reactions on a free nucleon target 
 %within the relevant energy range 
 are given by:  
 %I WOULD SIMPLIFY THE ARRAYS TO INCLUDE ONLY PION PRODUCTION THEN INCLUDE THE A STATEMENT LIKE "KAON, %HYPERON AND  RESONANCE DECAYS TO MORE MASSIVE STATES ARE ALSO POSSIBLE HOWEVER AT MUCH REDUCED RATES...
 \begin{eqnarray}
 \left.
\begin{array}{l}
  \nu_{l}(k) + n(p) ~~~\longrightarrow~~~ l^{-}(k^{\prime}) + p (p^{\prime}) , \\
  \bar{\nu}_{l}(k) + p(p) ~~~\longrightarrow~~~ l^{+}(k^{\prime}) + n (p^{\prime}) ,
  \end{array}\right\}~~~\mbox{(CC QE)}
 \end{eqnarray}

  \begin{eqnarray}
  \nu_{l}/\bar{\nu}_{l}(k) + N(p) ~~~\longrightarrow~~~ \nu_{l}/\bar{\nu}_{l}(k^{\prime}) + N (p^{\prime})~~~~(\textrm{NC \;\;elastic})
 \end{eqnarray}
 
 %\begin{comment}
  % \begin{eqnarray}
% \left.
%\begin{array}{l}
 % \nu_{l}/\bar{\nu}_{l}(k) + N(p) ~~~\longrightarrow ~~~ l^{-}/l^{+}(k^{\prime}) + N (p^{\prime}) +\end{comment} \pi(p_{\pi}), \\
  %\begin{comment}
  %\nu_{l}/\bar{\nu}_{l}(k) + N(p) ~~~\longrightarrow~~~ l^{-}/l^{+}(k^{\prime}) + N (p^{\prime}) + n \pi(p_{\pi}), \\
  %\nu_{l}(k) + N(p) ~~~~~~~\longrightarrow~~~ l^{-}(k^{\prime}) + N (p^{\prime}) + K^{+}(p_{K}), \\
  %\bar{\nu}_{l}(k) + N(p) ~~~~~~~\longrightarrow~~~ l^{+}(k^{\prime}) + N (p^{\prime}) + K^{-}(p_{K}), \\
  %\nu_{l}/\bar{\nu}_{l}(k) + N(p) ~~~\longrightarrow~~~ l^{-}/l^{+}(k^{\prime}) + Y (p^{\prime}) + K(p_{K}), \\
 % \nu_{l}/\bar{\nu}_{l}(k) + N(p) ~~~\longrightarrow~~~ l^{-}/l^{+}(k^{\prime}) + N (p^{\prime}) + \eta(p_{\eta}), \\
 % \bar{\nu}_{l}(k) + N(p) ~~~~~~~\longrightarrow~~~ l^{+}(k^{\prime}) + Y (p^{\prime}) , \\
 % \qquad \quad \vdots \qquad \qquad \quad \qquad \quad \vdots
  %\end{comment}
 % \end{array}\right\}~~~\mbox{(CC resonance)}\;\;\;\;\;\;
% \end{eqnarray}
 %\end{comment}
    \begin{eqnarray}
   \nu_{l}/\bar{\nu}_{l}(k) + N(p) ~~~\longrightarrow ~~~ l^{-}/l^{+}(k^{\prime}) + N (p^{\prime}) + m \pi(p_{\pi})~~(\textrm{CC ~resonance})
 \end{eqnarray}
  \begin{eqnarray}
   \nu_{l}/\bar{\nu}_{l}(k) + N(p) ~~~\longrightarrow ~~~ \nu_{l}/\bar{\nu}_{l}(k^{\prime}) + N (p^{\prime}) + m \pi(p_{\pi})~~(\textrm{NC ~resonance})
 \end{eqnarray}
 %(\text{NC \;\;elastic})
 %   \begin{eqnarray}
% \left.
%\begin{array}{l}
 % \nu_{l}/\bar{\nu}_{l}(k) + N(p) ~~~\longrightarrow ~~~ \nu_{l}/\bar{\nu}_{l}(k^{\prime}) + N (p^{\prime}) %+ \pi(p_{\pi}), \\
 % \nu_{l}/\bar{\nu}_{l}(k) + N(p) ~~~\longrightarrow~~~ \nu_{l}/\bar{\nu}_{l}(k^{\prime}) + N (p^{\prime}) %+ n \pi(p_{\pi}), \\
%  \nu_{l}/\bar{\nu}_{l}(k) + N(p) ~~~\longrightarrow~~~ \nu_{l}/\bar{\nu}_{l}(k^{\prime}) + Y (p^{\prime}) + K(p_{K}), \\
 % \nu_{l}/\bar{\nu}_{l}(k) + N(p) ~~~\longrightarrow~~~ \nu_{l}/\bar{\nu}_{l}(k^{\prime}) + N (p^{\prime}) %+ \eta(p_{\eta}), \\
 % \qquad \quad \vdots \qquad \qquad \quad \qquad \quad \vdots
 % \end{array}\right\}~~~\mbox{(NC resonance)}\;\;\;\;\;\;
% \end{eqnarray}

   \begin{eqnarray}
  \nu_{l}/\bar{\nu}_{l}(k) + N(p) ~~~\longrightarrow~~~l^{-}/l^{+}(k^{\prime}) + X (p^{\prime}) \;\;\;(\textrm{CC~DIS})
  \end{eqnarray}
    \begin{eqnarray}
  \nu_{l}/\bar{\nu}_{l}(k) + N(p) ~~~\longrightarrow~~~ \nu_{l}/\bar{\nu}_{l}(k^{\prime}) + X (p^{\prime})\;\;\;(\textrm{NC~DIS})
 \end{eqnarray}
 where the quantities in the parenthesis represent respective momenta carried by the particles, $N$ represents a proton or neutron, $\pi$ represents any of the three pion charge states depending upon  charge conservation, m represents number of pions in the final state and $X$ represents jet of hadrons in the final state.  Besides these production modes, kaon, hyperon, eta production and resonance decays to more massive states are also possible, however, at much reduced rates.
 
 In this review, the considered  signatures of $\nu_l$ and $\bar\nu_l$ interactions with nuclear targets are exclusively \emph {charged current} interactions yielding a charged lepton in the final state. In addition, we will be concentrating on the higher hadronic effective mass states that transition into and are within the deep-inelastic scattering regime.  This transition region includes higher effective mass resonant and non-resonant single and multi-pion production. Although the quasi-elastic~\cite{Sobczyk:2011bi} interaction
 and $\Delta$ resonance production~\cite{Paschos:2012va} 
 are also important in the few GeV region, they are not within the scope of this review. 
 
 As indicated, when a $\nu_l$ or $\bar\nu_l$ interacts with a nucleon bound in a nuclear target, nuclear medium effects become important. These nuclear medium effects are energy dependent and moreover different for each interaction mode. In resonant and non-resonant production nuclear effects of the initial state such as Fermi motion, binding energy, Pauli blocking, multi-nucleon  correlation effects have to be taken into account.
 %medium-ependent quenching dof coupling strengths and mesonic effects 
 In addition, final state interaction of  the produced nucleons and pions within the nucleus are also very important. 
 %These calculations appear as the nuclear models in Monte Carlo simulations and there is some tension between the data and these simulations.
 There are several theoretical calculations of these initial and final state nuclear medium effects in inelastic scattering where one pion is
 produced~\cite{Singh:1998ha}-\cite{Hernandez:2013jka}. However, as summarized in a recent white paper from NuSTEC~\cite{Alvarez-Ruso:2017oui}, there are much more limited studies of multi-$\pi$ resonant production and the other shallow inelastic scattering(SIS)  processes such as non-resonant $\pi$ production and the resulting interference of resonant/non-resonant states in the weak sector.  
 %These important points will be discussed in this review.
 
 The importance of non-resonant meson production and the resulting interference effects with resonant production is receiving renewed emphasis currently since there are efforts underway to produce more theoretically-based estimates~\cite{Hernandez:2016yfb, Kabirnezhad:2016nwu} of these processes rather than the phenomenological approach of extrapolating the DIS cross sections to lower hadronic mass W used in some MC generators~\cite{Kovarik:2015cma,Bodek:2002ps}.
Since it is not possible to experimentally distinguish resonant from non-resonant pion production, this kinematic regime as both the $Q^2$ of the non-resonant meson production and $W$ of the resonant region increase and transitions into the DIS region can only be studied in terms of inclusive production for example, by Morf\'{i}n et al.~\cite{Morfin:2012kn}, Melnitchouk et al.~\cite{Melnitchouk:2005zr}, 
Lalakulich et al.~\cite{Lalakulich:2006yn}, Christy et al.~\cite{Christy:2011cv}, and more recently by the Ghent group~\cite{Andreopoulos:2019gvw} that has employed Regge theory to describe this transition region.  This then is where low $Q^2$ non-resonant meson production in the SIS region transitions into higher $Q^2$ quark-fragmented meson production in the DIS region.  The need of improved understanding of $\nu_l/\bar\nu_l$-nucleus scattering cross sections in this transition region has generated considerable interest in studying Quark-Hadron Duality (duality) in the weak sector. 

 Duality has been studied in the electroproduction sector for both nucleon and nuclear targets and there is a body of evidence that 
%both Global and Local 
duality does approximately hold in this sector.
 %through the observation of parity violating asymmetry in the scattering of polarized electron from proton and deuteron targets at JLab~\cite{Wang:2014guo,Wang:2013kkc}.
 The few studies of duality in the weak sector have had to be based on theoretical models since no high-statistics, precise experimental data is available. These studies have not been encouraging suggesting that increased experiment and better modeling is required.  This also suggests caution in using the approach of simply extrapolating the DIS cross sections to lower hadronic mass $W$ used in some MC generators to estimate non-resonant $\pi$ as well as resonant multi-$\pi$ production in the SIS region.   If a form of duality is found to be valid for neutrino scattering, it can be 
  effectively used to theoretically describe the SIS/DIS transition region of $\nu_l/\bar \nu_l$-{\emph nucleon} scattering. 
  %The nuclear medium effects can be then incorporated in applying the Q duality to $\nu_l/\bar \nu_l$-nucleus scattering cross sections in the transition region. 

% ~\cite{SajjadAthar:2007bz,SajjadAthar:2009cr,Haider:2011qs,Haider:2012nf, Haider:2012ic, Athar:2013gya, Haider:2015vea,
%  Haider:2014iia,Haider:2016tev,Haider:2016zrk,Zaidi:2019mfd,Zaidi:2019asc} 

Increasing W and $Q^2$ of the interaction brings the regime of deep-inelastic scattering.  The definition of DIS is based upon the kinematics 
  %(Fig.\ref{fg:fig1mod}) 
 of the interaction products 
%and there is no precise way to distinguish the onset of the DIS region from the resonance region.
and is primarily defined with $Q^2 \ge 1.0$ GeV$^2$.  To further separate resonance produced pions from quark-fragmented pions a requirement of $ W \ge 2.0 $ GeV is made.  In the DIS region, charged lepton induced processes have been used to explore quark and gluon structures of nucleons and nuclei for quite some time.
It was assumed that the structure functions of a free nucleon would be the same as the structure functions of a nucleon within the nucleus environment. 
However, close to four decades ago EMC~\cite{Aubert:1983xm,Aubert:1986yn} performed experiments using a muon beam in the energy region of 120-280 GeV and measured cross sections from an iron target compared with the results on a deuterium target. It was found that the ratio of the cross section ${2\sigma_{Fe} \over A \sigma_D}$ is not unity in the DIS region.  This was surprising as this is the region where the underlying degrees of freedom should be quarks and gluons, while the deviation from unity suggested that nuclear medium effects were important.

In these ($\ell^\pm$-A) DIS interactions the deviation from 1.0 in the ratio of nuclear to nucleon structure functions as a function of $x_{Bjorken}$ ($\equiv x$), reflecting these nuclear medium effects, have been categorized in four regions: "shadowing" at lowest-x ($ \lessapprox 0.1$), "antishadowing" at intermediate x (0.1 to $\lessapprox 0.25$), the "EMC effect" at medium x (0.25 to $\lessapprox 0.7$) and "Fermi Motion effect" at high x ($\gtrapprox 0.7)$. Various attempts have been made both phenomenologically~\cite{Kovarik:2015cma},\cite{deFlorian:2003qf}-\cite{Khanpour:2016pph} 
as well as theoretically~\cite{Malace:2014uea}-\cite{Armesto:2006ph} 
 to understand these nuclear medium effects.  However, while the shadowing and Fermi motion regions are now better understood, there is still no community-wide, accepted explanation for antishadowing and the EMC effect.

Turning now to neutrinos, in the study of DIS neutrinos have significant importance over charged-leptons by having an ability to interact with 
particular quark flavors which help to understand the parton distributions inside the target nucleon. Hence, precise determination of weak structure functions ($F_{iA}^{WI}(x,Q^2$); $i=1,2,3,L$) is important. The nuclear effects in  neutrino DIS analyses had been assumed to be the same as for charged lepton-nucleus ($\ell^\pm$-A) DIS data. However there are now both theoretical and  experimental suggestions that the nuclear effects in the DIS region may be different for $\nu_l/\bar\nu_l$-nucleus interactions as there are contributions from the axial current in the weak sector and different valence and sea quark contributions for each observable. Therefore, an independent and quantitative understanding of the DIS nuclear medium effects in the weak sector is required.   

 The historical experimental study of neutrino-nucleus ($\nu$-A) scattering in the DIS region is summarized in sections~\ref{Subsec-BC} and~\ref{Subsec-Iron} and began during the bubble chamber era of the 1970's~\cite{Office:1973aa,Carmony:1976pw,Furuno:2003ng,Reinhard:1973rs,Snow:1971zz}.  It continued through the higher-statistics, mainly iron and lead experiments, of the 1990's using higher energy $\nu/\nub$ beams such as CDHSW~\cite{Berge:1989hr}, CCFR~\cite{Oltman:1992pq,Seligman:1997fe}, NuTeV~\cite{Tzanov:2005kr}.  Currently \href{http://minerva.fnal.gov}{MINER$\nu$A} at Fermilab is dedicated to the measurement of these cross section to better understand the nuclear medium effects and has taken data using the medium energy NuMI beam  ($<E_\nu> \sim 6$ GeV)
 in  $\nu_l$ as well as $\bar\nu_l$ modes with several nuclear targets $^{12}C$, $^{56}Fe$ and $^{208}Pb$ and the large central scintillator (CH) tracker.  
 %There are ongoing efforts to precisely measure neutrino-nucleus interaction cross sections  however, having to use low-energy neutrino beams optimized for neutrino oscillation experiments, .  For example, the $\nu_l/\bar\nu_l$ experiment  at Fermilab   using several nuclear targets He, C, O, Fe and Pb in the relevant $E_\nu$ range.
 In addition to the dedicated \minerva\ experiment there are the \href{https://t2k-experiment.org}{T2K} experiment in Japan as well as the \href{https://novaexperiment.fnal.gov}{NOvA} experiment in the USA, although
 primarily oscillation experiments, also currently contributing to cross section measurements. At these lower energies, SIS events dominate however DIS events still contribute to the event rates although with more limited kinematic reach. \footnote{For lower energy neutrinos, the \href{http://www-microboone.fnal.gov/}{MicroBooNE} experiment plus the future short baseline near detector \href{http://sbn-nd.fnal.gov/}{SBND} and far detector \href{http://icarus.lngs.infn.it} {ICARUS} in the Fermilab Booster neutrino beam will be measuring cross sections mainly in the E $\lessapprox$ 1 GeV region.   As the total cross section in this energy range is dominated by QE scattering and lower effective hadronic mass mainly $\Delta$ resonance production, we will not be explicitly covering this lower energy region.}  
 
 The method for testing the relevant nuclear models with  experimental results involve the 
 Monte Carlo (MC) generator the experiments employ. At present several MC generators have been developed like GiBUU~\cite{Buss:2011mx}, NuWro~\cite{Golan:2012wx}, GENIE~\cite{Andreopoulos:2009rq} and NEUT~\cite{Hayato:2009zz}   that are used within the experimental community.  These MC generators each have variations of a nuclear model, plus many other experiment dependent effects that are usually more accurately determined, involved in predicting what a particular experiment should detect. Comparing the predictions of these generators with the experimental measurements gives an indication of the accuracy of the nuclear model employed.  
 
 In addition to refining the nuclear model, the MC generator is also a necessary component for determining important experimental parameters 
 such as acceptance, efficiency and systematic errors. To perform all these functions the MC generators need production models for 
 each of the interactions they simulate as well as the nuclear model. For resonance production the (often modified) Rein-Sehgal model~\cite{Rein:1980wg} (R-S) or the more recent Berger-Sehgal (B-S) model~\cite{Berger:2007rq} is widely used. However 
 these models are limited to single pion production. It is essential to also study the full multi-pion production from nucleon resonances in the energy region of 1 - 10 GeV, where various resonances like
P$_{33}(1232)$, P$_{11}(1440)$, D$_{13}(1520)$, S$_{11}(1535)$, S$_{11}(1650)$, P$_{13}(1720)$,
etc. contribute. For most MC generators the DIS process is simulated using the Bodek-Yang model~\cite{Bodek:2002ps}. This DIS simulation is then extrapolated down into the SIS region.  The extrapolation is supposed to account for all non-resonant processes and resonant multi-$\pi$ production. 
  
  At the planned accelerator-based, long-baseline $\nu_l$-oscillation experiments such as the Deep Underground Neutrino Experiment(DUNE) using an argon target it is expected that more than 30$\%$ of the events would come from the DIS region and more than 50$\%$ of the events would come from the SIS(W $\ge M_\Delta$) plus DIS regions. Additionally the atmospheric $\nu_l$ studies in the proposed Hyper-K experiment 
(Hyper-Kamiokande), using water target, will also have significant SIS and DIS contributions.  It is  consequently important to have improved knowledge of nuclear medium effects in these lower-energy regions and therefore timely to revisit the present status of both the theoretical/phenomenological and experimental understanding of these scattering processes.

\begin{figure}
\begin{center}
\includegraphics[height=7.5 cm, width=12 cm]{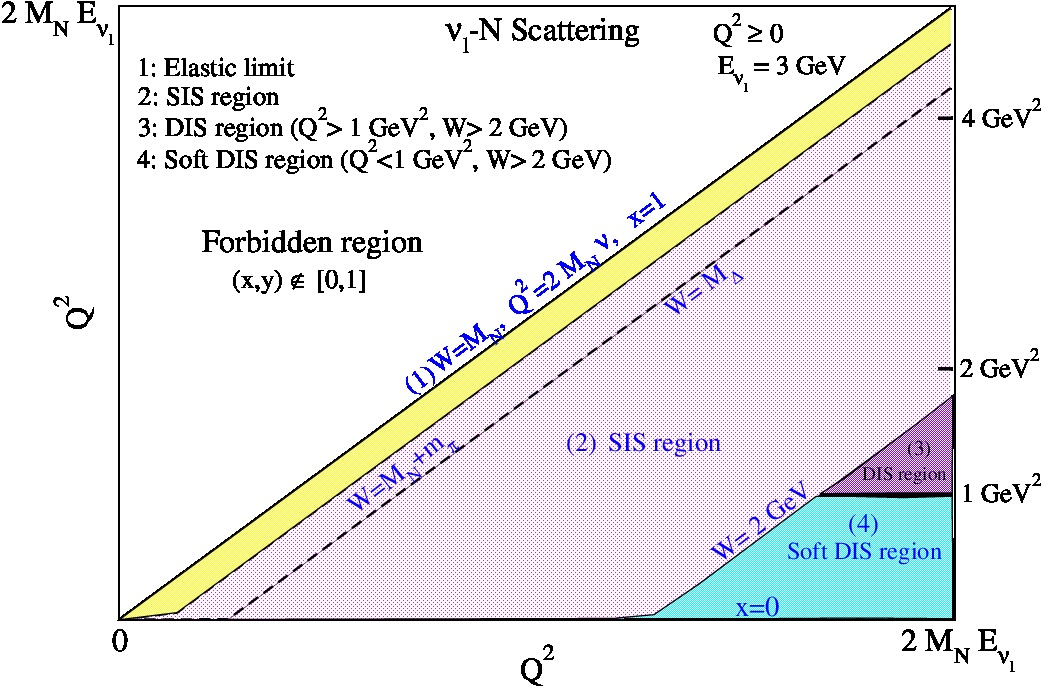}
\includegraphics[height=7.5 cm, width=12 cm]{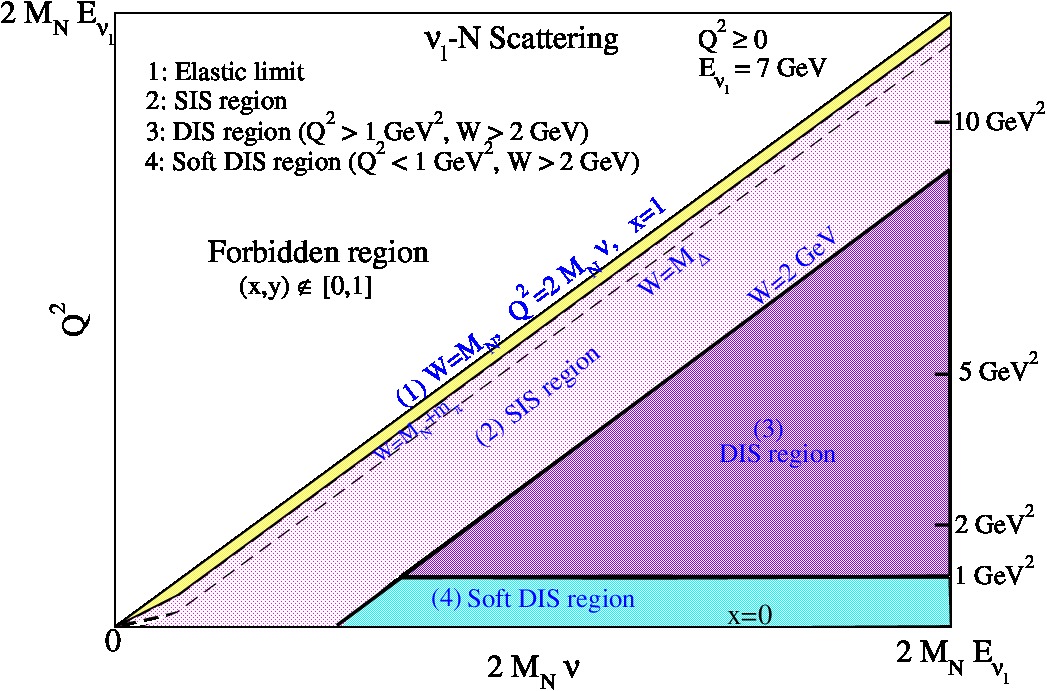}
 \end{center}
\caption{Allowed kinematical region for $\nu_l~-~N$ scattering in the ($Q^2, \nu$) plane for 
$E_\nu$=3 GeV(top panel) and $E_\nu$=7 GeV(bottom panel). The square of the invariant mass  is 
defined as $W^2=M_N^2+2M_N\nu-Q^2$ with the nucleon mass $M_N$ and the energy transfer $\nu$. The inelasticity is defined as $y=\frac{\nu}{E_\nu}=\frac{(E_\nu - E_l)}{E_\nu}$ and then the forbidden region in terms of
$x$ and $y$ is defined as $x,y~\notin~[0,1]$.
The elastic limit is $x=\frac{Q^2}{2M_N\nu}=1$  and, for this review, the SIS region has been practically defined as the
region for which $M_N+M_\pi \le W \le 2GeV$ and $Q^2 \ge 0$ covering both non-resonant and resonant meson production. The DIS region is defined as the region for which $Q^2 \ge 1~GeV^2$ and $W \ge 2~GeV$, and the 
Soft DIS region is defined as $Q^2 < 1 GeV^2$ and $W \ge 2~GeV$.
%(also part of SIS). 
Notice the yellow  band($M_N < W < M_N+M_\pi$), where we do not 
expect anything from $\nu-N$ scattering. However, this region becomes important when the scattering takes place with a nucleon within a nucleus due to the multi-nucleon correlation effect.}
 \label{fg:fig1mod}
\end{figure}

To summarize, using the $Q^2-\nu$ plane (Fig.\ref{fg:fig1mod}), one may define the relationship of the various regions like elastic ($W=M_N$), resonance ($M_N+M_\pi \le W \le 2$ GeV), DIS ($Q^2>1$ GeV$^2$, $W>2$ GeV) as well as the region of soft DIS ($Q^2<1$ GeV$^2$ and $W>2$ GeV). Soft DIS is where nonperturbative QCD must be taken into serious consideration and is yet to be fully explored. In addition to these categories, shallow inelastic scattering (SIS) is technically defined as non-resonant pion production with $Q^2<1$ GeV$^2$ and W $> M_N+M_\pi$.  It is apparent that the resonant and non-resonant pion production with $W<2$ GeV overlap and cannot be distinguished. For this review then, the practical definition of SIS is taken to include both non-resonant and resonant pion production and their interferences with $W<2$ GeV. In order to emphasis the neutrino energy dependence of different 
scattering processes, in Fig.\ref{fg:fig1mod} the variation of $Q^2-\nu$ plane is shown at $E_\nu=3$ GeV (upper panel) and $E_\nu=7$ GeV (lower panel).
As one moves away from the higher $W$ region, where DIS (that deals with the quarks and gluons) is the dominant process to the region of SIS (resonant + nonresonant processes
having hadrons as a degree of freedom), the boundary between these two regions is not well defined. In the literature, $Q^2 \ge 1$ GeV$^2$ has been chosen as the lower limit required to be 
interacting with the hadron's constituents. A kinematic constraint of $W\ge 2$ GeV is also applied to help distinguish the contributions from the resonance region and DIS. To better understand this transition from the resonance to DIS regions, the phenomenon of quark-hadron duality comes into play that basically connects the free and confined partons. 

% Move following commented-out to DIS section
%Another important aspect is that the medium modifications for weak structure functions may be different from the electromagnetic structure functions~\cite{Haider:2016zrk}. Furthermore, the nuclear dependence of the ratio of longitudinal to transverse  structure functions $R_{LA}(x,Q^2)$ has also been a topic of much experimental investigation in many experiments~\cite{Dasu:1993vk, Amaudruz:1992wn, Arneodo:1996ru, Gomez:1993ri} but no significant nuclear dependence has been reported. However, a recent analysis of these experiments shows a nontrivial nuclear dependence of $R_{LA}(x,Q^2) - R_{Lp}(x,Q^2)$ and its implications on the EMC ratio $F_{2A}(x,Q^2) \over F_{2N}(x,Q^2)$~\cite{Guzey:2012yk, Solvignon:2009it}. 

 The plan of the paper
 %, a detailed update of relevant sections of the NuSTEC White Paper~\cite{Alvarez-Ruso:2017oui},
 is the following. In section-\ref{nu-n}, we present in brief the theoretical formalism for the $\nu_l/\bar\nu_l$-
 nucleon scattering cross section including the QCD corrections.  In section-\ref{section3}, we describe in short the various phenomenological as well as  
 theoretical efforts to understand nuclear medium effects in weak interaction processes and compare the theoretical results of Aligarh-Valencia group
 ~\cite{SajjadAthar:2009cr}-\cite{Haider:2012nf},\cite{Haider:2016zrk}-\cite{Zaidi:2019asc} with experimental results.
 In section-\ref{section4} we cover the phenomenological and experimental treatment of the SIS region including a detailed examination of duality.  In section-\ref{section5} we present the phenomenological and experimental treatment of the DIS region.  In section-6 we present a comparison of theoretical and phenomenological (nuclear PDFs) predictions with existing higher-energy experimental results.  In section 7 we present our conclusions 
 on what is needed both theoretically and experimentally to improve our understanding of the physics of the SIS and DIS regions and our predictions for neutrino nucleus interactions with the lower neutrino energy and nuclei relevant for future oscillation experiments.
 %have been presented and discussed in section-\ref{section4}.
%  The plan of the paper is the following. In section-\ref{nu-n}, we present in brief the formalism for the $\nu_l/\bar\nu_l$-
%  nucleon scattering cross section, where we also mention the QCD corrections.  In section-\ref{section3}, we describe in short the various phenomenological as well as 
%  theoretical efforts to understand nuclear medium effects in weak interaction processes. Some of the theoretical results of Aligarh-Valencia group
%  ~\cite{SajjadAthar:2009cr,Haider:2011qs,Haider:2012nf,Haider:2016zrk,Zaidi:2019mfd,Zaidi:2019asc}  
%  have been presented and discussed in section-\ref{section4}.

\section{$\nu_l/\bar\nu_l$-Nucleon Scattering}\label{nu-n} 
\subsection{$\nu_l$-Nucleon Scattering: Shallow Inelastic Scattering}
\label{T-SIS}
\begin{figure}
\begin{center}
  \includegraphics[height=5 cm, width=6.8 cm]{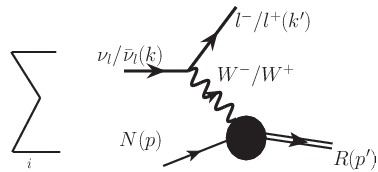}
\end{center}
 \caption{Diagrammatic representation of resonance excitations for $\nu_l(\bar\nu_l)~+~N~\rightarrow~l^-(l^+)~+~R$, where $R$ represents the different resonances 
 contributing to the hadronic current.}
 \label{review-fig2}
%  {Breview-fig2}
\end{figure}

   For the resonance production process 
 \begin{equation}
  \nu_l/\bar\nu_l(k)+N(p) \rightarrow l^-/l^+(k')+R(p')\nonumber
 \end{equation}
 the inclusive cross section is given as a sum of the
 individual contribution from the resonance excitations $R$, where $R=\Delta,~N^\ast$, etc. This is diagrammatically shown in Fig.\ref{review-fig2}. In the above relation, the quantities in the parenthesis
 are the four momenta of the corresponding particles.
The cross section for the resonance excitation of the individual resonance may be written as:
\begin{equation}\label{review-aa}
 \frac{d^2\sigma}{d\Omega'_l dE_l^\prime} \propto \frac{\it A(p^\prime)}{\sqrt{(k\cdot p)^2-m_l^2M_R^2}}~
 {L}_{\mu\nu} W_R^{\mu\nu}
\end{equation}
 where ${L}_{\mu\nu}$ is the leptonic tensor which is given by 
 \begin{equation}
  L_{\mu\nu}=k_\mu k'_\nu+k_\nu k'_\mu - k\cdot k' g_{\mu\nu} \pm i \epsilon_{\mu\nu\rho\sigma} k^\rho k'^\sigma
  \label{lepeq}
 \end{equation}
and $W_R^{\mu\nu}$ is the hadronic tensor corresponding to the $N(p)$ excitation of the resonance
 $R(p')$, which may be 
 schematically 
 given as
\begin{eqnarray}\label{review-ab}
 W_R^{\mu\nu}&=&\overline{\sum}\sum \langle R(p^\prime)|J^\mu|N(p)\rangle^\ast \langle R(p^\prime)|J^\nu|N(p)\rangle,\\
 A(p^\prime)&=&\frac{\sqrt{{p^\prime}^2}}{\pi}~\frac{\Gamma(p^\prime)}{({p^\prime}^2-M_R^2)^2~+~{p^\prime}^2\Gamma^2(p^\prime)},
\end{eqnarray}
 where $\Gamma(p^\prime)$ is the momentum dependent width and $M_R$ is the Breit-Wigner mass of the resonance. $\langle R(p^\prime)|J^\mu|N(p)\rangle$ corresponds 
 to the transition matrix element for the transition $N(p) ~\rightarrow~ R(p^\prime)$ induced by the current $J^\mu$. The transition matrix element 
 for the vector and the axial vector currents are characterized by the various transition form factors depending upon the spin
 of the excited resonance $R(p^\prime)$.  

 For example, in the case of the transition $N(p) ~\rightarrow~ R ^{\frac{3}{2}}(p^\prime)$, the general structure for the hadronic current of
 spin three-half resonance excitation is determined by the following equation 
\begin{eqnarray}\label{eq:had_current_3/2}
J_{\mu}^{\frac{3}{2}}=\bar{\psi}^{\nu}(p') \Gamma_{\nu \mu}^{\frac32} u(p), 
\end{eqnarray}
where $u(p)$ is the Dirac spinor for nucleon, ${\psi}^{\mu}(p)$ is the Rarita-Schwinger spinor for spin three-half resonance and 
$\Gamma_{\nu \mu}^{\frac32}$ has the following general structure for the positive(+) and negative(-) parity states : 
\begin{eqnarray}\label{eq:vec_3half_pos}
  \Gamma_{\nu \mu }^{\frac{3}{2}^+}= \left[{V}_{\nu \mu }^\frac{3}{2} - {A}_{\nu \mu }^\frac{3}{2}\right]  \gamma_5;  ~~~~~~~~~
  \Gamma_{\nu \mu }^{\frac{3}{2}^-}= {V}_{\nu \mu }^\frac{3}{2} - {A}_{\nu \mu }^\frac{3}{2} ,
\end{eqnarray}
where $V^{\frac32}(A^{\frac32})$ is the vector(axial-vector) current for spin three-half resonances, which are described in terms of ${C}^V_i$(${C}^A_i$) transition($N \rightarrow R$) form factors which are $Q^2$ dependent.  
 
 Similarly the hadronic current for the spin $\frac12$ resonant state is given by  
\begin{eqnarray}\label{had_curr_1/2}
J_{\mu}^{\frac{1}{2}}=\bar{u}(p') \Gamma_\mu^{\frac12} u(p), 
\end{eqnarray}
where $u(p)$ and $\bar u(p^\prime)$ are respectively, the Dirac spinor and adjoint Dirac spinor for spin $\frac{1}{2}$ particle and 
$\Gamma_\mu^{\frac12}$ is the vertex function which for the positive(+) and negative(-) parity states are given by 
\begin{equation}\label{eq:vec_half_pos}
  \Gamma_{\mu}^{{\frac12}+} = {V}_{\mu}^{\frac12} - {A}_{\mu}^{\frac12};~~~~~~~  \Gamma_{\mu}^{{\frac12}-} = \left[ {V}_{\mu}^\frac{1}{2} - {A}_{\mu}^\frac{1}{2}
  \right] \gamma_5
  \end{equation}
where $V_{\mu}^{\frac{1}{2}}$ represents the vector current and $A_{\mu}^{\frac{1}{2}}$ 
 represents the axial vector current. These currents are parameterized in terms of vector$(F_i(Q^2) (i=1,2))$ and axial 
vector$({g_1}(Q^2)$ and ${g_3}(Q^2))$ form factors.

Using the above prescription, the expression for the hadronic current is obtained and $W_R^{\mu\nu}$ in Eq.(\ref{review-ab}) is evaluated,
 which then is written in a form similar to Eq.~\ref{ch2:had_ten_N} i.e. in terms of
 $W_{jR}^{WI},~j=1-3$. Finally $W_{jR}^{WI}$ is related with the dimensionless structure functions $F_{jR}^{WI}$, following the same analogy as given in
 Eq.~(\ref{ch2:relation}), and the cross section(Eq.\ref{ch2:dif_cross}) is evaluated.
 
\begin{figure}
 \begin{center}
\includegraphics[height=9cm]{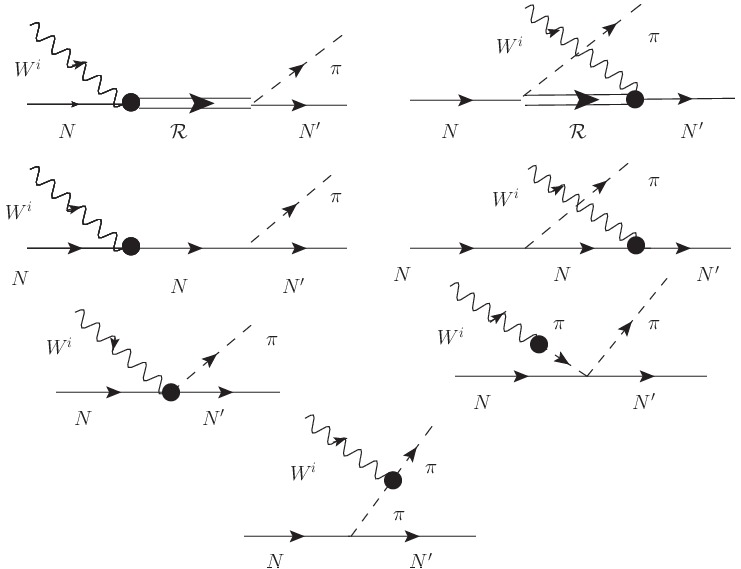}
\caption{Feynman diagrams contributing to the hadronic current corresponding to $W^{i} N \to N^{\prime} \pi^{\pm,0}$, where $(W^i 
\equiv W^\pm \; ; i=\pm)$ for charged current processes and $(W^i \equiv Z^0 \; ; i=0)$ for neutral current processes with $N,
N^{\prime}=p ~\rm{or}~ n$. First row represents the direct and cross diagrams for the resonance production where ${\cal R}$ 
stands for different resonances, second row represents the nucleon and cross nucleon terms while the contact and pion pole terms 
are shown in the third row while the last row represents the pion in flight term. The second, third and fourth rows represent non-resonant pion production.}
\label{fig:feynmann}
\end{center}
\end{figure}

  Besides the resonant terms, non-resonant terms also contribute to the scattering cross section. They are better known as background terms and play important role
  across the neutrino energy spectrum. These non-resonant
 background terms have contributions from the s-, t-, and u- channel Born terms, contact terms, meson in flight term, etc. There are various ways of including these
 terms like using non-linear sigma model, coupled channel approach, etc. For example, here we will briefly discuss the non-resonant background terms considered by \cite{Alam:2015gaa} obtained using non-linear sigma model. In the case of pion production, the non-resonant background terms involve five diagrams viz, direct nucleon pole~(NP), cross nucleon pole~(CNP), 
 contact term~(CT), pion pole~(PP) and pion in flight~(PF) terms (shown in Fig.\ref{fig:feynmann}), which are calculated using a chiral symmetric Lagrangian, obtained in the non-linear sigma model.
  
 The contributions from the different non-resonant background terms to the hadronic current are expressed as~\cite{Leitner:2008ue, Alam:2015gaa, Hernandez:2007qq}
% \footnotesize
\begin{eqnarray} \label{eq:background}
j^\mu\big|_{NP} &=& \mathcal{A}^{NP} \bar u({p}') 
 \not{k}_\pi\gamma_5\frac{\not{p}+\not{q}+M_N}{(p+q)^2-M_N^2+ i\epsilon}\left [V^\mu_N(q)-A^\mu_N(q) \right]  
u({p}),\nonumber \\
j^\mu\big|_{CNP} &=& \mathcal{A}^{CP} \bar u({p}') \left [V^\mu_N(q)-A^\mu_N(q) \right]
\frac{\not{p}'-\not{q}+M_N}{(p'-q)^2-M_N^2+ i\epsilon} \not{k}_\pi\gamma_5  u({p}), \nonumber \\
j^\mu\big|_{CT} &=& \mathcal{A}^{CT} \bar u({p}') \gamma^\mu\left (
  g_{1} f_{CT}^V(Q^2)\gamma_5 - f_\rho\left((q-k_\pi)^2\right) \right ) u({p}), \\
j^\mu\big|_{PP} &=& \mathcal{A}^{PP}f_\rho\left((q-k_\pi)^2\right) \frac{q^\mu}{M_\pi^2+Q^2}
  \bar u({p}')\ \not{q} \ u({p}),\nonumber \\
j^\mu\big|_{PF} &=& \mathcal{A}^{PF}f_{PF}(Q^2) \frac{(2k_\pi-q)^\mu}{(k_\pi-q)^2-M_\pi^2}
  2M_N\bar u({p}\,')  \gamma_5 u({p}\,),\nonumber
\end{eqnarray}
where $M_\pi$ is the mass of pion and $M_N$ is the nucleon mass. The constant factor 
$\mathcal{A}^i, i = $ NP, CNP, CT, PP and PF, are tabulated in Table--\ref{tab:born_para}. 
 For details see 
 Refs.\cite{Leitner:2008ue}, \cite{Hernandez:2007qq}-\cite{Sobczyk:2018ghy}.
 
\begin{table*}[h]
 \begin{center}
%   \vspace{1cm}
    \renewcommand{\arraystretch}{2.}
    \begin{tabular*}{150mm}{@{\extracolsep{\fill}}|c|ccc|ccc|}
%       \noalign{\vspace{-8pt}}
      \hline \hline
      Constant term $\rightarrow$    & \multicolumn{3}{c|}{$\mathcal{A}$(CC $\nu$)}&\multicolumn{3}{c|}{$\mathcal{A}$(CC 
      $\bar \nu$)}   \\ \hline
      Final states  $\rightarrow$  &   $p\pi^{+}$ & $n\pi^{+}$ & $p\pi^{0}$&
      $n\pi^{-}$ & $n\pi^{0}$ & $p\pi^{-}$ \\ \hline
      NP  &   0   & $\frac{-ig_1}{\sqrt{2}f_\pi}$&  $\frac{-ig_1}{2 f_\pi}$ &
              0   & $\frac{ig_1}{2f_\pi}$ & $\frac{-ig_1}
              {\sqrt{2} f_\pi}$ \\
      CP &   $\frac{-ig_1}{\sqrt{2}f_\pi}$  &0 & $\frac{ig_1}{2f_\pi}$  &
              $\frac{-ig_1}{\sqrt{2}f_\pi}$ & $\frac{-ig_1}{2f_\pi}$ &  0 \\
      CT  &   $\frac{-i}{\sqrt{2}f_\pi}$ &$\frac{i}{\sqrt{2}f_\pi}$ & $\frac{i}{ f_\pi}$ &
              $\frac{-i}{\sqrt{2}f_\pi}$ & $\frac{-i}{f_\pi}$  & $\frac{i}{\sqrt{2}f_\pi}$\\
      PP  &   $\frac{i}{\sqrt{2}f_\pi}$ & $\frac{-i}{\sqrt{2}f_\pi}$& $\frac{-i}{ f_\pi}$ &
              $\frac{i}{\sqrt{2}f_\pi}$&$\frac{i}{f_\pi}$& $\frac{-i}{\sqrt{2}f_\pi}$  \\
      PF  &   $\frac{-i g_1}{\sqrt{2}f_\pi}$ & $\frac{i g_1}{\sqrt{2}f_\pi}$ & $\frac{i g_1}{ f_\pi}$   &
              $\frac{-i g_1}{\sqrt{2}f_\pi}$ & $\frac{-i g_1}{f_\pi}$ & $\frac{i g_1}{\sqrt{2}f_\pi}$ \\
      \hline \hline
    \end{tabular*}
  \end{center}
    \caption{The values of constant term($\mathcal{A}^i$) appearing in Eq.~\ref{eq:background}, where $i$ corresponds to the nucleon 
    pole(NP), cross nucleon pole(CP), contact term(CT), pion pole(PP) and pion in flight(PF) terms. $f_\pi$ is pion weak decay 
    constant and $g_{1}$ is nucleon axial vector coupling.}
    \label{tab:born_para}
\end{table*}

The vector$(V^\mu_N(q))$ and axial vector$(A^\mu_N(q))$ currents for the NP and CNP diagrams, in the case of charged current 
interactions, are calculated neglecting the second class currents and are given by,
\begin{eqnarray}\label{eq:vec_curr}
V^\mu_N(q)&=&f_{1}^{V}(Q^2)\gamma^\mu + f_{2}^{V}(Q^2)i\sigma^{\mu\nu}\frac{q_\nu}{2M_N} \\
\label{eq:axi_curr}
A^\mu_N(q)&=& \left(g_{1}(Q^2)\gamma^\mu + g_{3}(Q^2) \frac{q^\mu}{M_N}  \right)\gamma^5,
\end{eqnarray}
where $f_{1,2}^{V}(Q^2)$ and $g_{1,3}(Q^2)$ are the vector and axial vector form factors for the nucleons. The isovector form 
factors \textit{viz.} $f_{1,2}^{V}(Q^2)$ are expressed as:
\begin{equation}\label{f1v_f2v}
f_{1,2}^V(Q^2)=F_{1,2}^p(Q^2)- F_{1,2}^n(Q^2), 
\end{equation}
where $F_{1}^{p,n}(Q^2)$ are the Dirac and $F_{2}^{p,n}(Q^2)$ are the Pauli form factors of nucleons. These form factors are, in 
turn, expressed in terms of the experimentally determined electric $G_E^{p,n}(Q^2)$ and magnetic $G_M^{p,n}(Q^2)$ Sachs form 
factors. 
 
On the other hand, the axial form factor($g_{1}(Q^2)$)  is generally taken to be of dipole form and is given by
\begin{equation}\label{fa}
g_1(Q^2)=g_{1}(0)~\left[1+\frac{Q^2}{M_A^2}\right]^{-2},
\end{equation}
where $g_{1}(0)$ is the axial charge and is obtained from the quasielastic $\nu_l$ and $\bar\nu_l$ scattering as well as from the 
pion electro-production data. We have used $g_{1}(0)=$ 1.267 and the axial dipole mass $M_A$=1.026 GeV, which is the world average 
value, in the numerical calculations. 

The next contribution from the axial part comes from the pseudoscalar form factor $g_{3}(Q^2)$, the determination of which is based 
on Partially Conserved Axial Current (PCAC) and pion pole dominance and is related to $g_{1}(Q^2)$ through the relation 
\begin{equation}\label{fp}
g_{3} (Q^2)=\frac{2M_N^2 \; g_{1} (Q^2)}{M_\pi^2+Q^2}.
\end{equation}

In order to conserve vector current at the weak vertex, the  two form factors \textit{viz.} $f_{PF}(Q^2)$ and $f_{CT}^{V}(Q^2)$ are 
expressed in terms of the isovector nucleon form factor as~\cite{Hernandez:2007qq}
\begin{equation}
 f_{PF}(Q^2) = f_{CT}^{V}(Q^2) = 2 f_{1}^V(Q^2).
\end{equation}
The $\pi \pi NN$ vertex has the dominant $\rho$--meson cloud contribution and following Ref.~\cite{Hernandez:2007qq}, we have 
introduced $\rho-$form factor $(f_{\rho}(Q^2))$ at $\pi \pi NN$ vertex and is taken to be of the monopole form:
\begin{equation}
 f_{\rho}(Q^2) = \frac{1}{1+Q^2/M_{\rho}^2}; \qquad \qquad \rm{ with } \; \qquad M_\rho = 0.776 \rm{ GeV}.
\end{equation}
In order to be consistent with the assumption of PCAC, $f_{\rho}(Q^2)$ has also been used with axial part of the contact term.

 The net hadronic current is then written as the sum of non-resonant and resonant contributions
 \begin{eqnarray}\label{had_curr_sum}
J_{\mu}=J_{\mu}^{NR}~+~J_{\mu}^{R}~e^{i\phi}, 
\end{eqnarray}
$\phi$ is the phase factor which tells us how the resonant channels
 add to the non-resonant contributions. Generally in numerical calculations $\phi$ is taken to be zero, that means these two are in the same phase and add up 
 coherently, however, in general this may not be necessarily true. 

$J_{\mu}^{NR}$ in Eq.\ref{had_curr_sum} gets the contribution from non-resonant diagrams shown in Fig.~\ref{fig:feynmann} as
\begin{equation}\label{nrback}
 J_{\mu}^{NR} = j_\mu\big|_{NP} + j_\mu\big|_{CNP} + j^\mu\big|_{CT} + j^\mu\big|_{PP} + j^\mu\big|_{PF},
\end{equation}
given in Eq.~(\ref{eq:background}).  For all the numerical calculations \cite{Alam:2015gaa} puts a constraint on W such that $M_N + M_{\pi} \le W \le 1.2~GeV$  while 
 evaluating $J_{\mu}^{NR}(W \le 1.2~GeV)$, which was due to the chiral limit.  Note this implies that the effect of the non-resonant contributions presented below are those contributions limited to $M_N + M_{\pi} \le W \le 1.2~GeV$ and do not include any of the additional non-resonant contributions between $1.2~GeV \le W \le 2.0~GeV$.  The non-resonant contribution in this missing W region could be significant since this contribution, along with any resonance plus interference contributions, must grow to transition into the total DIS inelastic cross section at W = 2 GeV.
 
 $J_{\mu}^{R}$
 has the contribution from spin $\frac{3}{2}$ and spin $\frac{1}{2}$ resonant states with positive or negative parity i.e. $J_{\mu}^{R}=J_{\mu}^{\frac{1}{2}}~+~J_{\mu}^{\frac{3}{2}}$. For the numerical evaluations they (\cite{Alam:2015gaa}) took the six low lying resonances contributing to one-pion production i.e.
 \begin{equation}
 J^{\mu}_{\rm{R}}=J^\mu_{P_{33}(1232)}~+~ J^\mu_{P_{11}(1440)}~+~J^\mu_{S_{11}(1535)}~+~J^\mu_{S_{11}(1650)}~+~J^\mu_{D_{13}(1520)}~+~J^\mu_{P_{13}(1720)},
 \end{equation}
 and the numerical results presented in \cite{Alam:2015gaa} for the total cross sections are for the three different cases, (i) with no cut on $W$,   an upper limit of $W$ as (ii) 1.4~GeV and (iii) 1.6~GeV, while evaluating $J_{\mu}^{R}$.
 
 For example, the authors of \cite{Alam:2015gaa} found that in the case of $\nu_\mu + p \rightarrow \mu^- + p + \pi^+$ induced reaction for a cut of $M_N + M_{\pi} \le W \le 1.2~GeV$  on $J_{\mu}^{NR}$ and no cut of $W$ on the resonance i.e. $J_{\mu}^{R}$, when the hadronic currents are added coherently(i.e. $\phi$=0), the main contribution to the total scattering cross section comes from $P_{33}(1232)$ resonance better known as the  $\Delta(1232)$ resonance and there is no contribution to the $p + \pi^+$ mode from the higher resonances 
 ($P_{11}(1440)$, $D_{13}(1520)$, $S_{11}(1535)$, $S_{11}(1650)$ and  $P_{13}(1720)$). It was also found in the case of $p~ +~ \pi^+$ production, that due to the presence of the non-resonant background terms i.e. $J_{\mu}=J_{\mu}^{NR}(W \le 1.2~GeV)~+~J_{\mu}^{\Delta}$, there is an increase in the cross section when compared with the results obtained using $\Delta(1232)$ term only in the hadronic current i.e. $J_{\mu}=J_{\mu}^{\Delta}$. This increase is about 12$\%$ at $E_{\nu_\mu} =1~GeV$ which becomes ∼ 8$\%$ at $E_{\nu_\mu} =2~GeV$.
 
 For $\nu_\mu + n \rightarrow \mu^- + n + \pi^+$ as well as $\nu_\mu + n \rightarrow \mu^- + p + \pi^0$ processes, there are contributions from the non-resonant background terms as well as other higher resonant terms, although $\Delta(1232)$ dominates. The net contribution to the total pion production due to the presence of the non-resonant background terms (i.e. $J_{\mu}=J_{\mu}^{NR}(W \le 1.2~GeV)~+~J_{\mu}^{\Delta}$) in $\nu_\mu + n \rightarrow \mu^- + n + \pi^+$ reaction results in an increase in the cross section of about 12$\%$ at $E_{\nu_\mu} =1~GeV$ which becomes 6$\%$ at $E_{\nu_\mu} =2~GeV$. When other higher resonances are also taken into account i.e. $J_{\mu}=J_{\mu}^{NR}(W \le 1.2~GeV)~+~J_{\mu}^{R}$, where $R$ also includes $\Delta$ as defined in Eq.\ref{had_curr_sum}, there is a further increase in the $n ~+~ \pi^+$ production cross section by about 40$\%$ at $E_{\nu_\mu} =1~GeV$ which becomes ∼ 55$\%$ at $E_{\nu_\mu} =2~GeV$. While in the case of $\nu_\mu + n \rightarrow \mu^- + p + \pi^0$ due to the presence of the non-resonant background terms the total increase in the $p + \pi^0$ production cross section is about 26$\%$ at $E_{\nu_\mu} =1~GeV$ which becomes ∼ 18$\%$ at $E_{\nu_\mu} =2~GeV$. Due to the presence of other higher resonances there is a further increase of about 35$\%$ at $E_{\nu_\mu} =1~GeV$ which becomes ∼ 40$\%$ at $E_{\nu_\mu} =2~GeV$. 
 
 When a cut of $W \le 1.4~GeV$(case-ii) or $W \le 1.6~GeV$ (case-iii) on the center of mass energy is applied on $J_{\mu}^{R}$,  then the over all cross section decreases.  
 %The presence of the non-resonant background terms ($M + m_{\pi} \le W \le 1.2~GeV$),results in an increase in the cross section for  the $\Delta$-dominated $\nu_\mu + p \rightarrow \mu^-  + p  + \pi^+$ process, which is about $10\%$ of the total scattering cross section at $E_{\nu_\mu}$= 1~GeV for both W cuts. For $\nu_\mu + n \rightarrow \mu^- + n  + \pi^+$ reaction this increase in the cross section is about $14\%$ at $E_{\nu_\mu}$=1~GeV which becomes $5\%$ at $E_{\nu_\mu}$=2~GeV. When other higher resonances i.e. $J_{\mu}=J_{\mu}^{NR}(W \le 1.2~GeV)~+~J_{\mu}^{R}$, for $n + \pi^+$ production are also taken into account there is a further increase in the cross section which is about $40\%$ at $E_{\nu_\mu}$=1~GeV. 
 The effect of these W cuts become apparent at higher neutrino energies where some energy dependence is observed when also considering higher resonances. For example with  $n + \pi^+$ production at $E_{\nu_\mu}$ = 2~GeV
 the increase in total cross section, compared to the non-resonant (W $\le 1.2~GeV$ ) + $\Delta$ cross section, is found to be $\sim 55\%$ for a cut $W \le 1.4GeV$  and $\sim 65\%$ for $W \le 1.6~GeV$. 
 
 It was observed that the inclusion of higher resonant terms lead to a significant increase in the cross section for $\nu_\mu + n \rightarrow \mu^- + n + \pi^+$ as well as $\nu_\mu + n \rightarrow \mu^- + p + \pi^0$ processes.
 Furthermore, it was also concluded that contribution from non-resonant background terms  with $W \le 1.2~GeV$ decreases with the increase in neutrino energy, while the total scattering cross section increases when other higher resonances were included in their calculations, although the $\Delta(1232)$ still dominates.  The net increase includes the contribution of the interference terms among the resonant and the non-resonant ($W \le 1.2~GeV$) contributions to the hadronic current.
  %While in the case of $\nu_\mu + n \rightarrow \mu^- + p + \pi^0$ process due to the presence of the non-resonant background terms i.e. $J_{\mu}=J_{\mu}^{NR}(W \le 1.2~GeV)~+~J_{\mu}^{\Delta}$ 
   %(with the constraint of $W$ on $J_{\mu}^{NR}$ i.e. $M + m_{\pi} \le W \le 1.2~GeV$) the total increase in the cross section is about $27\%$ at $E_{\nu_\mu}$=1~GeV for $W \le 1.4~GeV$ or $W \le 1.6~GeV$ cuts on $J_{\mu}^{\Delta}$.  Due to the presence of other resonances ($J_{\mu}=J_{\mu}^{NR}(W \le 1.2~GeV)~+~J_{\mu}^{R}$) there is a further increase of about 10$\%$ at $E_{\nu_\mu}$=1~GeV.  

 It must be pointed out that in electromagnetic interactions, phase dependence has been studied by a few groups, whereas in the weak interactions there is hardly any such study.
\subsection{$\nu_l$-Nucleon Scattering: Deep-Inelastic Scattering}
\begin{figure}
\begin{center}
\includegraphics[height=5 cm, width=13.5 cm]{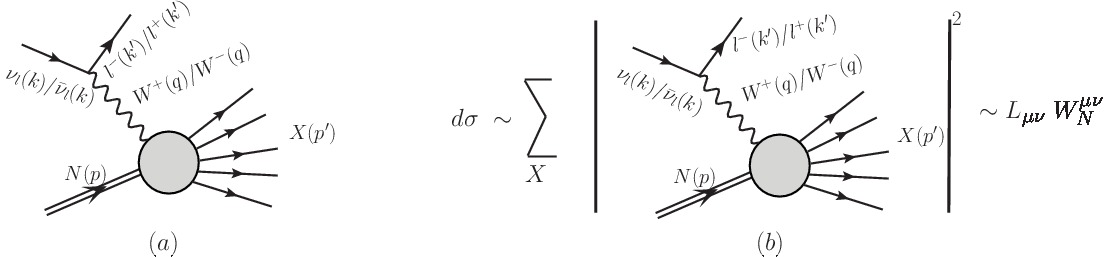}
\end{center}
\caption{(a) Feynman diagrams for the $\nu_l/\bar\nu_l$ induced DIS process. (b) $\nu_l({\bar\nu}_l) - N$ inclusive scattering where the 
 summation sign represents the sum 
over all the hadronic states such that the cross section($d\sigma$) for the deep inelastic scattering  
$\propto L_{\mu \nu} W_{N}^{\mu \nu}$.}
\label{fg:fig1}
\end{figure}

 The basic process for charged current DIS is given by(Fig.\ref{fg:fig1}a)
\begin{equation} 	\label{reaction}
\nu_l/\bar \nu_l(k) + N(p) \rightarrow l^{-}/l^{+}(k^\prime) + X(p^\prime),~l=e,~\mu,
\end{equation}
 where a $\nu_l/\bar\nu_l$ interacts with 
a nucleon($N$), producing a charged lepton($l$) and jet of hadrons($X$) in the final state. In the above expression $k$ and $k^\prime$ are the four momenta of incoming $\nu_l/\bar\nu_l$ and outgoing lepton respectively;
$p$ is the four momentum of the target nucleon and $p^\prime$ is the four momentum of final hadronic state $X$. This process is mediated by the 
exchange of virtual boson $W^\pm$ having four momentum $q(=k-k'=p'-p)$. The cross section for the inclusive scattering of a $\nu_l/\bar\nu_l$ from a nucleon target is
proportional to the leptonic tensor($L_{\mu \nu}$) and the hadronic tensor($W^{\mu \nu}_{N}$), where the hadronic tensor is obtained by summing over all the final 
states (Fig.\ref{fg:fig1}b). 

The double differential scattering cross section
evaluated for a nucleon target in its rest frame is expressed as:
\begin{equation} 	\label{ch2:dif_cross}
\frac{d^2 \sigma^{WI}_N}{d \Omega_l' d E_l^{\prime}} 
= \frac{{G_F}^2}{(2\pi)^2} \; \frac{|{\bf k}^\prime|}{|{\bf k}|} \;
\left(\frac{M_W^2}{q^2-M_W^2}\right)^2
L_{\mu \nu}
\; W^{\mu \nu}_{N}\,,
\end{equation}
where ${G_F}$ is the Fermi coupling constant, $\Omega_l'$, $E_l^\prime$ refer to the outgoing lepton and $-q^2=Q^2$ with $Q^2\ge 0$. 
 The expression of leptonic tensor $L_{\mu \nu}$ is given in Eq.\ref{lepeq} and the most general form of the hadronic tensor
$W_{N}^{\mu \nu}$ in terms of structure functions which 
depend on the scalars $q^2$ and $ p.q$, is given by
\begin{eqnarray}\label{ch2:had_ten_N}
W_{N}^{\mu \nu} &=&
\left( \frac{q^{\mu} q^{\nu}}{q^2} - g^{\mu \nu} \right) \;
W_{1N}^{WI}(\nu, Q^2)
+ \frac{W_{2N}^{WI}(\nu, Q^2)}{M_N^2}\left( p^{\mu} - \frac{p . q}{q^2} \; q^{\mu} \right)
 \nonumber\\
&\times&\left( p^{\nu} - \frac{p . q}{q^2} \; q^{\nu} \right)-\frac{i}{2M_N^2} \epsilon^{\mu \nu \rho \sigma} p_{ \rho} q_{\sigma}
W_{3N}^{WI}(\nu, Q^2) + \frac{W_{4N}^{WI}(\nu, Q^2)}{M_N^2} q^{\mu} q^{\nu}\nonumber\\
&&+\frac{W_{5N}^{WI}(\nu, Q^2)}{M_N^2} (p^{\mu} q^{\nu} + q^{\mu} p^{\nu})
+ \frac{i}{M_N^2} (p^{\mu} q^{\nu} - q^{\mu} p^{\nu})
W_{6N}^{WI}(\nu, Q^2)\,,
\end{eqnarray}
where $W_{iN}^{WI}(\nu, Q^2);~(i=1-6)$ are the nucleon structure functions and $\nu(=k^0-k^{'0})$ is the energy transfer.

In the limit $m_l \rightarrow 0$, the terms depending on $W_{4N}^{WI}(\nu,Q^2)$, 
$W_{5N}^{WI}(\nu,Q^2)$ and $W_{6N}^{WI}(\nu,Q^2)$ in Eq.~\ref{ch2:had_ten_N} do not contribute to the cross
section and DIS processes are described by the three nucleon structure functions $W_{1N}^{WI}(\nu, Q^2)$, $W_{2N}^{WI}(\nu, Q^2)$ and $W_{3N}^{WI}(\nu, Q^2)$. 
 Note that when compared to the electromagnetic process there is an additional structure function $W_{3N}^{WI}(\nu, Q^2)$
 due to parity violation in the case of weak interactions. 
 When $Q^2$ and $\nu$ become  large the structure functions $W_{iN}^{WI} (\nu,Q^2);~(i=1-3)$ are generally redefined in terms of the dimensionless nucleon
structure functions $F_{iN}^{WI} (x)$ as: 
\begin{equation}\label{ch2:relation}
\left.
\begin{array}{l}
M_N W_{1N}^{WI}(\nu, Q^2)=F_{1N}^{WI}(x),\\
\nu W_{2N}^{WI}(\nu, Q^2)=F_{2N}^{WI}(x),\\
\nu W_{3N}^{WI}(\nu, Q^2)=F_{3N}^{WI}(x).
\end{array}\right\}
\end{equation}
 $F_2^{WI}(x)$ at the leading order(LO) for 
$\nu_l$ and $\bar\nu_l$ induced processes on proton and neutron targets are given by assuming that the CKM matrix
 is almost unitary in its $2 \times 2$ upper left corner or equivalently 
that the heavy flavors bottom and top do not mix with the lightest ones such that:
\numparts
\begin{eqnarray}\label{f2parton}
 F_2^{\nu p}&=& 2 x \left[d(x) + s(x) + \bar u(x) + \bar c(x) \right],\\
  F_2^{\bar\nu p}&=& 2 x \left[ u(x) + c(x) + \bar d(x) + \bar s(x) \right]\\
 F_2^{\nu n}&=& 2 x \left[ u(x) + s(x) + \bar d(x) + \bar c(x) \right]\\
  F_2^{\bar\nu n}&=& 2 x \left[ d(x) + c(x) +\bar u(x) + \bar s(x) \right]
\end{eqnarray} 
\endnumparts
So, for an isoscalar nucleon (N) target assuming $s(x)=\bar s(x)$ and $c(x)=\bar c(x)$, we may write
\begin{eqnarray}\label{sfnuc}
F_2^{\nu N}(x)&=&F_2^{\bar\nu N}(x)\nonumber\\
&= & x\left[ u(x) + \bar u(x) + d(x)  + \bar d(x) + s(x) + \bar s(x) + c(x) + \bar c(x)\right]
\end{eqnarray}
The weak structure function $F_3^{WI}(x)$ at the leading order(LO) for $\nu_l$ and $\bar\nu_l$ interactions on the proton and neutron targets are given by
\numparts
\begin{eqnarray}\label{f3parton}
 xF_3^{\nu p}(x)&=& 2 x \left[d(x) + s(x) - \bar u(x) - \bar c(x) \right],\\
  xF_3^{\nu n}(x)&=&  2 x \left[ u(x) + s(x) - \bar d(x) - \bar c(x) \right],\\
 xF_3^{\bar\nu p}(x)&=& 2 x \left[ u(x) + c(x) - \bar d(x) - \bar s(x) \right],\\
  xF_3^{\bar\nu n}(x)&=&  2 x \left[ d(x) + c(x) - \bar u(x) - \bar s(x) \right]
\end{eqnarray} 
\endnumparts

\noindent and for an isoscalar nucleon target,
\begin{equation}\label{f3nuc}
  F_3^{\nu/\bar\nu N}(x) = \frac{F_3^{\nu/\bar\nu p}(x) + F_3^{\nu/\bar\nu n}(x)}{2}
\end{equation}
 The parton distribution functions (PDFs) (defined in Eqs.\ref{sfnuc} and \ref{f3nuc}) for the nucleon have been determined by various groups and they are known in the literature by the acronyms MRST~\cite{Martin:1998sq}, 
 GRV~\cite{Gluck:1998xa}, 
 GJR~\cite{Gluck:2007ck}, MSTW~\cite{Martin:2009iq}, ABMP~\cite{Alekhin:2016uxn}, ZEUS~\cite{Chekanov:2002pv}, HERAPDF~\cite{HERA}, 
 NNPDF~\cite{DelDebbio:2007ee}, CTEQ~\cite{Nadolsky:2008zw}, CTEQ-Jefferson Lab (CJ)~\cite{Accardi:2009br}, MMHT~\cite{Harland-Lang:2014zoa}, etc. In the present 
 work the numerical results are presented using CTEQ~\cite{Nadolsky:2008zw} and MMHT~\cite{Harland-Lang:2014zoa} nucleon parton distribution functions. 

 The weak structure function can be compared directly with the electromagnetic structure function $F_2^{EM}(x)$ 
\begin{eqnarray}
 F_2^{EM}(x)&=& \frac{F_2^{ep} + F_2^{en}}{2}= x\left[\frac{5}{18}(u(x) + \bar u(x)+d(x) \right.\nonumber\\
 &+& \left. \bar d(x)) + \frac{1}{9} (s(x)+\bar s(x)) + 
 \frac{4}{9}(c(x)+\bar c(x))  \right]~~~
\end{eqnarray} 
for an isoscalar nucleon target, by defining the ratio of electromagnetic to weak structure functions 
% \begin{footnotesize}
\begin{eqnarray}\label{tzanov_eq}
\frac{F_2^{EM}(x)}{F_2^{WI}(x)} &=& \frac{F_2^{eN}(x)}{F_2^{\nu/\bar\nu N}(x)}   = \it R^{EM/WI}(x) \nonumber\\
 &= &\frac{5}{18}\left[1 - \frac{3}{5} \frac{s(x) + \bar s(x) - c(x) -
\bar c(x)}{\sum (q(x) + \bar q(x))} \right],~~~~
\end{eqnarray}

%where the electromagnetic structure function  is given by,

\noindent and continue with the assumption $s(x) = \bar s(x) = c(x) = \bar c(x)$, the above expression reduces to
\begin{equation}
 F_2^{EM}(x) = \frac{5}{18} F_2^{WI}(x)
\end{equation}
In the quark parton model (QPM), where transverse momentum of partons is considered to be zero, the longitudinal structure function $F_L(x)$ is then also 0.  In this case, $F_1(x)$ is often expressed in terms of $F_2(x)$ using Callan-Gross relation, i.e.
\begin{equation}
 F_2(x) = 2 x F_1(x)
\end{equation} 
However, the modified QPM structure functions show a $Q^2$ dependence and partons possess a finite value of transverse momentum. Consequently, the longitudinal structure function has 
non-zero value leading to the violation of Callan-Gross relation which has also
been discussed in the literature~\cite{Whitlow:1991uw}-\cite{ Dasu:1988ms}.

The longitudinal structure function $F_{L}^{WI}(x,Q^2)$ is defined as
\begin{eqnarray}
% F_{L}^{WI}(x,Q^2) &=& {2 x \nu(1-x)M_N \over 4 \pi^2 \alpha} \sigma_L^{WI}(x,Q^2),\nonumber\\
F_{L}^{WI}(x,Q^2) &=& \left(1+\frac{4M_N^2 x^2}{Q^2}\right) F_{2}^{WI}(x,Q^2)-2xF_{1}^{WI}(x,Q^2),
\label{eq:fl}
\end{eqnarray}
where $F_{1}^{WI}(x,Q^2)$ is purely transverse in nature while $F_{2}^{WI}(x,Q^2)$ is an admixture of longitudinal and transverse components.
% where $\gamma^2=\left(1+\frac{4M_N^2 x^2}{Q^2}\right)$.
The ratio of longitudinal to transverse structure function $R_{L}^{WI}(x,Q^2)$ is given by
\begin{eqnarray}\label{rln}
 R_{L}^{WI}(x,Q^2) &=&  \frac{F_L^{WI}(x,Q^2)}{F_T^{WI}(x,Q^2)}={F_{L}^{WI}(x,Q^2) \over 2 x F_{1}^{WI}(x,Q^2)},\nonumber\\
  &=& \left(1+{4M_N^2 x^2 \over Q^2} \right)\frac{F_{2}^{WI}(x,Q^2)}{2xF_{1}^{WI}(x,Q^2)} - 1
\end{eqnarray}
% with
% \begin{eqnarray}\label{r2em}
%  R_{2,N}^{WI}(x,Q^2) = \frac{F_{2N}^{WI}(x,Q^2)}{2xF_{1N}^{WI}(x,Q^2)}.
% \end{eqnarray}

%  {\bf INTRODUCE AND DESCRIBE R = SIGL/SIGT HERE!}
A finite value of the ratio $R_{L}^{WI}(x,Q^2)$ has been measured in the $\nu_l/\bar\nu_l$ scattering by CCFR experiment~\cite{Yang:2001xc} in iron as well as several charged-lepton scattering
experiments~\cite{Whitlow:1991uw, Benvenuti:1989rh, Tao:1995uh} have also measured this ratio. In general it is expected that this ratio should be $A$ dependent and this dependence will be discussed in the later sections.  
  
{At low and moderate $Q^2$, structure functions show $Q^2$ dependence, therefore the above relation becomes:}
\begin{equation}\label{equal}
 F_2^{EM}(x,Q^2) = \frac{5}{18} F_2^{WI}(x,Q^2)
\end{equation}

\noindent Therefore, any deviation of $\it R^{EM/WI}(x,Q^2)=\frac{F_2^{EM}(x,Q^2)}{F_2^{WI}(x,Q^2)} $ from $\frac{5}{18}$ and/or any dependence on $x$, $Q^2$ will give information about the strange and charm quarks
distribution functions in the nucleon.

Now, we may write the differential scattering cross section~(Eq.\ref{ch2:dif_cross}) in terms of the dimensionless nucleon structure functions with respect to 
Bjorken scaling variable $x\left(=\frac{Q^2}{2 M_N \nu}\right)$ and the inelasticity 
$y\left(=\frac{\nu}{E_\nu}=\frac{E_\nu - E_l}{E_\nu}\right)$ as:
\begin{eqnarray}\label{d2sigdxdy_weak}
 { d^2\sigma_N^{WI} \over dx dy }&=& \frac{G_F^2 s}{2\pi} \left( \frac{M_W^2}{M_W^2+Q^2}\right)^2 \left[x y^2 F_{1N}^{WI}(x,Q^2) \right.\nonumber\\
 &+& \left. \left(1-y-{M_N x y  \over 2 E} \right) F_{2N}^{WI}(x,Q^2)\right.\nonumber\\
 &&  \left.  \pm x y\left(1-\frac{y}{2} \right)F_{3N}^{WI}(x,Q^2)\right],
 \end{eqnarray}
where the upper/lower sign is for $\nu_l/\bar\nu_l$ and $s=(p+k)^2$ is the center of mass energy squared.

 In the next subsection, the $Q^2$ evolution of nucleon structure functions from leading order to higher order terms as well as the non-perturbative effects such as target mass
 correction and higher twist effects important for  low and moderate $Q^2$ will be discussed.

\subsection{QCD Corrections}\label{free_effect}
\subsubsection{NLO and NNLO Evolutions}
According to the naive parton model (NPM), in the Bjorken limit structure functions depends only on $x$, i.e.
\begin{eqnarray}
F_{1N}(x,Q^2)  \longrightarrow_{x \to finite}^{[Q^2\to \infty, \nu \to \infty]} F_{1N}(x)\nonumber\\
F_{2N}(x,Q^2) \longrightarrow_{x \to finite}^{[Q^2\to \infty, \nu \to \infty]}F_{2N}(x)\nonumber
\end{eqnarray}
However, in QCD,  partons present inside the nucleon may interact among themselves via gluon exchange. The incorporation of contribution from gluon emission cause the $Q^2$ dependence 
of the nucleon structure functions, i.e. Bjorken scaling is violated. The $Q^2$ evolution of
structure functions is determined by the DGLAP evolution equation~\cite{Altarelli:1977zs} which is given by
\begin{footnotesize}
\begin{eqnarray*}
 \frac{\partial}{\partial ln Q^2}\left(\begin{array}{c}
                                   q_i(x,Q^2) \\
 g(x,Q^2)
                                 \end{array}\right)
&=&\frac{\alpha_s(Q^2)}{2 \pi}\sum_j \int_x^1\frac{dz}{z}\;
\left(
\begin{array}{c}
 P_{q_i q_j}(\frac{x}{y},\alpha_s(Q^2)) ~~~~P_{q_i g}(\frac{x}{y},\alpha_s(Q^2))\\
 P_{g q_j}(\frac{x}{y},\alpha_s(Q^2)) ~~~~ P_{g g}(\frac{x}{y},\alpha_s(Q^2))
\end{array}\right)\\
&\times&\left( \begin{array}{c}
 q_j(y,Q^2) \\
 g(y,Q^2)
 \end{array}\right),
\end{eqnarray*}
\end{footnotesize}

where $\alpha_s(Q^2)$ is the strong coupling constant, $q$ and $g$ are the quark and gluon density distribution functions, and $P(\frac{x}{y},\alpha_s(Q^2))$ are the splitting functions which are expanded in
power series of $\alpha_s(Q^2)$. Now, one may express the nucleon structure functions in terms of the convolution of coefficient function
($C_f\;;\;(f=q,g)$) with the density distribution of partons ($f$) inside the nucleon as
\begin{equation}\label{f2_conv}
 x^{-1} F_{i}^{WI}(x) = \sum_{f=q,g} C_{f}^{(n)}(x) \otimes f(x)\; ,
\end{equation}
where $i=2,3,L$, superscript $n=0,1,2,...$ for N$^{(n)}$LO and symbol $\otimes$ is the Mellin convolution. 
To obtain the convolution of coefficient functions with
parton density distribution, we use the following expression~\cite{Neerven}
\begin{equation}
 C_f(x)\otimes f(x) =\int_x^1 C_f(y)\; f\left(\frac{x}{y} \right) {dy \over y}
\end{equation}
This Mellin convolution turns into simple multiplication in the N-space.
The parton coefficient function are generally expressed as 
\begin{eqnarray}
 C_{f}(x,Q^2)=\underbrace{C_{f}^{(0)}}_{LO}+\underbrace{\frac{\alpha_s(Q^2)}{2\pi}C_{f}^{(1)}}_{NLO}+\underbrace{\left(\frac{\alpha_s(Q^2)}{2\pi}\right)^2\;C_{f}^{(2)}}_{NNLO}+...
\end{eqnarray}
% For example, we may write $F_{2N}(x)$ terms of coefficient function as
In the limit of $Q^2 \to \infty$, the strong coupling constant $\alpha_s(Q^2)$ becomes very small and therefore, the higher order terms such as next-to-leading order (NLO),
next-to-next-to-leading order (NNLO), etc., can be neglected in comparison to the leading order(LO) term.
But for a finite value of $Q^2$, $\alpha_s(Q^2)$ is large and next-to-leading order terms give a significant contribution followed by next-to-next-to-leading 
order term. The details of the method to incorporate QCD evolution are given in Refs.~\cite{ Neerven}-\cite{Ratcliffe:2000kp}. 
To calculate the structure functions, we use the NLO evolution of the parton distribution functions given in terms of the 
 power expansion in the strong coupling constant $ \alpha_s(Q^2)$. 
Following the works of Vermaseren et al.~\cite{Vermaseren} and van Neerven and Vogt \cite{Neerven}, the QCD corrections at NLO 
 for the evaluation of $ F_2 (x)$ structure function may be written as
\begin{equation}\label{eq:Fa-dec}
 x^{-1}  F_2(x) \: = \: C_{2,\rm ns}(x) \otimes q_{\rm ns}
 + \langle e^2 \rangle \left( C_{2,\rm q}(x) \otimes q_{\rm s}
                            + C_{2,\rm g}(x) \otimes g \right) \:\: ,
\end{equation}
and for the evaluation of $F_3(x)$ structure function, it may be written as
\begin{equation}
  F_3(x) \;=\; C_3(x) \otimes q_v(x)\;\;,
\end{equation}
where $q_{\rm s}$, $q_{\rm ns}$ and $q_{v}$ are respectively the flavor singlet, non-singlet and valence quark distributions, 
 $C_{2,q}(x)$ and  $C_{2,ns}(x)$ are singlet and non-singlet coefficient 
functions for the quarks, $C_{2,\rm g}(x)$ is the coefficient function for the gluons and $C_{3}(x)$ is the coefficient function for $F_3(x)$. The coefficient functions 
are defined in Refs.~\cite{ Neerven,Vermaseren,Moch:2008fj}. $\langle e^2 \rangle$ 
represents the average squared charge which is $\langle e^2 \rangle=\frac{5}{18}$ for four flavors of quarks in the case of EM interaction and $\langle e^2 \rangle=1$ for the weak interaction channel. 
%The Mellin convolution is given by
%\begin{equation}\label{Mellin_Convolution}
%[C\otimes f](x)=\int_x^1\frac{dy}{y}\;C(y)\;f\left(\frac{x}{y}\right)\,.
%\end{equation}

%START HERE ON FRIDAY.
\subsubsection{Target Mass Correction Effect:}
\label{TMC}
 The target mass correction (TMC) is a non-perturbative effect, which comes into the picture at lower $Q^2$. 
 At finite value of $Q^2$, the mass of the target nucleon and the quark masses modify the Bjorken variable $x$ with the light cone momentum fraction. For the massless quarks, the parton light cone momentum fraction is given by the Nachtmann variable $\xi$ which is related to the Bjorken variable $x$ as
 \begin{equation}
 \xi=\frac{2x}{1+\sqrt{1+\frac{4 M_N^2 x^2}{Q^2}}}.
 \end{equation}
 
 The Nachtmann variable $\xi$ depends only on the hadronic mass and will not have corrections due to the masses of final state quarks. However, for the massive partons, the Nachtmann variable $\xi$ gets modified to $\bar\xi$. These variables $\xi$ and $\bar\xi$ are related to the Bjorken variable as:
\begin{equation}
    \bar\xi=\xi \left( 1+\frac{m_q^2}{Q^2}\right)
\end{equation}
where $m_q$ is the quark mass. It is noticeable that the Nachtmann variable corrects the Bjorken variable for the effects of hadronic mass while the generalized variable $\bar\xi$ further corrects $\xi$ for the effects of the partonic masses~\cite{Schienbein:2007gr}.

TMC effect is associated with the finite mass of the target nucleon $M_N$ and is significant at low $Q^2$ and high $x$ ($x^2M_N^2/Q^2$ is large) which is an important region to determine the distribution of valence quarks. 
%TMC which are mostly relevant when . 
 The TMC effect involving powers of $1/Q^2$ are usually incorporated into the leading twist (LT) %{\bf LT not yet defined} 
term following the prescription of Refs.~\cite{Nachtmann:1973mr,Georgi:1976ve}. For a discussion of the impact of TMC see Ref.~\cite{Steffens:2012jx}. 

% Unfortunately, this kinematic region has not been much explored unlike the region of high $Q^2$ and low $x$. 
%  TMC effect is also known as ``kinematic higher twist effect''. In order to incorporate the target mass corrections, the Bjorken variable $x$ is replaced 
% by the Nachtmann variable $\xi$ which is defined as
% \begin{eqnarray}
%  \xi=\frac{2x}{1+\sqrt{1+\frac{4 M_N^2 x^2}{Q^2}}}
% \end{eqnarray}
% For a low value of momentum transfer squared($Q^2\approx M_N^2$), Nachtmann variable significantly deviates
% from the Bjorken variable. However, in the case of high momentum transfer square,
% %($Q^2\gg m_q^2$), 
% the Nachtmann variable reduces to the Bjorken variable, i.e., $\xi \longrightarrow x$. 
To incorporate the target mass corrections, Aligarh-Valencia group have followed the work of Schienbein et al.~\cite{Schienbein:2007gr}, where the
expressions of structure functions including TMC
effect are approximated as
\begin{eqnarray}\label{ftmc}
 F_{1N}^{TMC}(x,Q^2) &\approx& {x \over \xi \gamma}~F_{1N}(\xi)~\left(1~+~2r(1-\xi)^2 \right),\nonumber\\
 F_{2N}^{TMC}(x,Q^2) &\approx& {x^2 \over \xi^2 \gamma^3}~F_{2N}(\xi)~\left(1~+~6r(1-\xi)^2 \right),\nonumber\\
  F_{3N}^{TMC}(x,Q^2) &\approx& {x \over \xi \gamma^2}~F_{3N}(\xi)~\left(1~-~r(1-\xi)~ln\xi \right).
\end{eqnarray}
In the above expressions $r=\frac{\mu x \xi}{\gamma}$, $\mu=\left(\frac{M_N}{Q}\right)^2$ and $\gamma=\sqrt{1+\frac{4M_N^2 x^2}{Q^2}}$, respectively.

\subsubsection{Higher Twist Effect:}
\label{HT}
Similar to the TMC effect, there is another non-perturbative effect known as ``higher twist(HT) effect'' or ``dynamical higher twist effect''.
This effect involves the interactions of struck quark with other quarks via the exchange of gluons and it is suppressed by the power of $\left({1 \over Q^2} \right)^n$, where $n=1,2,....$.
This effect is also pronounced in the region of low $Q^2$ and high $x$ like the TMC effect but negligible for high $Q^2$ and low $x$. 

For lower values of $Q^2$, a few $GeV^2$ or less, non-perturbative 
phenomena could become important for a precise modeling of cross sections. 
%{\bf in addition to high-order QCD corrections~\cite{Alekhin:2007fh} do you want to mix in this unknown concept at this point?}. 
In the formalism of the operator product expansion (OPE)~\cite{Wilson:1969zs, Shuryak:1981kj}, unpolarized structure 
functions can be expressed in terms of powers of $1/Q^2$ (power corrections):  
\begin{equation}
F_{i}(x,Q^2) = F_{i}^{\tau = 2}(x,Q^2)
+ {H_{i}^{\tau = 4}(x) \over Q^2} 
+ {H_{i}^{\tau = 6}(x) \over Q^4} + .....   \;\;\; i=1,2,3, 
\label{eqn:ht}
\end{equation}
where the first term ($\tau=2$) is known as the twist-two or leading twist (LT) term, 
and it corresponds to the scattering off a free quark. 
 This term obeys the Altarelli-Parisi equations and is expressed in terms of PDFs.  It is responsible for the evolution 
of structure functions via perturbative QCD $\alpha_s(Q^2)$ corrections. 
% {\bf DEFINE WHAT THIS MEANS}. 
The HT terms with $\tau = 4,6$,\ldots
reflect the strength of multi-parton correlations ($qq$ and $qg$), and the HT corrections spoil the QCD factorization, so one 
has to consider their impact on the PDFs extracted in the analysis of low-$Q$ data.  
 The coefficients $ H_i(x)$ can only be determined in QCD as a result of the non-perturbative calculation. However, 
 due to their non-perturbative origin, current models can only provide a 
qualitative description for such contributions. The coefficients $ H_i(x)$ are usually determined via
reasonable assumptions from fits to the data~\cite{Alekhin:2013nda,Accardi:2016qay}. 

Existing information about the dynamical HT terms in lepton-nucleon structure functions is 
scarce and somewhat controversial. Early analyses~\cite{Miramontes:1989ni,Whitlow:1990gk} 
suggested a significant HT contribution to the longitudinal structure function $F_L(x)$.
% {\bf has this been defined?}.
The subsequent 
studies with both charged leptons~\cite{Virchaux:1991jc}-\cite{Alekhin:2002fv} and 
neutrinos~\cite{Kataev:1999bp} raised the question 
of a possible dependence on the order of QCD calculation used for the leading twist. 

A recent HT study~\cite{Alekhin:2007fh} including both charged lepton and $\nu_l/\bar\nu_l$ DIS data suggested that dynamic HT corrections affect the region of $Q^2 < 10$ GeV$^2$ and are 
largely independent from the order of the QCD calculation. However, the verification of QH duality at JLab implies a suppression of additional HT terms with respect to the average DIS behavior, 
down to low $Q^2\sim 1$ GeV$^2$~\cite{Niculescu:2000tk} with further details in section~\ref{Subsec-twist}. 
Furthermore, our  formalism suggests that as long as we demand $Q^2 \ge$ 1.0 $GeV^2$ and $W \ge $ 2.0 GeV, after the application of TMC there is no appreciable higher twist effect.

An empirical approach to take into account the effects of both kinematic and dynamical HT corrections on structure functions~\cite{Bodek:2003wd} is often implemented in MC generators.  This method is based upon LO structure functions (using GRV98 PDFs) in which the Bjorken variable $x$ is replaced by an adhoc scaling variable $\xi_w$ and all PDFs are modified by $Q$-dependent $K$ factors. The free parameters in the $\xi_w$ variable and in the $K$ factors are fitted to existing data. 

%An extrapolation of the HT terms on DIS structure functions to the transition and resonance region results in sizable corrections at low invariant masses $W<1.9$ GeV. . 

It is worth noting that the transition from the high $Q^2$ behavior of structure functions, 
well described in terms of perturbative QCD at leading twist, to the asymptotic limit for $Q^2\to 0$ defined by current conservation 
arguments in electroproduction, is largely controlled by the HT contributions. 
In this respect $\nu_l/\bar\nu_l$ interactions are different with respect to charged leptons, 
due to the presence of an axial-vector current dominating the cross sections at low $Q^2$ and the structure function does not go to zero as $Q^2\to 0$. 
The effect of the PCAC~\cite{Kopeliovich:2004px,Adler:1964yx} 
in this transition region can be formally considered as an additional HT contribution 
and can be described with phenomenological form factors~\cite{Kulagin:2007ju}. 
In the limit of $Q^2 \to 0$ for both charged leptons and neutrino scattering $F_T \propto Q^2$, while in the case of electromagnetic interaction $F_L \propto Q^4$
 and is dominated by the finite PCAC
contribution in the weak current. As a result, the ratio $R_L=F_L/F_T$
% {\bf Fel and FT have not yet been introduced!}
has a very different behavior in $\nu_l/\bar\nu_l$ scattering at 
small $Q^2$ values~\cite{Kulagin:2007ju} and this fact must be considered in the extraction of weak structure functions from the measured differential cross-sections. 
% {\bf need to show around eq 15 that there is a relation between F1 and F2 involving R}

\section{$\nu_l/\bar\nu_l$-Nucleus Scattering : Deep-Inelastic Scattering Theory} \label{section3}
After the EMC measurements in the early 1980s~\cite{Aubert:1983xm,Aubert:1986yn} and observation made by them henceforth named as the ``EMC effect'' that the ratio of $\frac{2F_2^A}{AF_2^D}$
 was not equal to 1.0 and was $x$ dependent, several other experiments were performed  
    by the different collaborations like 
    SLAC~\cite{Gomez:1993ri}, HERMES~\cite{Ackerstaff:1999ac}, BCDMS~\cite{Bari:1985ga, Benvenuti:1987az}, NMC~\cite{Arneodo:1995cs, Amaudruz:1995tq},
    JLab~\cite{Seely:2009gt}, etc.
    using nuclear targets, both moderate and heavy, for a wide range of Bjorken variable $x$($0 < x < 1$) 
    and four momentum transfer square $Q^2$, and the following observations were concluded from electroproduction experiments:
 \begin{itemize}
     \item although the shape of the effect does not change with mass number $A$, the strength of the nuclear medium effect increases with the increase in mass number $A$ and
     \item the functional form has very weak dependence on $Q^2$. 
 \end{itemize}
    The results for the nuclear medium effects on mass dependence $A$ were consistent with $log(A)$ and average nuclear density~\cite{Malace:2014uea}.   
    To understand nuclear medium effects on the structure functions, there are two broad approaches, one is the phenomenological approach involving determination of 
    the effective parton distribution of nucleons within a nucleus, and the other is theoretical 
    approach where dynamics of the nucleons in the nuclear medium is taken into consideration. The phenomenological approach will be presented in section~\ref{section4}.
 
 Theoretically many models have been proposed to study these effects on the basis of nuclear binding, nuclear medium modification including short range correlations in nuclei~\cite{Malace:2014uea}-\cite{Akulinichev1985}, 
 pion excess in nuclei~\cite{Bickerstaff:1989ch, Kulagin:1989mu, Marco:1995vb},\cite{ Ericson:1983um}-\cite{ Berger1987}, 
 multi-quark clusters~~\cite{Jaffe:1982rr}-\cite{Cloet:2005rt}, 
 dynamical rescaling~\cite{Nachtmann:1983py, Close:1983tn}, nuclear shadowing~\cite{Frankfurt:1988nt, Armesto:2006ph}, 
 etc. In spite of these efforts, no comprehensive theoretical/phenomenological understanding of the nuclear modifications of the  bound nucleon across the 
 complete range of $x$ and $Q^2$ consistent with the presently available experimental data exists~\cite{Arneodo:1992wf},\cite{Geesaman:1995}-\cite{Piller:1999wx}.
 In a recent phenomenological study Kalantarians et al.~\cite{Kalantarians:2017mkj} have made a comparison of electromagnetic vs weak nuclear structure functions ($F_{2A}^{EM}(x,Q^2)$ vs $F_{2A}^{WI}(x,Q^2)$)
 and found out that at low $x$ these two structure functions are different. Theoretically,  there have been very few calculations to study nuclear medium effects in the weak structure functions and moreover, there exists limited literature where explicitly a comparative study has been made\cite{Haider:2016tev, Zaidi:2019asc}. Therefore, 
 it is highly desirable to make a detailed theoretical as well as experimental studies of nuclear medium effects on the weak structure functions and compare the results with the EM structure functions for a wide range of $x$ and $Q^2$ for moderate as well as heavy nuclear targets.
  
  For the evaluation of weak nuclear structure functions not much theoretical efforts have been made except that of 
    Kulagin et al.~\cite{Kulagin:2007ju} and Athar et al.(Aligarh-Valencia group)~\cite{SajjadAthar:2007bz}-\cite{Zaidi:2019asc}.
    Aligarh-Valencia group~\cite{SajjadAthar:2007bz}-\cite{Zaidi:2019asc} has studied nuclear medium effects in the structure functions in a microscopic model 
 which uses relativistic nucleon spectral function to describe target nucleon momentum distribution incorporating the effects of Fermi
motion, binding energy and nucleon correlations in a field theoretical model. The spectral function that describes 
 the energy and momentum distribution of the nucleons in nuclei is
obtained by using the Lehmann's representation for the relativistic nucleon propagator and nuclear many body theory is used to calculate it
 for an interacting Fermi sea in the nuclear matter~\cite{FernandezdeCordoba:1991wf}. A local density approximation is then applied to translate these 
 results to a finite nucleus. Furthermore, the contributions of the pion and rho meson clouds in a many body
field theoretical approach have also been considered which is based on Refs.~\cite{Marco:1995vb,GarciaRecio:1994cn}. In the next subsection, the theoretical approach of 
Aligarh-Valencia group is discussed.

 \subsection{Aligarh-Valencia Formulation}\label{theo}
 It starts with the differential scattering cross section for the charged current inclusive $\nu_l/\bar\nu_l$-nucleus deep inelastic scattering process
\begin{equation}
 \nu_l / \bar\nu_l(k) + A(p_A) \rightarrow l^-/l^+(k') + X(p'_A),
\end{equation}
 written in analogy
with the charged current $\nu_l(\bar\nu_l)-N$ scattering discussed in section-\ref{nu-n},  
 by replacing the hadronic tensor for the nucleon i.e. $W^{\mu\nu}_N$ in Eq.\ref{ch2:dif_cross} with the 
 nuclear hadronic tensor $W^{\mu\nu}_A$:
\begin{eqnarray}\label{xecA}
{ d^2\sigma^{WI}_A \over d\Omega^\prime_l dE^\prime }&=& {G_F^2  \over (2\pi)^2} {|{\bf k^\prime}| \over|{ \bf k}|} {\left(M_W^2\over M_W^2+Q^2 \right)^2} 
L_{\mu\nu}^{WI} W^{\mu\nu}_A.
\end{eqnarray}
 and $W^{\mu\nu}_A$ is written in terms of the weak nuclear structure functions $W_{iA}^{WI}(\nu,Q^2)$ ($i=1,2,3$) as
\begin{eqnarray}
 \label{nuc_had_weak}
 W^{\mu\nu}_A &=& \left( \frac{q^{\mu} q^{\nu}}{q^2} - g^{\mu \nu} \right) \;
W_{1A}^{WI}(\nu,Q^2)
+ \frac{W_{2A}^{WI}(\nu,Q^2)}{M_A^2}\left( p_A^{\mu} - \frac{p_A . q}{q^2} \; q^{\mu} \right)\nonumber\\
&\times& \left( p_A^{\nu} - \frac{p_A . q}{q^2} \; q^{\nu} \right)
 \pm \frac{i}{2M_A^2} \epsilon^{\mu \nu \rho \sigma} p_{A \rho} q_{\sigma}~
W_{3A}^{WI}(\nu,Q^2),~~~~~~
\end{eqnarray}
% {\bf what is MA2??}
where $M_A$ is the mass and $p_A$ is the four momentum of the nuclear target and the  positive/negative sign is for the $\nu_l/\bar\nu_l$. The leptonic tensor in Eq.\ref{xecA} has
the same form as given in Eq.\ref{lepeq}. 
In the present work, the scattering process has been considered in the laboratory frame, where target nucleus is at rest($p_A=(p_A^0=M_A,{\bf p_A}=0)$). Therefore, one may define
\begin{eqnarray}
p_{_A}^\mu&=&(M_{_A},\vec 0),~~ \nonumber\\
% ,~~~~~\nu_A=\frac{p_{_A}\cdot q}{M_{_A}}=\frac{(p_{A}^0  q^0 - {\bf p_A \cdot q})}{M_{_A}}=q^{0},~~ \nonumber\\
x_A&=&\frac{Q^2}{2 p_A \cdot q}=\frac{Q^2}{2 p_{A}^0  q^0 } = \frac{Q^2}{2 A~M_N q^0}
\end{eqnarray}
However, the nucleons bound inside the nucleus are not stationary but they are continuously moving with finite momentum, i.e. $p=(p^0,{\bf p}\ne 0)$ and their motion corresponds to the 
Fermi motion. These nucleons are thus off shell. If we take the momentum transfer of the bound nucleon along the $z$-axis such that $q^\mu=(q^0,0,0,q^z)$ then Bjorken variable $x_N$ is given by
 \begin{equation}
 x_N = \frac{Q^2}{2 p \cdot q} = \frac{Q^2}{2 (p^0 q^0 - p^z q^z)}
\end{equation}
These bound nucleons may also interact among themselves via strong interaction and thus various nuclear medium effects are introduced which
play important roles in the different regions of the Bjorken variable $x$.
 %Fermi motion, binding energy, nucleon correlations, mesonic contributions, etc. .
  In the following subsections (\ref{spec} and \ref{mes}), these various nuclear medium effects like Fermi motion, binding, nucleon correlations,
isoscalarity correction and meson cloud contribution taken by Aligarh-Valencia group are discussed in brief.

\subsubsection{Fermi motion, binding and nucleon correlation effects:\;\;}\label{spec}
To calculate the scattering cross section for a neutrino interacting 
with a target nucleon in the nuclear medium to give rise to the process $\nu_l + N 
\rightarrow l^- + X$, we start off with a flux of neutrinos hitting a collection of target
 nucleons over a given length of time. Now a majority will simply pass
 through the target without interacting while a certain fraction will
 interact with the target nucleons leaving the pass-through fraction and
 entering the fraction of neutrinos yielding final state leptons and
 hadrons.  Here we introduce the concept of "neutrino self energy" that has a real and imaginary part. The real part modifies the lepton mass(it is similar to the delta mass or nucleon mass modified in the nuclear medium) while the imaginary part is related 
 to this fraction of interacting
 neutrinos and gives the total number of neutrinos that have participated in
 the interactions that give rise to the charged leptons and hadrons.
 The basic ingredients of the model are given in Appendix~\ref{app:self}-\ref{app:lda}.

 The neutrino self energy (Appendix~\ref{app:self}) is evaluated corresponding to the diagram shown in Fig.\ref{wself_energy} (left panel), 
 and the cross section for an element of volume $dV$ in the rest frame of the nucleus is 
related to the probability per unit time ($\Gamma$)
of the $\nu_l$ interacting with a nucleon bound inside a nucleus. $\Gamma dt dS$ provides probability times a differential of area ($dS$) which is nothing but the cross section
($d\sigma$)~\cite{Marco:1995vb}, i.e.
%   The cross section for an element of volume $dV$ in the nucleus is defined as~\cite{Marco:1995vb}:
\begin{equation}\label{defxsec}
d\sigma=\Gamma dt ds=\Gamma\frac{dt}{dl}ds dl=\Gamma \frac{1}{v}dV = \Gamma \frac {E_l}{\mid {\bf k} \mid}d^3 r,
% =\frac{-2m_l}{E_l({\bf k})} Im \Sigma (k)\frac{E_l({\bf k})}{\mid {\bf k} \mid}dV,
\end{equation}
where $v\left(=\frac{\mid {\bf k} \mid}{E_l}\right)$ is the velocity of the incoming $\nu_l$. The probability per unit time of the interaction of $\nu_l$ with the 
nucleons in the nuclear medium to give the final state is related to the imaginary part of the $\nu_l$ self energy as~\cite{Marco:1995vb}:
\begin{equation}\label{eqr}
-\frac{\Gamma}{2} = \frac{m_\nu}{E_\nu({\bf k})}\; Im \Sigma(k),
\end{equation}
where $\Sigma(k)$ is the neutrino self energy (shown in Fig.\ref{wself_energy} (left panel)).
By using Eq.\ref{eqr} in Eq.\ref{defxsec}, we obtain
\begin{equation}\label{eqq}
 d\sigma= \frac{-2m_\nu}{\mid {\bf k} \mid} Im \Sigma (k) d^3 r
\end{equation}
Thus 
to get $d\sigma$, we 
are required to evaluate the
imaginary part of neutrino self energy $Im \Sigma (k)$ which is obtained by following the Feynman rules:
\begin{figure}
\begin{center}
 \includegraphics[height=4.0 cm, width=11 cm]{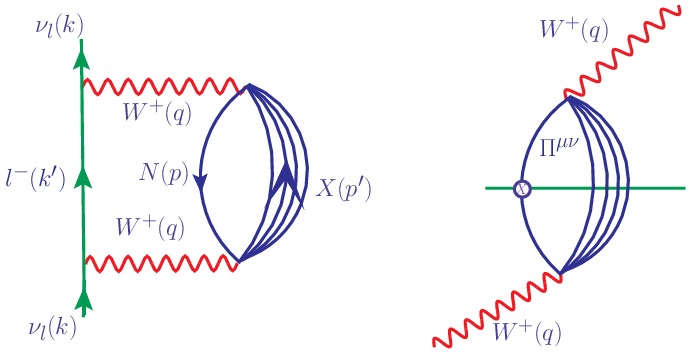}
 \end{center}
 \caption{Diagrammatic representation of the neutrino self-energy (left panel) and intermediate vector boson $W$ self-energy (right panel).}
 \label{wself_energy}
\end{figure}
% \begin{eqnarray}\label{self_weak}
% \Sigma(k)&=&\frac{i G_F}{\sqrt{2}} \int \frac{d^4q}{(2 \pi)^4} \frac{4 L_{\mu\nu}^{WI}}{m_l}\frac{1}{(k^{\prime 2}-m_l^2+i\epsilon)}\left(\frac{M_W}{q^2-M_W^2}\right)^2 \Pi^{\mu\nu}(q)\;,
% \end{eqnarray}
% where we have used the relation $\frac{g_W^2}{8 M_W^2}=\frac{G_F}{\sqrt{2}}$ and the properties of gamma matrices.
% Imaginary part of the neutrino self-energy may be obtained by following the Cutkowsky rules as
\begin{equation}\label{nu_imslf1}
Im \Sigma(k)=\frac{ G_F}{\sqrt{2}} {4 \over m_\nu} \int \frac{d^3 k^\prime}{(2 \pi)^4} {\pi \over E({\bf k^\prime})} \theta(q^0) \left(\frac{M_W}{Q^2+M_W^2}\right)^2\;Im[L_{\mu\nu}^{WI} \Pi^{\mu\nu}(q)]
\end{equation}
In the above expression, $\Pi^{\mu\nu}(q)$ is the $W$ boson self-energy, which is written in terms of the nucleon ($G_l$) and meson ($D_j$) propagators
(depicted in Fig.~\ref{wself_energy} (right panel)) following the Feynman rules and is given by
\begin{eqnarray}\label{wboson}
 \Pi^{\mu \nu} (q)&=& \left(\frac{G_F M_W^2}{\sqrt{2}}\right) \times \int \frac{d^4 p}{(2 \pi)^4} G (p) 
\sum_X \; \sum_{s_p, s_l} \prod_{i = 1}^{N} \int \frac{d^4 p'_i}{(2 \pi)^4} \; \prod_{_l} G_l (p'_l)\; \nonumber \\  
&\times&  \prod_{_j} \; D_j (p'_j)<X | J^{\mu} | N >  <X | J^{\nu} | N >^* (2 \pi)^4 \nonumber\\
&\times&\delta^4 (k + p - k^\prime - \sum^N_{i = 1} p'_i),\;\;\;
\end{eqnarray}
where $s_p$ is the spin of the nucleon, $s_l$ is the spin of the fermions in $X$, $<X | J^{\mu} | N >$ is the hadronic current for the initial state nucleon 
to the final state hadrons, index $l,~j$ are respectively, stands for the fermions and for the bosons in the final hadronic state $X$, and $\delta^4 (k + p - k^\prime - \sum^N_{i = 1} p'_i)$ ensures the conservation
of four momentum at the vertex.
The nucleon propagator $G(p)$ inside the nuclear medium
provides information about the propagation of the nucleon from the initial state to the final state or vice versa. 

\begin{figure}
\begin{center}
 \includegraphics[height=3.5 cm, width=10 cm]{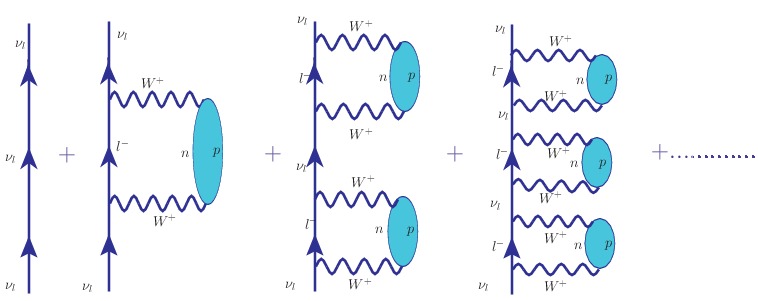}
 \end{center}
 \caption{Diagrammatic representation of neutrino self energy in the nuclear medium.}
 \label{n_self}
\end{figure}

 The relativistic nucleon propagator $G(p^0,p)$ in a nuclear medium is obtained by starting with 
 the relativistic free nucleon Dirac propagator $G^{0}(p^{0},{{\bf p}})$ which is written in terms of the
contribution from the positive and negative energy components of the nucleon described by the Dirac spinors
$u({\bf p})$ and $v({\bf p})$~\cite{Marco:1995vb,FernandezdeCordoba:1991wf}. 
 Only the positive energy contributions are retained as the negative energy contributions are suppressed. 
 In the interacting 
Fermi sea, the relativistic nucleon propagator is then written
in terms of the 
nucleon self energy $\Sigma^N(p^0,\bf{p})$ which is shown in Fig.\ref{n_self}. In nuclear many body technique, the quantity that contains all the information on
single nucleon properties is the nucleon self energy $\Sigma^N(p^0,\bf{p})$. For an interacting Fermi sea the relativistic nucleon propagator is written in terms of
the nucleon self energy and in nuclear matter the interaction is taken into account through Dyson series expansion. Dyson series expansion may be understood as 
the quantum field theoretical analogue of the Lippmann-Schwinger equation for the dressed nucleons, which is in principle an infinite series in perturbation theory.
%  However, this can be summed over and gives rise to the effective interaction in the medium
% using Dyson series expansion in terms of 
% nucleon self energy $\Sigma^N(p^0,\bf{p})$ which is shown in Fig.\ref{n_self}.
This perturbative 
expansion is summed in a ladder approximation as 
\begin{eqnarray}\label{gp1}
G(p)&=&\frac{M_N}{E({\bf p})}\frac{\sum_{r}u_{r}({\bf p})\bar u_{r}({\bf p})}{p^{0}-E({\bf p})}+\frac{M_N}{E({\bf p})}\frac{\sum_{r}u_{r}({\bf p})\bar
u_{r}({\bf p})}{p^{0}-E({\bf p})}\Sigma^N(p^{0},{\bf p})\nonumber\\
&&\times \frac{M_N}{E({\bf p})} \frac{\sum_{s}u_{s}({\bf p})\bar u_{s}({\bf p})}{p^{0}-E(P)}+..... \nonumber \\
&=&\frac{M_N}{E({\bf p})}\frac{\sum_{r} u_{r}({\bf p})\bar u_{r}({\bf p})}{p^{0}-E({\bf p})-\sum_{r}\bar u_{r}({\bf p})\Sigma^N (p^{0},{\bf p})u_{r}({\bf p})
\frac{M_N}{{E({\bf p})}}}\;\;\;
\end{eqnarray}
 The nucleon self energy $\Sigma^N(p^{0},{\bf p})$ is 
 spin diagonal, i.e., $\Sigma^N_{\alpha\beta}(p^{0},{\bf p})=\Sigma^N(p^{0},{\bf p})\delta_{\alpha\beta}$, where $\alpha$ and $\beta$ are spinorial indices. 
 The nucleon self energy $\Sigma^N(p)$ is obtained following the techniques of standard many body theory and is taken from 
Ref.~\cite{FernandezdeCordoba:1991wf, Oset:1981mk} which uses the nucleon-nucleon scattering cross section 
 and the spin-isospin effective interaction with random phase approximation(RPA) correlation as inputs. In this approach
the real part of the self energy of nucleon is obtained by means of dispersion relations using the expressions for the imaginary part 
 which has been explicitly calculated. The Fock term, which does not have imaginary part, does not contribute either to $Im \Sigma^N(p^{0},{\bf p})$ or to $Re \Sigma^N(p^{0},{\bf p})$ 
 through the dispersion relation and its contribution to $\Sigma^N(p^{0},{\bf p})$ is explicitly calculated and added to $Re \Sigma^N(p^{0},{\bf p})$~\cite{FernandezdeCordoba:1991wf}.
 The model however misses some contributions from similar terms of Hartree type which are independent of nucleon momentum $p$. 
 This semi-phenomenological model of nucleon self energy is found to be in reasonable agreement with those obtained 
in sophisticated  many body calculations and has been successfully used in the past to study nuclear medium effects in many processes
induced by photons, pions and leptons~\cite{Gil:1997bm, Gil:1997jg}.
 The expression for the nucleon self energy in the nuclear matter i.e. $\Sigma^N(p^0,{\bf{p}})$ is taken from Ref.~\cite{FernandezdeCordoba:1991wf}, 
 and the dressed nucleon propagator is expressed as
 \begin{small}
 \begin{eqnarray}\label{Gp}
G (p) =&& \frac{M_N}{E({\bf p})} 
\sum_r u_r ({\bf p}) \bar{u}_r({\bf p})
\left[\int^{\mu}_{- \infty} d \omega 
\frac{S_h (\omega, {\bf{p}})}{p^0 - \omega - i \eta}
+ \int^{\infty}_{\mu} d  \omega 
\frac{S_p (\omega, {\bf{p}})}{p^0 - \omega + i \eta}\right],\;\;~~~
\end{eqnarray}
\end{small}
where $S_h (\omega, {\bf{p}})$ and $S_p (\omega, {\bf{p}})$ are the hole
and particle spectral functions, respectively. $\mu=\epsilon_F+M_N$ is the chemical potential, $\omega=p^0-M_N$ is the removal energy and $\eta$ is the infinitesimal small quantity, i.e. $\eta \to 0$.   The spectral function and its properties have been discussed in brief in Appendix-\ref{spec_nuc} and Appendix-\ref{spec_prop}, respectively.
% Using Eqs.~\ref{gp1} and~\ref{Gp} above the expressions for 
% the hole and particle spectral functions can be easily obtained in terms of the nucleon self energy and they are given in Ref.\cite{FernandezdeCordoba:1991wf} 
% with their properties.

The cross section (Appendix-\ref{app:lda}) is then obtained by using Eqs.~\ref{eqq} and \ref{nu_imslf1} : 
\begin{equation}\label{dsigma_3}
\frac {d\sigma_A^{WI}}{d\Omega_l' dE_l'}=-\frac{G_F^2}{(2\pi)^2}\frac{|\bf{k^\prime}|}{|\bf {k}|}\left(\frac{M_W^2}{Q^2+M_W^2}\right)^2  \int  Im\left(L_{\mu\nu}\Pi^{\mu\nu}\right) d^{3}r, 
\end{equation}

Now by comparing the above equation with Eqs.\ref{xecA}, \ref{wboson} and \ref{Gp} the expression of the nuclear hadronic tensor for an isospin symmetric nucleus in terms of 
 the nucleonic hadronic tensor and spectral function, is obtained as~\cite{Haider:2015vea}
\begin{equation}\label{conv_WAa}
W^{\mu \nu}_{A} = 4 \int \, d^3 r \, \int \frac{d^3 p}{(2 \pi)^3} \, 
\frac{M_N}{E ({\bf p})} \, \int^{\mu}_{- \infty} d p^0 S_h (p^0, {\bf p}, \rho(r))
W^{\mu \nu}_{N} (p, q), \,
\end{equation}
where the factor of 4 is for the spin-isospin of nucleon and $\rho(r)$ is the charge density of the nucleon in the nucleus. In general, nuclear density have various phenomenological parameterizations known 
in the literature as the harmonic oscillator(HO) density, two parameter Fermi density(2pF),
modified harmonic oscillator (MHO) density, etc. The proton density distributions are obtained from the electron-nucleus scattering experiments, while the neutron densities
are taken from the Hartree-Fock approach~\cite{GarciaRecio:1991wk}. Thus the density parameters corresponds to the charge density for proton or equivalently the neutron matter density for neutron. Recently at the JLab, PREX and CREX collaborations~\cite{Hagen:2015yea}-\cite{prex-crex} have made efforts to directly measure the neutral  weak form factor of a few nuclei from  which the neutron rms radii of nuclei can be obtained. Further development in this area would be of great help to determine precisely the neutron form factor in nuclei for a broad mass range.

For a nonisoscalar nuclear target, the nuclear hadronic tensor is given by
\begin{equation}\label{conv_WAan}
 W^{\mu \nu}_{A} = 2 \sum_{\tau=p,n} \int d^3 r \int \frac{d^3 p}{(2 \pi)^3} 
\frac{M_N}{E ({\bf p})} \, \int^{\mu_\tau}_{- \infty} d p^0 S_h^\tau (p^0, {\bf p}, \rho^\tau(r))\;
W^{\mu \nu}_{N} (p, q), \,
\end{equation}
where $\mu_p(\mu_n)$ is the chemical potential for the proton(neutron). $S_h^p(\omega,{\bf p},\rho_p(r))$ and $S_h^n(\omega,{\bf p},\rho_n(r))$ are the 
 hole spectral functions for the proton and neutron, respectively, which 
 provide information about the probability distribution of finding a proton and neutron with removal energy $\omega$ and three momentum ${\bf p}$ inside the nucleus.
 
Now to evaluate the weak dimensionless nuclear structure functions by using Eq.(\ref{conv_WAa}), the appropriate components of 
nucleonic ($W^{\mu\nu}_N$ in Eq.\ref{ch2:had_ten_N}) and nuclear ($W^{\mu\nu}_A$ in Eq.\ref{nuc_had_weak}) hadronic tensors
along the $x,~y$ and $z$ axes are chosen. The dimensionless nuclear structure functions $F^{WI}_{iA}(x,Q^2)(i=1,2,3)$, following the analogy between the nucleon structure functions given in Eq.\ref{ch2:relation} and the nuclear structure functions $W_{iA}^{WI}(\nu,Q^2)$ are defined as
\begin{equation}\label{relation1}
\left.\begin{array}{l}
F_{1A}^{WI}(x_A,Q^2)=M_A W_{1A}^{WI}(\nu_A, Q^2),\nonumber\\
F_{2A}^{WI}(x_A,Q^2)=\nu_A~W_{2A}^{WI}(\nu_A,Q^2),\nonumber \\
F_{3A}^{WI}(x_A,Q^2)=\nu_A~W_{3A}^{WI}(\nu_A, Q^2),
      \end{array}
      \right\}
\end{equation}
where the energy transfer $\nu_A=\frac{p_{_A}\cdot q}{M_{_A}}=q^{0}$.

%  In the present work, the scattering process has been considered in the laboratory frame, where target nucleus is at rest(${\bf p_A}=0$) and the nucleons are moving with finite such that
% momentum(${\bf p}\ne 0$). 
% \begin{eqnarray}
% p_{_A}^\mu&=&(M_{_A},\vec 0),~~~~~\nu_A=\frac{p_{_A}\cdot q}{M_{_A}}=\frac{(p_{A}^0  q^0 - {\bf p_A \cdot q})}{M_{_A}}=q^{0},~~ \nonumber\\
% x_A&=&\frac{Q^2}{2 p_A \cdot q}=\frac{Q^2}{2 p_{A}^0  q^0 } = \frac{Q^2}{2 A~M_N q^0}
% \end{eqnarray}
%  Taking the momentum transfer of the bound nucleon along the $z$-axis such that $q^\mu=(q^0,0,0,q^z)$ and then Bjorken variable $x_N$ is given by
%  \begin{equation}
%  x_N = \frac{Q^2}{2 p \cdot q} = \frac{Q^2}{2 (p^0 q^0 - p^z q^z)}
% \end{equation}
By taking the $zz$ component of the hadronic tensors($W_{N}^{\mu \nu}$ of Eq.\ref{ch2:had_ten_N} and $W^{\mu\nu}_A$ of Eq.\ref{nuc_had_weak}), for a
nonisoscalar nuclear target the following expression is obtained~\cite{Haider:2011qs}:
  \begin{eqnarray} 
\label{had_ten151weakni}
F_{2A,N}^{WI}(x_A,Q^2)  &=&  2\sum_{\tau=p,n} \int \, d^3 r \, \int \frac{d^3 p}{(2 \pi)^3} \, 
\frac{M_N}{E_N ({\bf p})} \, \int^{\mu_\tau}_{- \infty} d p^0 ~S_h^\tau (p^0, {\bf p}, \rho^\tau(r)) \nonumber\\
&\times&\left[\left(\frac{Q}{q^z}\right)^2\left( \frac{|{\bf p}|^2~-~(p^{z})^2}{2M_N^2}\right) +  \frac{(p^0~-~p^z~\gamma)^2}{M_N^2}\times\right. \nonumber\\
&&\left.\left(\frac{p^z~Q^2}{(p^0-p^z~\gamma) q^0 q^z}+1\right)^2\right]  
\left(\frac{M_N}{p^0-p^z~\gamma}\right) F_{2\tau}^{WI}(x_N,Q^2).\;\;\;~~      
\end{eqnarray}

The choice of $xx$ components of the nucleonic(Eq.~\ref{ch2:had_ten_N}) and nuclear(Eq.~\ref{nuc_had_weak}) hadronic tensors lead to the expression of 
$F_{1A,N}^{WI}(x,Q^2)$ as
\begin{eqnarray}\label{f1ani}
 F_{1A,N}^{WI}(x_A,Q^2) &=& 2 \sum_{\tau=p,n} A M_N \int d^3r\int \frac{d^3 p}{(2 \pi)^3} 
\frac{M_N}{E_N ({\bf p})}  \int^{\mu_\tau}_{- \infty} d p^0 S_h^\tau (p^0, {\bf p}, \rho^\tau(r))\nonumber\\
&\times&\left[\frac{F_{1\tau}^{WI}(x_N,Q^2)}{M_N} + \left(\frac{p^x}{M_N}\right)^2 \frac{F_{2\tau}^{WI}(x_N,Q^2)}{\nu} \right]\;\;
\end{eqnarray}
in the case of nonisoscalar nuclear target.

Now by using the $xy$ components of the nucleonic(Eq.~\ref{ch2:had_ten_N}) and nuclear(Eq.~\ref{nuc_had_weak}) hadronic tensors in Eq.~\ref{conv_WAa},
the parity violating nuclear structure function is obtained as
\begin{eqnarray}\label{f3a_weak_noniso}
 F_{3A,N}^{WI}(x_A,Q^2) &=& 2 A \sum_{\tau=p,n} \int \, d^3 r \, \int \frac{d^3 p}{(2 \pi)^3} \, 
\frac{M_N}{E_N ({\bf p})} \, \int^{\mu_\tau}_{- \infty} d p^0 S_h^\tau (p^0, {\bf p}, \rho^\tau(r)) \nonumber\\
&\times&  \frac{q^0}{q^z}\left({p^0 q^z - p^z q^0  \over p \cdot q} \right)F_{3\tau}^{WI}(x_N,Q^2),\;\;~~
\end{eqnarray}
for a nonisoscalar nuclear target.

For an isoscalar target, the factor of 2 in Eqs. \ref{had_ten151weakni}, \ref{f1ani} and \ref{f3a_weak_noniso}, will be replaced by 4 and the contribution will come from the 
nucleon's hole spectral function $S_{h}(p^0, {\bf p}, \rho(r))$ instead of the individual contribution from proton and neutron targets 
in $S_h^\tau (p^0, {\bf p}, \rho^\tau(r));~(\tau=p,n)$.

The results obtained by using Eqs.~\ref{had_ten151weakni}, \ref{f1ani}, and \ref{f3a_weak_noniso} for a nuclear target
are labeled as the results with the spectral function(SF) only. 
\begin{figure}
\begin{center}
\includegraphics[height=3.5 cm, width=3 cm]{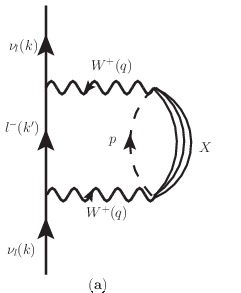}
 \includegraphics[height=3.5 cm, width=9 cm]{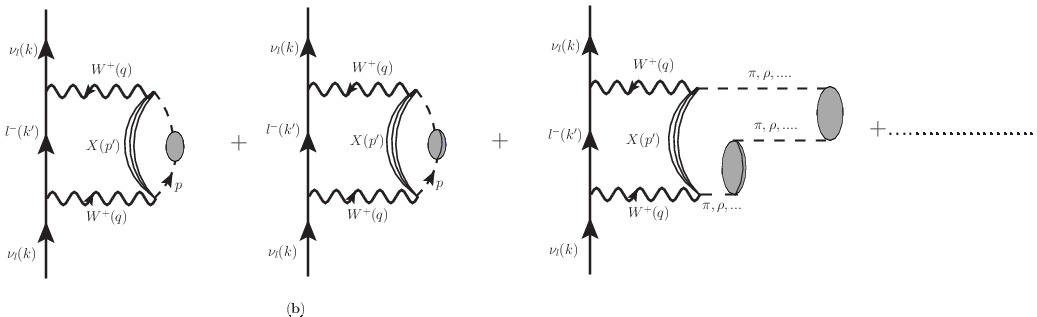}
 \end{center}
 \caption{Neutrino self energy diagram accounting for neutrino-meson DIS (a) the bound nucleon propagator is substituted with a meson($\pi$ or $\rho$) propagator 
 (b) by including particle-hole $(1p–1h)$, delta-hole $(1\Delta–1h)$,
 $1p1h-1\Delta1h$, etc. interactions.}
 \label{n_self-1}
\end{figure}
Furthermore, the nucleons bound inside the nucleus may interact among themselves via meson exchange such as $\pi,~\rho,$ etc. The interaction of intermediate vector boson 
 with these mesons
play an important role in the evaluation of nuclear structure functions. Therefore, the mesonic effect has been incorporated in the Aligarh-Valencia model and is discussed in
the next sub-subsection~\ref{mes}.

 \subsubsection{Mesonic effect}\label{mes}
 There are virtual mesons (mainly pion and rho meson) associated with each nucleon bound inside the nucleus. This mesonic cloud gets strengthened by the strong
 attractive nature of the 
 nucleon-nucleon interaction, which leads to a reasonably good probability of interaction of virtual bosons(IVB) with a meson instead of a 
 nucleon~\cite{Marco:1995vb,Kulagin:2004ie,  Ericson:1983um,LlewellynSmith:1983vzz}.
 Although the contribution from the pion
 cloud is larger than the contribution from rho-meson cloud, nevertheless, the rho contribution is non-negligible, and 
 both of them are positive in all the range of $x$.
 The mesonic contribution is smaller in lighter nuclei, while it becomes more pronounced in heavier nuclear
 targets and dominates in the 
 intermediate region of $x~ (0.2 < x < 0.6)$. 
  It may be pointed out that calculations performed with only the spectral function, result in a reduction in the nuclear structure function from the free 
 nucleon structure function. While the inclusion of mesonic cloud contribution leads to an 
 enhancement of the nuclear structure function, and it works in the right direction to explain the experimental
 data~\cite{Marco:1995vb,SajjadAthar:2009cr,Haider:2015vea}. 
 
 To obtain the contribution from the virtual mesons, the neutrino self energy is again evaluated using many body techniques
 ~\cite{Marco:1995vb}, 
 and to take into account mesonic effects a diagram similar to the one shown in Fig.\ref{wself_energy} is drawn, except that instead of a
nucleon now there is a meson which results in the change of a nucleon propagator by a meson
propagator. This meson propagator does not correspond to the free mesons as one lepton can not decay into another lepton, one pion and $X$
 but corresponds to the mesons arising due to the nuclear medium effects by using a modified meson propagator. These
mesons are arising in the nuclear medium through particle-hole $(1p–1h)$, delta-hole $(1\Delta–1h)$,
 $1p1h-1\Delta1h$, $2p-2h$, etc. interactions as depicted in Fig.\ref{n_self-1}.
 
 To evaluate the mesonic structure function $F_{2 A,a}^{WI}(x,Q^2)~~(a=\pi,\rho)$ the imaginary part of the meson propagator is used instead of spectral function,
 and the expression for $F_{_{2 A,a}}^{ WI}(x,Q^2), ~~(a=\pi,\rho)$ obtained by them~\cite{Haider:2011qs} is given by:
\begin{eqnarray} \label{pion_f21}
F_{_{2 A,a}}^{ WI}(x,Q^2)  &=&  -6 \kappa \int \, d^3 r  \int \frac{d^4 p}{(2 \pi)^4} 
        \theta (p^0) ~\delta I m D_a (p) \;2M_a~\left(\frac{M_a}{p^0~-~p^z~\gamma}\right)\nonumber \\
&\times&\left[\frac{Q^2}{(q^z)^2}\left( \frac{|{\bf p}|^2~-~(p^{z})^2}{2M_a^2}\right)  
+  \frac{(p^0~-~p^z~\gamma)^2}{M_a^2} \right.\nonumber\\
&& \left.\times\left(\frac{p^z~Q^2}{(p^0~-~p^z~\gamma) q^0 q^z}~+~1\right)^2\right] ~F_{_{2a}}^{ WI}(x_a)~~
\end{eqnarray}
where $\kappa=1$ for pion and $\kappa=2$ for rho meson, $x_a=-\frac{Q^2}{2p \cdot q}$, $M_a$ is the mass of pion or rho meson. $D_a(p)$ is the pion or rho meson propagator in the nuclear medium given by 
 \begin{equation}\label{dpi}
D_a (p) = [ p_0^2 - {\bf {p}}\,^{2} - M_a^2 - \Pi_{a} (p_0, {\bf p}) ]^{- 1}\,,
\end{equation}
with
\begin{equation}\label{pionSelfenergy}
\Pi_a(p_0, {\bf p})=\frac{f^2}{M_\pi^2}\;\frac{C_\rho\;F^2_a(p){\bf {p}}\,^{2}\Pi^*}{1-{f^2\over M_\pi^2} V'_j\Pi^*}\,.
\end{equation}
In the above expression, $C_\rho=1$ for pion and $C_\rho=3.94$ for rho meson. $F_a(p)={(\Lambda_a^2-M_a^2) \over (\Lambda_a^2 - p^2)}$ is the $\pi NN$ or $\rho NN$ form 
factor, $p^2=p_0^2~-~{\bf p}^2$, $\Lambda_a$=1~$GeV$ and $f=1.01$. For pion (rho meson), $V_j'$ is the longitudinal (transverse)
part of the spin-isospin interaction and $\Pi^*$ is the irreducible meson self energy that contains the contribution of particle-hole and delta-hole excitations. Various quark and antiquark PDFs parameterizations for pions are available in the literature such as given by Conway et al.~\cite{Conway:1989fs}, Martin et al.~\cite{Martin:1998sq}, Sutton et al.~\cite{Sutton:1991ay}, Wijesooriya et al.~\cite{Wijesooriya:2005ir}, Gluck et al.\cite{Gluck:1991ey}, etc. Aligarh-Valencia group have observed~\cite{Zaidi:2019mfd} that the choice of pionic PDFs parameterization would not make any significant difference in the event rates.
 In this work, the parameterization given by Gluck et al.\cite{Gluck:1991ey} has been taken into account for pions and for the rho mesons same PDFs as for the pions have been used.

The choice of $\Lambda_a=$ 1 GeV, $(a=\pi,\rho)$ have been fixed by Aligarh-Valencia group~\cite{SajjadAthar:2009cr, Haider:2015vea} to describe the nuclear
medium effects in electromagnetic nuclear structure function $F_{2A}^{EM}(x,Q^2)$ necessary to explain the data from JLab and 
other experiments performed using charged lepton scattering from several nuclear targets in the DIS region.

\subsubsection{Shadowing and Antishadowing effects}

Aligarh-Valencia group has taken the shadowing effect into account by following the works of Kulagin and Petti~\cite{Kulagin:2004ie,Kulagin:2007ju} who have used the original Glauber-Gribov multiple scattering theory.  In the case of $\nu_l/\bar\nu_l$ induced DIS processes, they have treated shadowing differently from the prescription applied in the case of electromagnetic structure functions~\cite{Kulagin:2004ie, Kulagin:2007ju}, due to the presence of the axial-vector  current  in  the $\nu_l$ interactions. 
The interference between the vector and the axial-vector currents introduces C-odd  terms  in  $\nu_l$  cross  sections,  which  are  described by structure function $F_3^{WI}(x,Q^2)$, and in their calculation of nuclear corrections, separate contributions to different structure functions according to their C-parity have been taken into account. This results in a different dependence  of nuclear effects on C-parity specially  in  the  nuclear  shadowing  region.  The same prescription has been adopted by the Aligarh-Valencia group.
 The Aligarh-Valencia group  points out that the inclusion of shadowing effect in the present model is not very comprehensive and more work is required. A review on the nuclear shadowing 
 in electroweak interaction has been done in \cite{Kopeliovich:2012kw}.
 
 %Although it may be pointed out that a change of $10\%$ in the shadowing and antishadowing treatment used in this work brings a change of only $1-2\%$ in the nuclear structure functions in a heavy nucleus like $^{208}Pb$ and is $<1\%$ in nuclei like $^{40}Ar$ and $^{56}Fe$ at low $x \sim 0.05-0.2$ and low $Q^2$, which becomes almost negligible at $x > 0.2$ and $Q^2 \ge 5 GeV^2$.

 \subsubsection{Isoscalarity Corrections}
 \label{iso_corrections}
 
In the case of heavier nuclear targets, where neutron number($N=A-Z$) is larger than the proton number($Z$) and their densities are also different, 
isoscalarity corrections become important. 
As most of the neutrino/antineutrino experiments are using heavy nuclear targets($N \ne Z$), phenomenologically the isoscalarity correction is taken into 
account by multiplying the experimental results with a correction factor defined as
\begin{eqnarray}\label{isocorr_em}
R_{A}^{Iso} &=& \frac{[{F_2^{\nu/\bar\nu p}+F_2^{\nu/\bar\nu n}] /2}}{{[Z F_2^{\nu/\bar\nu p} + (A-Z) F_2^{\nu/\bar\nu n}]/ A}}\;,~~
\end{eqnarray}
where $F_2^{\nu/\bar\nu n}$ are the weak structure functions for the proton and the neutron, respectively.

 \subsection{Results and Discussions}\label{section4}
  Aligarh-Valencia group have applied their model to study the effects of the nuclear medium on the electromagnetic structure 
 functions~\cite{SajjadAthar:2009cr,Haider:2015vea,Zaidi:2019mfd} as well as the weak structure functions~\cite{SajjadAthar:2007bz,Haider:2011qs,Haider:2012nf,Zaidi:2019asc} and have made
  a comparison between weak and electromagnetic nuclear structure functions for a wide range of $x$ and $Q^2$~\cite{Haider:2016zrk}. Furthermore an important effect, the isoscalarity 
  correction for the nonisoscalar nuclear targets has been studied by them (as discussed in \ref{spec} and \ref{mes}). They have applied their model to study medium effects in extracting $sin^2\theta_W$ 
 using the Paschos-Wolfenstein relation~\cite{Haider:2012ic}. This model has been applied successfully to study the Drell-Yan processes~\cite{Haider:2016tev} and 
 parity violating asymmetry with nuclear medium effects using polarized electron beam($\vec e$)~\cite{Haider:2014iia}.
 
\begin{figure}
\begin{center} 
     \includegraphics[height=0.35\textheight,width=0.75\textwidth]{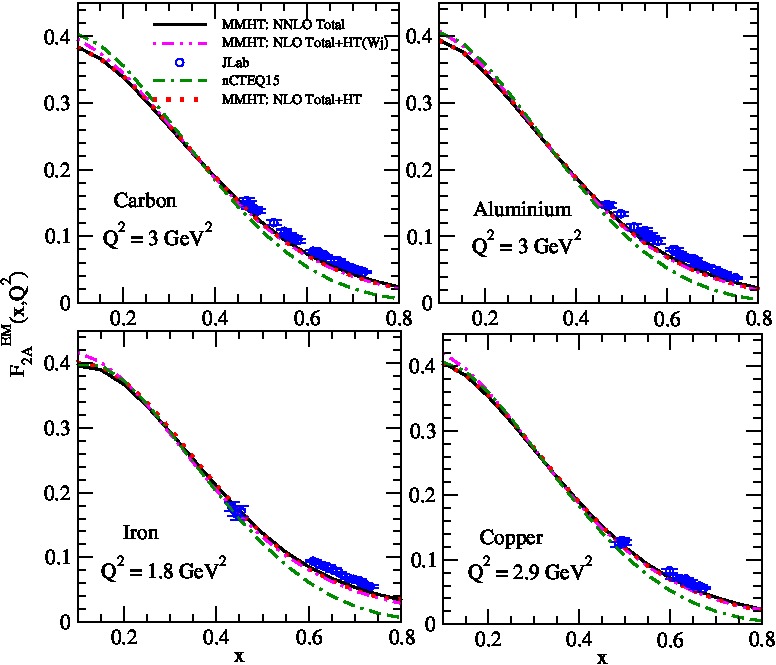}
 \caption{$F_{2A}^{EM}(x,Q^2)$ vs $x$ at different values of $Q^2$, in $^{12}$C, $^{27}$Al, $^{56}$Fe and $^{64}$Cu
 with the full model at NLO and NNLO using MMHT nucleon PDFs~\cite{Harland-Lang:2014zoa}. The results at NNLO are shown by solid 
 line and at NLO with the HT effect (renormalon approach~\cite{Dasgupta:1996hh, Stein:1998wr}) are shown by the dashed-double dotted line 
 using the pionic PDFs parameterization given by Wijesooriya et al.~\cite{Wijesooriya:2005ir} and by the dotted line for the parameterization
 of Gluck et al.~\cite{Gluck:1991ey}. The results are also obtained by using the nuclear PDFs parameterization given by nCTEQ group~\cite{Kovarik:2015cma}
 (double-dashed dotted line) and the experimental points are the JLab data~\cite{Mamyan:2012th}.}
 \label{figure5}
 \end{center}
\end{figure}

  This model describes the nuclear structure functions $F_{iA,N}^{WI}(x_A,Q^2)~(i=1-3)$ (defined in Eqs.~\ref{had_ten151weakni}, \ref{f1ani} and 
  \ref{f3a_weak_noniso}), 
 in terms of the nucleon structure functions $F_{iN}^{WI}(x_N,Q^2)$, convoluted with the spectral function which takes into account Fermi motion, 
 binding energy and nucleon 
 correlation effects followed by the mesonic and shadowing effects.  For the evaluation 
 of $F_{iN}^{WI}(x_N,Q^2)$ at the leading order(LO), free nucleon PDFs are used. Therefore, their numerical results do not use nuclear PDFs. 
 The results presented in this review are obtained using nucleon PDFs of MMHT~\cite{Harland-Lang:2014zoa} as well as CTEQ6.6 in the 
  MS-bar scheme~\cite{Nadolsky:2008zw}. $F_{iA,\pi}^{WI}(x,Q^2)$ and $F_{iA,\rho}^{WI}(x,Q^2)$ are the 
  structure functions giving pion and rho mesons contribution. 
  In the literature, various pionic PDFs parameterizations are 
  available and this work uses 
  %like that of Gluck et al.~\cite{Gluck:1991ey}, Wijesooriya et al.~\cite{Wijesooriya:2005ir},
 %Sutton et al.~\cite{Sutton:1991ay}, Conway et al.~\cite{Conway:1989fs}, etc. 
 %The numerical results presented here are 
  the pionic PDFs parameterization of Gluck et al.~\cite{Gluck:1991ey} as in Fig.~\ref{figure5}. Also for the comparison pion PDFs of Wijesooriya et al.~\cite{Wijesooriya:2005ir} have been used.
  To evaluate the nucleon structure functions in the kinematic region of low and moderate $Q^2$, where the higher order perturbative 
 corrections and the non-perturbative effects become important, PDFs evolution up to NNLO has been performed and included the effects of TMC and higher twist in the 
 numerical calculations. For the evolution of nucleon PDFs at the next-to-leading order(NLO) and next-to-next-to-leading order(NNLO) the works of Vermaseren et al.~\cite{Vermaseren:2005qc} and Moch et al.~\cite{ Moch:2008fj,Moch:2004xu} have been followed.
 The target mass correction effect has been included following the method of Schienbein et al.~\cite{Schienbein:2007gr}. 
 The dynamical higher twist correction has been taken into account following the methods of Dasgupta et al.~\cite{Dasgupta:1996hh} and 
 Stein et al.~\cite{Stein:1998wr} at NLO. 
  
The theoretical results obtained in the Aligarh-Valencia model~\cite{SajjadAthar:2007bz}-\cite{Zaidi:2019asc}
 are presented
 and compared with the experimental data wherever available. The first case is when the calculations are performed using the spectral function (SF) only
 and then the contribution from meson clouds as well as shadowing effect 
 are taken into account and this corresponds to the full model (Total) results as quoted by the authors~\cite{SajjadAthar:2007bz}-\cite{Zaidi:2019asc}. 
 The expression of total nuclear structure functions with the full theoretical model is given by
\begin{eqnarray}\label{f1f2_tot}
 F_{iA}^{WI}(x,Q^2)=F_{iA,N}^{WI}(x,Q^2)+F_{iA,\pi}^{WI}(x,Q^2)+F_{iA,\rho}^{WI}(x,Q^2)
 +F_{iA,shd}^{WI}(x,Q^2),\;\;\;\;\;
\end{eqnarray}
where $i=1-2$. $F_{iA,N}^{WI}(x,Q^2)$ are the nuclear structure function which has contribution from only the spectral function, $F_{iA,\pi/\rho}^{WI}(x,Q^2)$ take into 
account mesonic contributions. $F_{iA,shd}^{WI}(x,Q^2)$ has contribution from the 
shadowing effect which is given by
\begin{equation}\label{shad_sf}
 F_{iA,shd}^{WI}(x,Q^2) = \delta R_i(x,Q^2) \times F_{i,N}^{WI}(x,Q^2),
\end{equation}
where $\delta R_i(x,Q^2)$ is the shadowing correction factor for which Kulagin and Petti~\cite{Kulagin:2004ie} has been followed. 
In this model, the full expression
for the parity violating weak nuclear structure function is given by,
\begin{eqnarray}\label{f3_tot}
 F_{3A}^{WI}(x,Q^2)= F_{3A,N}^{WI}(x,Q^2) + F_{3A,shd}^{WI}(x,Q^2).
\end{eqnarray}
Notice that this structure function has no mesonic contribution and mainly the contribution to the nucleon structure function comes from the valence quarks distributions. 
For $F_{3A,shd}^{WI}(x,Q^2)$ similar definition has been used as given
in Eq.(\ref{shad_sf}) following the works of Kulagin et al.~\cite{Kulagin:2004ie}.
 
  First the results for the nuclear structure function($F_{2A}^{EM}(x,Q^2)$) in the case of electromagnetic interaction have been presented in
 Fig.\ref{figure5}, for the different nuclear targets like $^{12}C$, $^{27}Al$, $^{56}Fe$ and $^{64}Cu$~\cite{Zaidi:2019mfd}
  at moderate values of $Q^2$($1.8 \le Q^2 \le 3.0$ GeV$^2$) and compared with the available experimental results of the JLab~\cite{Mamyan:2012th}. 
    The nuclear targets are treated as isoscalar. For the evaluation of free nucleon structure functions, MMHT~\cite{Harland-Lang:2014zoa} parameterization has been used.
   The numerical results are shown for the full model using a similar expression as Eq.\ref{f1f2_tot} for the electromagnetic nuclear structure functions 
   $F_{2A}^{EM}(x,Q^2)$ with nucleon structure functions $F_{2N}^{EM}(x_N,Q^2)$:
  \begin{itemize}
   \item  at NNLO with mesonic PDFs of Gluck et al.~\cite{Gluck:1991ey} 
   \item  at NLO with HT effect and mesonic PDFs of Gluck et al.~\cite{Gluck:1991ey}
   \item at NLO with HT effect and mesonic PDFs of Wijesooriya et al.~\cite{Wijesooriya:2005ir}
  \end{itemize}
  It may be noticed from Fig.\ref{figure5} that the dependence of different pionic PDFs parameterizations have not much effect 
  on the evaluation of $F_{2A}^{EM}(x,Q^2)$. 
  Also the results obtained show that as long as  TMC is applied, NNLO is within a few percent of the results obtained at NLO with HT effect.  Further details of this interesting observation can be found in~\cite{Yang:1998zb}.
In literature, along with the free nucleon PDFs parameterizations, different nuclear PDFs are also available like AT12~\cite{AtashbarTehrani:2012xh}, nCTEQ15~\cite{Kovarik:2015cma}, 
EPPS16~\cite{Eskola:2016oht}, etc. Also, for the comparison, in this figure, the results obtained using nuclear PDFs of nCTEQ group~\cite{Kovarik:2015cma} has been shown. 
 It may be noticed that the theoretical 
 results obtained using the full model are reasonably in good agreement with the 
 nCTEQ results~\cite{Kovarik:2015cma} and show a good agreement with the JLab experimental data~\cite{Mamyan:2012th} in the 
 region of intermediate $x$. However, for $x>0.6$ and $Q^2 \approx 2~GeV^2$ they slightly underestimate the experimental results. 
 Since the region of high $x$ and low $Q^2$ is the transition region of nucleon resonances and DIS, the present theoretical results might indeed differ from the experimental data. 
 With the increase in $Q^2$, theoretical results show a better agreement with the experimental observations of 
 JLab~\cite{Mamyan:2012th} in the entire range of $x$. 
 
  \begin{figure}
\begin{center} 
     \includegraphics[height=0.28\textheight,width=0.75\textwidth]{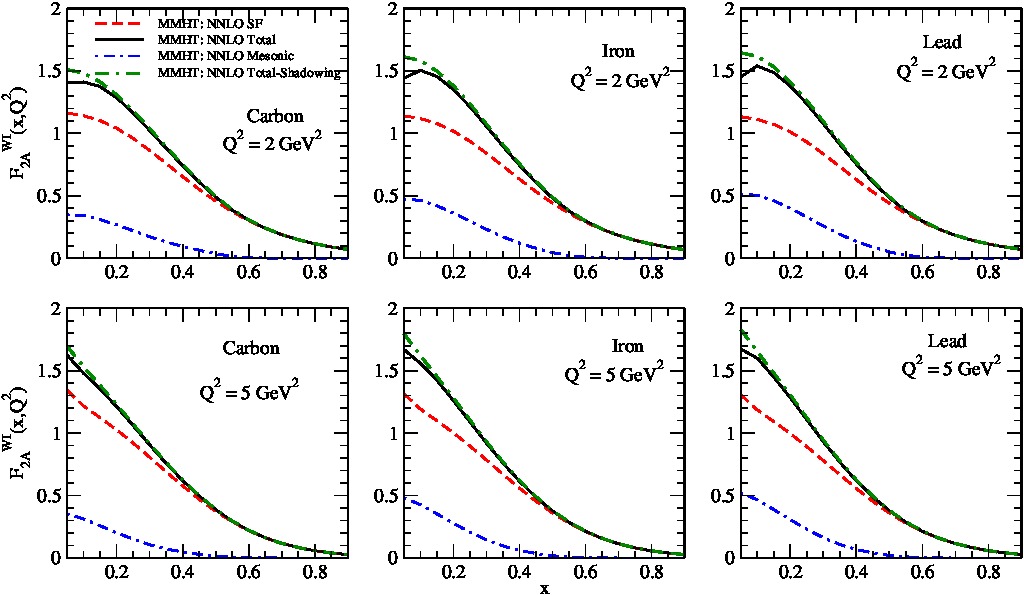}
 \caption{Results are shown for the weak nuclear structure function $F_{2A}^{WI}(x,Q^2)$ vs $x$ at $Q^2=2,~5~GeV^2$, in $^{12}$C, $^{56}$Fe and $^{208}$Pb 
 for {\bf (i)} only the spectral function (dashed line), {\bf (ii)} only the mesonic contribution (dash-dotted line) using Eq.\ref{pion_f21}, {\bf (iii)} the
 full calculation (solid line) using Eq.\ref{f1f2_tot} as well as {\bf (iv)} the double-dash-dotted line is the result without the shadowing and antishadowing effects. The numerical calculations have been performed at NNLO by using the MMHT~\cite{Harland-Lang:2014zoa} nucleon
 PDFs parameterizations.}
 \label{figurewk}
 \end{center}
\end{figure}
 Turning now to the weak interactions, in Fig.~\ref{figurewk}, the results are presented for $F_{2A}^{WI}(x,Q^2)$ vs $x$ for $^{12}C$, $^{56}Fe$ and $^{208}Pb$, for isoscalar nuclear targets,  at the different values of $Q^2$ chosen to reflect the current neutrino beam energies. The numerical results are obtained first by using the spectral function (dashed line), then we have included mesonic effect(dash-dotted line) and the final result by including the shadowing and antishadowing effects is shown by the solid line.   
 From the figure, it may be observed that the mesonic contributions result in an enhancement in the nuclear structure functions and
 is significant in the low and intermediate 
 region of $x$. Moreover, the effect is more pronounced at low $Q^2$ and becomes larger with the increase in mass number $A$. 
 For example, in comparison to the total
 contributions (solid line) in  carbon, the mesonic contribution at $x=0.1$  is found to be $24\%$ in iron which increases to $33\%$ in lead. With the increase in $x$(say $x=0.4$) the enhancement reduces to 
 $13\%$ and $18\%$ respectively and becomes almost negligible for $x\ge 0.6$ at $Q^2=2~GeV^2$.
 To depict the coherent nuclear effects(shadowing) which results in suppression of the structure functions at low $x$, the results without shadowing are shown with the double-dash-dotted line, and it may be observed that with the increase in mass number of the nuclear target($^{56}Fe$ vs $^{208}Pb$), the strength of suppression becomes larger.
%   For example, it reduces the total result by
%  $5\%$ at $x=0.1$ and $< 2\%$ at $x=0.2$ for $Q^2=2~GeV^2$, and this reduction becomes $3\%$ at $x=0.1$ and $< 1\%$ at $x=0.2$ for $Q^2=2~GeV^2$. 
   \begin{figure}
\begin{center} 
     \includegraphics[height=0.35\textheight,width=0.75\textwidth]{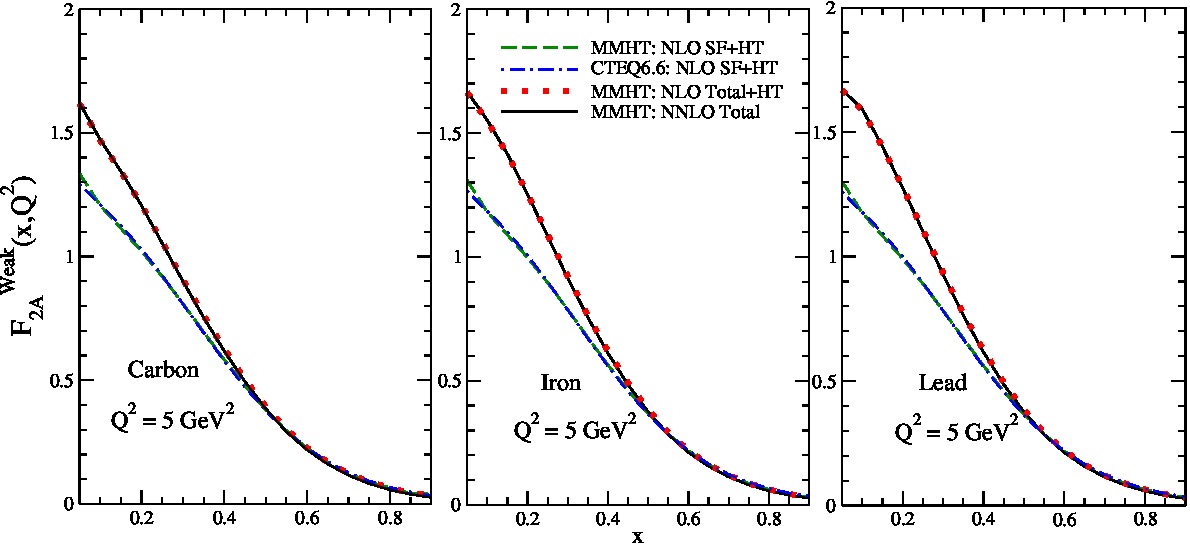}
 \caption{Results are shown for $F_{2A}^{WI}(x,Q^2)$ vs $x$ at a fixed $Q^2=5~GeV^2$, in $^{12}$C, $^{56}$Fe and $^{208}$Pb 
 for only the spectral function (dashed line) and for the full calculation (dotted line) at NLO with HT effect (renormalon approach~\cite{Dasgupta:1996hh, Stein:1998wr}) 
 using MMHT~\cite{Harland-Lang:2014zoa} nucleon PDFs parameterizations. 
 Solid line is the result of the full calculation at NNLO using MMHT PDFs parameterizations~\cite{Harland-Lang:2014zoa}. 
 Notice that the curves for NLO+HT and NNLO are almost the same implying equivalence of the two (NLO+HT and NNLO) for all $x$.
 The results at NLO obtained using only the spectral function with HT effect 
 are also compared with the corresponding results obtained using the CTEQ6.6~\cite{Nadolsky:2008zw} nucleon PDFs parameterization in the MS-bar scheme. All the nuclear targets are treated as isoscalar.}
 \label{figure6a}
 \end{center}
\end{figure}
In Fig.\ref{figure6a}, we present the results for $F_{2A}^{WI}(x,Q^2)$ in three different nuclear targets viz. $^{12}$, $^{56}Fe$ and $^{208}Pb$. These results are obtained with the spectral function(SF) as well as for the full model  using the nucleon PDFs  evaluated at NLO with higher twists(HT).
 To study the dependence of nuclear structure functions on the nucleon PDFs parameterization the numerical calculations have been performed by using the MMHT~\cite{Harland-Lang:2014zoa} as well as 
CTEQ6.6~\cite{Nadolsky:2008zw} nucleon PDFs parameterizations in the MS-bar scheme. From the figure, it may be observed that there is hardly any dependence of $F_{2A}^{WI}(x,Q^2)$ on the
 different choice of nucleon PDF parameterizations. 
 When the results using the  full prescription vs spectral function (with MMHT PDFs at NLO including the HT effect) are compared, we find the effect of mesonic contributions are quite significant in the region of present kinematic interest, which increases with the increase in the mass number. Also to observe the effect of PDFs evolution of the nucleon, on the nuclear structure functions the results are presented at NNLO using the full model and compared these results with results obtained at 
 NLO with HT effect. It may be observed that the results of NLO+HT is the same($<1\%$) when compared with the results obtained at NNLO in the entire region of $x$. For the detailed discussion, 
 please see the Refs.~\cite{Zaidi:2019mfd, Zaidi:2019asc}. 

\begin{figure}
\begin{center} 
     \includegraphics[height=0.28\textheight,width=0.75\textwidth]{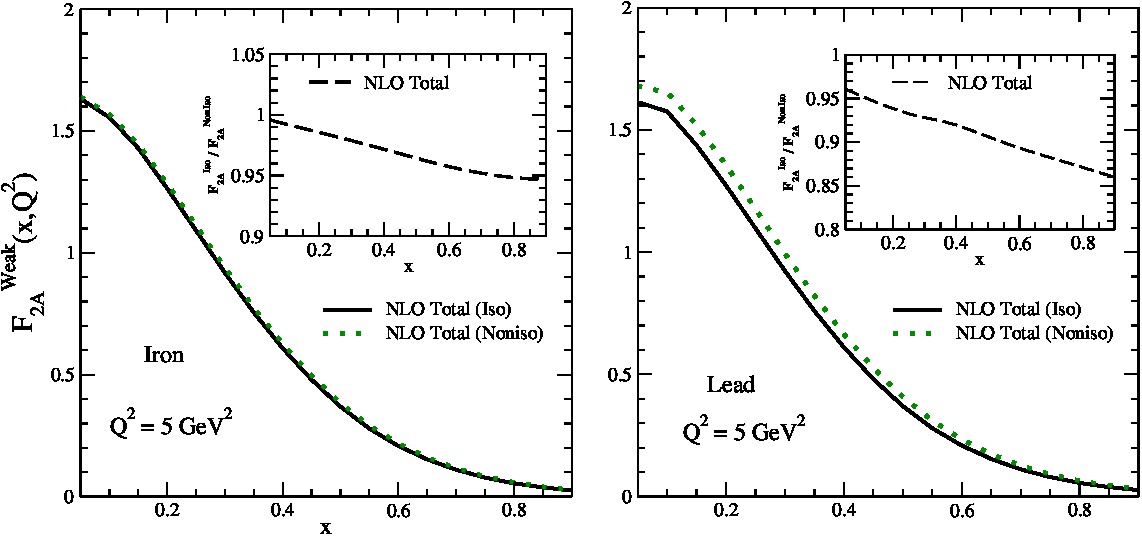}
 \caption{$F_{2A}^{WI}(x,Q^2)$ vs $x$ at fixed $Q^2=5~GeV^2$, in $^{56}$Fe and $^{208}$Pb
 with the full calculation at NLO treating $^{56}$Fe and $^{208}$Pb as isoscalar (solid line) and nonisoscalar (dotted line) targets.
%  Dotted line is the result of the full calculation at NLO treating $^{56}$Fe and $^{208}$Pb as nonisoscalar targets. 
These calculations are performed using CTEQ6.6~\cite{Nadolsky:2008zw} nucleon PDFs in the MS-bar scheme. 
 Inset in the figures depicts the ratio of nuclear structure functions $\frac{F_{2A}^{Iso}}{F_{2A}^{NonIso}}$ treating $^{56}$Fe and $^{208}$Pb as 
 isoscalar(Iso) and nonisoscalar(NonIso) nuclear targets using the full model. 
 }
 \label{figure6}
 \end{center}
\end{figure}

To study the effect of isoscalarity correction in nonisoscalar nuclear targets like $^{208}Pb$, 
 the Aligarh-Valencia model performs numerical calculations independently for isoscalar nuclear targets by normalizing the spectral function to 
 the number of nucleons($A$) using the nucleon density parameters and getting the correct binding energy (very close to the experimental values) of the nucleons in the nucleus which has been discussed in section-\ref{section3}. Similarly for the nonisoscalar nuclear targets, the spectral function is normalized to the proton number($Z$) using the proton density parameters, and the neutron numbers ($A-Z$) using the neutron density parameters. Fig.\ref{figure6} shows the isoscalarity vs nonisoscalarity effect,
 where the results are 
 presented at $Q^2=5~GeV^2$ for $F_{2A}^{WI}(x,Q^2)$ in $^{56}$Fe and $^{208}$Pb. 
%  The theoretical results for the isoscalarity corrections 
%  are also compared with the phenomenological results obtained by using Eq.(\ref{isocorr_em}) {\bf (not shown here)}. 
  In the inset of these figures, the isoscalarity effect has been explicitly shown by plotting the ratio
 $\frac{F_{2A}^{Iso}(x,Q^2)}{F_{2A}^{NonIso}(x,Q^2)}$ vs $x$ for the full theoretical model which deviates from unity in the entire range of $x$. This correction is $x$ as well as nuclear mass $A$ dependent, and becomes more pronounced with the increase in $x$ as well as with the increase in the nuclear mass number $A$.
 
       \begin{figure}
\begin{center}
\includegraphics[height=0.2\textheight,width=0.75\textwidth]{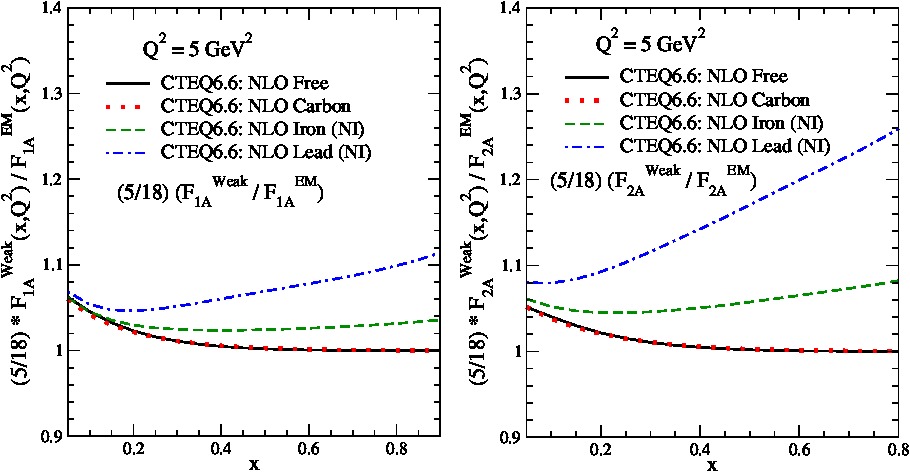}\\
\includegraphics[height=0.2\textheight,width=0.75\textwidth]{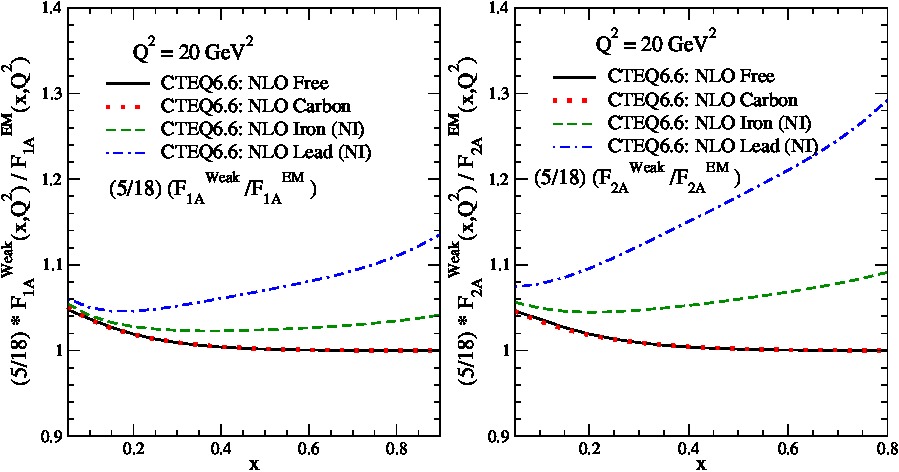}
\caption{Results for the ratio ${\it R^\prime}=\frac{\frac{5}{18} F_{i A}^{WI}(x,Q^2)}{F_{i A}^{EM}(x,Q^2)}; ~(i=1,2)$ are obtained
with the full model at NLO in $A=$ $^{12}$C, $^{56}$Fe and $^{208}$Pb at $Q^2=5~\mbox{and}~20$ GeV$^2$ by using the CTEQ6.6 nucleon PDFs in 
the MS-bar scheme~\cite{Nadolsky:2008zw}.
 The left figures are for $F_1(x,Q^2)$ and the right are for $F_2(x,Q^2)$.
The numerical results are obtained assuming $^{56}Fe$ and $^{208}Pb$ to be nonisoscalar target nuclei and are
 compared with the results obtained for the isoscalar free nucleon target.}
  \label{chapew_figure16}
  \end{center}
\end{figure}

In Fig.~\ref{chapew_figure16}, the variation of nuclear medium effects in the electromagnetic and weak interactions has been shown by using different nuclear targets.
%   To observe the variation of nuclear medium effects in the electromagnetic and weak interactions, the results are presented in Fig.~\ref{chapew_figure16}
%     for the ratio $R^\prime=\frac{\frac{5}{18} F_{iA}^{WI}(x,Q^2)}{F_{iA}^{EM}(x,Q^2)}$; ($i=1,2$) in 
%  various nuclei like 
%  $^{12}$C, $^{56}$Fe and $^{208}$Pb at $Q^2=5$ and $20$ GeV$^2$. 
%  These results are obtained assuming $^{56}Fe$ and $^{208}Pb$ to be nonisoscalar target nuclei and are
%  compared with the results obtained for the isoscalar free nucleon target. 
 It should be noticed from the figure that the ratio $R^\prime$ deviates from unity in the region of 
low $x$ even for the free nucleon case which implies the non-zero contribution from strange and charm quarks distributions. However, for $x\ge 0.4$, where 
the contribution of strange and charm quarks are almost negligible, the ratio approaches towards unity. Furthermore, if one assumes $s=\bar s$ and $c=\bar c$ then
in the region of small $x$, this ratio would be unity for an isoscalar nucleon target following the $\left(\frac{5}{18}\right)^{th}$-sum rule.
 One may also observe that for heavier nuclear targets like
 $^{56}Fe$ and $^{208}Pb$, this deviation becomes more pronounced. 
  This shows that the difference in charm and strange quark distributions could be
  significant in heavy nuclei. One may also notice that the isoscalarity corrections are different in $F_{1A}^{EM}(x,Q^2)$
  than in $F_{2A}^{EM}(x,Q^2)$ although the difference is small. 
%   For example, for $Q^2=5~GeV^2$ in $^{208}Pb$ it is found to be $5\%$ at $x=0.1$, $\approx 8\%$ at $x=0.6$ and 
%   $10\%$ at $x=0.8$ for $F_{1A}(x,Q^2)$, while in the case of $F_{2A}(x,Q^2)$ it becomes $8\%$, $12\%$ and $13\%$ at $x=0.1,~0.6$ and $0.8$, respectively.
  \begin{figure}
  \begin{center}
 \includegraphics[height=0.28\textheight, width =0.75\textwidth]{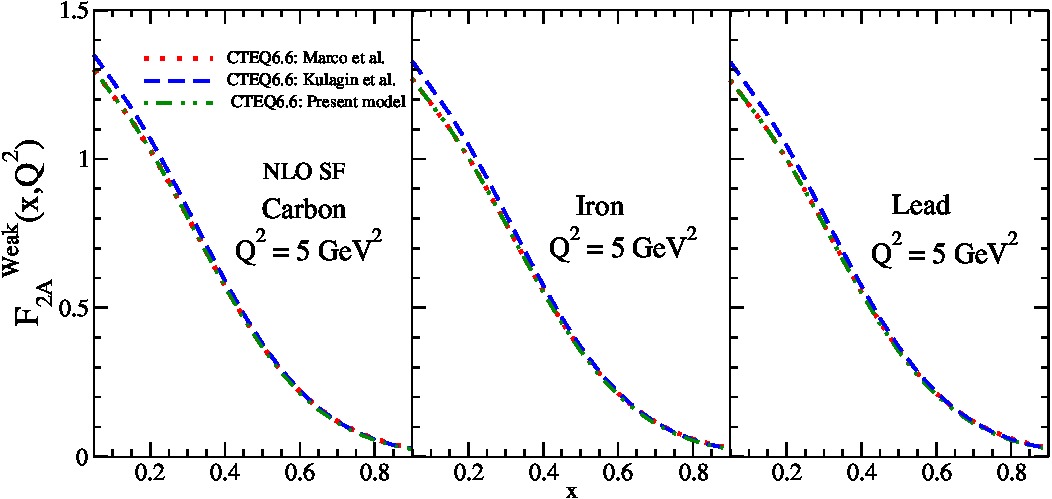}
 \end{center}
 \caption{Results for $F_{2A}^{WI}(x,Q^2)\;(A=$ $^{12}$C, $^{56}$Fe and $^{208}$Pb) vs $x$ are shown using models given by 
 Marco et al.~\cite{Marco:1995vb}(dotted line) in the Bjorken limit, Kulagin et al.~\cite{Kulagin:2007ju}(dashed line)
 and the present model (dashed-double dotted line) in the non-Bjorken limit to observe the model dependence of the spectral function at $Q^2=5~GeV^2$. Numerical results are evaluated
 at NLO by using CTEQ6.6~\cite{Nadolsky:2008zw} nucleon PDFs in the MS-bar scheme. All the nuclear targets are treated to be isoscalar here.}
 \label{figure7}
\end{figure}
    \begin{figure}
\begin{center} 
     \includegraphics[height=0.3\textheight,width=0.75\textwidth]{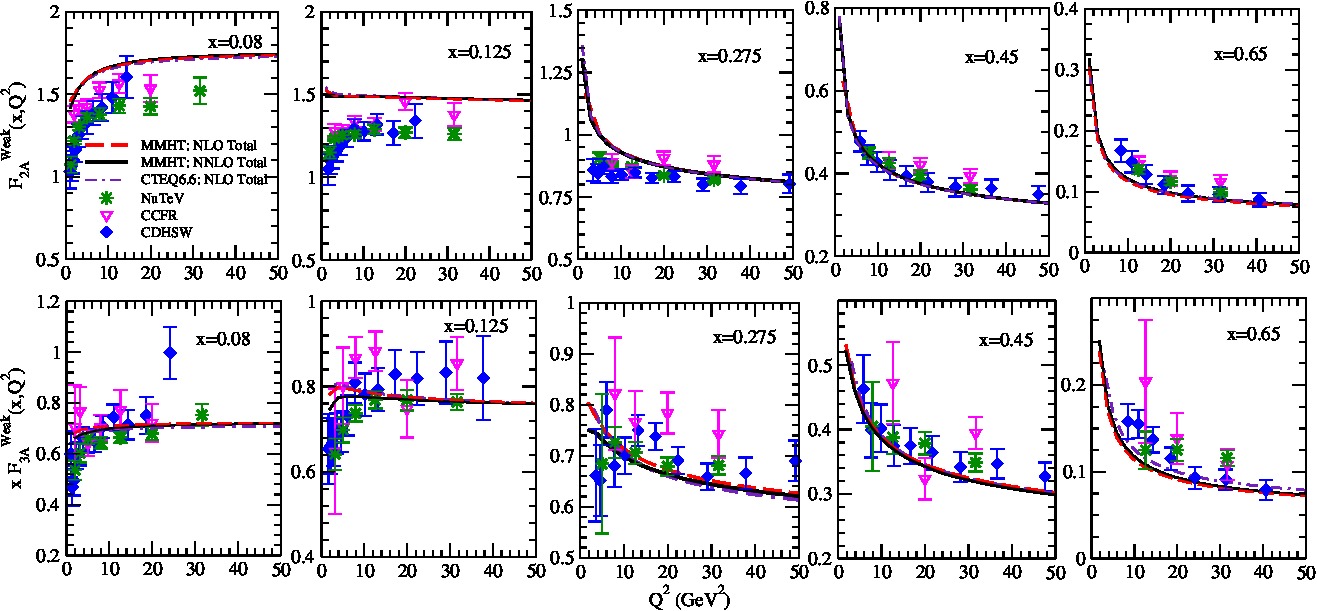}
 \caption{Results are presented for $F_{2A}^{WI}(x,Q^2)$ and $x F_{3A}^{WI}(x,Q^2)$ vs $Q^2$ in $^{56}$Fe using the full model at different values of $x$. The results are obtained by using 
 CTEQ6.6 nucleon PDFs at NLO in the MS-bar scheme (dotted line), MMHT at NLO(dashed line) and NNLO(solid line). 
 The experimental points are the data from CDHSW~\cite{Berge:1989hr} (solid diamond), CCFR~\cite{Oltman:1992pq} (empty triangle)  and NuTeV~\cite{Tzanov:2005kr} (star symbol) experiments.
  In the present case iron is treated as isoscalar nuclear target.}
    \label{figure8}
  \end{center}
 \end{figure}
%  The model dependence of spectral function has also been studied by Ruiz Simo and M.~J.~Vicente Vacas~\cite{RuizSimo:2008jk}.
 
%  There are various spectral functions available in the literature like that of Fernandez de Cordoba et al.~\cite{FernandezdeCordoba:1991wf}, Kulagin et al.~\cite{Kulagin:2007ju}, 
%  Ankowski et al.\cite{Ankowski:2007uy}, etc.
 In Fig.\ref{figure7}, the model dependence of the spectral function has been studied by using the different 
 spectral functions~\cite{Marco:1995vb,Kulagin:2007ju, FernandezdeCordoba:1991wf} available in the literature.
%  and the results of weak nuclear structure function
%  $F_{2A}^{WI}(x,Q^2)$; ($A=^{12}$C, $^{56}$Fe, $^{208}$Pb) vs $x$ at $Q^2=5~GeV^2$ has been 
%    obtained with only the spectral function and are presented (i) in the non-Bjorken limit and (ii) in the Bjorken limit~\cite{FernandezdeCordoba:1991wf,Marco:1995vb}, 
%    as well as using the spectral function of
%  Kulagin et al.~\cite{Kulagin:2007ju}. 
%   The numerical calculations have been performed at NLO by using the CTEQ6.6 nucleonic 
%  PDFs parameterization~\cite{Nadolsky:2008zw} in the MS-bar scheme. All the nuclear targets are treated to be isoscalar here.
  From the figure, it may be observed that the difference in the results obtained in the low $x$ and low $Q^2$ region vs Bjorken limit, is within 1$\%$ of each other.
  The results obtained by using the spectral function of Kulagin et al.~\cite{Kulagin:2007ju} show small difference even at low 
 $x$ and low $Q^2$ for the nuclei under consideration and the difference gradually becomes smaller with the increase in $x$ and $Q^2$. {\it Hence, it may be concluded that the 
 nuclear structure functions show  very little dependence on the choice of spectral function.}
%  In Ref.~\cite{RuizSimo:2008jk}, similar observations have been made for $^{40}Ca$ at $Q^2=20~GeV^2$. Furthermore, these authors have also studied 
%  the nuclear dependence by taking the spectral function of Ankowski et al.\cite{Ankowski:2007uy} and found very little dependence
%  on the choice of spectral 
%  function~\cite{RuizSimo:2008jk}.

 For the $\nu_l/\bar\nu_l$ scattering cross sections and structure functions high statistics measurements have been made by
 CCFR~\cite{Oltman:1992pq}, CDHSW~\cite{Berge:1989hr}, and NuTeV~\cite{Tzanov:2005kr} experiments by using iron as nuclear target. Experimentally, 
 the extraction of structure functions are done by using the differential scattering cross sections measurements. These experiments have 
 been performed in a wide range of $\nu_l/\bar\nu_l$ energies $20\le E_\nu \le 350~GeV$. Using Aligarh-Valencia formalism, Haider et al.~\cite{Haider:2011qs}
 have studied nuclear modifications for the $\nu_l/\bar\nu_l$ induced processes on iron target and compared their results with the available
 experimental data~\cite{Berge:1989hr, Oltman:1992pq, Tzanov:2005kr}.
  The theoretical results in Fig.\ref{figure8}, obtained by using the full model are compared with the available experimental data~\cite{Berge:1989hr, Oltman:1992pq, Tzanov:2005kr}.
%  The numerical calculations have been performed by using the nucleonic PDFs parameterizations of 
%  CTEQ6.6~\cite{Nadolsky:2008zw} in the MS-bar scheme at NLO and MMHT~\cite{Harland-Lang:2014zoa} at NLO as well as at NNLO. 
     \begin{figure}
\begin{center} 
     \includegraphics[height=0.3\textheight,width=0.75\textwidth]{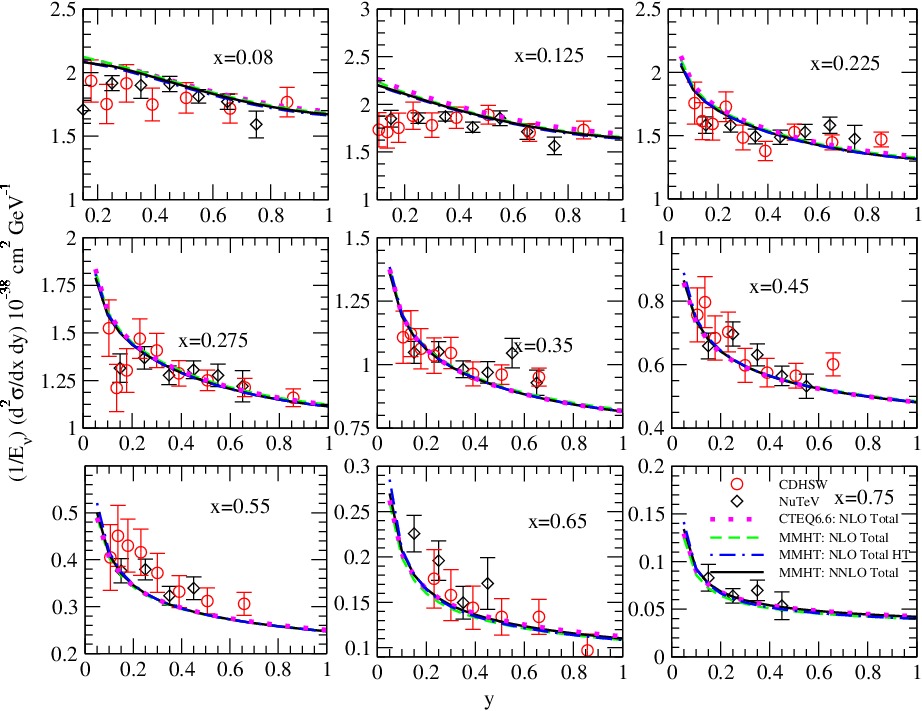}
 \caption{Results of the differential scattering cross section $\frac{d^2\sigma}{dxdy}$ vs $y$, at different $x$ for $\nu_\mu$ induced reaction on iron target at $E_{\nu_\mu}=65~GeV$ are shown.
 The results are obtained by using {\bf (i)}
 CTEQ6.6~\cite{Nadolsky:2008zw} nucleon PDFs at NLO in the MS-bar scheme (dotted line), {\bf (ii)} MMHT nucleon 
 PDFs~\cite{Harland-Lang:2014zoa} at NLO without (dashed line) and with the HT effect (renormalon approach: dashed-dotted line) as well as at NNLO (solid line).
 The experimental points are the data from CDHSW~\cite{Berge:1989hr} and NuTeV~\cite{Tzanov:2005kr} experiments. Here iron is treated as isoscalar target.}
 \label{figure44}
 \end{center}
 \end{figure}
%  The results are 
%  presented at the different values of $x$ and are compared with the available experimental data~\cite{Tzanov:2005kr,Oltman:1992pq,Berge:1989hr}.
These results differ from the experimental data in the region of low $x$ and low $Q^2$, however, with the 
 increase in $x$ and $Q^2$ they are found to be in reasonably good agreement.
 It is important to point out that the additional
 uncertainty(due to normalization) of $\pm 2.1\%$ has not been included in the NuTeV analysis~\cite{Tzanov:2005kr} and the experimental results
 also differ among themselves. 
 
 Moreover, the results obtained by using the CTEQ6.6~\cite{Nadolsky:2008zw} (in the MS-bar scheme) 
 and MMHT~\cite{Harland-Lang:2014zoa} PDFs parameterizations are consistent. 
 The numerical results evaluated at NNLO using MMHT nucleon PDFs parameterization~\cite{Harland-Lang:2014zoa} for $F_{2A}^{WI}(x,Q^2)$ show 
 a reasonably good agreement with the results evaluated at NLO while the results for $xF_{3A}^{WI}(x,Q^2)$ differ in the region of low $x$ and low $Q^2$. For 
 example, at $x=0.275(0.45)$, the difference is found to be $6\%(\approx 2\%)$ for $Q^2=1.8~GeV^2$ and $3\%(1\%)$ for $Q^2=5.8~GeV^2$. The detailed discussion of the nuclear medium effects for a wide range of $x$ is available
 in Ref.~\cite{Haider:2011qs}.
 
In Fig.\ref{figure44}, the results are presented for $\frac{d^2\sigma}{dxdy}$ vs $y$ at $E_{\nu_\mu}=65~GeV$ and keeping $Q^2>1~GeV^2$ with the full model. 
 The numerical calculations have been performed by using the 
 CTEQ6.6~\cite{Nadolsky:2008zw} nucleon PDFs in the MS-bar scheme at NLO as well as MMHT~\cite{Harland-Lang:2014zoa} nucleon PDFs at NLO without and with the HT
 effect (renormalon approach~\cite{Dasgupta:1996hh, Stein:1998wr}), and at NNLO. One may notice that in the present kinematical region, the  
 numerical results obtained for all the cases shown in the figure seems to be in agreement within a percent. This was expected because the perturbative QCD corrections 
 have inverse power dependence on $Q^2$. The experimental data show 
 a good agreement with the numerical results except in the region of low $x~(\le 0.275)$ and low $y~(\le 0.2)$, where theoretical results overestimate the experimental data.
 
 In section-\ref{comparison}, we also present the theoretical results using Aligarh-Valencia model and the phenomenological results of nCTEQnu for the differential cross section (i.e. $\frac{d^2\sigma_A}{dxdy}$) vs $y$ for different values of x for the incoming beam of energy E = 35 GeV for $\nu$ and $\nub$ scattering on Fe and Pb nuclear targets and compare them with the experimental results from NuTeV in Fe and CHORUS in Pb. These theoretical and the phenomenological 
 results are also compared with the experimental results from NuTeV and CDHSW in Fe at $E_{\nu, \nub}$=65 GeV and CHORUS in Pb at $E_{\nu, \nub}$=55 GeV.
 
 In section-\ref{prediction}, we have given predictions of the theoretical results using Aligarh-Valencia model and the phenomenological results of nCTEQnu for the differential cross section vs y in Fe, Pb and Ar at $E_{\nu, \nub}$=6.25 GeV as well as at $E_{\nu, \nub}$=2.25 GeV in Ar.
  These predictions may be useful in the analysis of MINERvA experiment being performed using Fe and Pb nuclear targets as well as the proposed DUNE experiment using liquid argon TPC.

\section{$\nu_l/\bar\nu_l$-Nucleus Scattering:  Shallow Inelastic Scattering Phenomenology} 
\label{section4}
%\section{Phenomenology and Experimental Results: Shallow Inelastic Scattering}
%\section{Shallow-Inelastic Scattering: Higher-W Resonances, Quark-Hadron Duality and Non-Perturbative QCD}
\label{Sec-SIS&Duality}
% 
%\subsection{Introduction}

Above neutrino quasi-elastic (QE) scattering in effective hadronic mass (W) comes the resonance region (RES) that starts with the $\Delta$ resonance followed by increasingly higher mass resonant states. These resonances sit atop a continuum of non-resonant $\pi$ production that starts at W = M + $m_{\pi}$. This resonant plus non-resonant $\pi$ production region transitions into the deep-inelastic scattering (DIS) region, where interactions occur on quarks, at a border kinematically defined for most experiments as  W $\ge$ 2.0 GeV and $Q^2 \ge 1 ~GeV^2$.  
The non-resonant pion production under all resonances is the very intriguing kinematic region  referred to technically as the shallow-inelastic scattering (SIS) region.  
%Although the production of exclusive resonances in the SIS region is an important topic, it is not the covered explicitly in this study.  
However, since it is not possible to experimentally distinguish resonant from non-resonant pion production or the interference between them, for this review SIS is practically defined as the sum of pion production processes contributing to inclusive scattering with W $\le$ 2.0 GeV.  Subsequent investigations are made into how models distinguish resonant from non-resonant production. 
%which, in turn, is considered in terms of quark-hadron duality and non-perturbative QCD effects.  
In particular the portion of SIS above the well-studied $\Delta$ has been minimally studied both experimentally and theoretically with neutrino scattering and is a very practical challenge which must be faced in all MC event generators. 

The challenge of this higher W SIS region theoretically was recently summarized by Nakamura~\cite{Andreopoulos:2019gvw} when discussing the dynamical coupled-channel approach to the resonances beyond the $\Delta$.  In the $\Delta$(1232) region: there is only a single resonance that dominates $\pi$ production; the non-resonant contribution is much smaller than the $\Delta$ and is well controlled by chiral perturbation theory; and the only decay channel that must be considered is $\pi$N.  For the region beyond the $\Delta$(1232) up to W $\lessapprox$ 2 GeV: no single resonance dominates and several comparable resonances overlap; the non-resonant contribution is comparable to the resonant contribution; the $\pi$N and $\pi\pi$N are comparable and strongly coupled as well as higher mass meson-baryon channels.

In addition to the individual pion production model approach that, eventually, must cover both resonance and non-resonance single- and multiple-pion channels, there is the intriguing alternative treatment of the SIS region by the GiBUU group based on nuclear transport theory~\cite{Leitner:2009ke}. While the current MC simulation programs treat the initial interaction and the subsequent final state interaction of produced hadrons independently, the GiBUU framework attempts to model the full space-time evolution of particles from the initial through final state interactions and emphasize that the initial and final state interactions should not be treated independently.  Using this framework they have predicted both resonant and non-resonant pion production within the SIS region~\cite{Mosel:2017nzk}.  The results of the GiBUU model will be presented in the discussion of duality.

An initial anomaly to note is that in some current Monte Carlo (MC) event simulators/generators ``DIS'' is defined as ``anything but QE and RES'', instead of the usually expressed kinematic condition on the effective hadronic mass such as $W>2$~GeV with $Q^2>1$~GeV$^2$.
Notice moreover that RES in these simulators is limited to 1$\pi$ production.  This suggests that such a MC generator definition of ``DIS'' must contain all non-resonant pion production as well as resonant multi-$\pi$ production.  This MC "DIS definition then includes a contribution from the kinematical region $Q^2<1$~GeV$^2$, which is certainly outside of the applicability of the genuine DIS formalism and consequently perturbative QCD.  Thus the MC definition of DIS contains also part of what we define as the SIS region. For this review "DIS" refers to the original kinematical definition of DIS.

This higher-W SIS region between the $\Delta$ resonance and DIS has been quite intensively  studied experimentally in electron/muon-nucleon (e/$\mu$-N) interactions and somewhat less thoroughly in e/$\mu$-nucleus (e/$\mu$-A) scattering. The studies of e/$\mu$-N interactions in this kinematic region have been used to test the hypothesis of quark-hadron duality (hereafter "duality"). Duality, as we shall see, relates the average of inclusive production cross sections in this SIS region to extrapolated results from the better known DIS region.  To further define the concept of duality, consider that perturbative QCD is well defined and calculable in terms of asymptotically free quarks and gluons, yet the process of confinement ensures that it is hadrons, pions and protons, that are observed.    One speaks the language of quarks/gluons in the DIS region and, as W decreases, transitions to speak the language of hadrons in the SIS region that includes both resonant and non-resonant pion production.  Duality can then be considered as a conceptual experimental bridge between free and confined partons.   It is important to note that the understanding of this SIS region is important for long-baseline oscillation experiments.  As has been mentioned, in the future DUNE experiment~\cite{Abi:2020evt}, more than 50\% of the interactions will be in these SIS and DIS regions with W above the mass of the $\Delta$ resonance.  

\subsection{Quark-Hadron Duality}
\label{Subsec-Duality}

Historically in the 1960’s the concept of what was to become "duality" began with the total pion-proton cross sections being compared with Regge fits to higher energy data.  It was concluded that low-energy hadronic cross sections on average could be described by the high-energy behavior.  In the 1970's Poggio, Quinn and Weinberg \cite{Poggio:1975af} suggested that higher energy inclusive hadronic cross sections, appropriately averaged over an energy range, should approximately coincide with the cross sections calculated using quark-gluon perturbation theory.  This directly implied that the physics of quarks and gluons could describe the physics of hadrons.  

Finally, also in the 1970's, Bloom and Gilman~\cite{Bloom:1970xb} defined duality by comparing the structure functions obtained from inclusive electron-nucleon DIS scattering with resonance production in similar experiments and the observation that the average over resonances is approximately equal to the leading twist (see~\ref{TMC}) contribution measured in the DIS region. This seems to be valid in each resonance region individually as well as in the entire resonance region when the structure functions are summed over higher resonances. That is the DIS scaling curve extrapolated down into the resonance region passes through the average of the "peaks and valleys" of the resonant structure.  In this picture, the resonances can then be considered as a continuing part of the behavior observed in DIS.  This would suggest there is a connection between the behavior of resonances and QCD, perhaps even a common origin in terms of a point-like structure for both resonance and DIS interactions.  Along this line it has been conjectured that there may exist two component duality where the resonance contribution and background contribution to the structure functions in the resonance excitation region corresponds respectively to the valence quarks, and the sea quarks contribution in structure functions in the DIS 
region~~\cite{Melnitchouk:2005zr}. However, these observations are to be verified by model calculations as well as by the experimental data when they become available with higher precision.  Currently, the observation of duality in charged-lepton scattering has the following main features~\cite{Lalakulich:2009zza}:
 \begin{itemize}
 \item the resonance region data oscillate around the scaling DIS curve  
  \item the resonance data are on an average equivalent to the DIS curve  
  \item the resonance region data moves towards the DIS curve with the increase in $Q^2$.
 \end{itemize}

As more data with better precision become available on inclusive lepton scattering from nucleons and nuclei a verification of QH duality with sufficient accuracy will provide a way to describe lepton-nucleon and lepton-nucleus scattering over the entire SIS region.  Significantly, if duality does hold for neutrino nucleon interactions, it would be possible to extrapolate the better-known neutrino DIS structure into the SIS region and give an indication of how well current event simulators are modeling the SIS region.  If the application of duality to our event generators can help us with this understanding it should be explored.

\paragraph{Duality and Charged-lepton Scattering}

\begin{figure}[h]
\begin{center}
\includegraphics[width=0.55\textwidth]{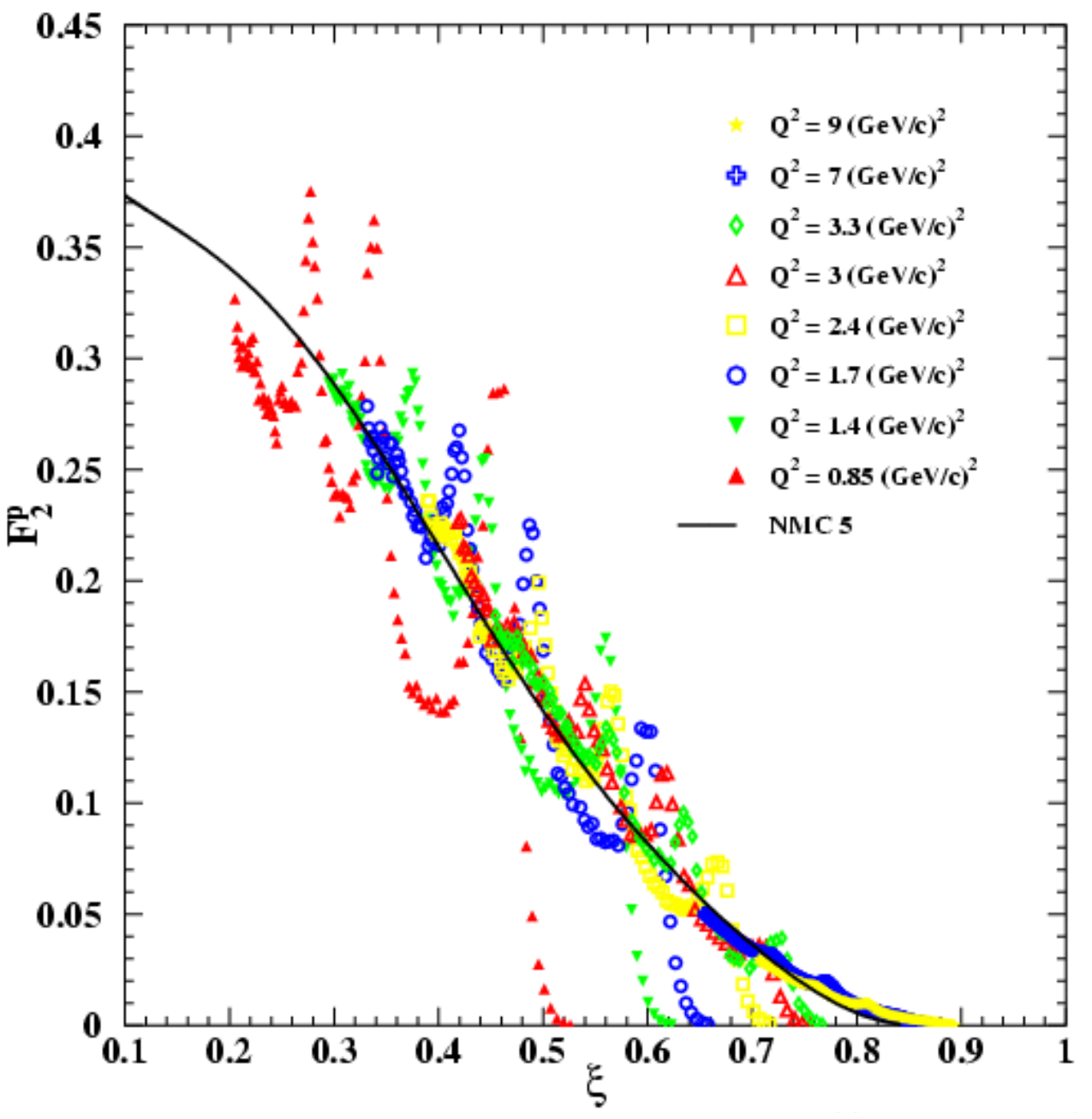}
\caption{Figure from~\cite{Niculescu:2000tk}. Comparison of $F_2^p$ from the series of resonances measured by E94-110 vs the Nachtmann variable $\xi$ (see below) at the indicated $Q^2$ compared to the extrapolated DIS measurement from the NMC collaboration at 5 $GeV^2$ }
\label{eN_duality94110}
\end{center}
\end{figure}

By the early 2000's there was considerable accumulation of charged lepton DIS studies at multiple laboratories with nucleon structure functions well measured over a broad range in $x,Q^2$, (x$_{Bjorken} \equiv x$).  Many experimental tests had supported the success of QCD and a new examination of duality with Jefferson Lab resonant production experiments was begun.  An early Jefferson Lab measurement (E94-110)  ~\cite{Malace:2009kw} showed that global duality was clearly observed for $Q^2 \ge 0.5 ~GeV^2$, as can be seen in Fig.~\ref{eN_duality94110}, with resonances following the extrapolated DIS curve.  

The experimental and theoretical study of duality proceeded relatively smoothly for e-N and even for e-A interactions and there are now \emph{visual} suggestions that duality holds for $F_2^{n,p,N}, F_1^p, F_L^p$, and $F_2^{D,C,Fe,Au}$. 
 
%Duality has also been observed to work well for the "EMC" effect (to be described in section ....) for W $ > W_\Delta$.  This also suggests an interesting observation that the EMC effect persists well below the DIS threshold that will be examined later. ADD FIGURE.  

However, with the much more accurate Jefferson Lab data, it was thought that there should be an improved method to test duality precisely.  A possible solution is to quantify the degree to which duality is satisfied by defining the ratio of integrals of structure functions, over the same $\xi$ interval, from the resonance (RES) region and DIS region.  To keep the same $\xi$ interval in the higher W DIS region compared to the lower W RES region requires a different $Q^2$ for the RES and DIS regions, thus the indexing of $Q^2$ in the ratios.  This method tests local duality within the integrals limits.  For perfect local quark-hadron duality the value of the ratio would be 1.0.
\begin{equation}
{\cal I_j} (Q^2_{RES}, Q^2_{DIS}) = \frac{\int_{\xi_{min}}^{\xi_{max}} d\xi F_j^{RES}(\xi, Q^2_{RES})}
{\int_{\xi_{min}}^{\xi_{max}} d\xi F_j^{DIS}(\xi, Q^2_{DIS})}\
\label{dratio}
\end{equation}

\hspace{150em} $\xi(x, Q^2) = \frac{2x}{1+\sqrt{1+4x^2M_N^2/Q^2}}$ 
\\
%For perfect local quark-hadron duality the value of I would be 1.0. 

\begin{figure}[h]
\begin{center}
\includegraphics[width=0.55\textwidth]{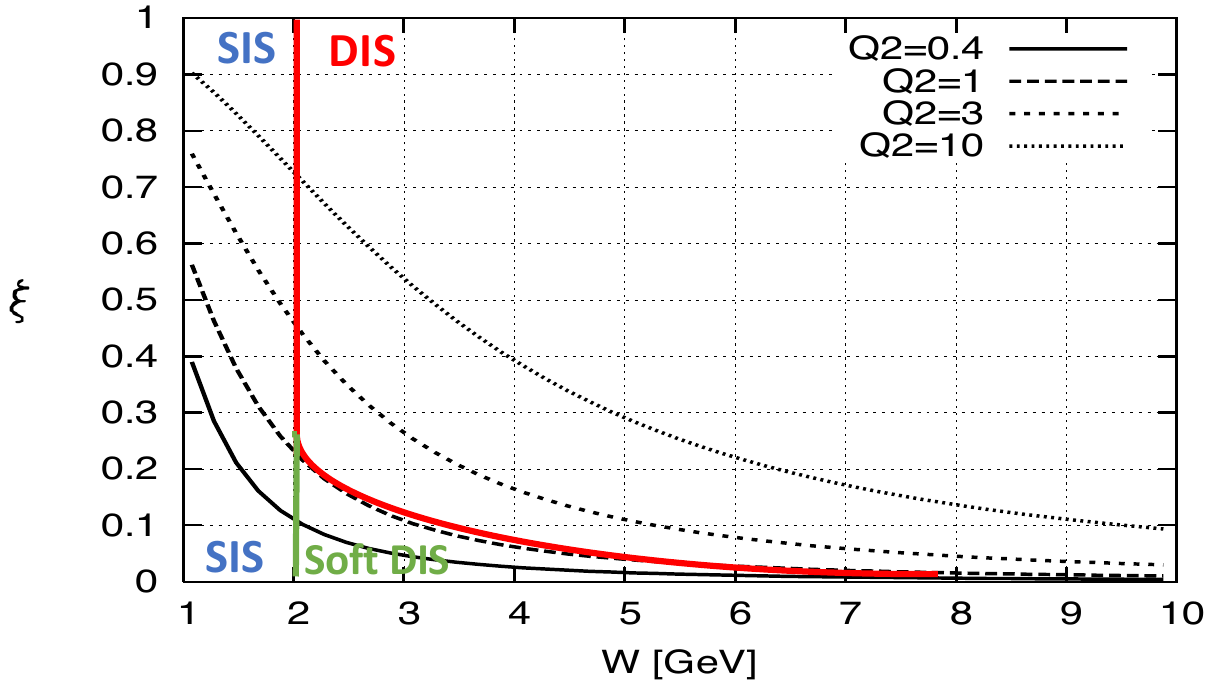}
\caption{Dependence of $\xi$ on W for specific values of $Q^2$. The interplay of these three variables with the kinematic regions of SIS, DIS and soft DIS are shown}.
\label{Fig-XivsW}
\end{center}
\end{figure}

The integrals use the Nachtmann variable: 
$\xi(x, Q^2)$ to account for target mass effects (TMC,section~\ref{TMC})   % = \frac{2x}{1+\sqrt{1+4x^2M^2/Q^2}}$  
and the integration over the resonance region is defined as typically $W_{min}=M_N+M_\pi$ and $W_{max} = 2.0$~GeV, which for a given $Q^2$ yields $\xi_{min}~and~\xi_{max}$. $F_j^{RES}$ is defined theoretically in section~\ref{T-SIS} and experimentally it is determined from the total inelastic cross section in the SIS region.
Fig.~\ref{Fig-XivsW} demonstrates the relationship between $\xi , Q^2$ and W corresponding to the SIS and DIS regions.  As an example, note that the $\xi$ range  of the open red triangles at 3 $GeV^2$ in Fig.~\ref{eN_duality94110}  cover the range $0.42 \le \xi \le 0.75$ that can now be directly related to a corresponding W range ($1.1 \lessapprox W(GeV) \lessapprox 2.0$),  the SIS region, using the 3 $GeV^2$ curve in this Fig.~\ref{Fig-XivsW}.

%In the case of DIS, the value of $Q^2_{DIS}$ is much larger and as a consequence the integral over $\xi$ runs over a quite different region in $W$.

\begin{figure}[h]
\begin{center}
\includegraphics[width=0.85\textwidth]{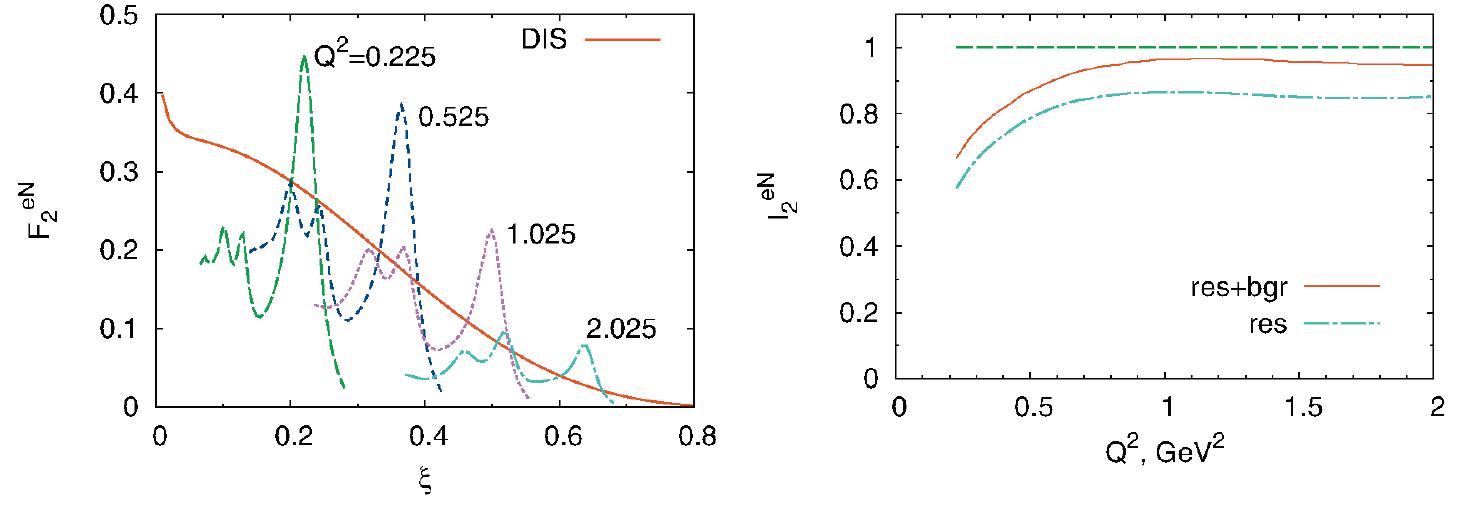}
\caption{Figure from ~\cite{Lalakulich:2009zza} illustrating duality for the isoscalar nucleon structure function calculated within the GiBUU electroproduction model. (Left) $F_2^{eN}$ as a function of $\xi$, for values of $Q^2$ indicated on the spectra, compared with the DIS QCD-fit results for $F_2^{eN}$ over the same $\xi$ range but at $Q^2$ = 10 GeV$^2$. (Right) Ratio $I_2^{eN}$ of the integrated F$_2$ in the resonance region to the integral over this DIS QCD fit to high $Q^2$ data. The $Q^2$ along the abscissa is the $Q^2$ involved in computing the limits $ \xi_{min} = \xi(W_1, Q^2)$ and $\xi_{max} = \xi(W_2, Q^2)$ of the integration of the numerator of $I_2^{eN}$. 
 }
\label{fig-I2eN}
\end{center}
\end{figure}

Using this new measure of agreement with quark-hadron duality for eN scattering a Giessen-Ghent collaboration~\cite{Lalakulich:2009zza} used the GiBUU model~\cite{Leitner:2009ke} that had been shown to reproduce the full range of Jlab e-nucleon resonance results covering the SIS kinematic region.  They found that, significantly, one {\emph must} include the non-resonant as well as the resonant contributions to the integral over the SIS region to improve the agreement with quark-hadron duality as shown in Fig.~\ref{fig-I2eN}.

\begin{figure}[h]
\begin{center}
\includegraphics[width=0.45\textwidth]{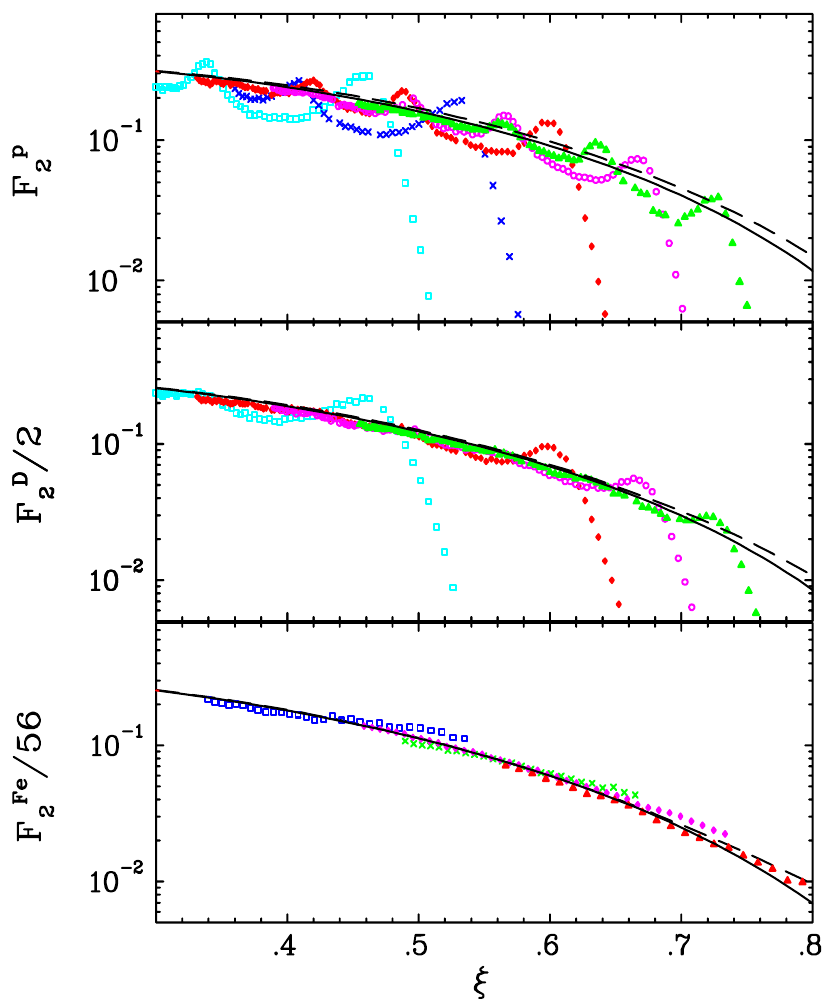}
\caption{Figure from~\cite{Arrington:2003nt}. The $F_2$ structure function per nucleon as a function of $\xi$ for (top to bottom) ep, eD and eFe. For the H and D data the quasielastic data has been removed while for the Fe data a cut of $W^2 \ge 1.2 GeV^2$ has been applied to remove the quasielastic peak.  The curves are the MRST and NMC DIS QCD fits with nuclear effect for e Fe applied}
\label{fig-Duality-FeData}
\end{center}
\end{figure}

When now considering nuclear as opposed to nucleon targets in e/$\mu$ scattering, the results of duality studies are not as straightforward.  When using nuclei, the Fermi motion of the bound nucleons within the nucleus serves to average (smear) the production of resonances over $\xi$ so that visual evaluation of duality should be more obvious in nuclei.  This concept is supported in Fig.~\ref{fig-Duality-FeData} that shows how the resonances that are clearly visible for e+p interactions are somewhat less defined in e+D interactions and essentially smoothed out completely for e+Fe interactions (where the curve has been modified for the EMC effect).  The curves for each are the MRST and NMC fits from the DIS region and, indeed, visual agreement with duality is apparent. The phenomenon of duality has now been observed in multiple  experiments on e-N and e-A scattering~\cite{ Malace:2014uea,Niculescu:2000tk, Mamyan:2012th, Malace:2009kw},\cite{ Niculescu:2015bla}-\cite{Niculescu:2015wka}.

\begin{figure}
\begin{picture}(500,150)(0,0) %   \graphpaper[20](0,0)(500,160)
 \put(50,0){\includegraphics[width=0.45\textwidth]{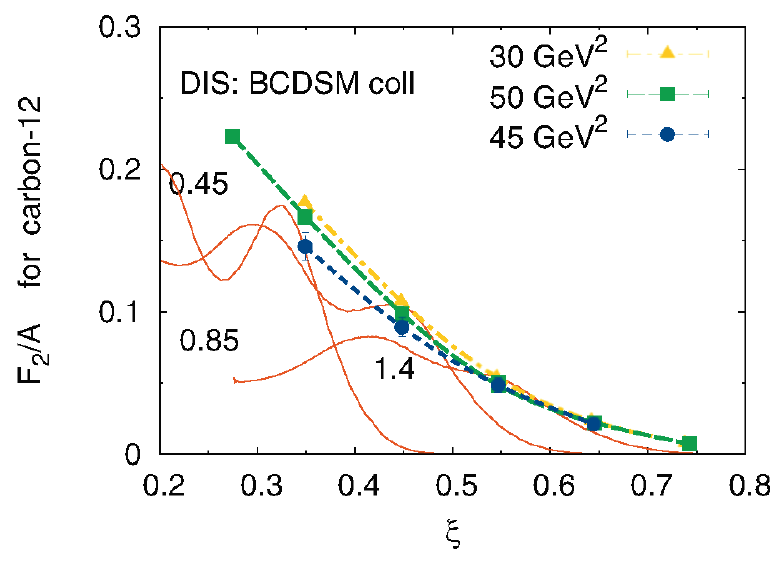}}
\put(250,0){\includegraphics[width=0.46\textwidth]{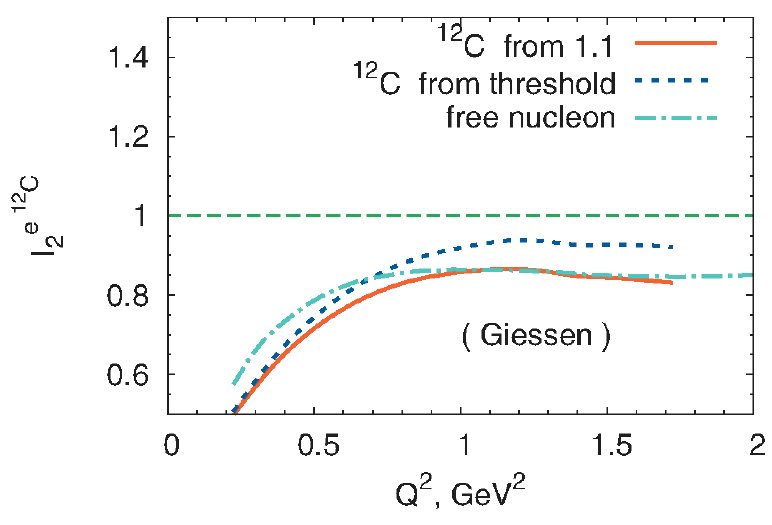}}
%\put(114,0){$(a)$} \put(374,0){$(b)$} 
\end{picture} 
\caption{ 
Figure from~\cite{Lalakulich:2009zza}: (Left) $F_2^{eC}$ as a function of $\xi$, for values of $Q^2$ indicated on the spectra, compared with the BCDMS data for $F_2^{eN}$ at the given $\xi$  at $Q^2$ = 30, 45 and 50 GeV$^2$. (Right) Ratio $I_2^{eC}$ of the integrated F$_2$ in the resonance region within the Giessen~\cite{Mosel:2017nzk} model to the integral over the DIS QCD fit to BCDMS high $Q^2$ data. The results are displayed for two choices of the lower limit for the integral of the numerator: W = 1.1 GeV (solid line) and "threshold" that takes into account the Fermi motion within the C nucleus (dotted line). For comparison, the ratio $I_2^{eN}$ for the free nucleon (dash-dotted line)is shown.
}
\label{fig:FI2eC}
\end{figure}

In contrast to these last comparisons that used {\em data} and seem to be clearly consistent with duality, Fig.~\ref{fig:FI2eC} (left) displays the result of using theoretical/phenomenological models, namely the Giessen (GiBUU) model for the resonance plus non-resonance contributions to the $F_2$ structure functions for a carbon target.  The model predictions for resonance production at several $Q^2$ values are compared to experimental data obtained by the BCDMS collaboration~\cite{Feltesse:1981zy} in muon-carbon scattering in the DIS region ($Q^2 \approx 30 - 50~GeV^2 $) that are shown as experimental points connected by smooth curves.  Due to the Fermi motion of the target nucleons, the peaks from the various resonance regions, which were clearly seen for the nucleon target, are hardly distinguishable for the carbon nucleus. The same effect was clearly demonstrated in Fig.~\ref{fig-Duality-FeData}.

The results for the ratio $I_2^{eC}$ are shown in Fig.~\ref{fig:FI2eC} (right). The curve for the isoscalar free-nucleon case, without including the non-resonant background, is the same as in Fig.~\ref{fig-I2eN}. One can see that the carbon curve obtained by integrating "from threshold" that takes into account Fermi motion of the nucleons within the carbon nucleus,  lies above the one obtained by integrating "from 1.1 GeV".  Recall that the flatter the curve is and the closer it gets to one, the higher the accuracy of local duality would be.  The GiBUU model in the SIS region emphasizes the importance of initial bound nucleon kinematics as well as non-resonant pion production being included in the calculations.
 
\begin{figure}[h]
\begin{center}
\includegraphics[width=0.55\textwidth]{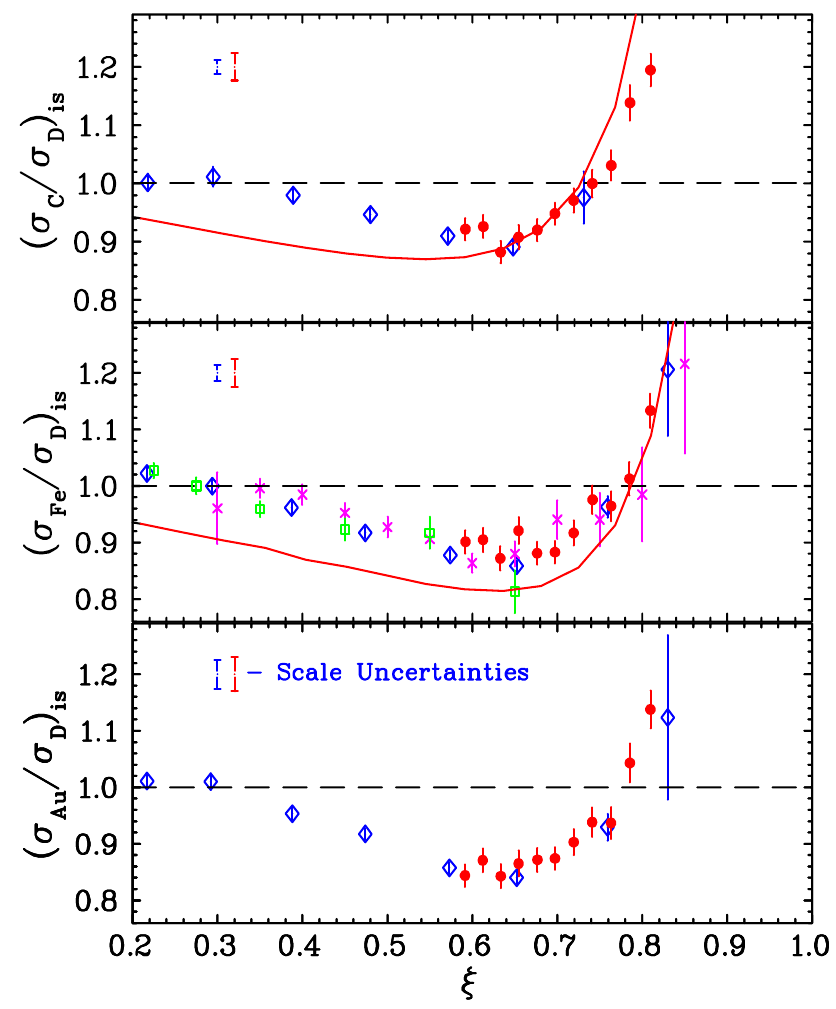}
\caption{Figure from~\cite{Arrington:2003nt} demonstrating the EMC effect in the resonance region.  The solid circles are Jefferson Lab data taken in the resonance region ($1.2 \le W^2 \le 3.0 GeV^2 ~and ~Q^2 = 4 GeV^2$ ) while all other data points are from DIS experiments. The red curve is a prediction of the EMC effect from reference~\cite{Gross:1991pi}}
\label{fig-EMCReso}
\end{center}
\end{figure}

%Duality for the isoscalar nucleon Fj"^^ structure function calculated within GiBUU model. (Left) F2^ as a function of ^, for Q = 0.225,0.525,1.025 and 2.025 GeV (indicated on the spectra), compared with the leading twist parameterizations at Q^ = 10 GeV . (Right) Ratio if^ of the integrated F2^ in the resonance region to the leading twist functions.
A further rather surprising and significant indication of duality in eA scattering can be found in~\cite{Arrington:2003nt} when discussing duality and the "EMC effect".  The EMC effect, to be covered in the next section, was previously thought to be a phenomena restricted to purely DIS kinematics.  However as shown in Fig.~\ref{fig-EMCReso} data covering the x-range of  the EMC effect measured in the resonance region intermixes seamlessly with EMC effect data taken in the DIS region. This intriguing result is further demonstration of quark-hadron duality with the nuclear structure functions in the resonance region exhibiting similar behavior as in the DIS region.  It can also be interpreted as a further suggestion that to have entered the DIS region is signified by $Q^2$ no matter what  W is involved.

\paragraph{Duality and Neutrino Scattering}

The experimental study of duality with neutrinos is much more restricted since the measurement of resonance production by $\nu$-N interactions is confined to rather low-statistics data obtained in hydrogen and deuterium bubble chamber experiments from the 70's and 80's.  Attempting to study duality with experimental $\nu$-A scattering is also limited due to very limited results above the $\Delta$ resonance in the SIS region. A recent NuSTEC workshop (\href{https://indico.cern.ch/event/727283/overview}{NuSTEC SIS/DIS Workshop})~\cite{Andreopoulos:2019gvw} concentrating on this SIS region with neutrino-nucleus interactions emphasized the considerable problems facing the neutrino community in this transition region.  Since there are no high-statistics experimental data available across the SIS region, $\nu$-N and $\nu$-A scattering duality studies are by necessity limited to theoretical models. Yet even the theoretical study of $\nu$-N/A duality is sparse with only  several full studies in the literature \cite{Lalakulich:2006yn,Lalakulich:2009zza},\cite{Sato:2003rq}-\cite{Gross:1991pi}.  This is troublesome since modern $\nu$ interaction simulation efforts can not then compare their results with duality predictions for $\nu$-N as they do for $\ell^\pm$-N interactions for confirmation.

\begin{figure*}
\begin{picture}(500,300)(0,0) %   \graphpaper[20](0,0)(500,160)
 \put(20,153){\includegraphics[width=0.3\textwidth]{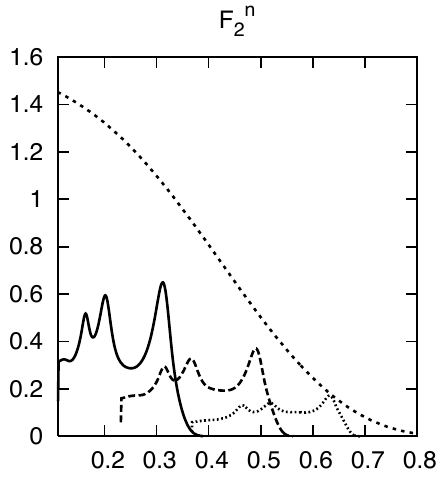}}
\put(170,150){\includegraphics[width=0.3\textwidth]{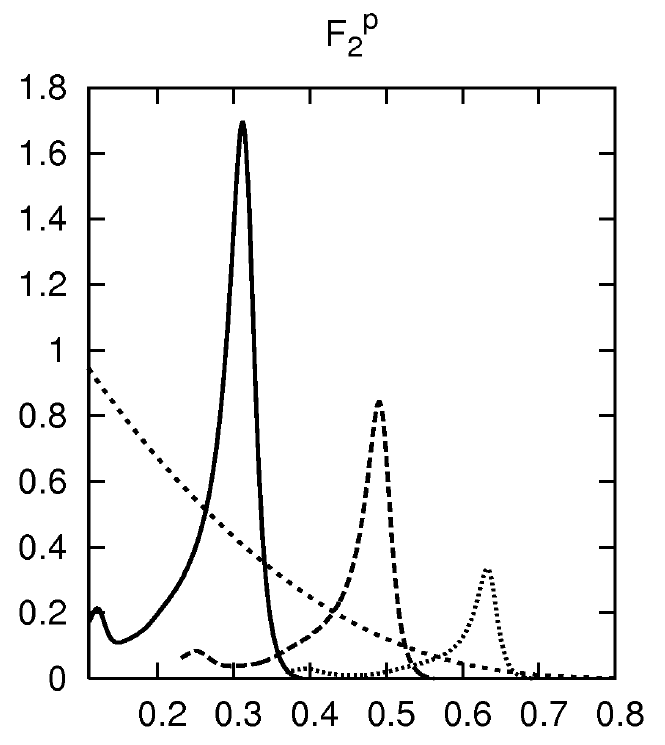}}
\put(320,146){\includegraphics[width=0.3\textwidth]{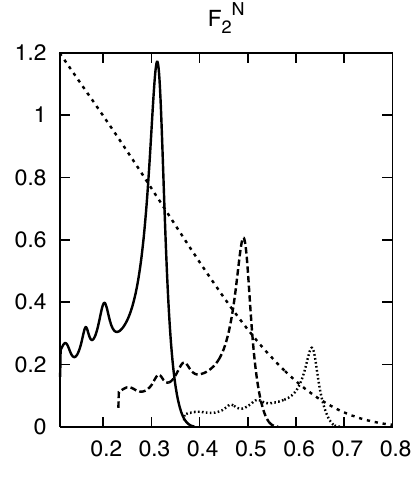}}
\put(90,140){$\xi$} \put(240,140){$\xi$}  \put(390,140){$\xi$}
\put(70,10){\includegraphics[width=0.85\textwidth]{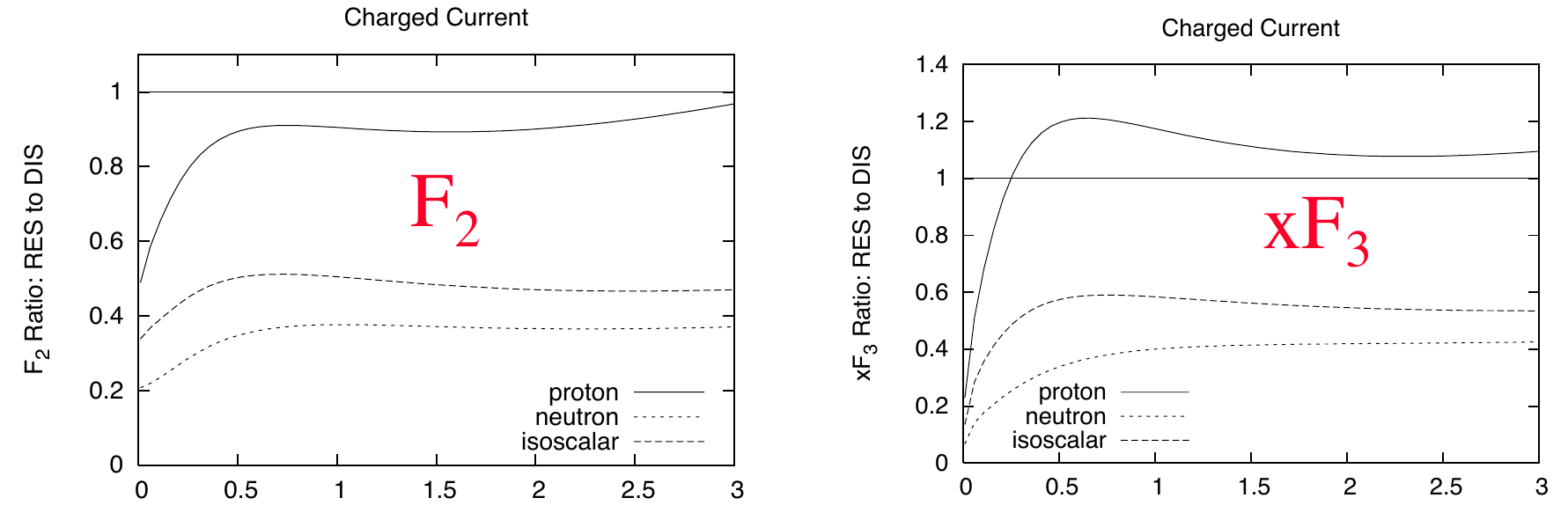}}
\put(165,0){$Q^2$} \put(360,0){$Q^2$}
\end{picture} 
\caption{Figure from~\cite{Graczyk:2005uv}: (upper) Comparison of the Rein-Sehgal F$_2$ structure functions vs $\xi$ for neutron, proton and the isoscalar nucleon target at Q$^2$ = 0.4, 1 and 2 GeV2 with the appropriate DIS scaling functions at Q$^2$ = 10 GeV$^2$. (lower) Ratio $I_{2,3}^{\nu}$ of the integrated F$_2$ (left) and xF$_3$ (right) in the resonance region to the integral over the DIS LO QCD fit at $Q^2$ = 10 GeV$^2$.
%On the left s $F_2^n$ vs $\xi$ in the middle $F_2^p$ vs $\xi$ and on the right $F_2^N$ vs $\xi$.
}
\label{n&p_duality}
\end{figure*}
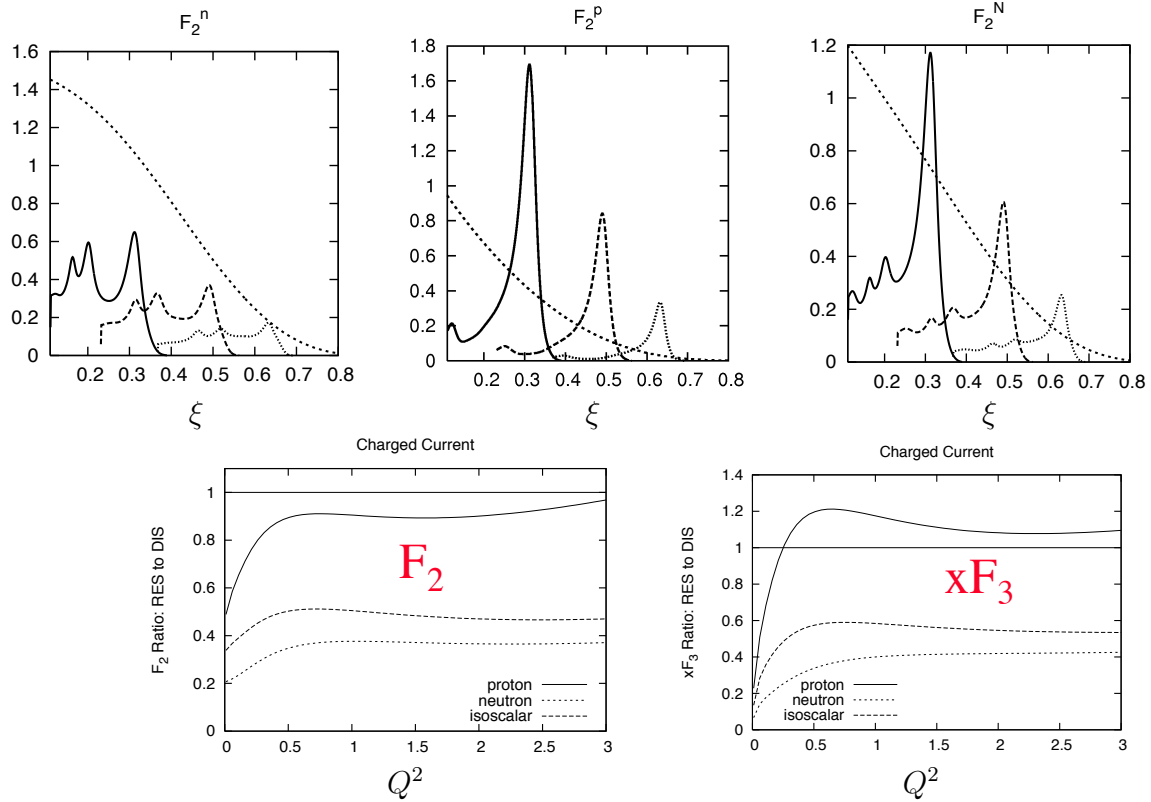

An early neutrino nucleon duality study~\cite{Graczyk:2005uv} by the Wroclaw group used the Rein-Sehgal model, which, as mentioned, is commonly used in current MC event generators for neutrino nucleon 1-$\pi$ resonance production. The study suggested that within the original R-S model for $\nu$-N 1-$\pi$ production across the SIS region, local duality is definitely not satisfied for neutron targets somewhat better for isoscalar targets and best, although not great, for proton target as shown in Fig.~\ref{n&p_duality}. This reflects the fact that {\it resonance production off a proton dominates the resonance region while in the DIS region $\nu$-n scattering dominates the DIS cross section}. Other analyses such as~\cite{Lalakulich:2009zza} observe a much smaller disagreement of duality for neutrons however it is still a disagreement.

%\put(0,0){\includegraphics[width=0.45\textwidth]{newfig9a_v3}}
%\put(260,0){\includegraphics[width=0.45\textwidth]{newfig9b_v3}}
%\put(114,0){$(a)$} \put(374,0){$(b)$} \end{picture} 

%The conclusions of this extended duality analysis for CC $\nu N$ interactions is that, as illustrated in Fig.~\ref{fig_Jan_RS-Int}: for the whole resonance region $(M + m_\pi \le W \le$ 2 GeV) and for $Q^2 \ge 0.5 GeV^2$  duality is satisfied {\em only for CC proton target reaction} and at best to the ∼ 20\% level;  there is also CC local duality in the vicinity of the $\Delta$ resonance for an isoscalar target.

%\begin{figure}[h]
%\begin{center}
%\includegraphics[width=0.55\textwidth]{Jan_RS-Int}
%\caption{Figure from~\cite{Graczyk:2005uv}: The integral ~\Eref{dratio} for CC interactions in the R-S model for resonances off proton (solid lines), neutron (dotted lines) and isoscalar target (dashed lines).}
%\label{fig_Jan_RS-Int}
%\end{center}
%\end{figure}

\begin{figure}[h]
\begin{center}
\includegraphics[width=0.85\textwidth]{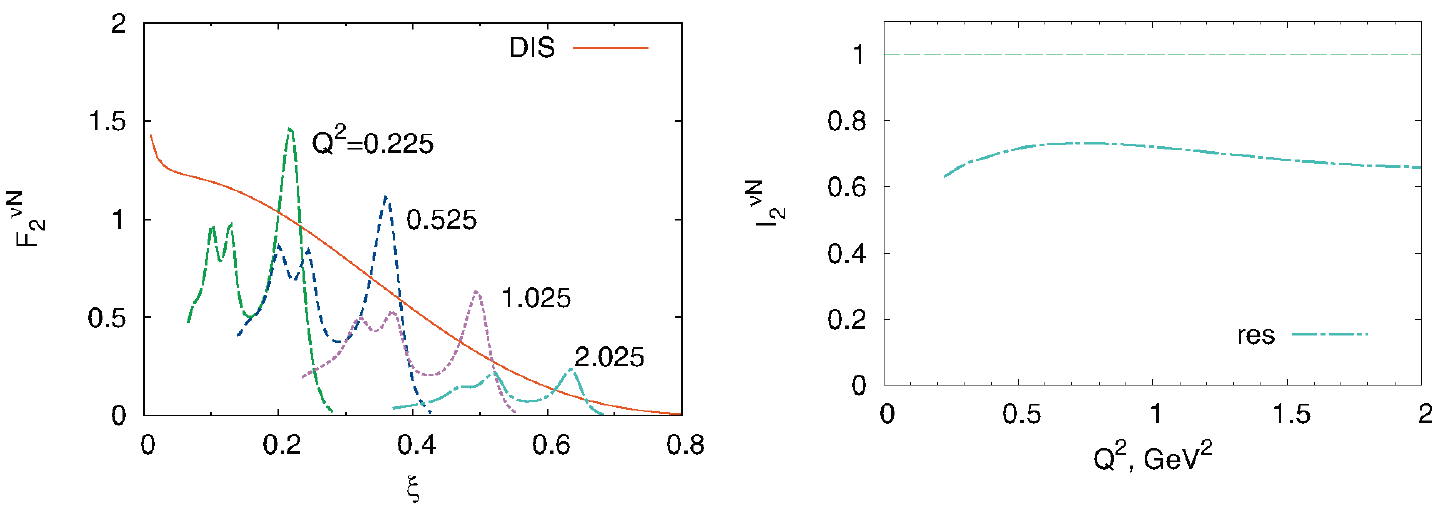}
\caption{Figure from~\cite{Lalakulich:2009zza}: Duality for the isoscalar nucleon F$_2^{\nu N}$ structure function calculated within the GiBUU model. (Left) F$_2^{\nu N}$ in the resonance region at different $Q^2$ indicated on the spectra as a function of $\xi$ compared with the leading twist parametrizations at Q$^2$ = 10 GeV$^2$ . (Right) From ~\Eref{dratio}  the ratio I$_2^{\nu N}$ of the integrated F$_2^{\nu N}$ in the resonance region to the DIS leading twist functions}
\label{fig_F2&I2nuN}
\end{center}
\end{figure}

Note, this group~\cite{Graczyk:2005uv} emphasized that the R-S model treatment of the non-resonant background, important for the quantitative evaluation of duality, is not very satisfactory. The significance of this non-resonant pion contribution has also been emphasized by the previously cited work of the Giessen-Ghent collaboration~\cite{Lalakulich:2009zza,Leitner:2009ke} that examined duality with $\nu$-N/A scattering as well as e-N/A scattering.  Using the GiBUU model in the resonance region (defined as $W<2$~GeV)
%with its emphasis on the importance of careful consideration of the non-resonant contribution to the pion production model (determined by fitting to the data) 
the value of $I_2^{\nu N}$ in ~\Eref{dratio} using the resonance contribution only (no non-resonance production included), even for the isoscalar nucleon, is  about 70\% as shown in Fig.~\ref{fig_F2&I2nuN} consistent with the importance of correctly accounting for the \emph{non-resonant} pion contribution.  This result can be directly compared to the earlier analysis of the Rein-Sehgal model that yielded a result for the integral of order 50\%  that did include the Rein-Sehgal estimate of the non-resonant pion contribution. 
%Again, in general for neutrinos, the resonance structure functions for proton are much larger than for neutron and in the case of DIS structure functions the situation is opposite.  
These results are, obviously,  model dependent but a general tendency is that {\em for larger W, DIS structure functions are much larger than the resonance contribution at lower W} and that the non-resonant contribution cannot be neglected. This general conclusion should be kept in mind for consideration of simulation programs treating the SIS region.

\begin{figure}[h]
\begin{center}
\includegraphics[width=0.85\textwidth]{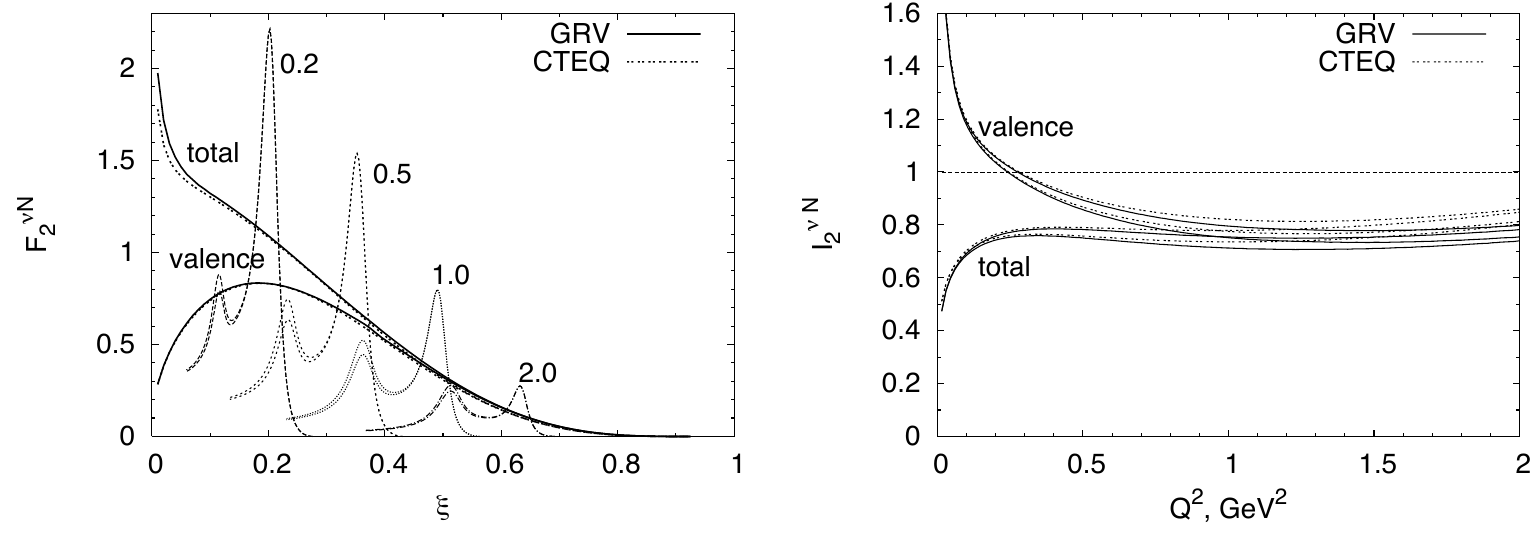}
\caption{Figure from~\cite{Lalakulich:2006yn}: Duality for the isoscalar nucleon structure function F$_2^{\nu N}$. (Left) F$_2^{\nu N}$ for resonances at Q$^2$ indicated on the spectra as a function of $\xi$, compared with the leading twist parametrizations (valence and total) at Q$^2$ = 10 GeV$^2$ . (Right) From ~\Eref{dratio}  the ratio I$_2^{\nu N}$ of the integrated F$_2^{\nu N}$ in the resonance region to the leading twist functions. The upper (lower) integrated ratios correspond to different $Q^2$ behavior of the axial form factors}
\label{fig_Valence-duality}
\end{center}
\end{figure}

%Other models for $\nu$ nucleon scattering that include higher-W multi-pion production as well as alternative non-resonant pion production models come to different conclusions.   as shown in 
From  Fig.~\ref{n&p_duality} and Fig.~\ref{fig_F2&I2nuN} there is a noticeable decrease at low values of $\xi$ of the integral ratio below $\approx Q^2 \le 0.5 ~GeV^2$.  This behavior resembles the fall-off of the valence quarks ($xF_3$) and was noted by several studies including~\cite{Lalakulich:2006yn,Graczyk:2005uv}. This led to the idea of two-component duality, which was originally proposed by Harari and Freund~\cite{Harari:1968jw,Freund:1967hw}. It essentially relates resonance production of pions with the valence quark component and non-resonant pion production with the sea quark component of the structure functions. This concept was tested via e-N interaction~\cite{Niculescu:2000tk} studies indicating that the $F_2$ structure function averaged over resonances at low values of $\xi$ ($\lessapprox$ 0.3) behaves like the valence quark contribution to DIS scaling as in Fig.~\ref{fig_Valence-duality}. 

This suggests the very intriguing concept that if overall duality is satisfied and the resonance contribution is dual to the valence DIS contribution, then the non-resonant background could be dual to the sea quark contribution.   This, in turn, suggests that duality could be used to guide a model for non-resonant pion production background.  

\begin{figure}[h]
\begin{center}
\includegraphics[width=0.85\textwidth]{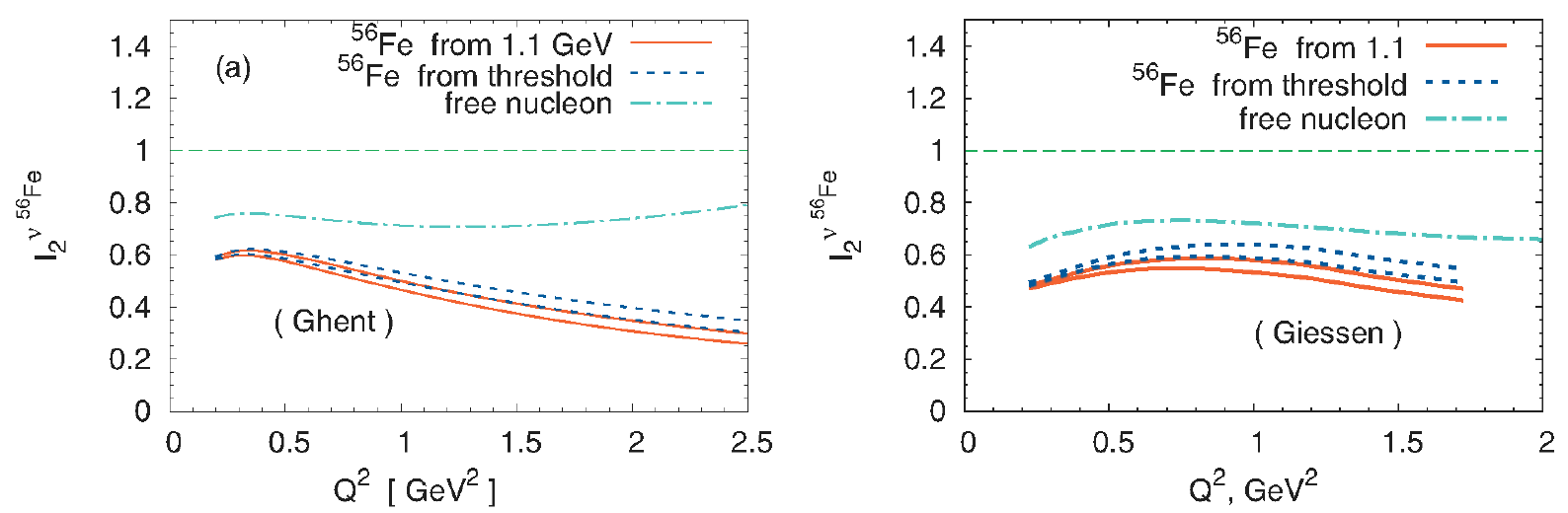}
\caption{Figure from~\cite{Lalakulich:2009zza}: Ratio I$_2^{\nu Fe}$ for iron calculated within the Ghent~\cite{Praet:2008zz} (left) and Giessen~\cite{Mosel:2017nzk}(right) models. For Fe the results are displayed for two choices of the lower limit of the numerator in the integral of ~\Eref{dratio}: W = 1.1 GeV (solid line) and "threshold" that takes into account the Fermi motion within the Fe nucleus (dotted line). For each of these two choices they used two sets of DIS data in determining the denominator of the integral I, one at Q$^2_{DIS}$ = 12.59 GeV$^2$ and the other at 19.95 GeV$^2$.  The ratio I$_2^{\nu N}$ for the free nucleon (dash-dotted line) is shown for comparison}
\label{fig_I2nuFe}
\end{center}
\end{figure}

Duality in the case of neutrino \emph{nucleus} interactions has been studied, again theoretically, in \cite{Lalakulich:2008tu}. In particular both the GiBUU and Ghent groups have used their respective resonance models to evaluate duality.  The main difference in the models is that GiBUU~\cite{Kaskulov:2011pr}
 uses a resonance model that includes single- and multi-$\pi$ decays plus heavier decay states while the Ghent model~\cite{Gonzalez-Jimenez:2016qqq} concentrates on 1$\pi$ decays but extended up to high effective masses using Regge trajectories.
%The results as in Fig.~\ref{fig_I2nuFe} from that reference suggest problems with applying duality to this process, particularly for non-isotropic nuclei such as Pb or even Fe or Ar.  Furthermore, when attempting to apply duality to any \nu-nucleus experimental results one has to be extremely cautious
%
%The $Q^2$ along the abscissa in Fig.~\ref{fig_I2nuFe} is the $Q^2$ involved in computing the limits $ \xi_{min} = \xi(W_1, Q^2)$ and $\xi_{max} = \xi(W_2, Q^2)$ of the integration of the numerator of $I_2^{\nu Fe}$.  Refer to the figure caption for further details of the figure. 
They observed as in Fig.~\ref{fig_I2nuFe} that the\emph{ computed} integrated resonance strength is about half of the measured DIS one. Contrary to the free nucleon case, where the ratios $I_i(Q^2)$ are at the level of 0.8-0.9, they found for nuclei such as Fe ratios of 0.6 for electro-production and 0.4 for neutrino production. This points towards a scale dependence in the role of the nuclear effects. It could suggest that nuclear effects act differently at lower $Q^2$ (resonance regime) than at higher $Q^2$ (DIS regime).
In this analysis the contributions of the non-resonant background was ignored. It was stressed that for more detailed investigations of duality a theoretical or phenomenological model for the non-resonant background across the entire resonance region will be required.  The inadequacy of the treatment of non-resonant meson production in current neutrino event simulators has been emphasized by recent studies~\cite{Rodrigues:2016xjj} that found that the non-resonant background evaluated from bubble chamber data is considerably smaller than the estimates in GENIE.  Note that subsequent experimental studies have preliminarily suggested that this large reduction in the GENIE non-resonant pion estimate is essentially only for the W-region around the $\Delta$ and the GENIE prediction for the higher-W SIS region may be valid. 

%The authors of~\cite{Lalakulich:2008tu} concluded:
%\begin{itemize}
% \item the ratio, $I_2^{\nuFe}$, is significantly smaller than 1 for all  %$Q^2$
%    \item the ratio is even smaller than $I_2^{\nu N}$ for the free nucleon
%\end{itemize}
However, these studies suggest the need for care in using duality to verify the strength of contributions of $\nu$-N scattering in the SIS region and, particularly, for considering the interpretation of duality with $\nu$-A scattering for nuclei with large excess neutron content.

%Among the few theoretical studies that have been done for $\nu$-N (note: not $\nu$-A ) scattering, some have  used the Rein-Sehgal model that is commonly used in current MC event generators. 
%If the resonance region is confined to $W<1.6$~GeV the dualityas defined in Eq.~(\ref{dual_ratio}) is satisfied at the 75-80\% level. If the resonance region is defined as $W<2$~GeVthe value of the integral in Eq.~\ref{dual_ratio} is only about 50\%. These results are to some extent model dependent but a general tendency is that for larger W, DIS structure functions are much larger than the resonance contribution, as clearly seen from Fig. 3 in   \cite{2Lalakulich:2006yn}   and Fig. 7 in \cite{Graczyk:2005uv}.  

%There has been an attempt to modify the tests of duality in $\nu$-N scattering by introducing a two component duality hypothesis that separates the resonance contribution as displaying duality with the valence quarks and the non-resonant background displaying duality with the sea. Investigation done within the Rein Sehgal model with $W<2$~GeV revealed no signature of two component duality. 

\subsection{Duality and the Transition to Perturbative QCD: "1 / Q$^2$" Effects}
\label{Subsec-twist}

In the calculation of the DIS integral for the denominator  of ~\Eref{dratio}, the LO or NLO \emph{leading twist} (see~\ref{HT}) perturbative QCD fit to high Q$^2$ data or, if an experimental measurement of F$_2$ was used, it was taken from higher-$Q^2$ measurements.  The important feature was that no higher twist "1 / Q$^2$" effects were included in the evaluation of the integral denominator of the ratio.  This being the case, the observation from Fig.~\ref{fig-I2eN} that the agreement with duality is quite close to complete is a suggestion that there are minimal additional higher twist effects in the DIS data or needed in the DIS theoretical expression as long as target mass correction (TMC) are included through the use of $\xi$
%it is evaluated for $Q^2 \ge 10 GeV^2$.  

Considering these conclusions, it could be possible to learn about possible higher twist effects by observing violations of duality for e/$\mu$ nucleon data at lower $Q^2$.  Current neutrino experiments are constrained by their lower-energy neutrino beams to the lower $Q^2$ edge of the DIS region where possible higher twist effects could then be a real complication of the analysis.  On the other hand, improved knowledge of higher twist contributions and how these contributions are exhibited as non-resonant pion production could provide a better understanding of the transition from perturbative to non-perturbative QCD, from the SIS to DIS regions. Improved  determination of the higher-twist effects should then be a goal of current and future analyses. 

There have been several studies investigating the link between duality and higher twist effects~\cite{Ji:1994br}-\cite{Melnitchouk:2011ab}.  In the earlier study~\cite{Ji:1994br} the authors emphasize the ability to use duality to determine higher twist contributions from structure function data in the resonance region by using moments (in x) of the structure function $F_2$. These higher moments in x emphasizing the contributions from increasingly higher x regions where higher twist effects are supposed to be larger. %For example, in the integral over x of the structure function $F_2(x,Q^2)$, they are able to determine that the ratio of the higher twist contribution to leading-twist contributions at  $Q^2 = 2 ~GeV^2$ is order 10\%.  The ratio of higher- to leading-twist contributiuons grows rather rapidly as the index of the moment increases thereby  emphasizing higher and higher x regions.  In~\cite{Melnitchouk:2005tn} the author examines the size of twist-4 effects using moments of the spin-dependent structure functions to suggest that higher twists are small for $Q^2 \ge 1 GeV^2$.  

\begin{figure}[h]
\begin{center}
\includegraphics[width=0.55\textwidth]{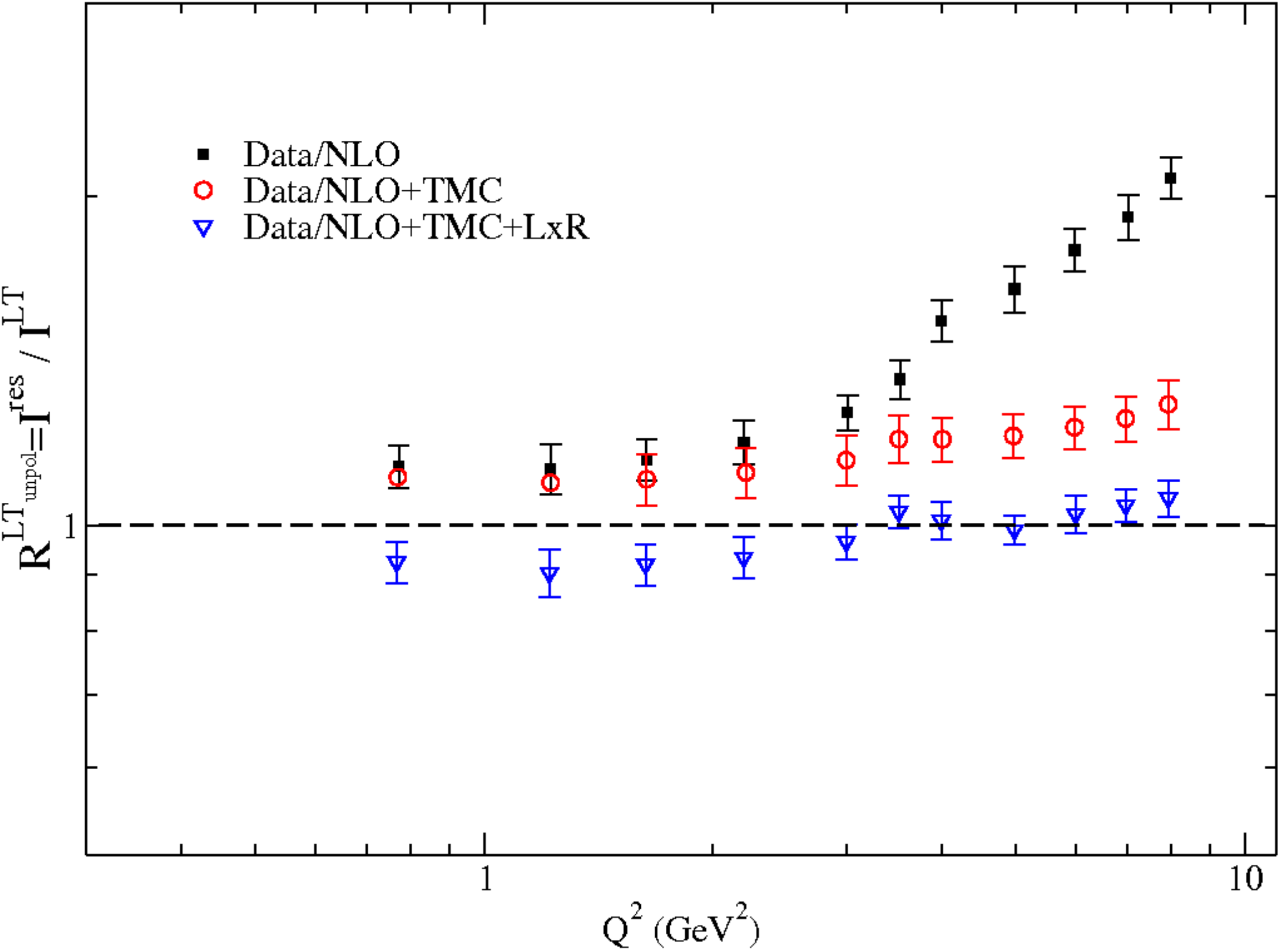}
\caption{Modified figure from~\cite{Fantoni:2005ip}: Ratio between the integrals of the measured structure functions and the calculated ones plotted as a function of Q$^2$ showing the effect of adding consecutively the TMC and large-x re-summation to the straight leading twist NLO QCD expression based on PDF fits.}
\label{fig-FBL-fig1L}
\end{center}
\end{figure}

The authors of~\cite{Bianchi:2003hi, Fantoni:2005ip} first examined duality in structure functions~\cite{Bianchi:2003hi} and then used the techniques developed in this study to understand the interplay of duality and higher twist\cite{Fantoni:2005ip}.  This study used a combination of nucleon data from Jefferson Lab and SLAC to form the numerator in a ratio of integrals similar to ~\Eref{dratio}.  The denominator is taken from dynamical parametrizations coming from free nucleon parton distribution functions.  Target-mass effects %(TMC - REFER TO SA SECTION ON TMAC)
 are then introduced and, in a separate step, they also include the large-x re-summation effects.  These re-summation effects, essentially, reduce the exaggerated Q$^2$-dependent suppression of F$_2$ as x approaches 1., which, in essence, adds strength to F$_2$ at large x and increases the integral.  The results of their study is shown in Fig.~\ref{fig-FBL-fig1L} and 
supports their conclusion that with the addition of the TMC and the inclusion of the large-x re-summation there is little space left for additional (1 / Q$^2$) higher twist effects for $Q^2 \ge 1.0 GeV^2$ .  This is then a quantitative exercise showing the power of using duality to better understand the need for both additional kinematical (TMC) and dynamical higher twist(HT) terms added onto the leading twist perturbative QCD expression for DIS structure functions.

 %The experimental extraction of higher-twist effects in the analysis of DIS structure functions is usually performed with an additional term multiplying the leading twist expression.  This additional term contains a constant depending on x divided by Q$^2$ such that, for example, $F_2^{LT+HT} = F_2^{LT} (1+C(x)/ Q^2)$.  The authors of~\cite{Fantoni:2005ip} analysed the Jefferson Lab and SLAC data using this expression and found the results shown in Fig.~\ref{fig-FBL-fig2L} that, as in Fig.~\ref{fig-FBL-fig1L}, clearly shows how the sequential inclusion of the TMC and large-x resummation strongly diminishes the need for further higher twist corrections.

% \begin{figure}[h]
%\begin{center}
%\includegraphics[width=0.55\textwidth]{FBL-fig2L}
%\caption{Figure from~\cite{Fantoni:2005ip}: The HT coefficient$(1+C(x)/ Q^2)$ as a function of x showing the effect of adding TMC and large-x resummation to the extraction.  For comparison the values for this HT coefficients obtained in~\cite{Alekhin:2002fv} using DIS data and the effect of TMC are shown.}
%\label{fig-FBL-fig2L}
%\end{center}
%\end{figure}
  
\subsection{Neutrino Simulation Efforts in the SIS region} 
\label{Subsec-BYmodel}
%Noting the disparity in simulations of the SIS region in Fig.~\ref{fig:SIDISwithGens} 
For an informative comparison of current simulation efforts in the SIS and DIS regions with emphasis on NEUT refer to Bronner's presentation in~\cite{Andreopoulos:2019gvw}. It is interesting to note that there is a common thread among MC event generators in attempts to bridge the transition from the SIS to DIS regions.  As a practical procedure for addressing this SIS region in contemporary neutrino event generators, such as GENIE, Bodek and Yang~\cite{Bodek:2010km} introduced a model (BY) that is used to bridge the kinematic region between the Delta resonance and DIS.  This BY model is also used by GENIE and other event generators to describe the DIS region as well and this application will be considered in subsequent sections. The model was developed using results from~\emph{electron-nucleon} inelastic scattering cross sections. 
%It uses leading order parton distribution functions and introduces a new scaling variable $\xi_w$ to include the effects of dynamic higher twist effects through a modified target mass correction.  
The model incorporates the GRV98~\cite{Gluck:1998xa} LO parton distribution functions replacing the variable x with their $\xi_w$ scaling variable to include the effects of dynamic higher twist effects through a modified target mass correction.  These modified parton distribution functions are used to describe data at high $Q^2$ and down to 0.8 GeV$^2$.  Below $Q^2$ = 0.8 GeV$^2$ they take the GRV98 LO PDFs to get the value of $F_2 (x, 0.8~GeV^2$) and multiply it by quark-flavor dependent K factors to reach lower $Q^2$ and W.
  
The BY model then compares the results obtained with the above procedure with the expectations of the duality concept as demonstrated with e/$\mu$-\emph{nucleon} inelastic scattering (see section~\ref{Subsec-Duality}).  They find their predictions to be consistent with the average of charged-lepton nucleon initiated resonance production from the $\Delta$ peak to the start of DIS and therefore consistent with duality.  Note that the predictions of this procedure are meant to also include the {\emph non-resonant meson production} in this region.  The steps to expand their predictions from e/$\mu$-\emph{nucleon} to e/$\mu$-\emph{nucleus}, is described in~\cite{Bodek:2003wd}.
In brief, a model for deuterium nuclear effects is used to produce e/$\mu$-\emph{deuterium} from e/$\mu$-\emph{nucleon} and then the measured ratio as a function of x  of e/$\mu$-\emph{Fe} to e/$\mu$-\emph{deuterium} is used to predict  e/$\mu$-\emph{Fe} so that this procedure is then valid for Fe targets only.
  
The BY procedure for $\nu$-N/A scattering is described in~\cite{Bodek:2010uf}.  They use the same GRV98LO, $\xi_w$ with K-factor approach as used for charged-lepton scattering but have quite different techniques for evaluating the factors since the axial-vector contribution, involving an axial K-factor, and the additional structure function ($xF_3$) of neutrino scattering must be considered.  For very high-E$_\nu$ and high-$Q^2$ both the vector and axial K-factors are expected to be 1.0 and the expressions for $F_2$ and $xF_3$ are straightforward.  Since the vector part of $F_2$ goes to 0 at $Q^2$ = 0 while the axial component does not, their approach to low-$Q^2$ must account for this difference in the vector and axial components of $F_2$.  They furthermore account for the differences in higher order QCD effects and scaling violations in $F_2$ and $xF_3$ at low-$Q^2$ and end up with expressions for $F_2^{vector}$,  $F_2^{axial}$ and $xF_3$ that they then use to predict neutrino nucleon interactions below the DIS region. 

\paragraph{Transition from SIS to DIS}
The B-Y expressions for this lower-W behavior, which, significantly, includes their estimate of non-resonant meson production, is expected to seamlessly blend with the straightforward expressions for $F_2$ and $xF_3$ they predict in the DIS region. They then have mimicked the concept of duality but based the extrapolation from the $\Delta$ to DIS on the described components of their model. It is important to note that this transition region is goverened by non-resonant pion production that, at low $Q^2$, is described in models quoted earlier.  This non-resonant pion production then becomes DIS as $Q^2$ becomes high enough to allow scattering off partons within the nucleus.  GENIE, employing the B-Y model, estimates the sum of non-resonant pion as well as multi-pion resonant production with this extrapolation from DIS.  GENIE then uses the AGKY hadronization model (see~\ref {Subsec-hadronization}) that for these lower values of W employs the KNO multiplicity model, to predict production of single and multi-pi events.  There is unfortunately very limited experimental data to compare with their predictions in this low-$Q^2$, low-W region.  However, an indication of a possible problem, an overestimation of the prediction in the region of $\Delta$ production, could be drawn from the  earlier quoted recent studies~\cite{Rodrigues:2016xjj} that found the non-resonant background to be considerably smaller than the estimates in GENIE that come from the BY model prediction of the average strength of the $F_2$ and $xF_3$ in this region.  

Whether this result can be related to the duality approach of extrapolating $F_2$ and $xF_3$ from the DIS regime down into the resonant region for neutrino scattering has not been explicitly considered.  However, we did learn~(Section \ref{Subsec-Duality}) that, from all models considered, such an extrapolation could indeed lead to overestimating this contribution in the lower-W resonant region.  That is a smooth extrapolation of the strength in the DIS region tends to overshoot the model predictions for the strength in the SIS region.  However an important point to recall is that many of the current models for resonance production include either no or very simple, approximate models for non-resonant pion production.

\subsection{Results and Discussion} 
\label{Subsec-SISResults}

\begin{figure}[h]
\begin{center}
\includegraphics[width=0.65\textwidth]{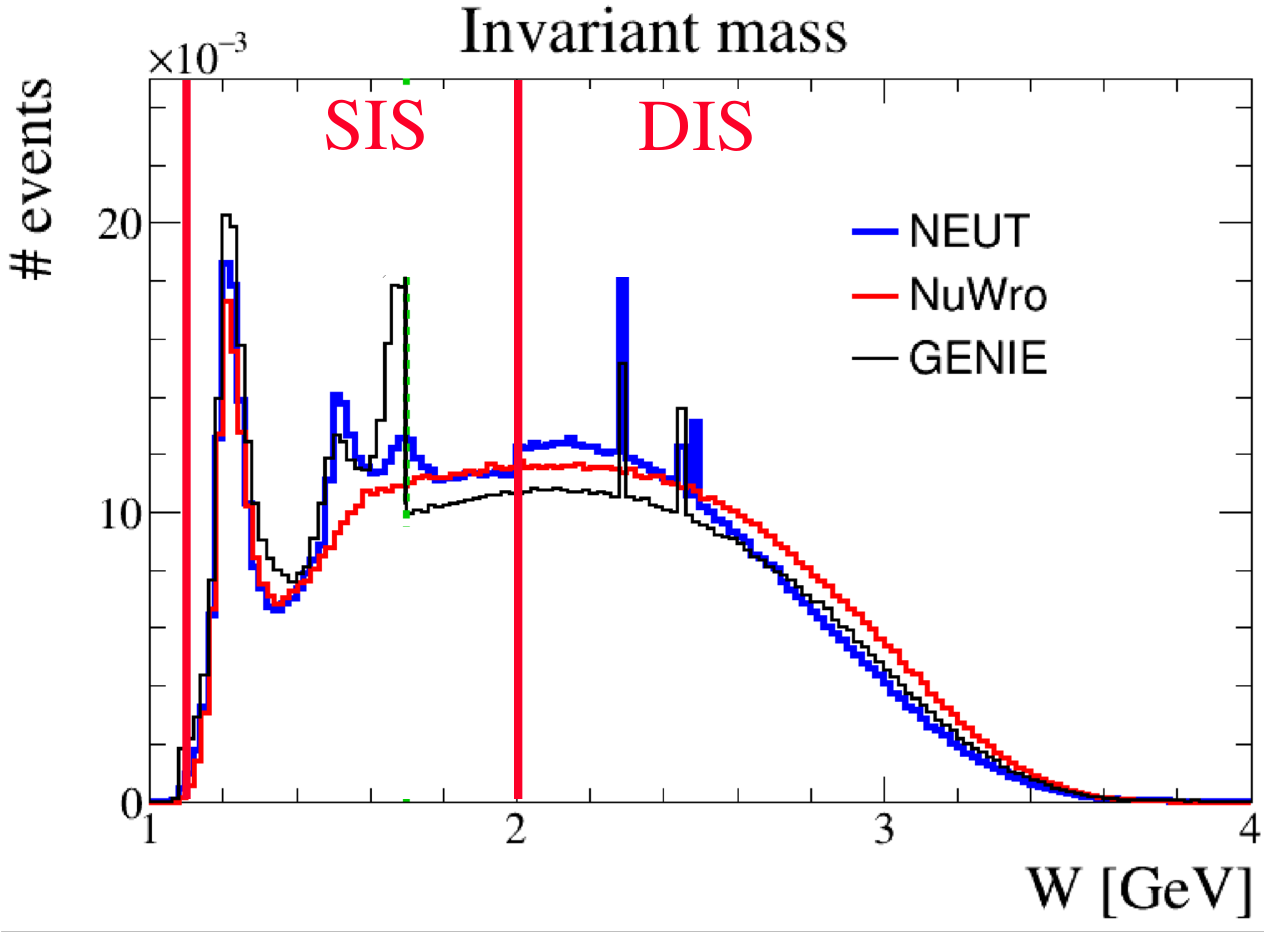}
\caption{Figure adapted from Bronner ~\cite{Andreopoulos:2019gvw} showing a comparison of the predictions of the community's then current simulation programs (NEUT 5.4.0, GENIE 2.12.10 and NuWro 18.02.1) over the range of W encompassing the SIS and DIS regions.  The predictions are for a 6 GeV neutrino on Fe.}
\label{fig:SIDISwithGens}
\end{center}
\end{figure}

The Shallow Inelastic Scattering region, particularly the higher-W transition to the Deep-Inelastic Scattering region in $\nu/\nub$ nucleon/nucleus scattering has been scarcely studied theoretically or experimentally. In particular the evolution of low-$Q^2$ non-resonant single pion production to low-$Q^2$ non-resonant multi-pion production and, as $Q^2$ increases, DIS pion production needs much more attention.  The lack of knowledge of this region is reflected in the disparity in the current predictions by the community's simulation programs as displayed in Fig.~\ref{fig:SIDISwithGens}.

There have been multiple studies of the $\Delta$ resonance region (W $\le$ 1.4 GeV), however only restricted studies by the \minerva ~ experiment including somewhat higher W single and  multi-pion production ( W $\le$ 1.8 GeV)~\cite{Stowell:2019zsh} and nothing for the interesting transition to DIS at even higher W.  
%\bibitem{Stowell:2019zsh} 
%  P.~Stowell {\it et al.} [MINERvA Collaboration],
  %``Tuning the GENIE Pion Production Model with MINERvA Data,''
 % arXiv:1903.01558 [hep-ex].
 
%These disagreements can have multiple sources; the current hydrogen and deuterium based bubble chamber results are low statistics with questionable systematics, the modified Rein-Sehgal model used by GENIE to generate resonances is rather dated, the GENIE-BY model used to generate the non-resonant meson production in the SIS region has been shown to be questionalble and finally the final state interaction model that yields/or-not pions in the final state measured by experiments is simply a best estimate cascade model.

As shown in this section, the application of duality seems to be quite different for e/$\mu$-N interactions  and $\nu/\nub$-N interactions. A brief summary would conclude that:
\begin{itemize}
 \item F$_2^{ep,en}$ - for e/$\mu$-N scattering qualitative and quantitative duality {\emph is observed}    
 \item F$_2^{\nu p,\nu n}$ - for $\nu/\nub$-N scattering duality is roughly observed for the average nucleon [(n+p)/2] but duality is {\emph not observed} for neutrons and protons individually. 
  \item For electroproduction with nuclei it is a different story. The quantitative evaluation of duality in e-A is not as good as with e-N.
   \item For $\nu$-A interactions it is not clear at all how duality works, particularly with nuclei having an excess of neutrons.
 \end{itemize}

The challenge of addressing duality with neutrinos is that in general in the SIS region the resonance structure functions for proton are much larger than for neutrons and in the case of deep-inelastic scattering the opposite is the situation. This does support the observation that if  duality is observed at all with neutrinos it is with the average nucleon [(n+p)/2].

However there is a more fundamental concern regarding the whole concept of testing duality experimentally. Can one really test duality if both the "DIS" and "SIS" regions are not  experimentally accessible at identical kinematics?  For example, Fig.~\ref{fig:SIDISwithGens} represents a neutrino energy typical for  the \minerva\ experiment and there is very limited range of W above the 2 GeV DIS cutoff available for any comparison to the SIS region.  
%If it is then necessary to test duality by comparing experimental SIS measurements with phenomenological DIS projections,
Furthermore, although there may be limited contributions of higher twist for lower-x and $Q^2$ structure functions, when including inclusive cross sections over all x and $Q^2$ leading twist alone may not be sufficient.  
Thus different extrapolations will give you better or worse agreement between the extrapolated "DIS" part and the measured SIS part.  There is a need for careful consideration of exactly what experimental tests can be made to test duality with neutrino nucleus interactions

%\minerva is investigating this possibility by using $\nu/\nub$ to measure the inclusive W, $\xi$ dependent cross sections for the full SIS region up to the start of the DIS region off carbon, with its equal number of protons and neutrons.  This can then be compared with measurements of the DIS region and both with expectations of duality.  

This also strongly suggests that rather than only experimental tests of duality  we should encourage a closer examination of just how well the current neutrino simulation event generators, GENIE, NEUT and NuWro obey duality in their treatment of the basic input, $\nu/\nub$ isoscalar nucleon scattering.

\section{$\nu_l/\bar\nu_l$-Nucleus Scattering: Deep-Inelastic Scattering Phenomenology} 
\label{section5}
%\section{Phenomenology and Experimental Results: Deep-Inelastic Scattering} \label{Sec-ExpDIS}
%\subsection{Introduction}
Neutrino scattering plays an important role in the QCD analysis of deep-inelastic scattering since the weak current has the unique ability to "taste" only particular quark flavors resolving the flavor of the nucleon's constituents: $\nu$ interacts with $d$, $s$, $\ubar$ and $\cbar$ while the $\overline{\nu}$ interacts with $u$, $c$, $\dbar$ and $\sbar$.  This significantly enhances the study of parton distribution functions and complements studies with electromagnetic probes. However, as helpful as this ability of the weak-interaction may be, it should be again emphasized that all high-statistic neutrino experiments have had to use heavier nuclear targets. This means the PDFs extracted from these experiments are for nucleons in the nuclear environment and are thus {\it nuclear} parton distribution functions nPDF. As will be shown, there is considerable difference between these A-dependent nPDFs and the free nucleon PDFs. Furthermore, since the relevant nuclear effects could involve multiple nucleon scattering as in shadowing or scattering from correlated nucleon pairs as possibly in the EMC effect these nPDFs might better be considered {\it nuclear} nPDFs and not necessarily the PDFs of single bound nucleons.  

%With the exceptionally large statistical sample and more manageable systematic uncertainties expected in current and future neutrino experiments, these experiments can, in principle, dramatically improve the isolation of individual nPDFs by measuring the full set of $\nu$ and $\bar \nu$ structure functions.  That is, for the first time, independently isolate all the nuclear structure functions $F_1^{\nu N_A}(x,Q^2)$, $F_1^{\bar \nu N_A}(x,Q^2)$, $F_2^{\nu N_A}(x,Q^2)$, $F_2^{\bar \nu N_A}(x,Q^2)$ $xF_3^{\nu N_A}(x,Q^2)$ and $xF_3^{\bar \nu N_A}(x,Q^2)$.  %Then by taking differences and sums of these structure functions, specific parton distribution functions in a given $(x, Q^2)$ bin can in turn be highlighted. 
%Extracting this full set of structure functions will rely on the $y$-variation (y = the ratio of hadron energy to incoming neutrino energy = $E_H / E_\nu)$ of the structure function coefficients in the expression for the cross-section. In the helicity representation, for example:

%\begin{eqnarray}
%\frac{d^2 \sigma^\nu}{dx d\qsq} &=& \frac{G^2_F}{2 \pi x}
%\Bigl[\frac{1}{2}
%\left( F_2^\nu (x,\qsq) + xF_3^\nu (x,\qsq) \right) + %\nonumber \\
%& & \frac{(1-y)^2}{2}
%\left(F_2^\nu (x,\qsq) - xF_3^\nu (x,\qsq) \right) - 
%2 y^2 F_L^\nu x,\qsq) \Bigr].
\label{eq:pnp1}
%\end{eqnarray}

%\noindent By analyzing the data as a function of $(1-y)^2$ in a given $(x, \qsq)$ bin for $\nu$ and $\bar \nu$, all six structure functions could be extracted.  Note however that high statistics and especially rigorous control of systematic errors, particularly those associated with the energy-dependent neutrino flux, are required to accurately extract the six structure functions.
%\footnote{Note that for this type of parton distribution function study anti-neutrino running and minimal flux-associated systematics will be essential.}

Historically the study of the DIS region and first tests of (nuclear) QCD with neutrinos were actually the primary goals of early experiments with higher energy neutrino beams.  However, the current focus on neutrino-oscillation studies, with the need to emphasize lower neutrino energies to maximize oscillations, has led to limiting the possibility to explore the full DIS region in such experiments.  For example, the future DUNE experiment with the huge statistics expected in the near detectors should allow an interesting study of DIS albeit in a rather limited kinematic range suggesting an interesting and necessary study of the non-pQCD / pQCD transition region in the nuclear environment as well as the lower-Q, lower-W DIS region.
%more precise determinations of $\nu$ A cross sections on various targets.

\subsection{Early Bubble Chamber DIS Results}
\label{Subsec-BC}
Neutrino scattering experiments have been studying QCD with DIS for over four decades.     The early pioneers in these studies were the bubble chambers such as the Gargamelle heavy liquid bubble chamber~\cite{Office:1973aa} normally filled with heavy freon $CF_3Br$, while the smaller ANL~\cite{Carmony:1976pw} and  BNL~\cite{Furuno:2003ng} chambers as well as the much larger BEBC~\cite{Reinhard:1973rs} at CERN and the 15' chamber~\cite{Snow:1971zz} at FNAL were normally filled with hydrogen or deuterium and occasionally mixed with heavier nuclei such as Ne  or using heavier liquids such as propane. With these bubble chambers, initial studies of QCD behavior with the axial vector current were undertaken by multiple collaborations.  

%Experimental measurements in the DIS kinematic region require (anti)neutrino beams of relatively high energies, several GeV and higher. 
These chambers using hydrogen or deuterium targets offered an ideal tool to probe the structure of the free nucleon and measure the very important fundamental production cross sections essential as input to modern neutrino scattering simulation programs. Unfortunately, the overall statistics was quite limited and totally insufficient for contemporary needs such as the vital input for modern event generators GENIE, NEUT and NuWro\footnote{There is consideration within the neutrino community to attempt to correct this insufficiency of free nucleon data with an H/D experiment using the high-intensity DUNE LBNF neutrino beam.}.
%(e.g. only about 9,000 $\bar \nu$ and 5,000 $\nu$ events were collected by BEBC on hydrogen~\cite{Allasia:1985hw}). 

%Aside from a low statistics study with the 15' chamber at FNAL ~\cite{Barish:1978cb} presenting an analysis of scaling-variable distributions in deep-inelastic $\numubar$-p  interactions,
Most of the early studies of QCD with bubble chambers were performed by CERN experiments.  An example of these early CERN studies is the publications~\cite{Bosetti:1982yy}  that showed the results of a combination of the lower energy Gargamelle ($CF_3Br$) PS run with the higher energy narrow-band beam BEBC Ne-H exposure.  
Note this analysis was performed before the discovery of the DIS x-dependent nuclear effects suggesting that simply combining the two experiments, using different nuclei, without considering these nuclear effects could have been problematic.  However with the BEBC run using a 73\% molar Ne/H mix the difference in nuclear effects between the Gargamelle and BEBC runs would have been smaller than the errors on the data.  
The point of these early CERN $\nu$ experiments was to perform first measurements of $\Lambda_{QCD}$ with neutrinos and to better understand the influence of non-perturbative effects such as target mass and higher-twist in a quantitative comparison of results with QCD.  

%The end results were:
%\begin{itemize}
%\item A quantitative test of the Callan-Gross relation $2xF_1 / F_2$ = 1.0 with neutrinos.  Using the differential cross section with respect to y they determined that this ratio, integrated over all x and $Q^2$  was $0.94 \pm .09 stat \pm (.07 systematic)$.  
%\item The structure functions and their moments exhibited deviations from Bjorken scaling which agreed qualitatively with the x and $Q^2$ dependences predicted by both LO and NLO QCD calculations.
%\item From their fits that included a higher-twist (1/$Q^2)$ term, the $\chi^2$ values and the comparison of the data with and without higher-twist they concluded that for $Q^2 \geq 1.0  GeV^2$  both the pure QCD form and the form including a  higher-twist term describe the data well.
%\end{itemize}
%%%%%%%%

\subsection{Massive Neutrino Scattering Detectors: DIS Results and QCD}
\label{Subsec-Iron}
%\subsection{High-Statistics Experimental Measurements} 
%\label{Subsec-HighStat}
%(${\cal{O}}(10^7)$ events)
 The first higher statistics $\nu$ and $\bar \nu$ nucleus measurements were performed by massive nuclear target detectors like CDHS(W) - iron~\cite{Berge:1987zw} and CHARM/CHARM II - marble/glass~\cite{DeWinter:1989zg}. These early experiments were followed by the CCFR~\cite{Seligman:1997fe} and NuTeV~\cite{Tzanov:2005kr} - iron experiments and the CHORUS - lead experiment ~\cite{KayisTopaksu:2011mx, Onengut:2005kv}.   As opposed to the high resolution of the earlier low statistics bubble chamber experiments, most of these experimental measurements using heavy nuclear targets could not resolve details of the hadronic shower and  concentrated on the inclusive $\nu$ and $\bar \nu$ cross section measurements.  

Even the contemporary MINOS oscillation experiment, with the requisite low energy $\nu/\nub$ beams, had to concentrate on total cross section measurements on iron~\cite{Adamson:2009ju} since the rather limited experimental resolution in the measurement of hadron energy resulted in poor x resolution. No extraction of x - $Q^2$  dependent differential cross sections was undertaken.

The NOMAD experiment was one of the first modern, finer-grained  experiments with an opportunity for high resolution measurements of exclusive states ~\cite{Wu:2007ab}. However NOMAD has yet to release their measurements of the inclusive cross sections and structure functions off the various nuclei in their experiment.  

The latest results come from the \minerva\ experiment that has measured charged current (CC)  $\nu$-A DIS cross sections on polystyrene, graphite, iron and lead targets both in the lower energy (LE) NuMI neutrino beam~\cite{DeVan:2016rkm,Mousseau:2016snl}, and, more recently, in the somewhat higher energy (ME) beam, which enabled increased statistics and a wider kinematic range.   

%Very limited information is currently available on nuclear modifications of cross sections and structure functions in (anti)neutrino inelastic interactions. The first measurement of nuclear effects was performed by the BEBC bubble chamber experiment from the ratio of neon and deuterium targets~\cite{Allport:1989vf}, providing a suggestion of nuclear shadowing at small $x_{Bj}$ and $Q^2$ values.  The high statistics measurement of nuclear effects in DIS took several more decades to accomplish. The MINER$\nu$A experiment has presented the results from their low-energy analysis of multiple nuclear targets. They show the differential scattering cross section in the form of ratios $\frac{d\sigma^i}{dx}/\frac{d\sigma^{CH}}{dx}$,~i=C, Fe, and Pb~\cite{Mousseau:2016snl} reflecting nuclear effects directly.

\begin{figure*}[h]
\begin{center}
\includegraphics[width=0.65\textwidth]{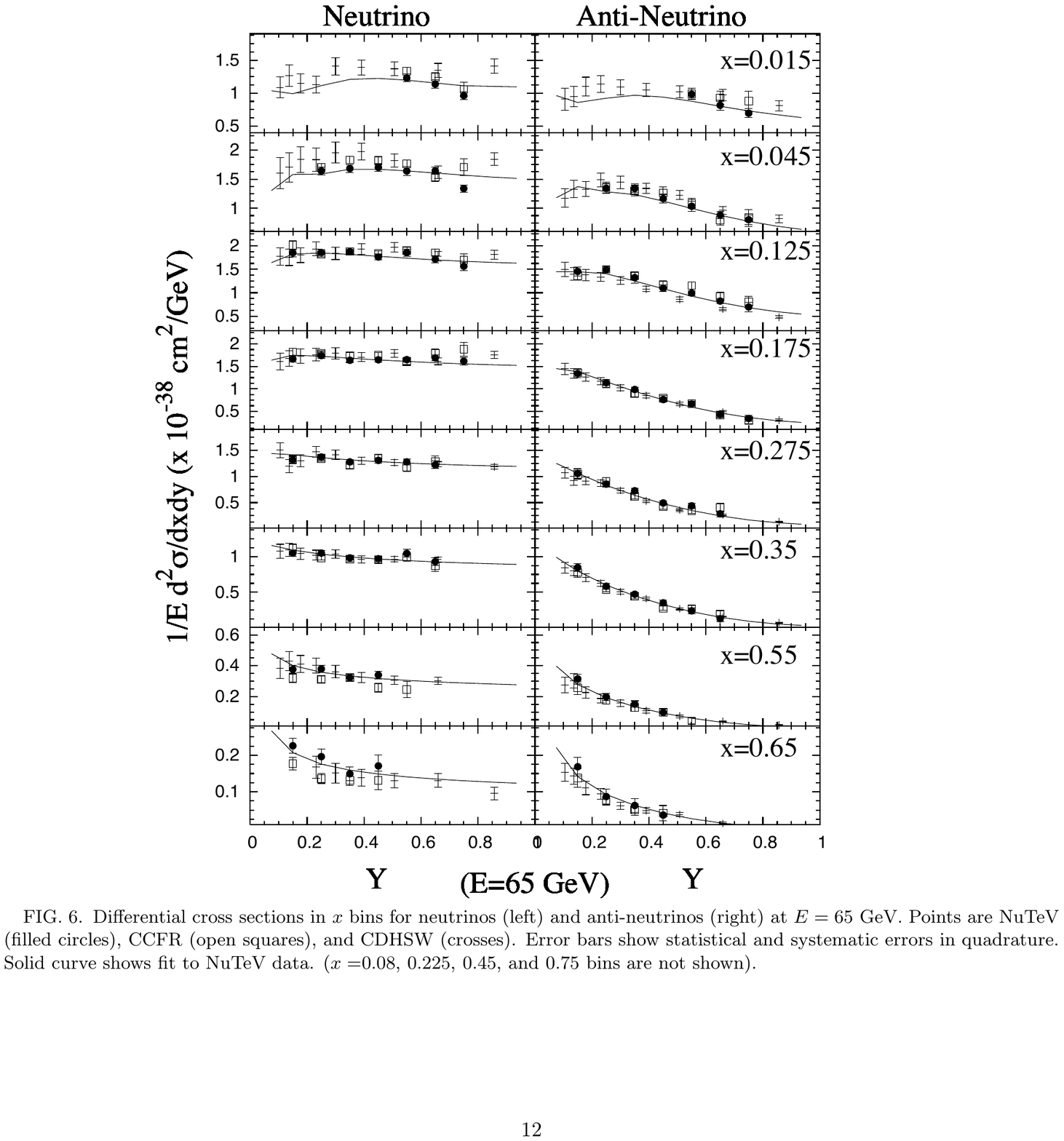}
\caption{Figure from~\cite{Tzanov:2005kr}.
Differential cross sections as a function of y in x bins for neutrinos and anti-neutrinos at E = 65 GeV. Points are NuTeV (filled circles), CCFR (open squares), and CDHSW (crosses). Error bars show statistical and systematic errors in quadrature. Solid curve shows fit to NuTeV data. } 

\label{fig:NuTeV-dsig65}
\end{center}
\end{figure*}

\begin{figure*}[h]
\begin{center}
\includegraphics[width=0.65\textwidth]{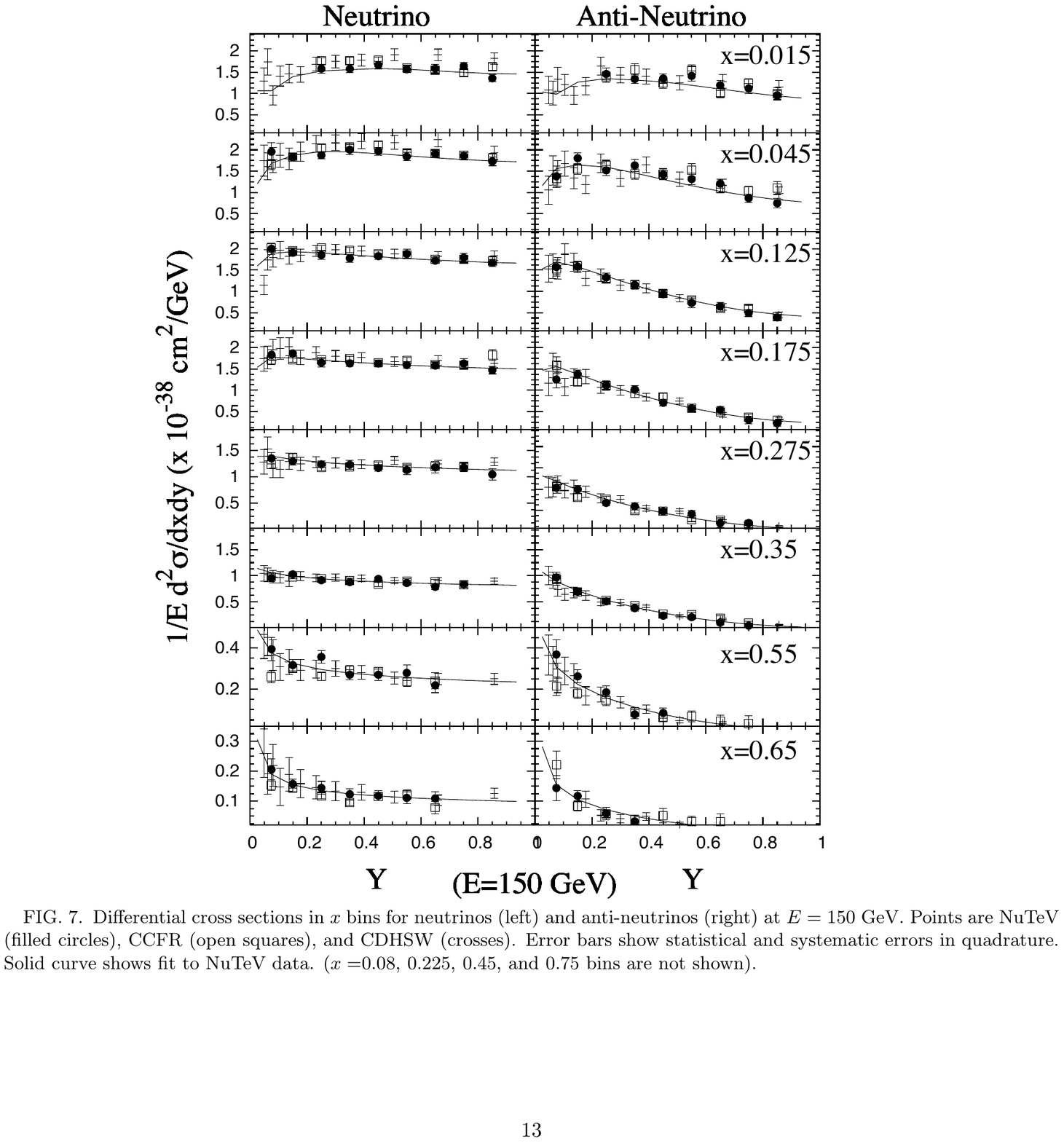}
\caption{Figure from~\cite{Tzanov:2005kr}.
Differential cross sections as a function of y in x bins for neutrinos and anti-neutrinos at E = 150 GeV. Points are NuTeV (filled circles), CCFR (open squares), and CDHSW (crosses). Error bars show statistical and systematic errors in quadrature. Solid curve shows fit to NuTeV data. } 

\label{fig:NuTeV-dsig150}
\end{center}
\end{figure*}

%
%The difficulty, of course, is that modern neutrino oscillation experiments demand high statistics which means that the neutrinos need massive nuclear targets to acquire these statistics.  This, in turn, complicates the extraction of free nucleon PDFs and demands nuclear correction factors that scale the results on a massive target to the corresponding result on a nucleon target. 
%
Without NOMAD results and \minerva\ ME results still pending, the latest high-statistics dedicated studies of QCD using neutrino scattering come from the NuTeV  \cite{Tzanov:2005kr} and CCFR~\cite{Seligman:1997fe} experiments off Fe as well as the CHORUS ~\cite{KayisTopaksu:2011mx,Onengut:2005kv} experiment off Pb.  The NuTeV experiment was a direct follow-up of the CCFR experiment using nearly the same detector as CCFR but with a different neutrino beam and analysis methods.  The NuTeV experiment accumulated over 3 million  $\nu$  and $\nub$ events in the energy range of 20 to 400 GeV off a mainly Fe target. The data were then corrected for QED radiative effects~\cite{Bardin:1981ft} and the charm production threshold.  A comparison of the NuTeV differential cross section results with those of CCFR and CDHSW are shown in %Fig.~\ref{fig:NuTeV-dsig},
Fig.~\ref{fig:NuTeV-dsig65} and Fig.~\ref{fig:NuTeV-dsig150} for two different beam energies. The importance of these directly measured cross sections as opposed to assumption-based extracted structure functions will be emphasized in subsequent sections describing the extraction of nuclear parton distributions.

% \begin{figure*} [h]
% \begin{picture}(500,300)(0,0) %   \graphpaper[20](0,0)(500,160)
%  \put(50,0){\includegraphics[width=0.45\textwidth]{NuTeV-dsig65}}
% \put(250,0){\includegraphics[width=0.45\textwidth]{NuTeV-dsig150}}
% %\put(114,0){$(a)$} \put(374,0){$(b)$} 
% \end{picture} 
% \caption{ 
% Figure from~\cite{Tzanov:2005kr}.
% Differential cross sections as a function of y in x bins for neutrinos and anti-neutrinos at E = 65 GeV (left) and 150 GeV (right). Points are NuTeV (filled circles), CCFR (open squares), and CDHSW (crosses). Error bars show statistical and systematic errors in quadrature. Solid curve shows fit to NuTeV data. 
% }
% \label{fig:NuTeV-dsig}
% \end{figure*}

\begin{figure*} [h]
\begin{picture}(500,300)(0,0) %   \graphpaper[20](0,0)(500,160)
 \put(50,0){\includegraphics[width=0.46\textwidth]{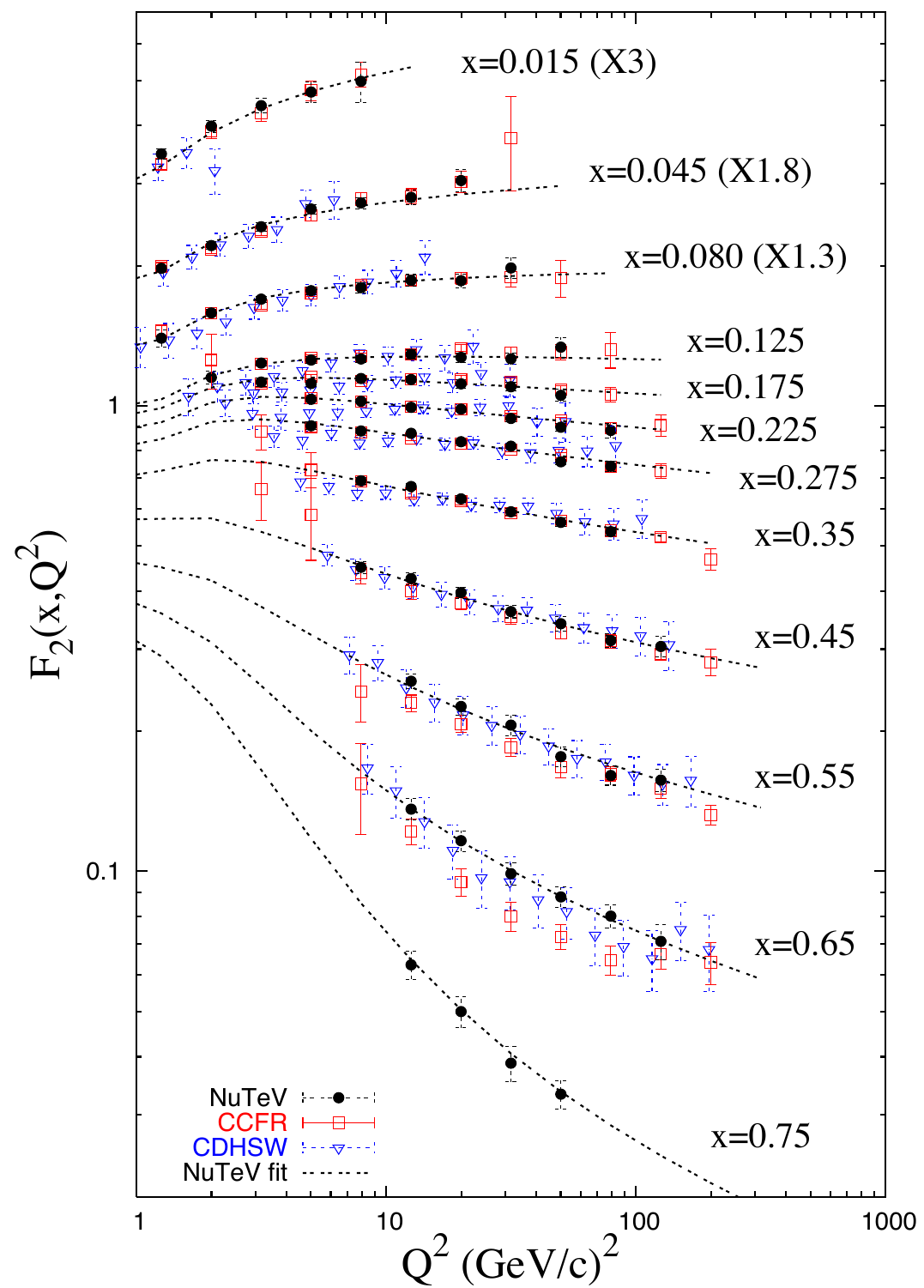}}
\put(250,0){\includegraphics[width=0.45\textwidth]{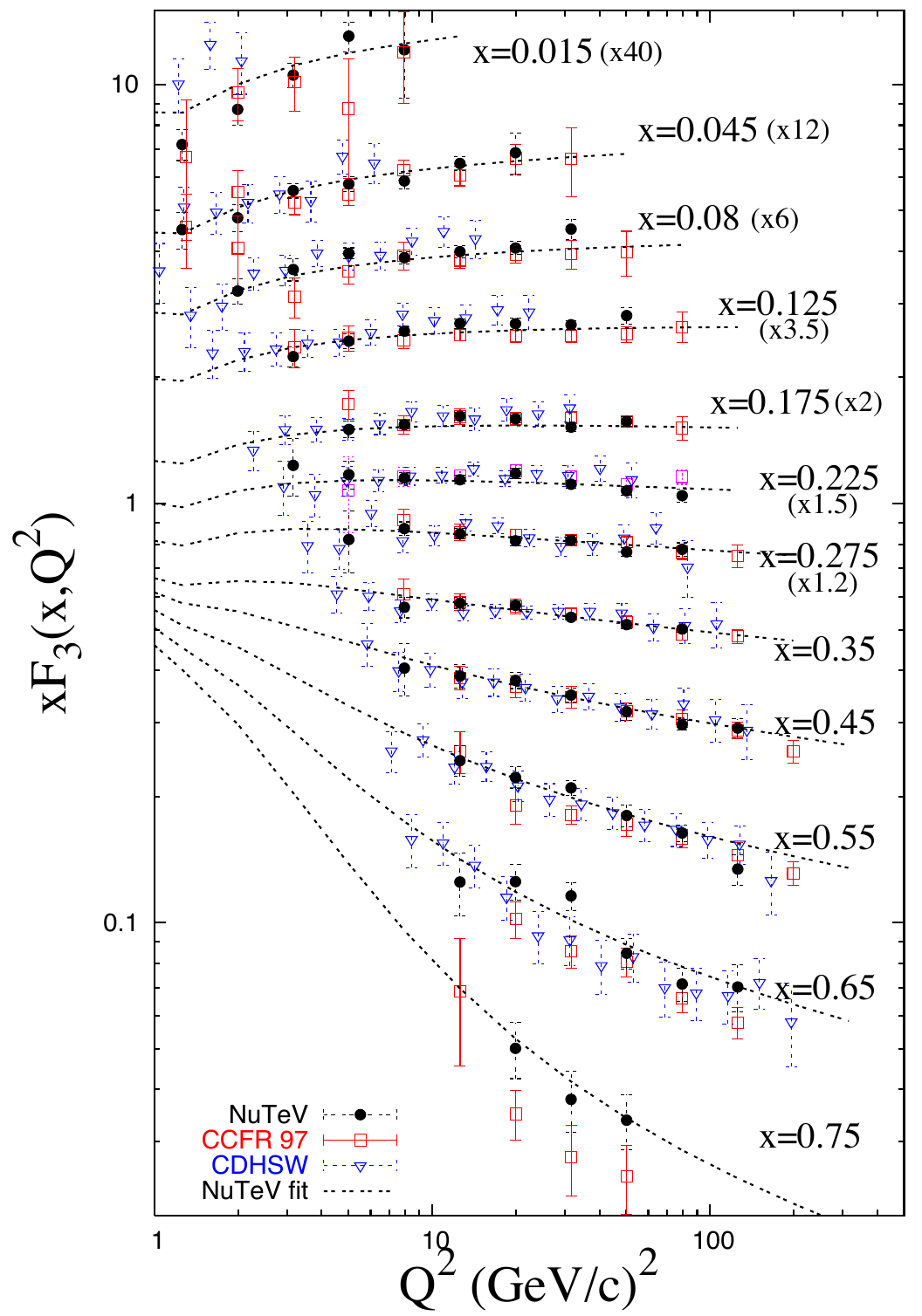}}
%\put(114,0){$(a)$} \put(374,0){$(b)$} 
\end{picture} 
\caption{Figure from~\cite{Tzanov:2005kr}.
NuTeV measurement of $F_2$ (left) and x$F_3$ (right) structure functions  (solid circles) compared with previous $\nu$-Fe results; CCFR (open circles) and CDHSW (triangles). The data are on iron and corrected to an isoscalar target and for QED radiative effects. The curve show the NuTeV model.
}
\label{fig:NuTeV+F23}
\end{figure*}

%\noindent 
A comparison of the NuTeV structure functions $F_2(x,Q^2)~and~xF_3(x,Q^2)$ derived from these cross sections with those from CCFR and CDHSW are shown in Fig.~\ref{fig:NuTeV+F23}. The main point is that the NuTeV structure function $F_2$ agrees with CCFR $F_2$ for values of $x \leq$  0.4 but is systematically higher, agreeing more with CDHSW, for larger values of $x$ culminating at $x \simeq$ 0.65 where the NuTeV result is $\simeq$ 20\% higher than the CCFR result.  Although the reason for this difference at high-x was not initially understood, it was finally traced to the difference of the magnetic field maps of the two experiments (that resulted is a shift of the muon energy scales of the two experiments),  the different cross section models used by NuTeV and CCFR and NuTeV’s improved muon and hadron energy smearing models.

\begin{figure*}[h]
\begin{center}
\includegraphics[width=0.65\textwidth]{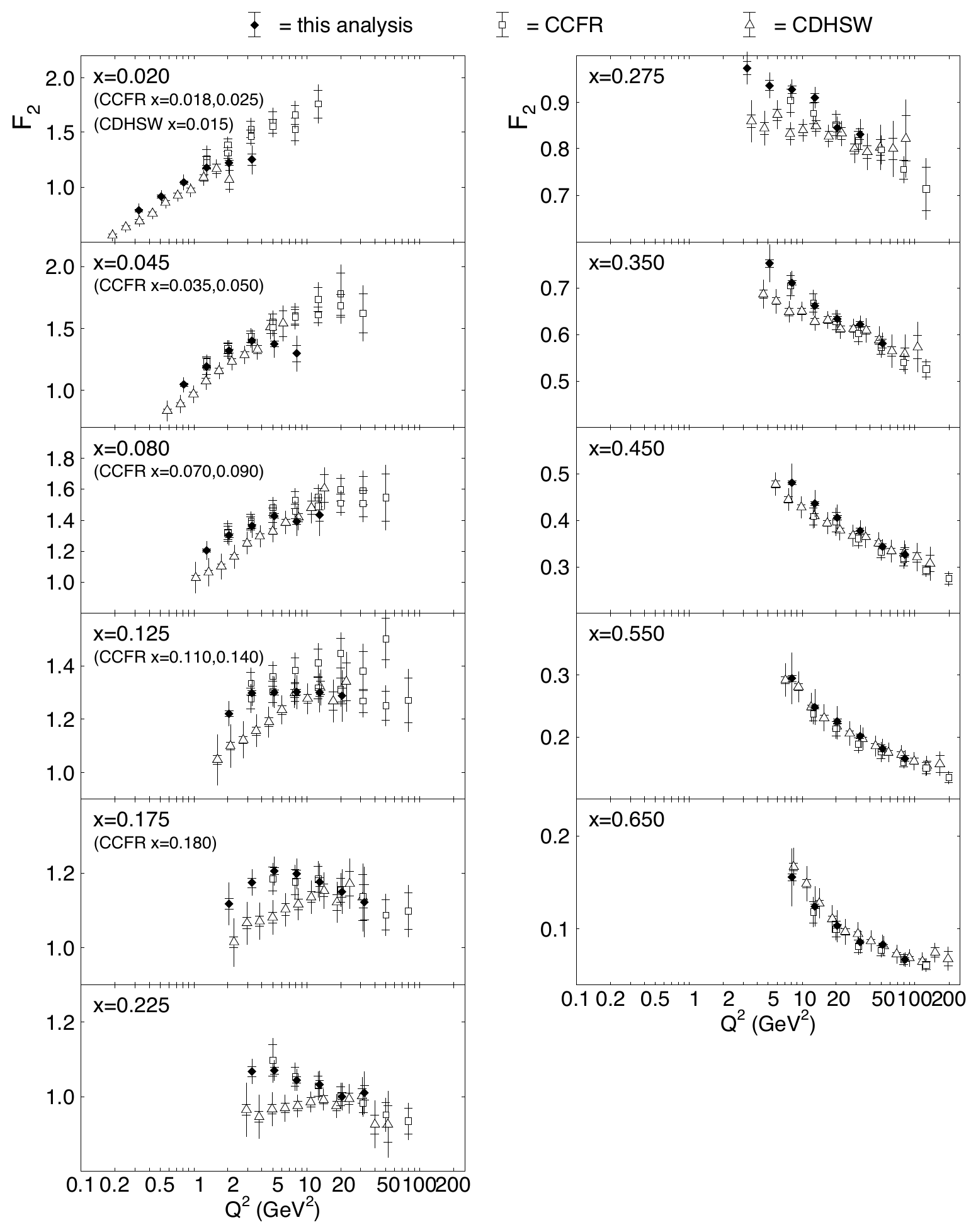}
\caption{Figure from~\cite{Onengut:2005kv} CHORUS~\cite{Onengut:2005kv} measurement of $F_2$ structure functions off Pb (solid circles) compared with previous $\nu$ Fe results from CCFR (open circles) and CDHSW (triangles).} 

\label{fig:CHORUS_F2}
\end{center}
\end{figure*}

\begin{figure*}[h]
\begin{center}
\includegraphics[width=0.65\textwidth]{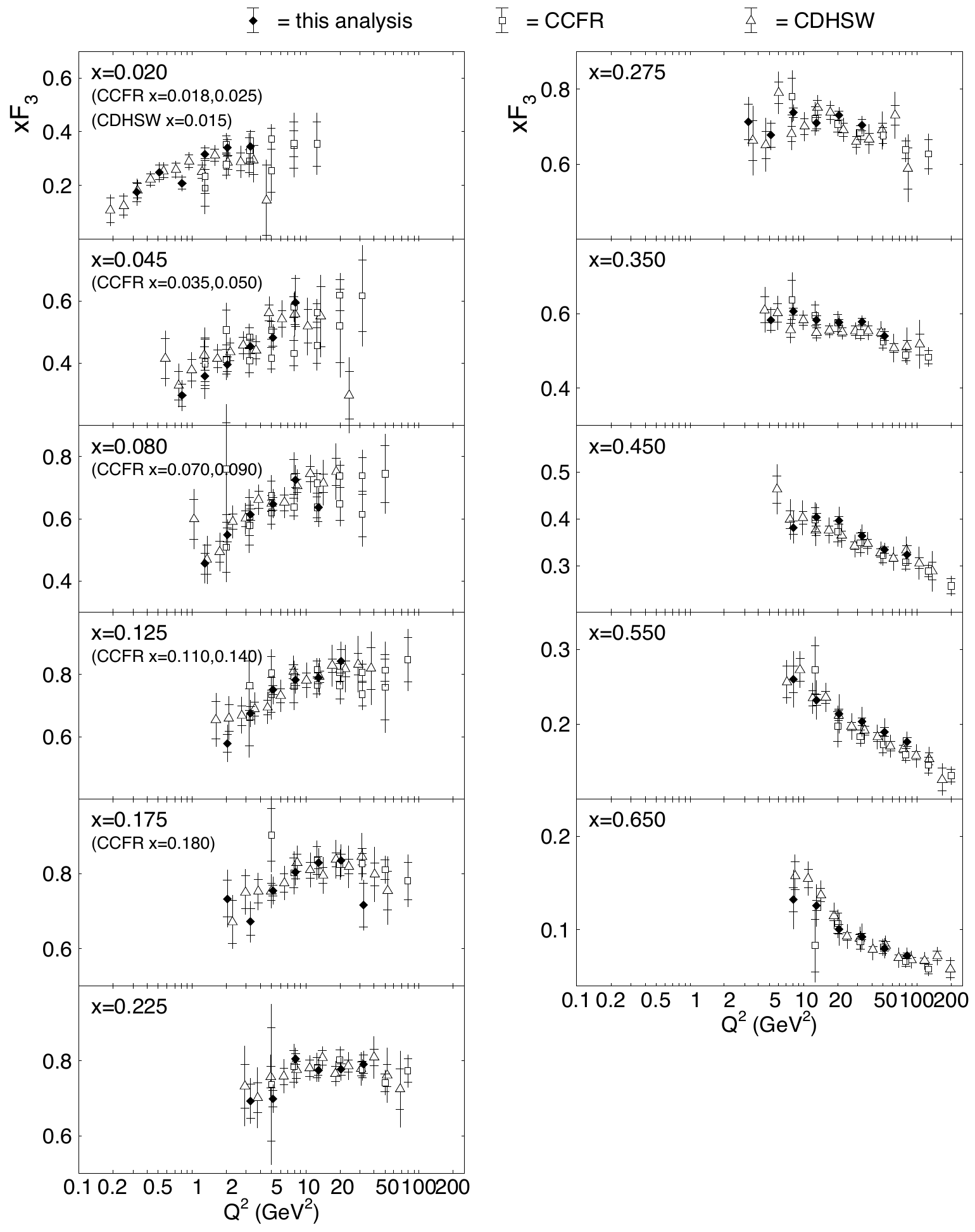}
\caption{Figure from~\cite{Onengut:2005kv} CHORUS~\cite{Onengut:2005kv} measurement of $xF_3$ structure functions off Pb (solid circles) compared with previous $\nu$ Fe results from CCFR (open circles) and CDHSW (triangles).} 

\label{fig:CHORUS_xF3}
\end{center}
\end{figure*}

Providing input on lead targets, the CHORUS detector was comprised of a high-resolution lead/scintillator calorimeter coupled with a large acceptance muon spectrometer for neutrino interactions in the calorimeter. The experiment used higher purity sign-selected neutrino and anti-neutrino beams to measure double differential cross-sections, in different bins of the neutrino energy, with minimal model-dependence. It is these cross sections that were used in the extraction of nuclear parton distributions.  From the differential cross sections the structure functions $F_2$ and x$F_3$ were extracted and are shown in %Fig.~\ref{fig:CHORUS_F23},
Fig.~\ref{fig:CHORUS_F2} and Fig.~\ref{fig:CHORUS_xF3}
along with the $\nu$-Fe results of CCFR and CDHSW.

% %\begin{figure*} [h]
% \begin{picture}(500,250)(0,0) %   \graphpaper[20](0,0)(500,160)
%  \put(50,0){\includegraphics[width=0.45\textwidth]{CHORUS_F2}}
% \put(250,0){\includegraphics[width=0.45\textwidth]{CHORUS_xF3}}
% \put(114,0){$(a)$} \put(374,0){$(b)$} 
% \end{picture} 
% \caption{ Figure from~\cite{Onengut:2005kv} CHORUS~\cite{Onengut:2005kv} measurement of $F_2$ (left) and x$F_3$ (right) structure functions off Pb (solid circles) compared with previous $\nu$ Fe results from CCFR (open circles) and CDHSW (triangles). }
% \label{fig:CHORUS_F23}
% \end{figure*}

\subsection{The Need for Nuclear Correction Factors}
\label{Subsec-NucCor}
%%%%%%%%%%%%%%%%%%%%
%%%%%%%%%from shadowing review

%The difficulty, of course, is that modern neutrino oscillation experiments demand high statistics which means that the neutrinos need massive nuclear targets to acquire these statistics.  This, in turn, complicates the extraction of free nucleon PDFs and demands nuclear correction factors that scale the results on a massive target to the corresponding result on a nucleon target. 
That the need for high statistics neutrino experiments resulted in the use of heavy nuclear targets eventually introduced significant complications in the attempt to extract {\it free nucleon} PDFs with these neutrino results.   
The goal of combining the many DIS experimental results on heavy nuclei ranging from C to Pb was thought not to be a problem in that the PDFs of nucleons in the nuclear environment were assumed to be the same as the free nucleon.  However, this was determined not to be the case with charged-lepton-nucleus ($\ell^\pm$-A) DIS data that dominated the early study of the nuclear effects in DIS measurements.  In the early '80s, the European Muon Collaboration~\cite{Aubert:1983xm} found that the per-nucleon structure functions $F_{2}$ for iron and deuterium were not only different but also that this difference changed as a function of $x$. This intriguing result initiated an over decade long series
 of follow-up  experiments from~\cite{Bodek:1983qn} up through~\cite{Adams:1995is} 
 %\cite{Dasu:1993vk, Amaudruz:1995tq, Arneodo:1995cs, Arneodo:1995cq,Arneodo:1996rv,nmc,hermes} 
to investigate the nuclear modifications of this ratio, $R[F_{2}^{\ell A}] = (F_{2}^{\ell A} /A) / (F_{2}^{\ell D}/2))$, over a wide range of nuclear targets with atomic number $A$.
These experiments established that  in the  scattering off nucleons within a nucleus in the deep-inelastic region with
$Q^{2}\geq1\ {\rm GeV}^{2}$, the ratio of cross section per nucleon in nuclei to that in deuterium varies considerably in the kinematic range from relatively small $x$ $\sim10^{-2}$ to large $x$ $\sim0.8$.  This led to the need of x-dependent  "nuclear correction factors" (NCF) that scale the results of scattering off nucleons in the nuclear environment to the corresponding result on a free nucleon target. These NCFs were then determined for charged-lepton nucleus scattering.
The behavior of these NCFs, the ratio $R[F_{2}^{\ell A}]$,
can be divided into four regions: 

\begin{itemize}
\item the shadowing region - $R[F_{2}^{\ell A}]$ $\leq$ 1 for x $\lessapprox$ 0.1,
\item the antishadowing region - $R[F_{2}^{\ell A}]$ $\geq$ 1 for 0.1 $\lessapprox$ x $\lessapprox$ 0.25 ,
\item the EMC effect - $R[F_{2}^{\ell A}]$ $ \leq$ 1 for 0.25 $\lessapprox$ x $\lessapprox$ 0.7,
\item and the Fermi motion region - $R[F_{2}^{\ell A}]$ $\geq$ 1 for x $\gtrapprox$ 0.7.
\end{itemize}

\noindent  with no single inclusive model able to explain the nuclear modifications across the whole x range. 

The shadowing suppression at small $x$ is the topic of a rigorous review \cite{Kopeliovich:2012kw}.  Nuclear shadowing had been predicted long before it was observed experimentally in lepton-nucleus interactions. Glauber~\cite{glauber:1959} was the first to suggest that 
a shadowing effect would be due to successive interactions of the impinging object with nucleons in the nucleus.  On the order of 15 years later Gribov~\cite{Gribov:1968jf} suggested that shadowing could be given in terms of elementary diffractive scattering cross sections. Then, at the turn of the century, Strikman and Frankfurt~\cite{Frankfurt:2003zd} generalized the ideas 
of Glauber and Gribov leading to, when combined with the factorization theorem, a QCD leading twist (LT) model,  which again incorporates rescattering of intermediate states.

In most current models, the origin of the shadowing effect is related to the hadronic fluctuations of the intermediate vector boson. This resolved hadronic component of the
IVB will coherently interact several times with the different nucleons in the nucleus$-$multiple scattering. These multiple scatters destructively interfere resulting in a reduction of the corresponding cross sections$-$shadowing. While the basis of the explanation with multiple scattering models is common, phenomenologically there is considerable variation in the details of
application from model to model. The hadronic component of the IVB may be given a partonic structure like in the dipole model~\cite{Kopeliovich:1981pz} or modeled as a superposition of hadronic states like vector meson dominance, or some combination of both approaches.  The models for shadowing were initially developed for charged lepton nucleus scattering, thus the vector current. More
recent studies~\cite{Armesto:2006ph, Fiore:2005bp, Fiore:2005yi} and explicitly~\cite{Kopeliovich:2012kw} based on the dipole model clearly demonstrates that there is a  difference in the shadowing response for the vector and axial vector currents. This is because the electromagnetic and weak interactions take place through the interaction of photons and $W^\pm/Z$ bosons, respectively, with the target hadrons.  Considering the large difference in mass, the hadronic fluctuation processes of photons and $W^\pm/Z$ bosons could be quite different.

 An additional difference in shadowing between the electromagnetic and weak processes is that sea quarks play an important role in this region of low $x$.  The role they play is quite different in the case of  the two processes. For example, the sea quark contribution, though small, is not same for $F_2^{EM}(x,Q^2)$ and $F_2^{WI}(x,Q^2)$ even at 
the free nucleon level and could evolve differently in a nuclear medium. 
% A comparative study of mesonic contributions in the case of electromagnetic and weak structure functions  
% should also be made in view of some recent work done for the DIS cross sections~\cite{Bodek:2010km, Haider:2011qs, prc85}.
% should also be made in view of some recent work in this direction~\cite{Bodek:2010km, Haider:2011qs, prc85}.
 Therefore, a microscopic understanding of the difference between $F_{2A}^{EM}(x,Q^2)$ and $F_{2A}^{WI}(x,Q^2)$ will be very instructive for studying the nuclear
 medium effects in DIS processes as emphasized  at the NuInt15~\cite{Andreopoulos:2019gvw} workshop.
 
Note that with the well-accepted explanation of shadowing involving hadronic fluctuations of the vector boson into quark-antiquark pairs it is important to emphasize, as mentioned earlier, that the nuclear PDFs associated with the low - x shadowing region are not necessarily the PDFs of a single bound nucleon but rather of multiple nucleons in the nuclear environment.
%Note that the currently available experimental data from charged lepton-nucleus DIS scattering have strongly correlated values of $x$ and $Q^2$.  With the requirement that $Q^2$ be large enough in order to apply perturbative QCD (e.g.,~$Q^2 > 1$ GeV$^2$),  the majority of available experimental data covers a relatively limited region in x, particularly for the study of nuclear shadowing iin the small x region.  
%The review by  Geesaman, Saito and Thomas~\cite{Geesaman:1995yd}  summarized the experimental situation in 1995, and  since then there has been limited amounts of new data on shadowing from charged-lepton nucleus scattering in the DIS region.   

The anti-shadowing region is theoretically less well understood but might be explained by the application of momentum, charge, and/or baryon number sum rules. There is work currently underway to follow up on an earlier study~\cite{Brodsky:2004qa} that suggests anti-shadowing is the {\it constructive} interference analog of the shadowing effect. These authors also suggest that anti-shadowing is not universal but rather quark-flavor dependent~\cite{Brodsky:2014hia}, which also suggests the idea of antishadowing is different depending on the interaction being examined.

The modifications at medium $x$ (the so-called {}``EMC effect") are still lacking a convincing, community-accepted explanation, but have often been described as nuclear binding and medium effects \cite{Geesaman:1995yd}.  It has also been shown~\cite{Arrington:2003nt} that this "EMC effect" persists at lower  W in the resonance/transition region albeit at higher $Q^2$ suggesting this is not a purely high-W DIS effect. Along this line, there is now growing quantitative evidence connecting the EMC effect with bound nucleons in short-range correlated (SRC) states~\cite{Hen:2014vua}. This would suggest this effect is not for all nucleons within a nucleus but is exhibited only for nucleons bound in multi-nucleon correlated states.

\begin{figure}[h]
\begin{center}
\includegraphics[width=0.45\textwidth]{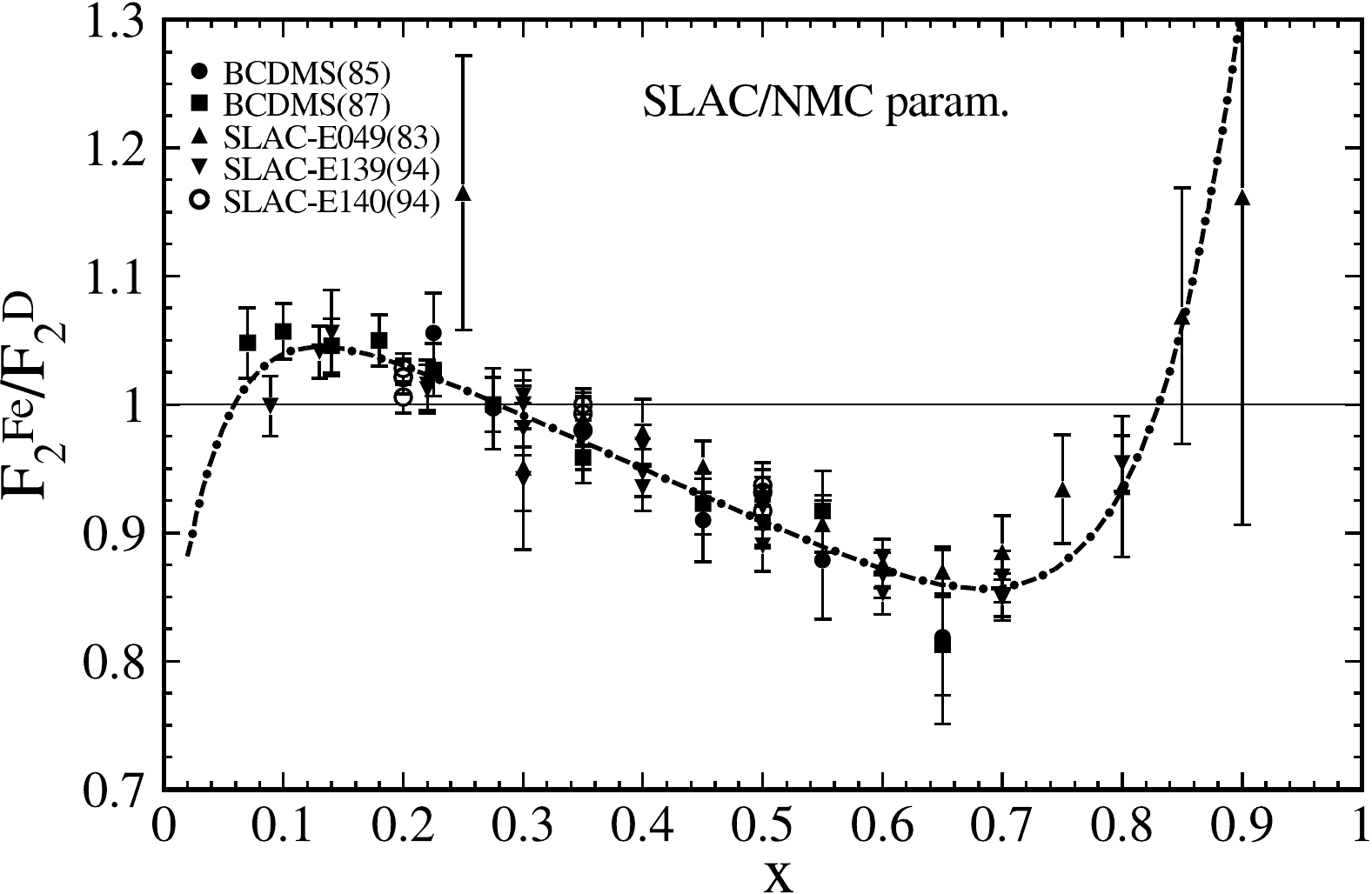}
\caption{Figure from~\cite{Schienbein:2009kk}. Nuclear correction factor, $F_{2}^{Fe}/F_{2}^{D}$, as a function of
$x$. The parametrized curve is compared to SLAC and BCDMS data \cite{Gomez:1993ri,Bari:1985ga,Benvenuti:1987az,Dasu:1993vk,Bodek:1983qn}. 
%The details of the shadowing regime are shown in the following Figure ~\ref{fig:gst-3} and Figure ~\ref{fig:gst-4} for other nuclei since Fe data in the shadowing region did not survive other applied cuts.
}
\label{fig:slac}
\end{center}
\end{figure}

With these qualifications, the evidence for nuclear effects in {\emph charged-lepton} nucleus scattering can be summarized in Fig.~\ref{fig:slac}, which displays the $F_{2}^{Fe}/F_{2}^{D}$ structure function ratio, as measured by both the SLAC e-A and the BCDMS $\mu$-A collaborations. The SLAC/NMC curve is the result of an A-independent parametrization fit to calcium (providing measurements in the shadowing region) and iron charged-lepton nucleus DIS data \cite{Gomez:1993ri,Dasu:1993vk,Owens:2007kp}. 

This SLAC/NMC curve has often been used as the standard nuclear correction factor (NCF) to convert data from a nuclear target to a free-nucleon target for both charged-lepton and neutrino interactions. However concern about the validity of the assumption that the NCF was the same for both charged-lepton and neutrino interactions actually started with a comparison of NuTeV, CCFR and CHORUS results with theory/phenomenology predictions based on charged-lepton scattering results.

\begin{figure*}[h]
\begin{center}
\includegraphics[width=0.65\textwidth]{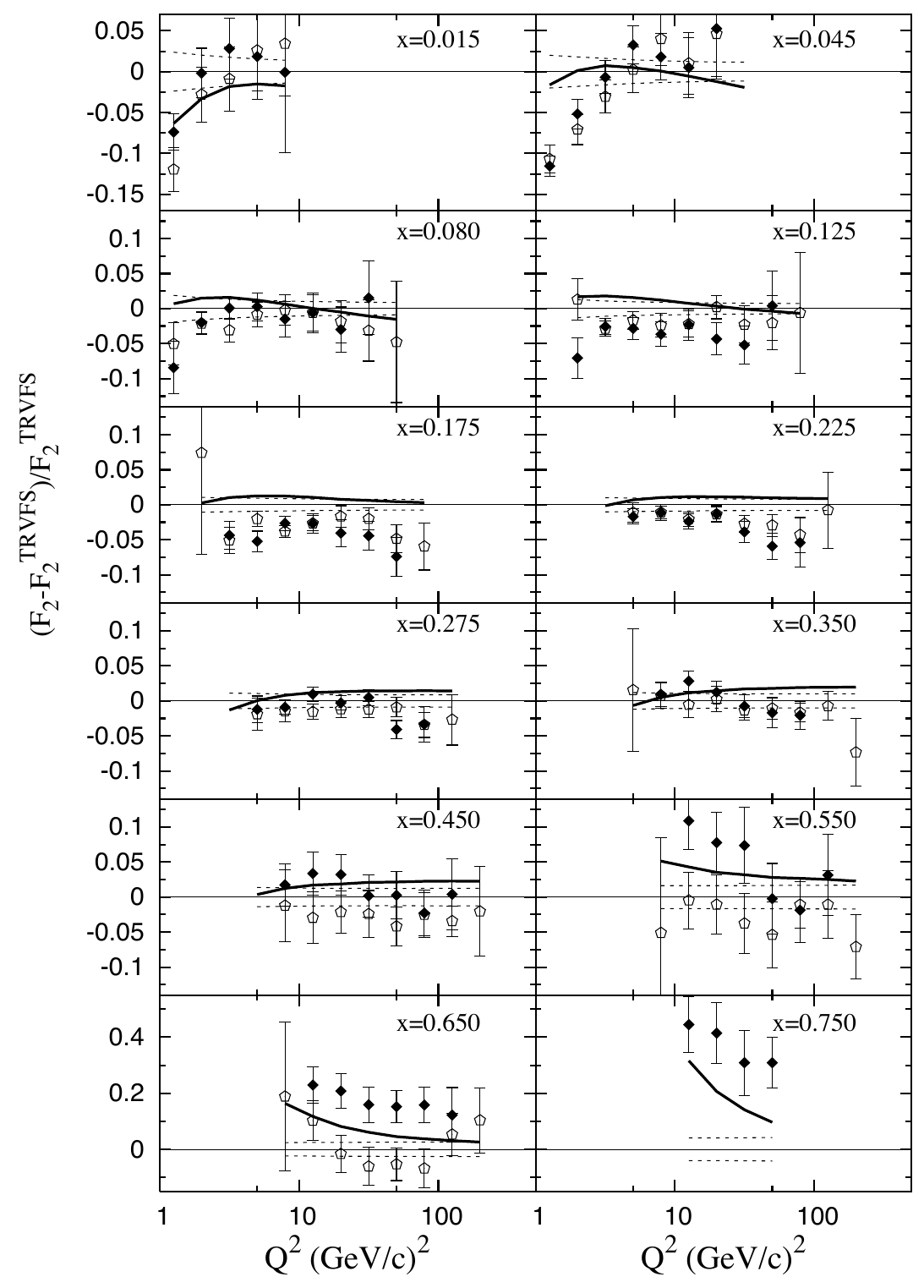}
\caption{Figure from~\cite{Tzanov:2005kr}. A comparison of the measurements of the $F_2$ structure functions by NuTeV (solid dots) and CCFR (open circles) and the predictions from the global PDF fits of the CTEQ collaboration (CTEQ5)  \cite{Lai:1999wy}
(solid line) and TRVFS(MRST2001E) ±1σ (dashed lines). The results are normalized to the Thorne-Roberts variable-flavor scheme (TRVFS) NLO QCD model that used the MRST2001 NLO PDFs~\cite{Thorne:2001gp}.} 

\label{fig:NuTeVThryF2}
\end{center}
\end{figure*}

\begin{figure*}[h]
\begin{center}
\includegraphics[width=0.85\textwidth]{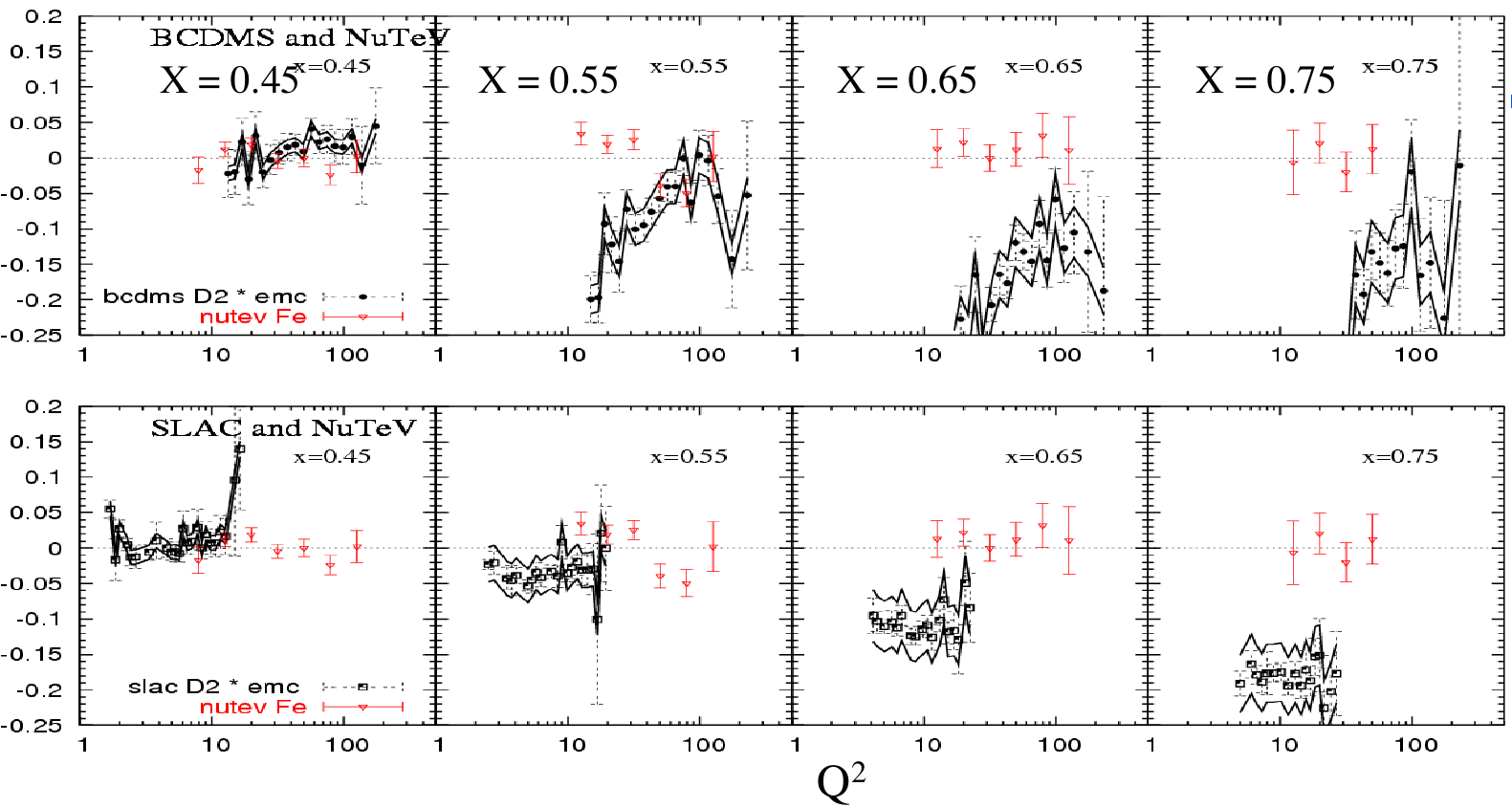}
%\vspace{-1.5cm}
\caption{A further examination of the high-x region of Figure~\ref{fig:NuTeVThryF2} showing the behavior of the NuTeV structure function $F_2$ compared to deuterium measurements from BCDMS and SLAC corrected for the measured (charged lepton) EMC effect on Fe.}

\label{fig:NuTevF2lowhi}
\end{center}
\end{figure*}

\begin{figure*}[h]
\begin{center}
\includegraphics[width=0.65\textwidth]{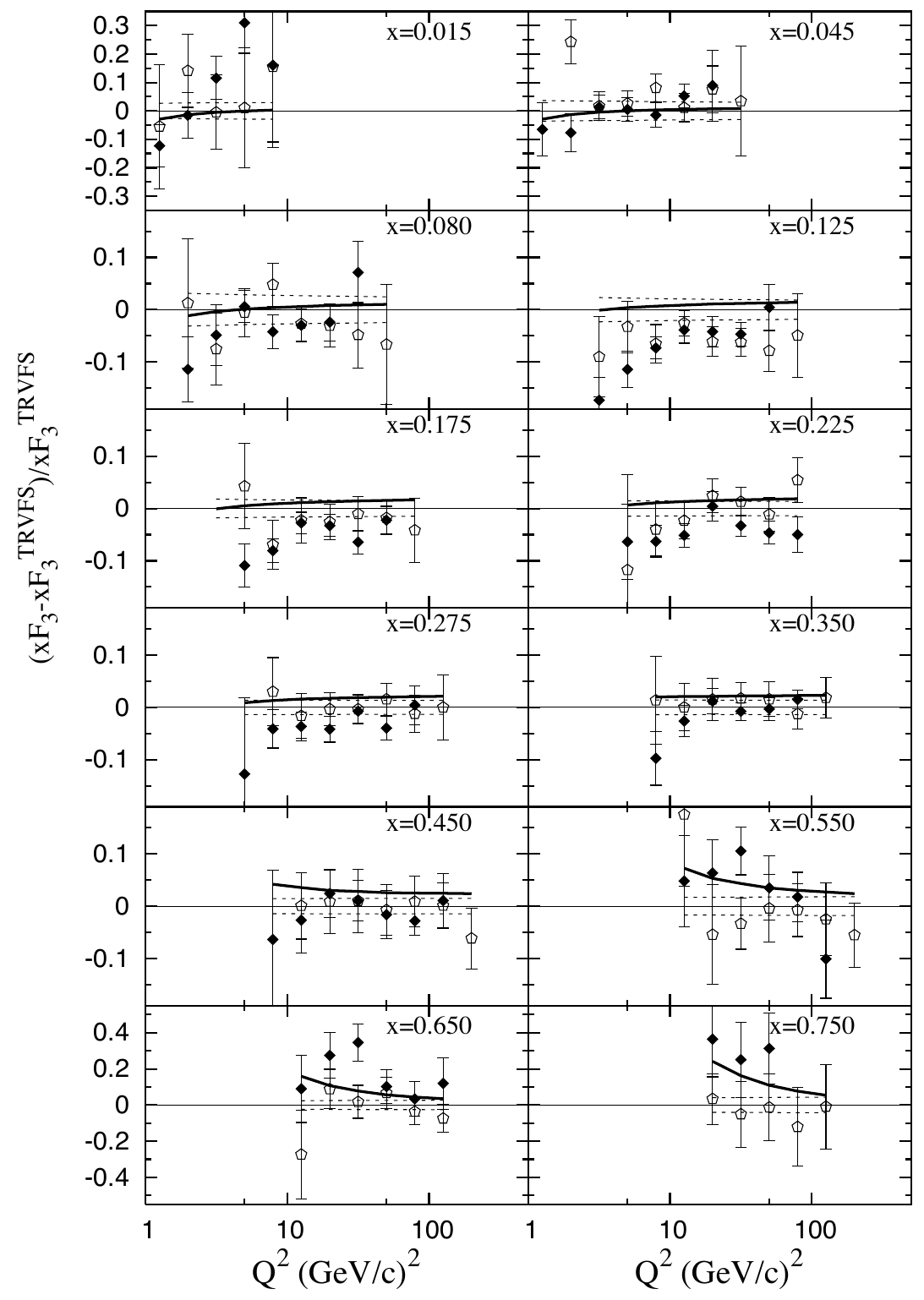}
\caption{Figure from~\cite{Tzanov:2005kr}. A comparison of the measurements of the x$F_3$ structure functions by NuTeV (solid dots) and CCFR (open circles) and the predictions from the global PDF fits of the CTEQ collaboration (CTEQ5)  \cite{Lai:1999wy}
(solid line) and TRVFS(MRST2001E) ±1σ (dashed lines). The results are normalized to the Thorne-Roberts variable-flavor scheme (TRVFS) NLO QCD model that used the MRST2001 NLO PDFs~\cite{Thorne:2001gp}.} 

\label{fig:NuTeVThryxF3}
\end{center}
\end{figure*}

A comparison of the NuTeV results with those of CCFR and the then current predictions of the major free-nucleon PDF-fitting collaborations CTEQ and MRST  ~\cite{Lai:1999wy},\cite{ Thorne:2001gp} are shown in, %Fig.~\ref{fig:NuTeVThryF23} ,
Fig.~\ref{fig:NuTeVThryF2} and Fig.~\ref{fig:NuTeVThryxF3} and, with emphasis on the $F_2$ high-x region, in Fig.~\ref{fig:NuTevF2lowhi}. The CTEQ and MRST curves (labeled as "TRVFS" that used the MRST2001E parton distribution functions) in Fig.~\ref{fig:NuTeVThryF2} and Fig.~\ref{fig:NuTeVThryxF3} %Fig.~\ref{fig:NuTeVThryF23} 
are corrected for nuclear target effects using the $Q^2$-independent \emph {charged-lepton} nuclear correction factors~\cite{deFlorian:2011fp, Conrad:1997ne}, target mass effects~\cite{Georgi:1976ve} and QED radiative effects.  Fig.~\ref{fig:NuTevF2lowhi} emphasizes high-x behavior of these neutrino structure functions compared to the charged-lepton derived structure functions by comparing the NuTeV results with the BCDMS and SLAC measured deuterium structure functions corrected for the measured {\emph charged-lepton} Fe EMC effect.  

It is important to emphasize the observation that NuTeV structure functions agree with the e/$\mu$-based theoretical calculations for $0.30 \le x \le 0.5$. However, for x $\le$ 0.08 both NuTeV and CCFR measure quite different $Q^2$-dependence than the charged-lepton-based theoretical predictions while for $0.08 \le x \le 0.3$ both NuTeV and CCFR results tend to be somewhat lower than the charged-lepton-based predictions.  At high-x,  $\ge$ 0.50 both NuTeV and CCFR results are systematically higher than the charged-lepton-based theoretical predictions. 

The conclusion of the  NuTeV collaboration was that their results suggest neutrino scattering favors \emph {smaller} nuclear effects compared to charged-lepton scattering\footnote{From Ref.~\cite{Tzanov:2005kr} "NuTeV perhaps indicates that neutrino scattering favors smaller nuclear effects at high-x than are found in charged-lepton scattering."}. It was then not a complete surprise that challenges were found when attempting to combine these $\nu(\nub)$-Fe results  with $\ell^\pm$-Fe and then, using $\ell^\pm$-A nuclear correction factors, to combine both with scattering results from free nucleons in global fits.  To further test for this suggested difference in charged-lepton and neutrino NCFs, the nuclear parton distribution functions were extracted independently by the nCTEQ collaboration for charged-lepton-based and neutrino-based event samples.

%TO BE INCLUDED IN LATER SECTION - At small x, theoretical calculations specifically for $\nu$ nucleus scattering suggest that in the shadowing region the nuclear correction for neutrinos has $Q^2$ dependence~\cite{Qiu:2004qk}. The standard nuclear correction obtained from a fit to charged lepton data implies a suppression of $\approx 10\%$ independent of $Q^2$ at x = 0.015, while for x = 0.015 reference~\cite{Qiu:2004qk} finds a suppression of 15\% at $Q^2 = 1.25 GeV^2$ and a suppression of 3.4\% at $Q^2 = 8.0 GeV^2$. This effect improves agreement with data at low-x.

% \begin{figure*}[h]
% \begin{picture}(500,300)(0,0) %   \graphpaper[20](0,0)(500,160)
%  \put(50,0){\includegraphics[width=0.46\textwidth]{NuTeVThryF2}}
% \put(260,0){\includegraphics[width=0.45\textwidth]{NuTeVThryxF3}}
% %\put(114,0){$(a)$} \put(374,0){$(b)$} 
% \end{picture} 
% \caption{Figure from~\cite{Tzanov:2005kr}. A comparison of the measurements of the $F_2$ (left) and x$F_3$ (right) structure functions by NuTeV (solid dots) and CCFR (open circles) and the predictions from the global PDF fits of the CTEQ collaboration (CTEQ5)  \cite{Lai:1999wy}
% (solid line) and TRVFS(MRST2001E) ±1σ (dashed lines). The results are normalized to the Thorne-Roberts variable-flavor scheme (TRVFS) NLO QCD model that used the MRST2001 NLO PDFs~\cite{Thorne:2001gp}.  
% %The model predictions have been corrected for target mass and, most significantly, for nuclear effects {\em assuming these corrections are the same for charge-lepton and neutrino interactions}
% }
% \label{fig:NuTeVThryF23}
% \end{figure*}

\subsection{Nuclear Parton Distribution Functions}
\label{Subsec-nPDF}
 
%{\bf The paragraph mentioned below has common features from section 5.1 first para}

It is obvious from Fig.~\ref{fig:slac} that the structure function of nucleons within a nucleus are different from the free nucleon structure functions.  Assuming that both free nucleons and nucleons in nuclei can be described with parton distribution functions (PDFs),  this suggests that the PDFs for a nucleon within the nuclear environment (nuclear parton distribution functions - nPDFs) will be different than those of the free nucleon. The partonic structure of these nucleons within a nucleus must reflect the nuclear environment and, as has been mentioned, in some regions of x can better be considered as "effective" nPDFs representing the interaction with multiple nucleons within the nucleus. Consequently, the nucleus cannot simply be considered as an ensemble of Z free protons PDFs and (A-Z) free neutron PDFs.  

Currently the analyses of both free nucleons and  nucleons within a nuclear environment are based on the same factorization theorems ~\cite{Collins:1985ue}-\cite{Collins:1998rz} that do not in any way consider the relevant nuclear environment. The PDFs of a free proton are extremely well studied with several global analyses of free proton PDFs regularly updated~\cite{Harland-Lang:2014zoa,Alekhin:2013nda,Stirling:2005td},\cite{Ball:2014uwa}-\cite{Jimenez-Delgado:2014twa}. Nuclear PDFs have been determined by several groups ~\cite{Hirai:2007sx}-\cite{deFlorian:2011fp}, \cite{Schienbein:2009kk} using global fits to experimental data that include, mainly, deep inelastic scattering and Drell-Yan lepton pair production on nuclei. However the fits can also include information from the LHC when nuclear ions are accelerated.

%The standard DGLAP equation is used to evolve quantities from one Q to another.

Our knowledge of nuclear PDFs is much less advanced than the free nucleon PDFs due to both theoretical and experimental limitations\footnote{The discussions and methods of the nCTEQ collaboration as presented in detail in the publication K. Kovarık et al.
%A. Kusina, T. Jezo, D. B. Clark, C. Keppel, F. Lyonnet, J. G. Morf\'{\i}n, F. I. Olness, J.F. Owens, I. Schienbein and J. Yu
~\cite{Kovarik:2015cma} are the basis for this section and will serve as an example of the process of determining nuclear PDFs.}. 
%As already indicated, it is assumed that the QCD factorization theorems are also valid for global nuclear PDF analyses of lepton-nucleus processes.  
For example, consider that there is a contribution to nuclear PDFs coming from  x $\ge$ 1.0 (expected to be rather small) that is mainly due to short-range correlated nucleon pairs allowed with nuclear targets but not currently included in the fits to nPDfs. 
%\footnote{Notably there is a fairly well established connection between x $\ge$ 1.0 events and SRC nucleon pairs.}  
Allowing this restriction, the nuclear proton nPDFs are assumed to have the same evolution equations and obey the same sum rules as the free proton PDFs. However, 
%the theoretical description of nuclear induced hard processes is more challenging due to the complex nuclear environment. %It is only in this context that the universal nuclear parton distributions ($f_i^A $(x, Q)) can be defined.  
the nuclear PDFs must account for nuclear effects such as shadowing, anti-shadowing and the EMC effect at leading twist.   
%Although these named nuclear effects are \emph{not} higher twist effects, 
Although higher twist contributions had been expected to be enhanced in a nucleus~\cite{Qiu:2003vd,Accardi:2003qn} due to the scattering of the outgoing partons through the nuclear medium, the theoretical analysis presented at the start of this review found that, on the contrary, as long as target mass effects are included the need for additional dynamical higher twist contributions is quite small.
%these higher twist effects can either be  included in the analysis or eliminated by kinematic cuts.  

The other challenge with nuclear PDFs is the lack of precise experimental data. Currently, the experimental constraints on the nPDFs for any single nucleus, except iron, are quite limited. Since for a global multi-nucleus fit, data from multiple nuclei must be included simultaneously in the fits,  
%more assumptions have to be made. Since the nuclear effects are clearly dependent on the number of nucleons within the nucleus, 
the non-trivial nuclear A dependence of the PDFs must be considered  by including a parmetrization of the A-dependence. The constraints on this parametrization are only as strong as the accuracy of the data in the fit.%{\bf---- Can we delete this part ?}

In spite of these challenges, as long as the fit was charged-lepton-based and \emph{the more accurate $\numu$-A DIS data were not used in these fits} the existing global nPDF analyses generally led to a reasonable description of the data confirming this picture.
There are essentially three types of global fits to determine the nPDFs:

\begin{itemize}
  \item Those that fit a multiplicative correction factor to apply to the free nucleon PDFs. 
\begin{center}
$f_{i}^{(p/A)}(x,Q) = R_i(x,Q,A) f_{i}^{free~proton}(x,Q)$ 
\end{center}
This method was used by the groups that pioneered the extraction of nPDFs~\cite{Hirai:2007sx}-\cite{deFlorian:2011fp}.
\item An attempt was made to use a convolution method~\cite{deFlorian:2003qf} to isolate the nPDFs.
\item And finally the method of native nuclear PDFs extracted using the same procedure as the free nucleon PDFs.
\end{itemize}

It is this last method employed by the \href{https://ncteq.hepforge.org}{Nuclear CTEQ Collaboration(nCTEQ)} group, a subgroup of the full CTEQ collaboration, that will be used as an example to describe the extraction of nPDFs in more detail.

%\paragraph{The nCTEQ Determination of the Nuclear PDFs}

In the nCTEQ framework~\cite{Kovarik:2015cma}, the parton distributions of the nucleus are constructed as:
\begin{equation}
f_{i}^{(A,Z)}(x,Q) = \frac{Z}{A}f_{i}^{p/A}(x,Q) + \frac{A-Z}{A}f_{i}^{n/A}(x,Q),
\label{eq:nucleusPDF}
\end{equation} 
%{\bf where $Z$ is number of protons and $A$ number of protons plus neutrons in the nucleus.--- no need, we have already define Z and A}  
Isospin symmetry is used to construct the PDFs of a neutron in the nucleus, $f_{i}^{n/A}(x,Q)$, by exchanging up- and down-quark distributions from those of the proton.

The parametrization of individual parton distributions are similar in form to that used in the free proton CTEQ
fits~\cite{Owens:2007kp,Pumplin:2002vw,Stump:2003yu}
and takes the following form at the input scale $Q_0$:

\begin{eqnarray}
%\begin{split}
xf_{i}^{p/A}(x,Q_{0}) = c_{0}\,x^{c_{1}}(1-x)^{c_{2}}e^{c_{3}x}(1+e^{c_{4}}x)^{c_{5}} \nonumber \\
%{\rm for}\quad &   
i=u_{v},d_{v},g,\bar{u}+\bar{d},s+\bar{s},s-\bar{s},\nonumber \\[2mm]
\frac{\bar{d}(x,Q_{0})}{\bar{u}(x,Q_{0})}
= c_{0}\,x^{c_{1}}(1-x)^{c_{2}}+(1+c_{3}x)(1-x)^{c_{4}}.
%\end{split}
\label{eq:param2}
\end{eqnarray}

The input scale is chosen to be the same as for the free proton fits~\cite{Owens:2007kp,Stump:2003yu}, namely $Q_0=1.3$ GeV and the DGLAP equation is used to evolve to higher Q. There is currently on-going discussions within the nCTEQ collaboration on adjusting $Q_0$ to a lower value to better reflect the $Q^2$ range of current neutrino experiments.

As in the other available nuclear PDFs~\cite{Hirai:2007sx}-\cite{deFlorian:2011fp},
nuclear targets are characterized by their atomic mass number $A$. However, in contrast to
those groups that derive a multiplicative factor to apply to the free proton PDFs, in the nCTEQ analysis the additional $A$ dependence is introduced directly to the $c$-coefficients $c_{k}\to c_{k}(A)$ in Eq. ~\ref{eq:param2}.
%\begin{equation}
%\begin{split}
%c_{k}\to c_{k}(A) \equiv c_{k,0}+c_{k,1}\left(1-A^{-c_{k,2}}\right),
 %\\
%\quad k=\{1,\ldots,5\}.
%\end{split}\label{eq:Adep}
%\end{equation}
The $c_{k}(A)$ are defined such that for $A=1$ one recovers the underlying PDFs of a free proton that are described in~\cite{Owens:2007kp} and which have the advantage of minimal influence from nuclear data.
%It is important to emphasize again that this form of parameterizing the nPDF coefficients specifically for a given nucleus A is different than other nPDF fits that result in a ratio of the nPDFs to the free nucleon PDFs.

\paragraph{nCTEQ nPDFs for a nucleus $A$ without including $\nu A$ results as input}

The data currently used in this global fit for nPDFs are from charged lepton DIS, and Drell-Yan lepton pair production experiments and are subject to the following cuts:
\begin{itemize}
\item DIS: $Q>2$ GeV and $W>3.5$ GeV,
\item DY: $M \ge 2$ GeV,\\
(where $M$ is the invariant mass of the produced lepton pair)  
%
%\item  $\pi^0$ production: $p_T>1.7$ GeV.
\end{itemize}
 These cuts are considerably more restrictive than other nuclear PDF analyses with the goal of limiting the importance of both kinematic and dynamic higher twists in the fit.  
 %With these cuts the number of data points used in the fit, 740 total, are 414 $F_2^A / F_2^D$, 202 $F_2^A / F_2^A^'$ and 92 Drell-Yan production.  

\begin{figure}[th]
\centering{}
\includegraphics[width=0.48\textwidth]{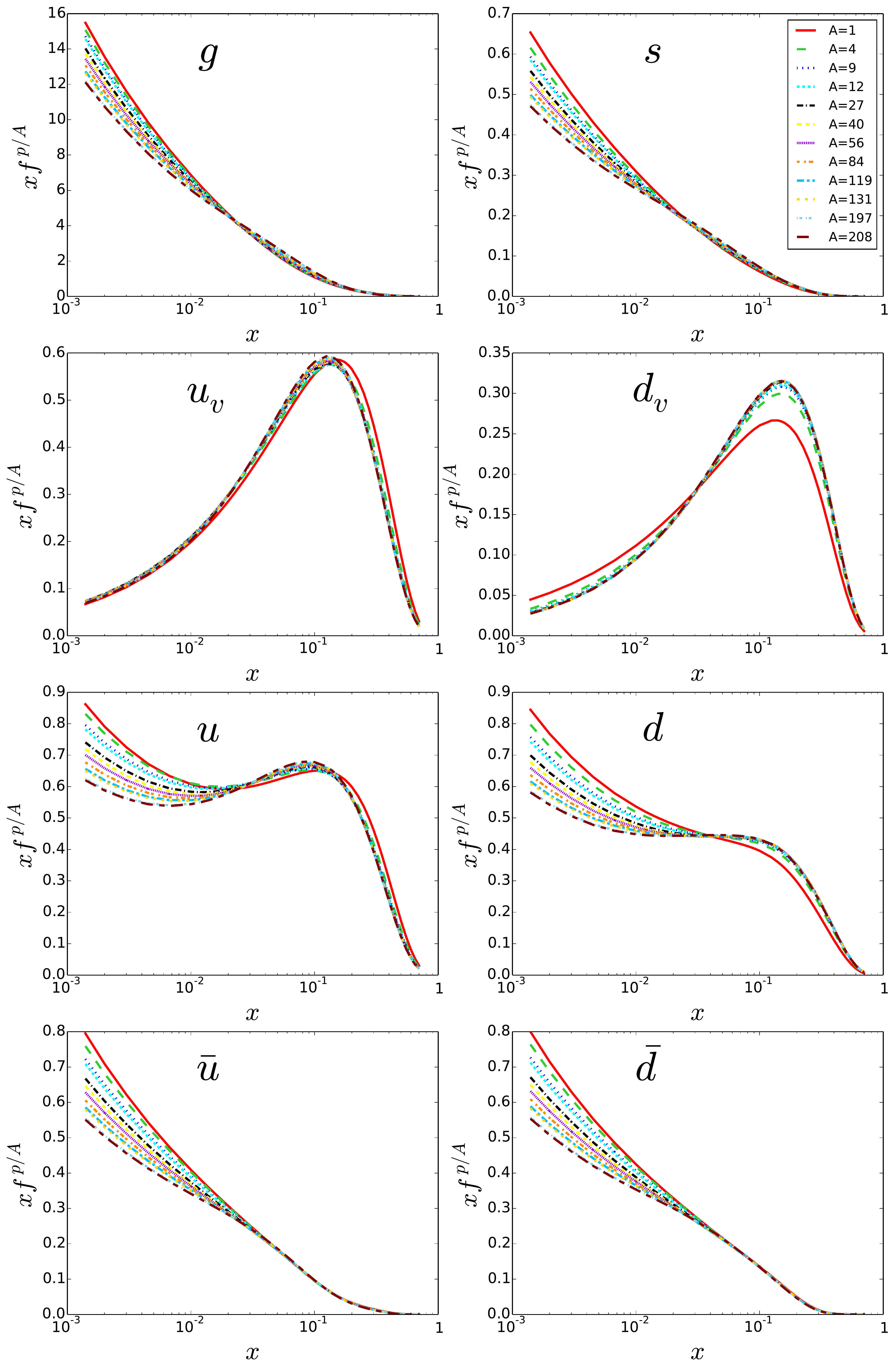}
\caption{Figure from~\cite{Kovarik:2015cma}. The A-dependence of the nCTEQ nuclear proton PDFs at the scale $Q=10$ GeV 
for a range of nuclei from the free proton ($A=1$) to lead ($A=208$).
}
\label{fig:compar_PDFs_diff-nuc}
\end{figure}

The results of this nCTEQ fit (labeled nCTEQ15 in the literature) yield the $A$-dependence of the
various nPDF flavors of a proton in nucleus A illustrated in Fig.~\ref{fig:compar_PDFs_diff-nuc} where the  central fit predictions for a range of nuclear $A$ values from $A=1$ (proton) to $A=208$ (lead) are displayed.
%----------------

%----------------
\begin{figure*}[th]
\centering{}
%\includegraphics[clip,width=0.48\textwidth]{nCTEQ15-PDFs-ratio_Q1.pdf}
%\quad{}
\includegraphics[width=0.55\textwidth]{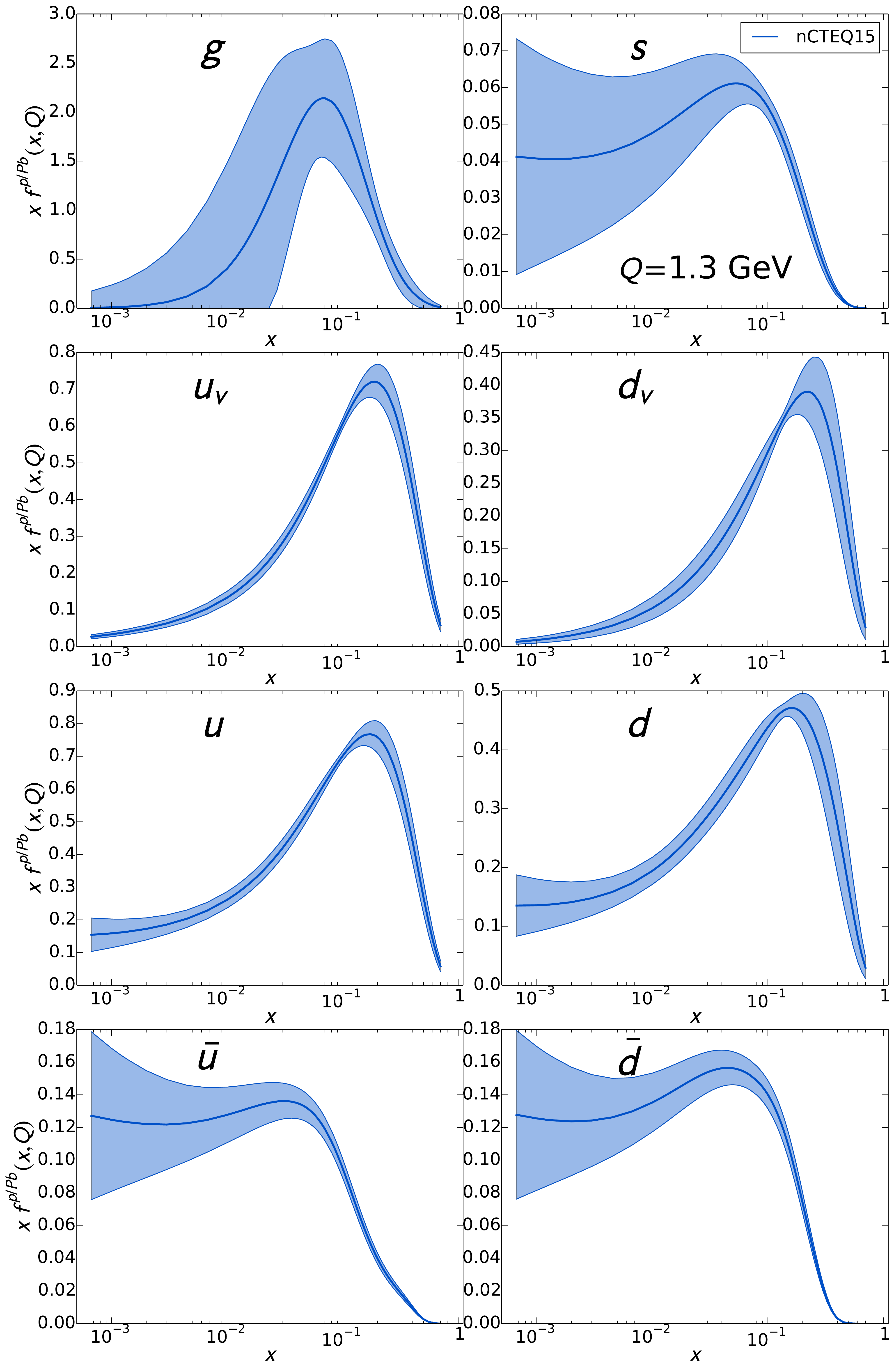}
\caption{Figure from~\cite{Kovarik:2015cma}. Results of the nCTEQ fit displaying the
actual PDFs for a proton in lead at the $Q_0$ scale of $Q=1.3$ GeV.}
\label{fig:nCTEQ15g_1}
\end{figure*}
%----------------

%When examining the $A$-dependence we observe 
%that as we move to larger $A$ the gluon and sea-distributions
%$\{g, \bar{u}, \bar{d}, s \}$ decrease at small $x$ values. 
%This trend is also present for the $\{u, d\}$ PDFs.
%On the other hand, the $A$-dependence of $\{u_v, d_v\}$ distributions
%is reduced relative to the other flavor  components.

Fig.~\ref{fig:nCTEQ15g_1} shows the nPDFs ($f^{p/Pb}$) for a proton  in a lead nucleus
at the input scale $Q=Q_0=1.3$ GeV.  The uncertainty bands arising from 
the error PDF sets  based upon the Hessian method with tolerance criterion are also shown.
Note that the uncertainty bands for $x\le 10^{-2}$ and $x\ge 0.7$ are
not directly constrained by data but only by the momentum and baryon number sum rules.

It should be emphasized that these "nCTEQ15" nuclear PDFs fit described here did not contain any input $\nu/\nub$-A scattering results.  The next section will describe the nCTEQ approach to determining the nPDFs for neutrino nucleus scattering.

\subsection{Nuclear Correction Factors for Neutrino Nucleus Scattering}
\label{Subsec-nuNucCor}
%$F_2^\nu(x,Q^2)$ and  $F_2^{\bar\nu}(x,Q^2)$
%Since very limited information was available on nuclear modifications of cross sections and structure functions in $\nu(\nub)$ nucleus inelastic interactions in the earliest global fits of PDFs, the nuclear correction factors Fig.~\ref{fig:slac} determined from charged-lepton nucleus interactions were simply assumed to apply also to $\nu A$ scattering.  
The first attempt at measuring nuclear effects, yielding a nuclear correction factor, with $\nu$ was performed by the BEBC bubble chamber experiment from the ratio of neon and hydrogen targets~\cite{Allport:1989vf} in the mixed Ne-H filling of the chamber. The measurement provided a suggestion of nuclear shadowing at small $x$ and $Q^2$ values, however, the large associated errors of these lower statistics measurements precluded any careful comparison with charged-lepton results.   Consequently, in earlier QCD global fits of nucleon PDFs  that attempted to include neutrino nuclear DIS data, the charged-lepton nuclear correction factors (Fig.~\ref{fig:slac}) were simply applied to neutrino nucleus scattering results as well.  

It was immediately noted that these early attempts to include neutrino-nucleus DIS scattering data, corrected with charged-lepton NCFs, introduced such tension in the shadowing region at low-x in global QCD fits that the low-x neutrino data was simply excluded in these early CTEQ nucleon PDF global analyses.  In more recent examinations of higher-x parton distribution functions, carried out by the CTEQ collaboration~\cite{Owens:2007kp, Owens:2007zz}, indications began to accumulate that the nuclear correction factors for neutrino nucleus scattering {\em not only in the shadowing region} could indeed be different than those for charged-lepton nucleus scattering.  A conclusion already voiced and quoted by the NuTeV collaboration 
%\footnote{"NuTeV perhaps indicates that neutrino scattering favors smaller nuclear effects at high-x than are found in charged-lepton scattering." ~\cite{Tzanov:2005kr}}.

A study to check these indications was then initiated by the nCTEQ collaboration to extract the neutrino nuclear correction factor $F_2^{\nu A}(x,Q^2)$ / $F_2^{\nu N}(x,Q^2)$.  The same procedure used to determine the correction factor for charged lepton nucleus scattering that resulted in the SLAC/NMC curve, was used. \footnote{It should  be apparent that the rather restrictive $Q_0$ and DIS minimal $Q^2$ and W cuts from the charged-lepton-based fits when applied to neutrino scattering results would rule out most contributions from contemporary neutrino nucleus experiments and are thus also being carefully reconsidered.} To apply this procedure  to $\nu$-A scattering, there were several data sets considered.  The earliest is the CDHSW $\nu$-Fe data followed by the CCFR $\nu$-Fe data, the NuTeV $\nu$-Fe data and finally the CHORUS $\nu$-Pb data. The weights of these data sets in the combined fit were dictated by the errors on the data. The NuTeV $\nu$-Fe and CHORUS  $\nu$-Pb  data had associated full covariant error treatment of the data, yielding maximal discriminatory power of the data.   The weight of the CDHSW and CCFR data, with their errors calculated via the sum of the squares of statistical and systematic errors, when combined with the NuTeV and CHORUS data with their full covariant error matrix for the fit, was greatly reduced.  Furthermore, even though both the NuTeV and CHORUS data sets have full covariant error matrices, the relatively small NuTeV errors with respect to the CHORUS errors enabled the NuTeV data points to dominate the combined fit.

An additional input to the fits was the NuTeV and CCFR di-muon data~\cite{Goncharov:2001qe} off Fe, which are sensitive to the strange quark content of the nucleon in the nuclear environment of Fe.   However, no other data such as charged-lepton  nucleus ($\ell^{\pm}$A) and DY were used.  Because the neutrinos alone do not have the power to constrain all of the PDF components, a minimal set of external constraints~\cite{Schienbein:2007fs} also had to be employed and some of these external assumptions do indeed affect the behavior of the fit parton distributions at small x - the shadowing region.  These include the Callan-Gross relation ($F_2^{\nu A} = 2 x F_1^{\nu A}$) as well as use of the assumption s = $\overline s$ and c = $\overline c$.  In subsequent fits of neutrino data, the results of the NuTeV analysis~\cite{Mason:2007zza} of the s-$\overline s$ asymmetry will be included.
%%\cite{Mason:2007zz}
%\bibitem{Mason:2007zz} 
 % D.~Mason {\it et al.} [NuTeV Collaboration],
 % Phys.\ Rev.\ Lett.\  {\bf 99}, 192001 (2007).
 % doi:10.1103/PhysRevLett.99.192001
  %%CITATION = doi:10.1103/PhysRevLett.99.192001;%%
  %117 citations counted in INSPIRE as of 13 Oct 2019
  
\begin{figure*}
\begin{picture}(500,155)(0,0) %   \graphpaper[20](0,0)(500,160)
 \put(20,0){\includegraphics[width=0.45\textwidth]{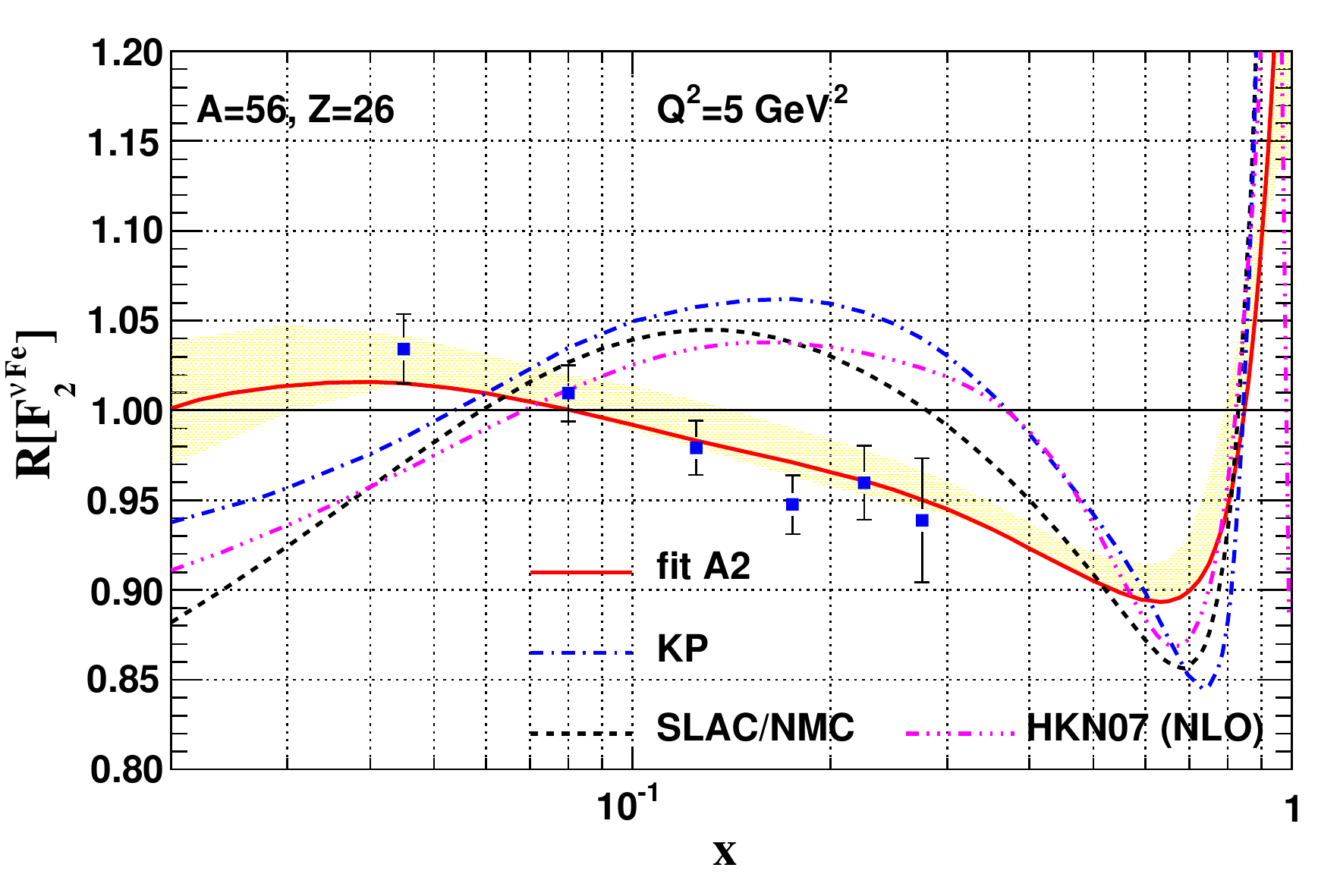}}
\put(240,0){\includegraphics[width=0.45\textwidth]{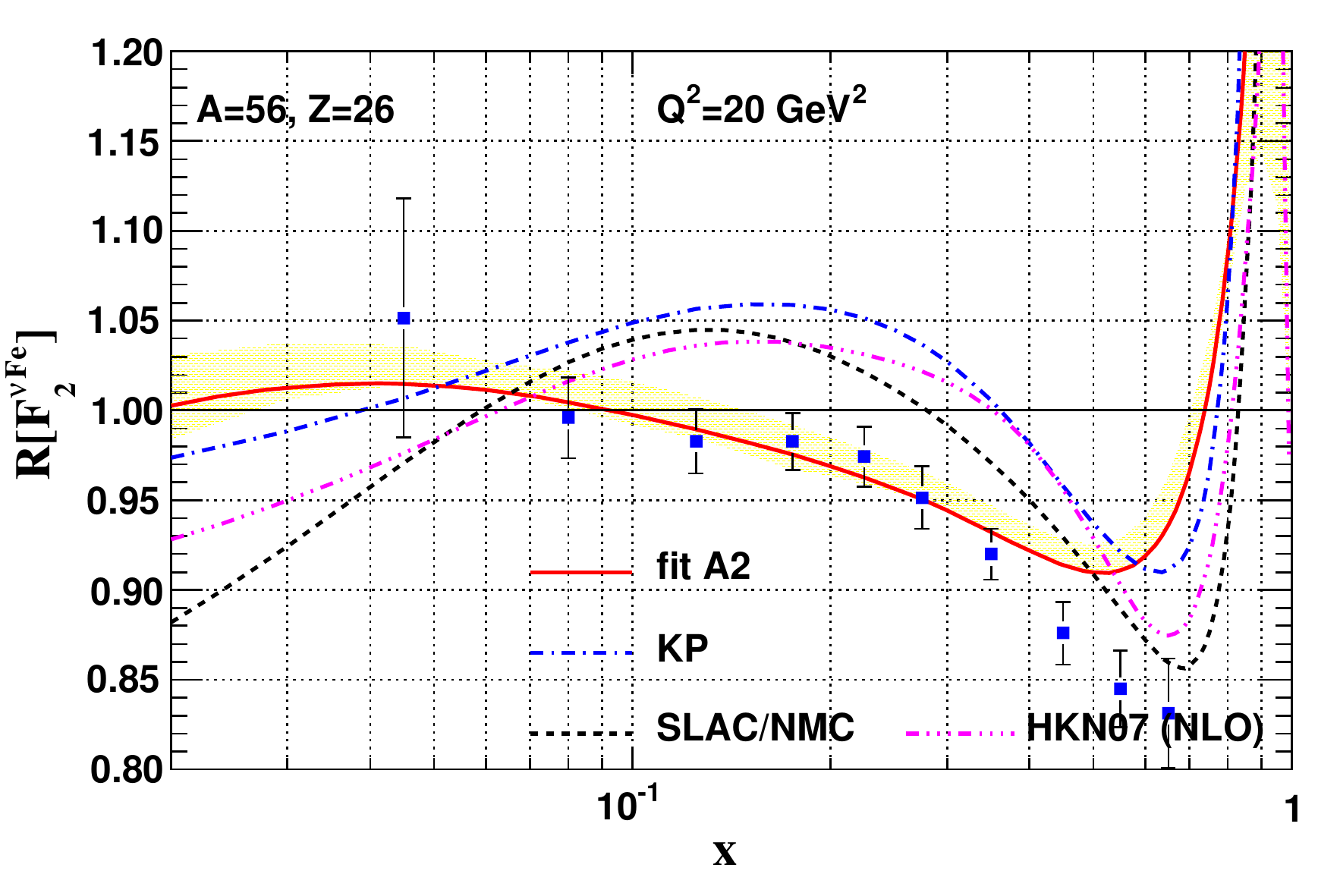}}
\put(134,0){$(a)$} \put(354,0){$(b)$} \end{picture} 
\caption{Figure from~\cite{Schienbein:2009kk}. 
Nuclear correction factor $R$ for the structure function $F_2$ in charged current $\nu Fe$ scattering at  a)~$Q^2=5~GeV^2$ and b)~$Q^2=20~GeV^2$.  The  solid curve shows the result of the nCTEQ analysis of NuTeV differential cross sections (labeled fit A2), divided by the results obtained with the reference fit (free-proton) PDFs;  the uncertainty from the A2 fit is represented by the yellow band. Plotted also are NuTeV data points of the average $F_2$ to illustrate the consistency of the fit with the input points. For comparison the correction factor from the Kulagin--Petti (KP) model \protect\cite{Kulagin:2006dg} (dashed-dot line),  from the Hirai, Kumano, Nagai (HKN07) fit  \protect\cite{Hirai:2007sx} (dashed-dotted line), and the SLAC/NMC parametrization, Fig.~\ref{fig:slac} (dashed line) of the charged-lepton nuclear correction factor are also shown.
We compute this for $\{A=56,Z=26\}$.
}
\label{fig:fig6a}
\end{figure*}

\begin{figure}
\begin{picture}(500,155)(0,0) %   \graphpaper[20](0,0)(500,160)
 \put(20,0){\includegraphics[width=0.45\textwidth]{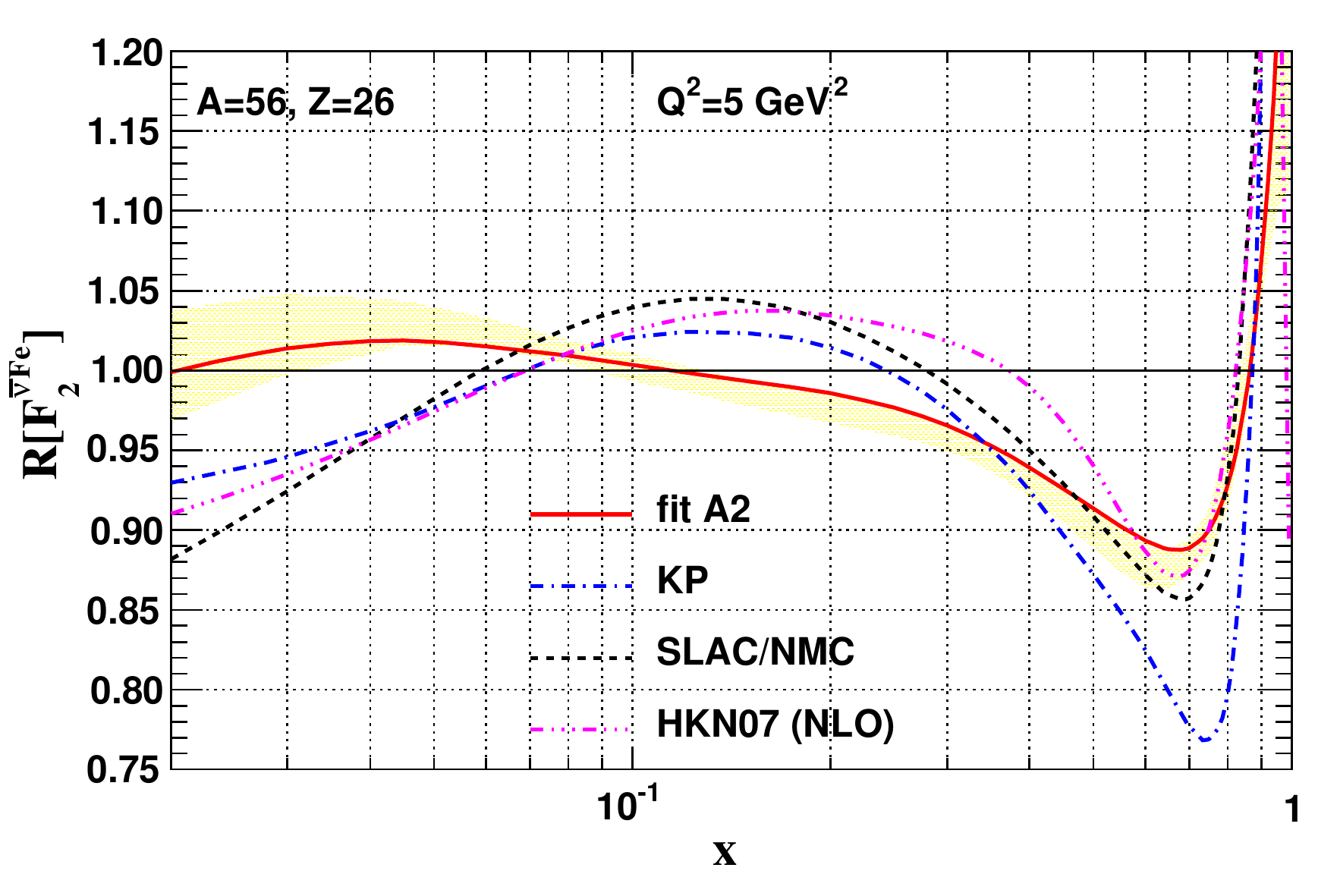}}
\put(240,0){\includegraphics[width=0.45\textwidth]{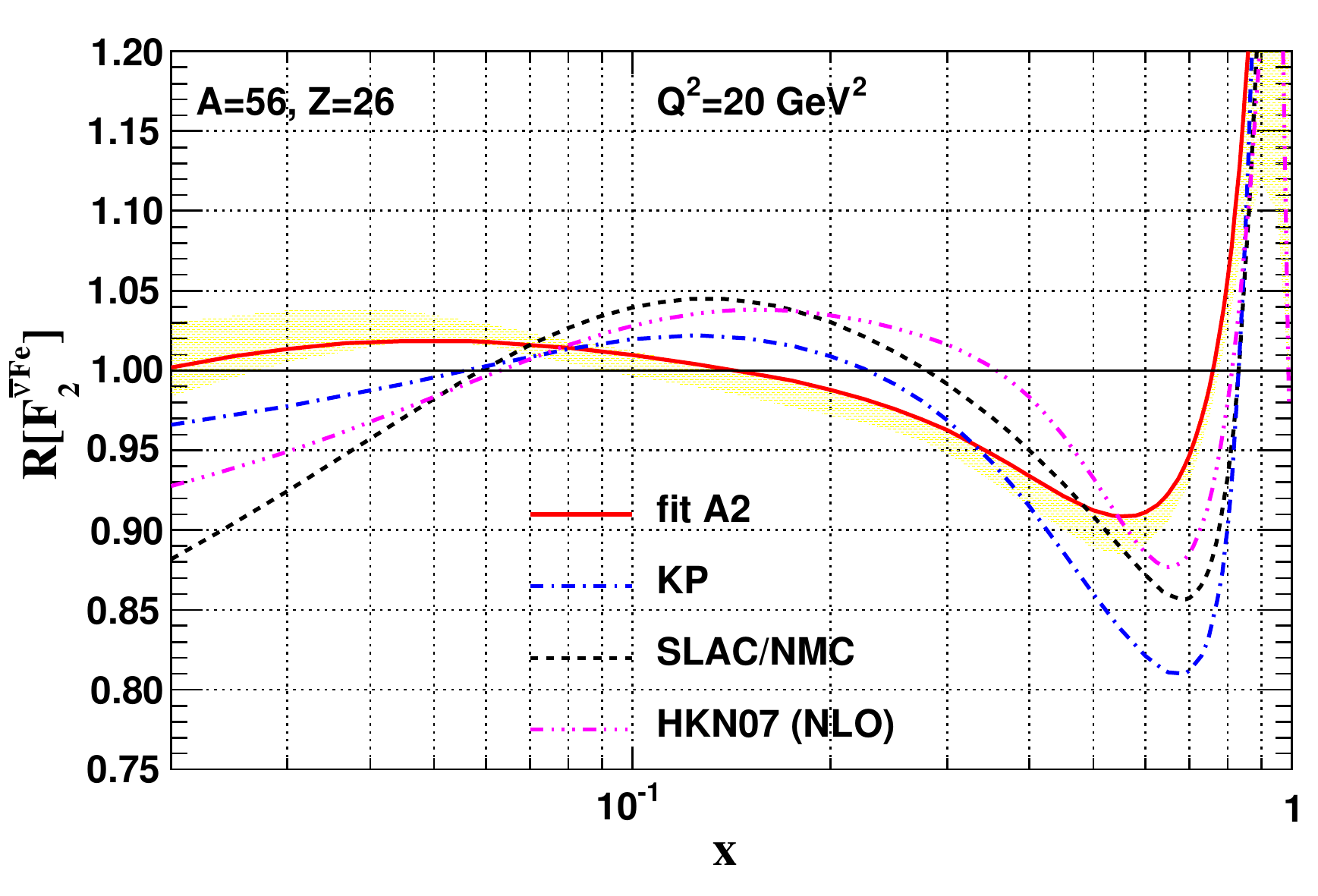}}
\put(134,0){$(a)$} \put(354,0){$(b)$} \end{picture} 
\caption{
The same as in Figure~\protect\ref{fig:fig6a} for $\overline{\nu} Fe$
scattering.
%We compute this for $\{A=56,Z=26\}$.
}
\label{fig:fig6b}
\end{figure}

It is important to note that the nCTEQ fit was made directly to the NuTeV and CHORUS measured double differential cross sections in order to extract the set of nPDFs of the nucleon in the nucleus.  The fit did {\em not} use the extracted NuTeV and CHORUS {\em structure function} results of the average value of $F_2(x, Q^2)$, which contains all the nuclear-dependent assumptions made to extract them such as, presumably A-dependent, $R_{em}(\sigma^{em}_L / \sigma^{em}_T)$ being used instead of $R_{weak}(\sigma^{WI}_L / \sigma^{WI}_T)$ and $\Delta~xF_3$.  The extracted nPDFs were then taken in ratio to the  free-nucleon PDFs~\cite{Owens:2007kp} to form the individual values of the nuclear correction factor R for a given x and $Q^2$.  It is also important to note that these free-nucleon PDFs that were used in the denominator of the nuclear correction factors were a special fit to ensure that any data involving nuclear targets was minimally involved.  These fits were performed separately for neutrino and anti-neutrino - not the average of both - as shown in Fig.~\ref{fig:fig6a} for $\nu$--$Fe$  and in Fig.~\ref{fig:fig6b} for $\bar\nu$-$Fe$.

%Since the difference between $F_2(\nu A)$ and $F_2(\bar\nu A)$ is small, 
%the consistency of these extracted values of $F_2(x, Q^2)$ with the measured average values from NuTeV can be shown. 
It was also possible to combine the fitted neutrino nPDFs to form the average of $F_2(\nu A)$ and $F_2(\bar\nu A)$ for a given x, $Q^2$ to compare directly with the NuTeV published values of this quantity.  This was also performed by nCTEQ  and results can be found in~\cite{Kovarik:2012pm}.
These studies by nCTEQ~\cite{Kovarik:2012zz} have shown a strong indication that there is indeed a difference between the $\ell^\pm$ A and the $\nu A$ nuclear correction factors.  An analysis by the HKN~\cite{Nakamura:2016cnn} group also finds some inconsistencies between $\nu(\nub)$ and charged-lepton data and most recently, a direct comparison~\cite{Kalantarians:2017mkj} of $F_2^{\nu Fe}$ with $F_2^{\ell^{\pm} Fe}$  structure functions observed a clear ($\approx{20\%}$) difference between $\nu(\nub)$ and charged lepton scattering off Fe for the structure functions at low x. 

%It is also of course possible to combine these fitted nPDFs to form the individual values of the average of $F_2(\nu A)$ and $F_2(\bar\nu A)$ for a given x and $Q^2$, to compare directly with the NuTeV published values of this quantity.  This was done and the nCTEQ  results are shown in Fig.~\ref{fig:F2lowQ}.
%and those of $\bar\nu$--$Fe$ are shown in Fig.~\ref{fig:fig6b}. 
%Note the ratios shown are using the published NuTeV values for the average value of $F_2(\nu A)$. 
%and   $F_2(\bar\nu A)$. 

\subsection{Comparison of the $\ell^{\pm}A$ and $\nu A$ Nuclear Correction Factors}
\label{Subsec-e_nuNucCor}

%In addition, the data sets from all available experiments consistently suggest that in the small $x<0.08$ region (anti)neutrino cross sections are significantly higher than predictions obtained by a simple re-scaling of the charged lepton cross sections. 

Certainly there are similarities in the general shape of the nCTEQ $\nu A$ and the SLAC/NMC (charged-lepton) nuclear correction factors. However the magnitude of the effects and the $x$-region where they apply are quite different.  The nCTEQ $\nu A$ fits confirm the earlier impression from the NuTeV collaboration  that the size of the nuclear corrections affecting the NuTeV data are not as strong as those obtained from charged lepton scattering. 

The nCTEQ $\nu$-A NCFs are noticeably "flatter" than the SLAC/NMC curve, especially at lowest and moderate-$x$ where the differences are significant.
%These results suggest that the size of the nuclear corrections extracted from the NuTeV data are smaller than those obtained from charged lepton scattering (SLAC/NMC).  
In the $\overline\nu$ case, these differences are smaller but persist across the full x range.  The nCTEQ collaboration emphasize that both the charged-lepton-based and neutrino-based results come directly from global fits to the data.  Other than the assumptions stated earlier, there is no model involved.  They further suggest that this difference between the results of charged-lepton and neutrino DIS is reflective of the long-standing {}``tension'' between the light-target charged lepton data and the heavy-target neutrino data in the historical global PDF fits \cite{Botts:1992yi,Lai:1994bb} particularly at small x. These nCTEQ results further suggest that the tension is not only between charged-lepton \emph{light-target} data and neutrino heavy-target data, but also between neutrino and charged-lepton \emph{heavy-target} data as well.  In other words a difference between charged-lepton ($\ell^{\pm}$-A) and the neutrino ($\nu$-A) nuclear correction factors when comparing the same A.

The general trend is that the anti-shadowing region is shifted to smaller $x$ values, and any decrease at low $x$ is minimal at $Q^2 = 5 ~GeV^2$ where shadowing is clearly observed in $\ell^{\pm}$-A scattering.  The fit to $\nu$-A in the shadowing region gradually approaches the charged-lepton fit with increasing $Q^{2}$.  However, the slope of the fit approaching the shadowing region from higher x, where the NuTeV measured points and the nCTEQ fit are consistently below the charged-lepton Fe fit, make it difficult to reach the degree of shadowing evidenced in charged-lepton nucleus scattering at even higher $Q^2$.  

There is indeed shadowing observed in $\nu$-A scattering however at lower $Q^2$ than the $5 ~GeV^2$ of the general comparison above.  This only heightens the difference between $\nu$-A and $\ell^\pm$-A nuclear correction factors.  Referring to Fig.~\ref{fig:NuTeVThryF2} it can be clearly seen that NuTeV and CCFR data favor a significant trend toward increased shadowing as $Q^2$ decreases down to $\approx$ 1.0 ~GeV$^2$.  This could suggest significant shadowing in the  regime of modern neutrino experiments with their low E$_\nu$ dominated beams. This point will be addressed shortly (see \ref{DIS-rescon}).

%as shown in Fig.~\ref{fig:F2hiQ}

%Note the ratios shown are using the published NuTeV values for the average value of $F_2(\nu A)$. 
%and   $F_2(\bar\nu A)$. 
%Although the neutrino fit has general features in common with the charged-lepton parameterization, the magnitude of the effects and the $x$-region where they apply are quite different.  The present results are noticeably flatter than the charged-lepton curves, especially at low- and moderate-$x$ where the differences are significant.  The comparison between the nCTEQ fit, that passes through the NuTeV measured points, and the charged-lepton fit is also very different in the lowest-x (shadowing), lowest-$Q^2$ region and gradually approaches the charged-lepton fit with increasing $Q^2$.   However, the slope of the fit approaching the shadowing region from higher x where the NuTeV measured points and the nCTEQ fit are consistently below the charged-lepton A fit, make it difficult to reach the degree of shadowing evidenced in charged-lepton nucleus scattering at even higher $Q^2$. 
%as shown in Fig.~\ref{fig:F2hiQ}

%For the nCTEQ analysis, the contrast between the charged-lepton ($\ell^{\pm}A$) case and the neutrino ($\nu A$) case is striking.   

Concentrating on these interesting differences found by the nCTEQ group, if the nuclear correction factors for the $\ell^{\pm}$-A and $\nu$-A processes are indeed different there are several far-reaching consequences. For example, what happens to the concept of "universal parton distributions".  To maintain the universality of nuclear parton distributions is there an additional term in the factorization ansatz needed to reflect the response of the nuclear environment to vector and axial vector probes?

\begin{figure}[h]
\begin{center}
\includegraphics[width=0.99\textwidth]{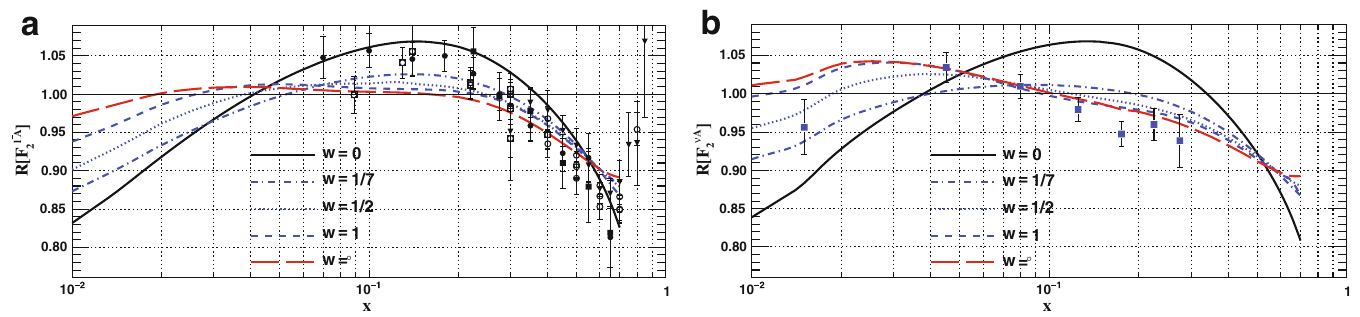}
\caption{Figure from~\cite{Kovarik:2010uv}. Predictions for the compromise fits for a) $\ell^{\pm}$Fe + DY  on the left and b)  $\nu Fe$ on the right for the indicated weight w as a function of x at $Q^2$ = 5 GeV$^2$.  }
\label{fig:nul-combo}
\end{center}
\end{figure}

\begin{figure}[h]
\begin{center}
\includegraphics[width=0.6\textwidth]{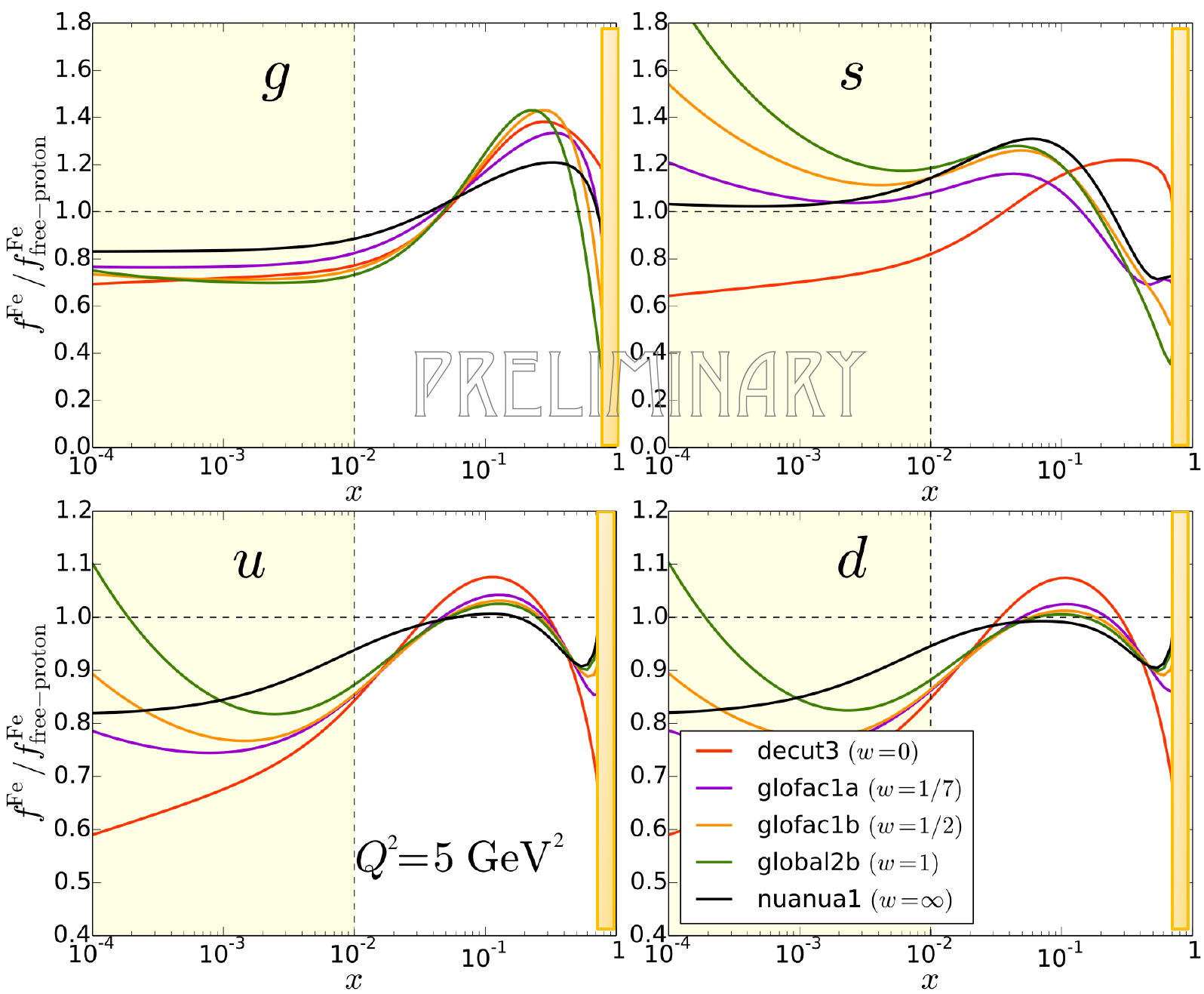}
\caption{Figure from~\cite{Andreopoulos:2019gvw}. Predictions for selected nuclear parton distributions in Fe for the indicated weight w as a function of x at $Q^2$ = 5 GeV$^2$. The main comparison is for the w=0, pure electroproduction and w = infinity, pure neutrino scattering. The shaded areas are where no appreciable date was available}
\label{fig:Kusina-nPDFs-wdep}
\end{center}
\end{figure}

Considering these possible significant consequences, the nCTEQ group performed a unified global analysis~\cite{Kovarik:2012zz} of the $\ell^{\pm}$-A, DY, and $\nu$-A data to determine if it would be possible to obtain a {}``compromise'' solution including both $\ell^{\pm}$-A and $\nu$-A data. They used a hypothesis-testing criterion based on the $\chi^{2}$ distribution that can be applied to both the total $\chi^{2}$ as well as to the $\chi^{2}$ of individual data sets.  Noting the large difference in the number of involved data points ($\ell^{\pm}$-A + DY) (708) and the $\nu$-A (3134), they introduced a weight (w) applied to the neutrino data sample that allowed adjustment for this rather large difference between the samples. With w = 0, only the $\ell^{\pm}$-A + DY was fit, w = 1 was a straight fit to both the $\ell^{\pm}$-A + DY and the $\nu$-A samples while w = $\infty$ was a pure $\nu$-A fit.  The results of the fit are displayed in Fig.~\ref{fig:nul-combo} and the corresponding w-dependent nuclear parton distribution functions are shown in Fig.~\ref{fig:Kusina-nPDFs-wdep}

It was concluded by these authors that it was {\em not possible} to accommodate the data from $\nu$-A and $\ell^{\pm}$-A DIS 
by an acceptable combined fit.  That is, when investigating the results in detail, the tension between the $\ell^{\pm}$-Fe and $\nu$-Fe data sets permits {\em no possible compromise fit} which adequately describes the neutrino DIS data along with the charged-lepton data and, consequently, $\ell^{\pm}$-Fe and $\nu$-Fe have different nuclear correction factors.

A compromise solution between $\nu$-A and $\ell^{\pm}$-A data can be found {\em only} if the full correlated systematic errors of the $\nu$-A data are {\em not} used and all the statistical and systematic errors are combined in quadrature thereby neglecting the information contained in the correlation matrix.  
%In other words the larger errors resulting from combining statistical and all systematic errors in quadrature reduces the discriminatory power of the fit such that the difference between $\nu$-A and $\ell^{\pm}$-A data are no longer  evident. 
This conclusion underscores the fundamental differences~\cite{Kovarik:2012zz} of the nCTEQ analysis with some of the other contemporary analyses~\cite{deFlorian:2011fp,Paukkunen:2013grz} using different statistical methods.  These other analyses suggest the $\nu$-A and $\ell^{\pm}$-A DIS data can be statistically consistent and relates the discrepancies to possible energy-dependent fluctuations of the NuTeV analysis. In particular they cite non-negligible differences in the absolute normalization between different neutrino data sets that, they claim, are large enough to prevent a tension-free fit to all data simultaneously.

On the other hand, a difference between $\nu$-A and $\ell^{\pm}$-A is not completely unexpected, particularly in the shadowing and antishadowing regions, and has previously been discussed in the literature \cite{Qiu:2004qk,Brodsky:2004qa,Brodsky:2014hia}.  The charged-lepton processes occur (dominantly) via $\gamma$-exchange, while the neutrino-nucleus processes occur via $W^{\pm}$-exchange. Since, as was stated, a (simplified) explanation of shadowing is that hadronic fluctuations of the vector boson interact coherently (like a "pion") off multiple nucleons in the nucleus and the interactions interfere destructively, the different nuclear shadowing corrections could simply be a consequence of the differing propagation of the hadronic fluctuations of the intermediate bosons (photon, $W$) through dense nuclear matter.  Perhaps the shadowing difference is due to the difference in vector boson masses, the W-boson is a much more localized probe than the photon.  The difference in antishadowing could indeed be a consequence of the quark-flavor dependence of antishadowing proposed by~\cite{Brodsky:2014hia}.

In particular, theoretical calculations~\cite{Qiu:2004qk} specifically for $\nu$ nucleus scattering suggest that at small x in the shadowing region the nuclear correction for neutrinos, as opposed to charged leptons, does have a rather strong $Q^2$ dependence. The standard nuclear correction obtained from a fit to charged lepton data implies a suppression of $\approx 10\%$ for iron compared to deuterium independent of $Q^2$ at x = 0.015.  While for x = 0.015  reference~\cite{Qiu:2004qk}  finds a suppression of 15\% at $Q^2 = 1.25 ~GeV^2$ and a suppression of 3.4\% at $Q^2 = 8.0 ~GeV^2$. This predicted  effect improves agreement with NuTeV data at low-x.  In addition, this definite $Q^2$ dependence of the $F_2$ structure function on Fe  at low x is supported by the predictions of the model of reference~\cite{Ducati:2006vh} shown in Fig. 5 of that reference.

Furthermore, since the structure functions in neutrino DIS and charged lepton DIS are distinct observables with different parton model expressions, it is not surprising that the nuclear correction factors would not be exactly the same.  What is, however, unexpected is the degree to which the $R$ factors differ between the structure functions $F_{2}^{\nu Fe}$ and $F_{2}^{\ell^\pm Fe}$. In particular the lack of evidence for shadowing in neutrino scattering at $Q^2$ = 8.0 ~$GeV^2$ down to $x\sim0.02$ is quite surprising.  
%How these surprising differences in R are directly related to different sets of nPDFs is shown in  Fig.~\ref{fig:PDFs(w)} that shows the set of nPDFs  corresponding to each value of w in the fits to the combined $F_{2}^{\nu Fe}$ and $F_{2}^{\ell^\pm Fe}$.  The comparison of the red (pure  $F_{2}^{\ell^\pm Fe}$) and the black (pure  $F_{2}^{\nu Fe})$ are particularly striking in the low-x shadowing region and the EMC region.  As the figure caption summarizes, there is little data especially for $\nu Fe$ below x = 0.1 and above x = 0.7.

%\begin{figure}[h]
%\begin{center}
%\includegraphics[width=0.99\textwidth]{PDFs(w)}
%\caption{nuclear PDFs resulting from the fits to $F_{2}^{\nu Fe}$ and $F_{2}^{\ell^\pm Fe}$  with the indicated weight w as a function of x at $Q^2$ = 5 GeV$^2$.  In addition to a lack of data below x = 0.1, there is limited data above x = 0.7 that can be use in the fit.}
%\label{fig:PDFs(w)}
%\end{center}
%\end{figure}

Should subsequent experimental results and analyses confirm the rather substantial difference between charged-lepton and neutrino scattering in the shadowing region at low-$Q^2$ it is interesting to speculate on the possible cause of the difference.  A  study of EMC~\cite{Aubert:1983xm}, BCDMS~\cite{Feltesse:1981zy} and NMC~\cite{Amaudruz:1995tq} \& data by a Hampton University - Jefferson Laboratory collaboration~\cite{Guzey:2012yk}  suggests that anti-shadowing in charged-lepton nucleus scattering may be dominated by the longitudinal structure function $F_L$.  As a by-product of this study, their figures hint that shadowing in the data of $\mu$-A scattering is being led by the transverse cross section with the longitudinal component crossing over into the shadowing region at lower x compared to the transverse.   

As summarized earlier, in the low-$Q^2$ region, the neutrino cross section is dominated by the longitudinal structure function $F_L$ via axial-current interactions since $F_T$ vanishes as  $Q^2 \rightarrow$ 0 similar to the behavior of charged lepton scattering.  If the results of the NuTeV analysis are verified, one contribution to the different behavior of shadowing at low-$Q^2$ demonstrated by $\nu$-A and $\ell$-A, in addition to the different hadronic fluctuations in the two interactions, could be due to the different mix of longitudinal and transverse contributions to the cross section of the two processes in this kinematic region.

Another hypothesis of what is causing the difference between neutrino and charged-lepton shadowing results comes from Guzey et al.~\cite{Guzey:2012yk} who speculates that at low x, low-$Q^2$  
%the value of y is close to unity and 
the neutrino interactions primarily
 probe the down and strange quarks.  This is very different than the situation with charged-lepton scattering where the contribution from down and strange quarks are suppressed by a factor of 1/4 compared to the up and charm. Therefore, the discrepancy between the observed nuclear shadowing in $\ell^{\pm}$-Fe total cross section at small $x$ and shadowing in total $\nu$-Fe cross section could be caused by the absence of nuclear shadowing of the strange quark nuclear parton distributions as extracted from the neutrino-nucleus data or even the poor knowledge of the strange-quark distribution in the free-nucleon that affects the neutrino-nucleus ratio more than the charged-lepton.  These suggestions are not inconsistent with the results shown in Fig.~\ref{fig:Kusina-nPDFs-wdep} that indicate no shadowing of the strange quark for neutrino scattering off Fe with the nCTEQnu nPDFs determined with $\nu$-A scattering data.
 %INSERT FIGURE OF NCTEQNU STRANGE QUARK DISTRIBUTION
 %$\nu$-A and $\ell^{\pm}$-A

It is worth repeating to emphasize that this difference in nPDFs depending on whether extracted from ($\nu/\nub$-A)-based or ($\ell^\pm$-A)-based interactions is a suggestion of non-universal nuclear parton distributions.  A way to salvage this concept of universal parton distributions could be to modify factorization to include consideration of the type of interaction in the nuclear environment. 

\subsection{Hadronization of Low Energy $\nu$-A Interactions}
\label{Subsec-hadronization}

Current and particularly the future DUNE long baseline oscillation experiments,
have neutrino energies up to $\lessapprox$ 10 GeV.  For such a broad range of neutrino energy, they will have to use information from the hadronic system in order to estimate the actual $E_\nu$ of an event and estimate the  backgrounds to their signal topologies.  Specific models for quasi-elastic and one-pion resonance production are available.  However, for example in the GENIE simulation program, multi-pion production through resonance decay and all non-resonant pion production are grouped together under the name GENIE "DIS" and the multiplicity of a given event is chosen through models that describe the hadronization of the initial hadronic component of the interaction. They will then need models that describe the initial state hadronization of the hadronic shower that is then followed by final state interactions of these produced hadrons. A good survey of the current hadronization models now in use within the community can be found in section seven of~\cite{Andreopoulos:2019gvw}. 

In the DIS region these hadonization models describe the formation of hadrons in inelastic interactions and are characterized by non-perturbative fragmentation 
functions (FF), which in an infinite momentum frame can be interpreted as probability distributions 
to produce a specific hadron of type $h$ with a fraction $z$ of the longitudinal momentum 
of the scattered parton. These universal fragmentation functions can not be easily calculated 
but can be determined phenomenologically from the analysis of high-energy scattering data 
\footnote{An example of a recent study of pion and kaon FF in $e^+e^-$ collisions can be found in Ref.~\cite{Sato:2016wqj} while the FF for charmed hadrons ($D,D_s,\Lambda_c$) in $\nu_l$ DIS interactions were
studied in Ref.~\cite{Samoylov:2013xoa}.}. 

Modern event generators often use the LUND string fragmentation model~\cite{Andersson:1983ia, Andersson:1987pr}, as implemented 
in the PYTHIA/JETSET~\cite{Sjostrand:1993yb}
packages, to describe the hadronization process. 
This model results in a chain like production of hadrons with
%with local compensation of quantum numbers. T
%The original partons are associated with the endpoints of a massless relativistic string to approximate a linearly confining color flux tube, while gluons are associated with energy and momentum carrying kinks on the string. The production rate of the created $q\bar q$ pairs leads to a Gaussian spectrum of the transverse momentum $p^2_\perp$ for the produced hadron,
an associated FF providing the probability that a 
given ratio $z$ between the hadron energy and the energy transfer is selected. 
The PYTHIA/JETSET implementation of this LUND model is controlled by many free parameters, 
which can be tuned to describe the data. A detailed study of the PYTHIA fragmentation 
parameters with $\nu$ data~\cite{Kuzmin:2013tza} 
from proton and deuterium targets was performed in Ref.~\cite{Katori:2014fxa}. 
In particular, the various parameter sets determined by the HERMES experiment were 
used within the GENIE event generator obtaining predictions in agreement with the 
measured hadron multiplicities. 

An independent tuning of the JETSET fragmentation 
parameters was performed in Ref.~\cite{Chukanov16} with NOMAD data from 
exclusive strange hadron production and inclusive momentum and angular distributions 
in $\nu$-C DIS interactions. However, as has been noted, in $\nu$-nucleus interactions 
the hadrons originating from the primary interaction can re-interact inside the nucleus. 
These final state interactions must, therefore, be taken into account in the determination 
of the effective fragmentation parameters from the observed final state hadrons. 

Since the physics of the LUND hadronization model is not applicable at lower values of the invariant mass $\approx{W<3 GeV}$, a better description of the data has been achieved with  a phenomenological description of the hadronization process in which the average hadron multiplicities are parametrized as linear functions of $\log W$ for each channel. This Koba-Nielsen-Olesen (KNO) 
scaling law~\cite{Koba:1972ng} can then be used to relate the 
dispersion of the hadron multiplicities at different invariant masses. 
%with a universal scaling function parameterized in terms of the Levy function. 
Both the averaged hadron multiplicities and the KNO functions are usually tuned from $\nu$ bubble chamber data. 

The challenge faced by the neutrino simulation programs is how to bridge the transition from the KNO procedure used at low W to the PYTHIA/JETSET LUND-based model at higher W. To do this the GENIE~\cite{Andreopoulos:2009rq} generator uses the hybrid AGKY 
approach~\cite{Yang:2009zx}, which has a gradual transition from the KNO hadronization 
model to PYTHIA in the region $2.3 \leq W \leq 3.0$ GeV and allows the average multiplicities to be continuous as a function of $W$. 
%However, since PYTHIA underestimates the dispersions at low $W$ with respect to bubble chamber data, the AGKY model is characterized by some discontinuities of the topological cross-sections in the hadronization transition region. 
The NEUT~\cite{Hayato:2009zz} generator has a more abrupt transition for the hadronization process, 
using KNO for $W<2$ GeV and PYTHIA for $W>2$ GeV.
The NuWro~\cite{Golan:2012wx} generator tuned both the average multiplicities and the 
corresponding dispersions to the available bubble chamber data in order to achieve 
continuous topological cross-sections. All three generators, GENIE, NEUT and NuWro, tune 
the average hadron multiplicities and dispersions from bubble chamber data. 

Before addressing specific hadronization techniques, it is important to again emphasize that not only do some generators effectively use these hadronization models within the DIS region, they also use these models to produce multi-pion resonant and all non-resonant mesons multiplicities in the resonance region. This mechanism then also provides the main contribution for multi-meson production in the resonance region.

\subsubsection{The AGKY Hadronization Model}
An excellent overview of this topic can be found in~\cite{Katori:2016blu}. The authors cover the full spectrum of available treatments of this topic as they apply to hadronization in the lower-W kinematic region.

\begin{figure}[tb]
\begin{center}
\includegraphics[width=0.75\textwidth]{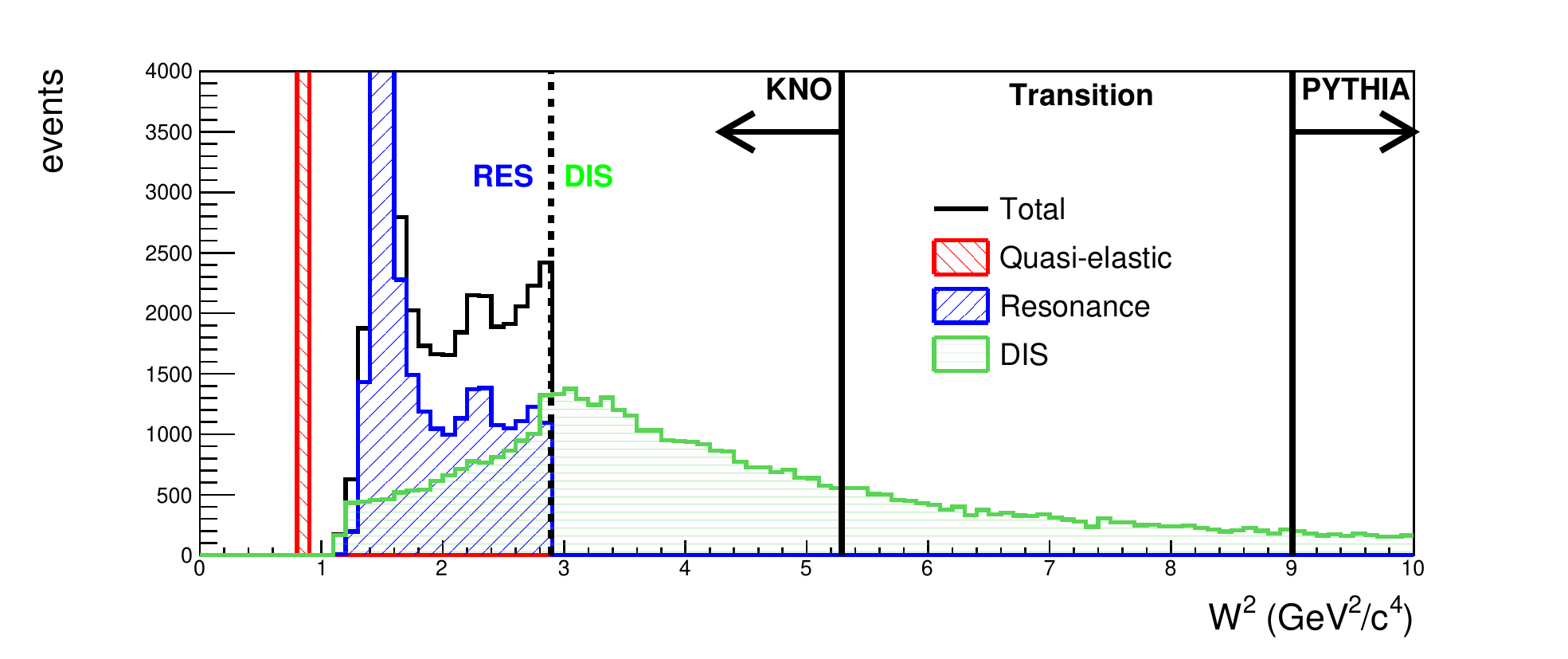}
\caption{ From reference~\cite{Katori:2016blu}.
$W^2$ distribution of $\numu$-water target interactions in GENIE showing the quasi-elastic scattering, the 
resonance interactions, and the DIS region.  
The $W$ distribution is further split into the three regions, KNO scaling-based model only region, PYTHIA only region, and the transition between the two regions used in the AGKY model.
}
\label{fig:KNOtoPYTHIA}
\end{center}
\end{figure}

The model used by the GENIE simulation program, the AGKY (initials of the main author's names - Andreopoulos, Gallagher, Kehayias and Yang) hadronization model~\cite{Yang:2009zx}, was developed for the MINOS experiment.
The model is split into three $W$ regions shown in Fig.~\ref{fig:KNOtoPYTHIA} with the AGKY model used to cover the hadronization of the GENIE DIS (horizontal hatched curve) in the figure. Also, as mentioned earlier, the so-called DIS region in GENIE extends to the low $W^2$ resonance region to describe non-resonant pion production as well as resonant multi-pion production in the resonance region. 

\begin{figure}[tb]
\begin{center}
\includegraphics[width=0.45\textwidth]{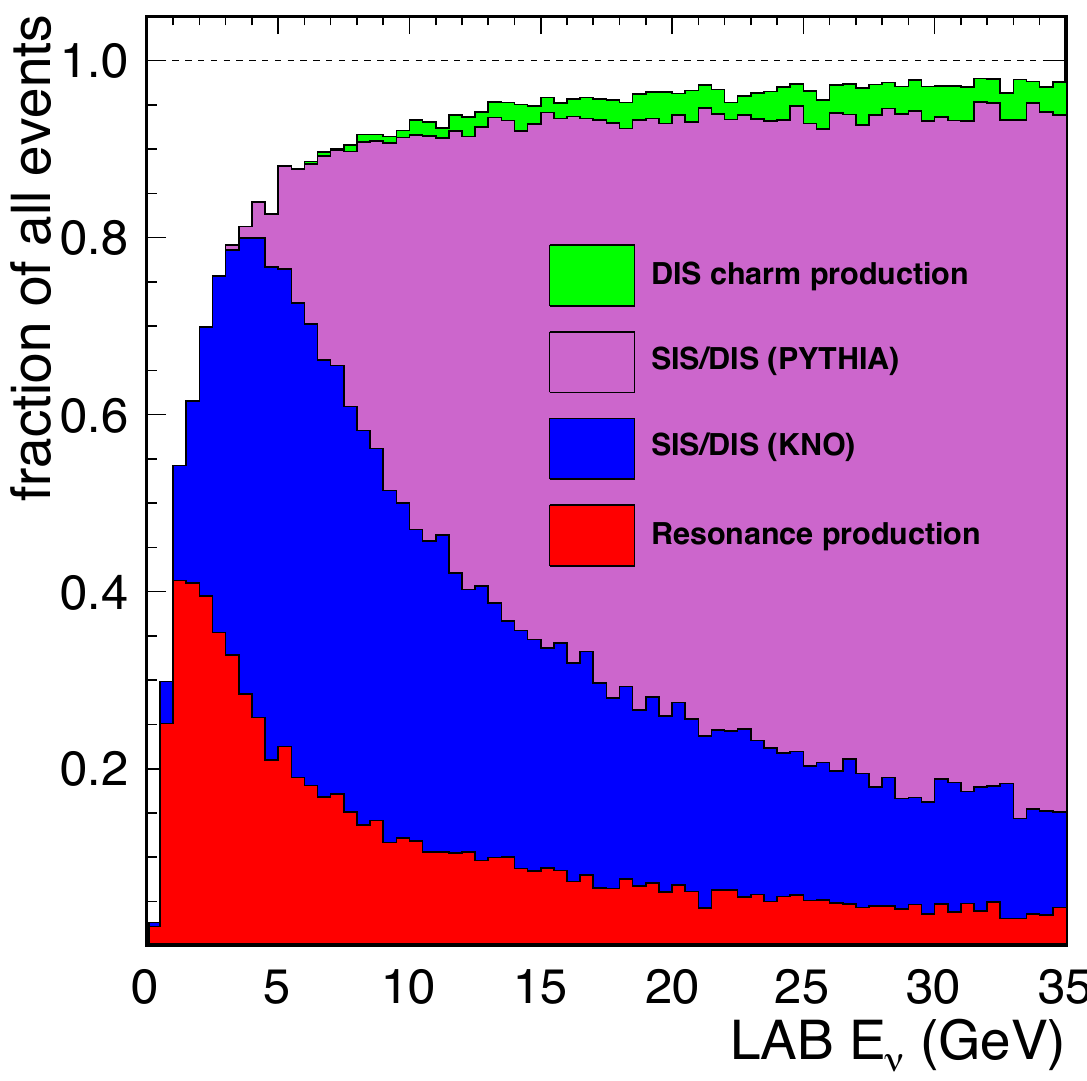}
\caption{ From Andreopoulos presentation in
reference~\cite{Andreopoulos:2019gvw}.
The figure presents the division of events coming from the GENIE 1-$\pi$ resonance model and using the AGKY model to generate events as a function of $E_\nu$ in GENIE.
}
\label{fig:Costas-GENIE-HAD}
\end{center}
\end{figure}

At lower $W$ $\le$ 2.3 GeV, a phenomenological description based on the Koba-Nielsen-Olesen (KNO) scaling law is used~\cite{Koba:1972ng} to simulate the hadron multiplicity of each interaction.  As $W$ increases beyond 2.3 GeV, the AGKY model gradually transitions from this KNO model to PYTHIA~\cite{Sjostrand:2019zhc} which is used for $W$ $\ge$ 3.0 GeV. This transition from the solely KNO to the solely PYTHIA region is based on the value of $W$.  As $W$ increases  the fraction of events hadronized using the PYTHIA model increases while the fraction using KNO decreases linearly. 
PYTHIA is a standard hadronization tool for higher energy physics experiments used by  neutrino interaction generators for hadronization at the relatively higher $W$ region.  Whether PYTHIA can be applied to such low W and resulting low multiplicities is not at all clear.
Refer to~\cite{Katori:2016blu} for further details of the KNO and PYTHIA models.

The actual results of the application of the AGKY model within GENIE is shown in Fig.~\ref{fig:Costas-GENIE-HAD}.  It is evident that already with an $E_\nu$ of $\approx{3~GeV}$ the meson multiplicities are coming more from KNO determination than from the GENIE 1-$\pi$ model.  It is also important to restate that such a procedure suggests that the KNO model is being used to govern non-resonant pion production rather than the explicit calculation of the relevant theory involved in the process.

\subsubsection{FLUKA: NUNDIS}

The FLUKA neutrino event generator is called NUNDIS that  describes the neutrino-nucleon interactions from Quasi Elastic through resonance production and into Deep Inelastic Scattering. Hadronization is performed with the FLUKA models based on the LUND string models, for details see Sala's summary in~\cite{Andreopoulos:2019gvw} and\cite{Battistoni:2013tra} from which much of this description has been drawn. 
%Important for current neutrino experiments are the treatments of low-mass events that require special treatment and are taken into account in FLUKA. 
They find that for very low mass situations standard hadronization has to be replaced by what they refer to as a "phase space explosion". This treatment has proven to be important for the correct simulation of single-pion production in neutrino interactions.  Although traditionally  associated only to resonance production, FLUKA finds the DIS contribution to the single-pion channel is significant and an important contribution to the one-pi channel in $\nu$-nucleon scattering.

%\begin{figure}[tb]
%\begin{center}
%\includegraphics[width=0.75\textwidth]{FLUKA-onepi}
%\caption{ From Salas presentation inreference~\cite{Andreopoulos:2019gvw}.The contribution of resonance and DIS scattering to the one-pi channel in $\nu$-nucleon scattering.}
%\label{fig:FLUKA-onepi}
%\end{center}
%\end{figure}

\begin{figure}[tb]
\begin{picture}(500,155)(0,0) %   \graphpaper[20](0,0)(500,160)
 \put(20,15){\includegraphics[width=0.4\textwidth]{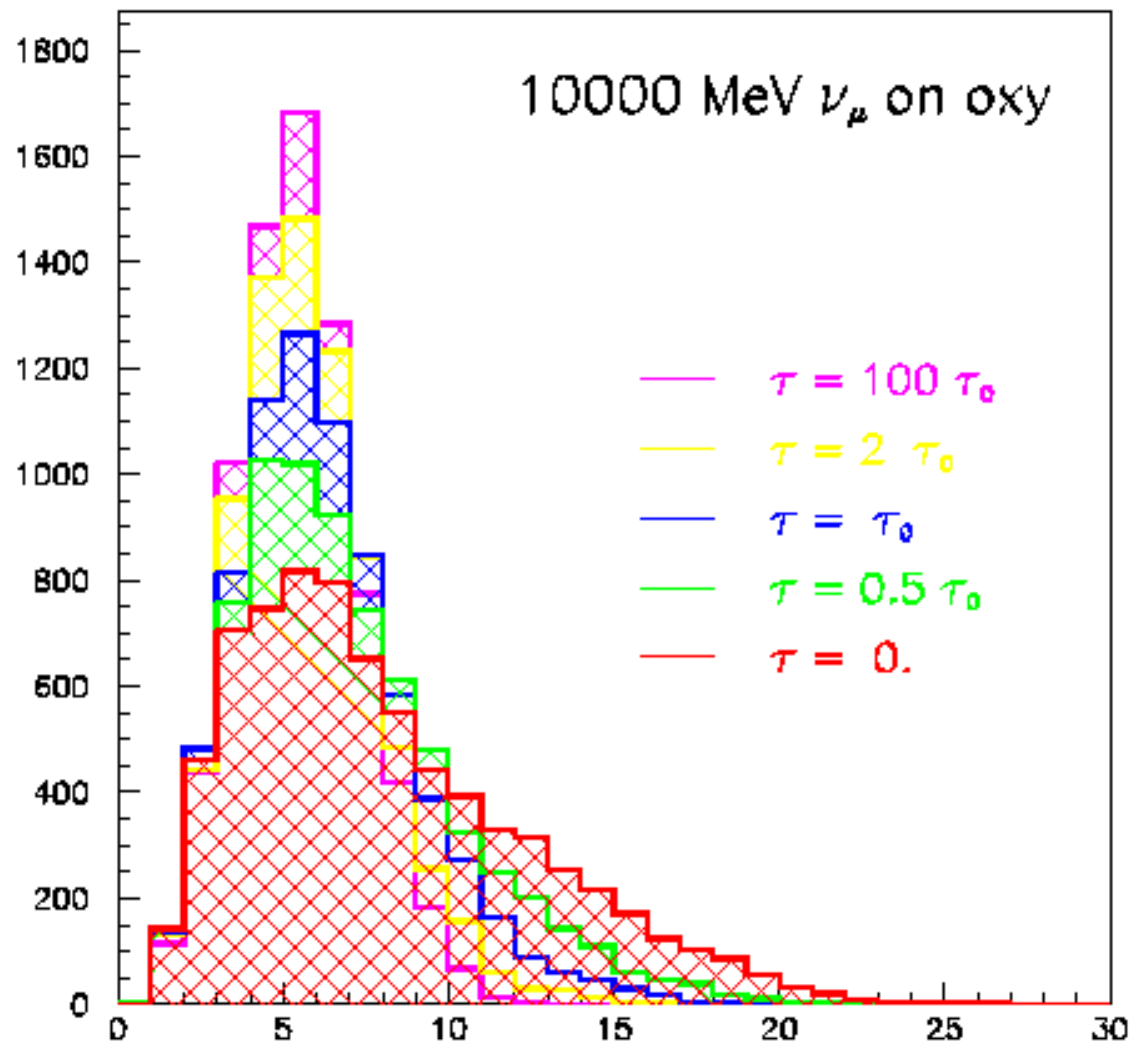}}
\put(220,0){\includegraphics[width=0.4\textwidth]{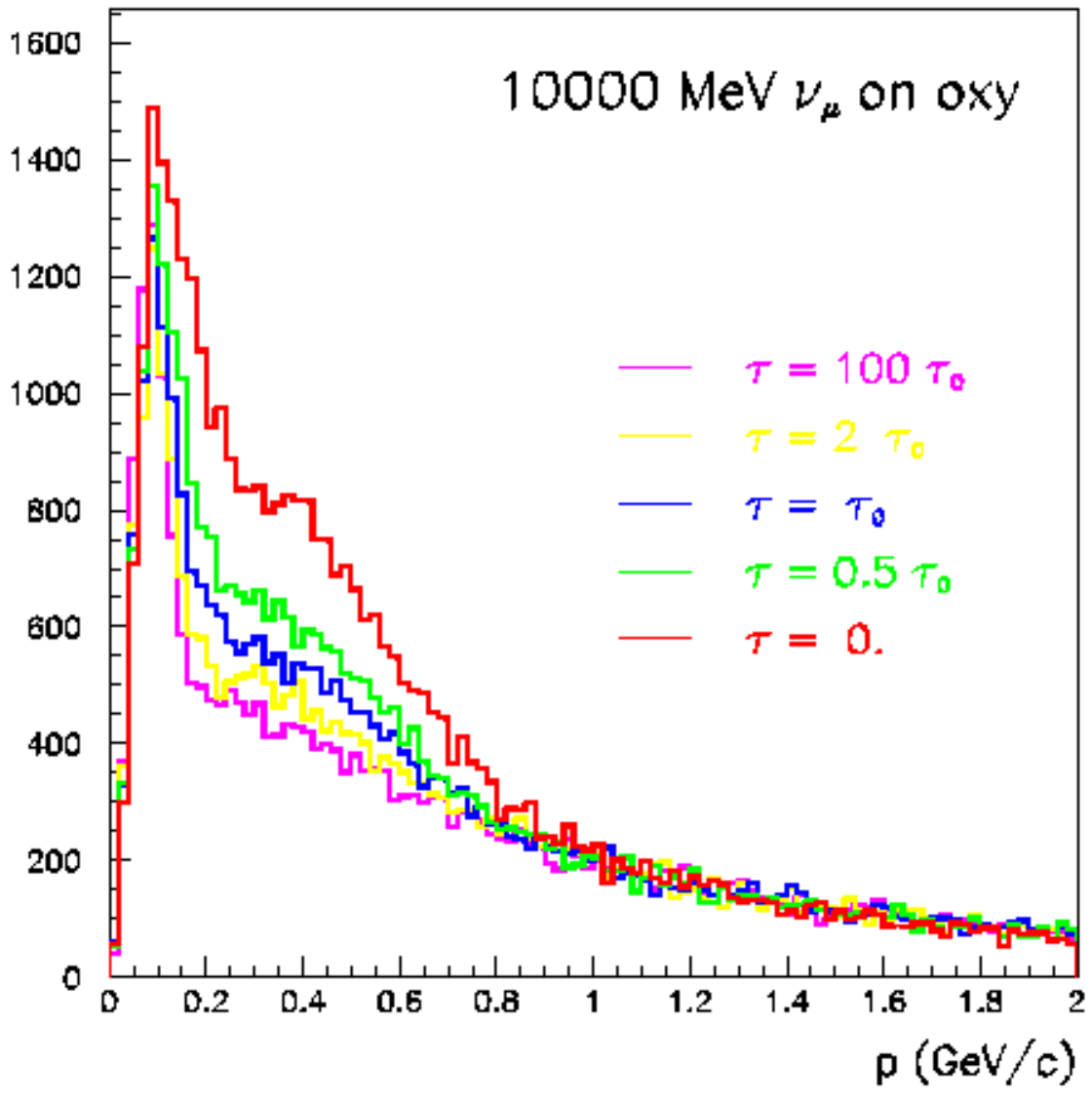}}
\put(100,5){\footnotesize {Multiplicity}}
%\put(374,0){$(b)$} 
\end{picture} 
\caption{
From Sala presentation in
reference~\cite{Andreopoulos:2019gvw}. The figures emphasize the relative change in distributions as a function of formation length with the vertical axis a arbitrary number of events.
The resulting dependence of the event multiplicities (left) and particle momentum distributions (right) are from FLUKA for a 10 GeV neutrino on oxygen when the formation length is varied over a wide range.}
\label{fig:FLUKA-Form}
\end{figure}

Important for FLUKA, and included in  GENIE, is the introduction of Formation Zone that can be understood as hadrons emerging from an inelastic interaction that require some time before beginning strong interactions with the surrounding environment. This has the effect of allowing certain hadrons to escape any final state interactions within the nucleus. Formation Zone is then important to correctly model hadronic interactions as is illustrated in ~Fig.~\ref{fig:FLUKA-Form} that shows the effect on both event multiplicities and the momentum spectra of these secondaries when the considered formation zone is varied.  For Formation zone set to 0 - no formation zone - the produced hadrons within the shower can immediately interact within the nucleus thus the average number of hadrons leaving the nucleus is largest and the average momentum of these hadrons is the smallest.  As the formation zone increases more of the hadrons leave the nucleus without interacting and the average multiplicty decreases with the average mometum of the hadrons increasing.

\subsection{Results and Discussion}
\label{DIS-rescon}
%\lessapprox  \gtrapprox

Although it has been emphasized that neutrino DIS scattering could be a particularly rich source for for flavor separation in determining free proton parton distribution functions, a serious problem in the neutrino community is the very poor state of knowledge of $\nu$-{\emph free nucleon} interactions.  There are presently only low-statistics bubble chamber results from the 1970's and 1980's that have relatively large statistical and systematic errors.  This severely limits the influence of neutrino scattering in free nucleon PDFs.  That these rather imprecise results are then used as the start of neutrino interaction simulations by the current community's event generators is also a matter of real concern. In addition, this also forces the determination of the denominator of nuclear correction factors for neutrino experiments to use a phenomenological estimate of $\nu$-free nucleon cross sections and structure functions formed from free nucleon PDFs.

%{\bf The content in paragraph mentioned below has already been mentioned earlier. Can we make it short ?}
Turning then to neutrino nucleus scattering, the NuTeV $\nu$-Fe and CHORUS $\nu$-Pb experiments are the most recent high-statistics DIS experiments that  have published double-differential $\nu/\nub$-A scattering cross sections as well as very detailed studies of systematic errors.   To be able to combine these NuTeV and CHORUS $\nu$-A results with other experiments in global fits of free-nucleon PDFs, a way of converting   $\nu$-Fe/Pb to $\nu$-nucleon -  nuclear correction factors -  had to be determined. 

Using the results from these experiments, nuclear effects of charged current deep inelastic $\nu$-A scattering were studied by the nCTEQ collaboration in the frame-work of a $\chi^2$ analysis and, in particular, a  set of iron nuclear correction factors for iron structure functions was extracted.  Comparing these results with structure function correction factors for $\ell^\pm$-Fe scattering it was determined that the neutrino correction factors differ in both shape and magnitude from the correction factors for $\ell^\pm$-Fe scattering. 

This difference, although not unexpected theoretically especially in the shadowing and antishadowing regions, is not universally seen by all groups examining nPDFs of neutrinos.  It is imperative that we carefully consider these contrasting results and gain an understanding of the $\nu$-A nuclear correction factors.  The nCTEQ study  of the $\nu$-Fe and $\nu$-Pb nPDFs provides a foundation for a general investigation that can address this topic.  However the results from a much wider variety of nuclear targets in a neutrino beam, able to access DIS kinematics, will be needed to definitively answer this question. 

%There is one experiment currently analysing data and a future neutrino experiment that, in principle, will have the small statistical and systematic errors needed to contribute to the  study of the $\nu$-A nuclear correction factor in the DIS region.
%The high statistics measurement of nuclear effects in DIS took several more decades to accomplish. The MINERνA experiment has presented the results from their low-energy analysis of multiple nuclear targets. They show the differential scattering cross section in the form of ratios dσi /dσCH , i=C, Fe, and Pb [22]
The \minerva\ neutrino-nucleus scattering experiment at Fermilab~\cite{Drakoulakos:2004gn}, a collaboration of high-energy and nuclear physicists, is currently analyzing data performing a systematic study of neutrino nucleus interactions.  The overall goals of the experiment are to measure absolute exclusive and inclusive cross-sections and study nuclear effects in $\nu$ - A  interactions with He, C, O, Fe and Pb nuclear targets.   

\begin{figure}[h]
\begin{center}
\includegraphics[width=0.99\textwidth]{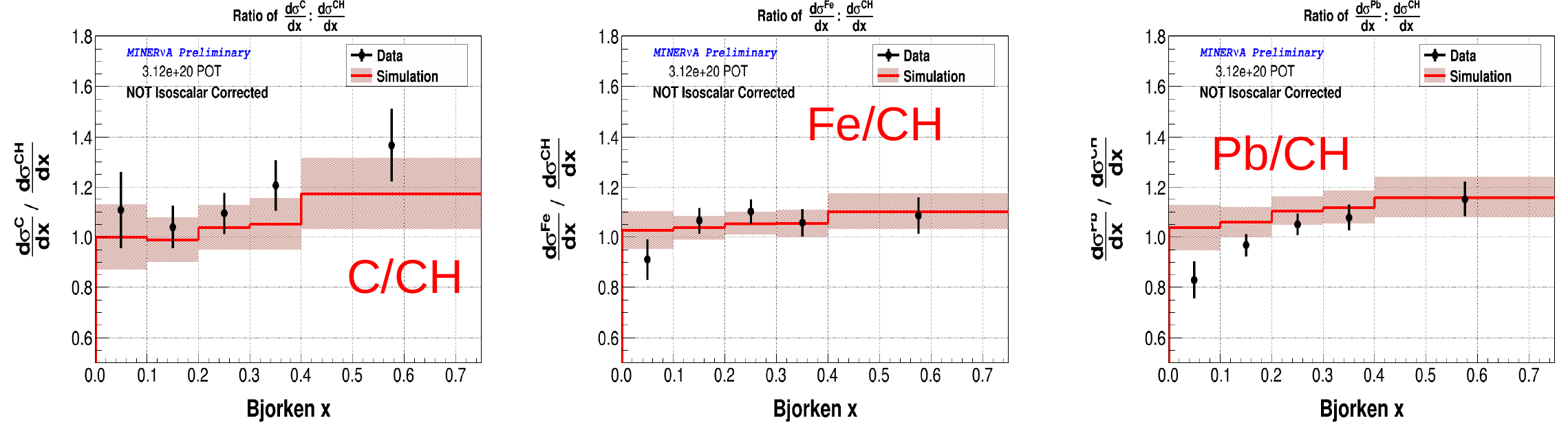}
\caption{Figure from~\cite{Mousseau:2016snl}. The ratios of the total DIS cross section on C (left), Fe (center) and Pb (right) to scintillator (CH) as a function of x. Data are drawn as points with statistical uncertainty and simulation as lines. The total systematic error is drawn as a band around the simulation in each histogram. The experimental results and simulations are not isoscalar corrected}
\label{fig:LE-ratios}
\end{center}
\end{figure}

For QCD oriented studies \minerva\ is pursuing systematic studies of the resonance-to-DIS (SIS) transition region and the lower-Q$^2$ DIS region. The \minerva\ experiment has finished both their low-$E_\nu$ (LE) exposure and their somewhat higher energy (ME) exposure that yielded a much higher fraction of DIS events with a considerably broader kinematic range than the lower energy data and is currently being analyzed. 
\minerva\ used the low-energy (LE) NuMI beam to initiate a first study of the DIS cross sections off the \minerva\ suite of nuclear targets and published~\cite{Mousseau:2016snl} the cross section ratios of target A to the nominal scintillator (CH) of the main tracker as shown in Fig.~\ref{fig:LE-ratios}.  

\begin{figure}[h]
\begin{center}
\includegraphics[width=0.85\textwidth]{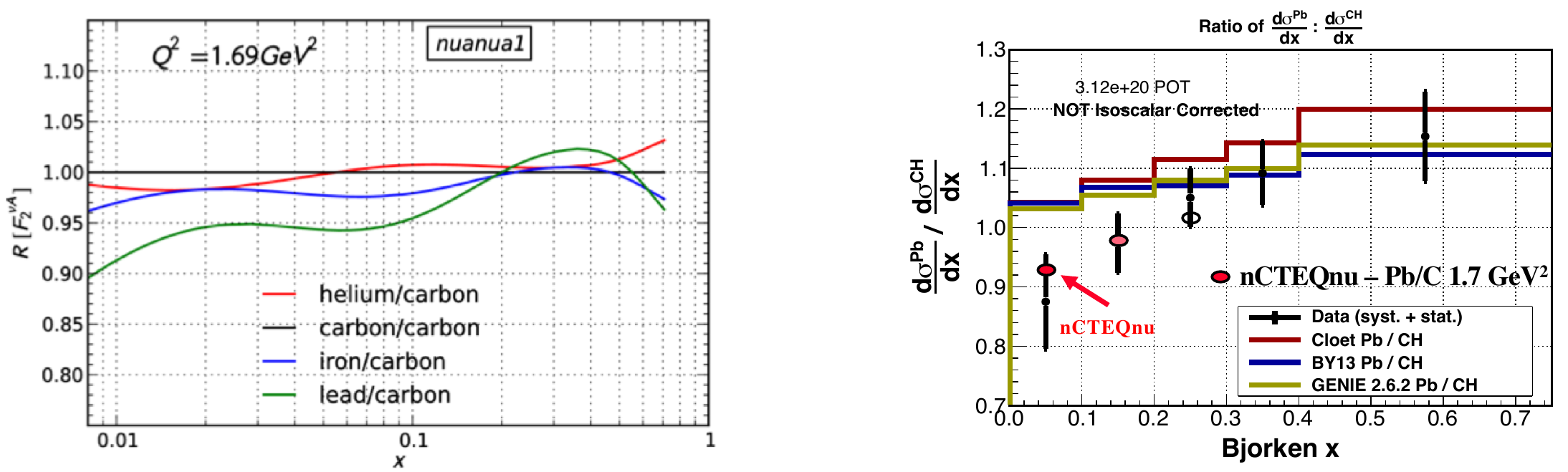}
\caption{(left) The x-dependent predictions for the ratios A/C of the structure function $F_2$ at $Q^2$ = 1.7 GeV$^2$ using the nuclear parton distributions determined from neutrino scattering -- nCTEQnu. (right) As in Fig.~\ref{fig:LE-ratios} the measured DIS cross section ratio of Pb/CH as a function of x from \minerva\ (data points) and various parametrizations of x- dependent nuclear effects~\cite{Bodek:2010km, Cloet:2006bq, Andreopoulos:2015wxa} as well as the predictions based on the nCTEQnu nPDFs.  The error bars on the data are the combined statistical and systematic uncertainties.}
\label{fig:nCTEQnu-Pb}
\end{center}
\end{figure}

These results can be compared to the predictions of nCTEQ nuclear PDF sets, namely the nCTEQnu nPDFs based on neutrino nucleus DIS scattering data.  The predictions of the extracted neutrino-based nuclear PDFs can be seen in Fig.~\ref{fig:nCTEQnu-Pb} (left) that shows the predicted ratios using these neutrino-based nCTEQnu nuclear PDFs at a $Q^2$ of 1.7 GeV$^2$. This is roughly the average $Q^2$ of the lowest x bin and close to the average of the neighboring x-bin in the cross section ratio of Pb to CH.  Fig.~\ref{fig:nCTEQnu-Pb} (right) displays the \minerva\ measured values for the x-dependent cross section ratios of Pb to CH compared to several current models for this ratio, based on charged-lepton nuclear effects, as well as the predictions of nCTEQnu ($\nu$-A) nuclear parton distributions. Although this is not the ratio of $F_2$ as the figure on the left, in this small x region the contribution of $xF_3$ is small so the cross section is dominated by $F_2$. In the lowest x bin. With the data having an approximate $Q^2$  of 1.8 GeV$^2$, the nCTEQnu predictions can be read off the plot to the left. Certainly the associated uncertainties are significant, however the measured points do favor the nCTEQnu predictions that reflect the low-x, low-$Q^2$ results of the NuTeV, CCFR and CHORUS results.

While these results are suggestive they are certainly not the statistically significant result needed to resolve this question. It is important that further experimental result with well-controlled errors are pursued to determine the neutrino nuclear correction factors over a wide range of A.  While the \minerva\ experiment is now addressing this question with a somewhat higher beam energy with targets of C, water, Fe and Pb, in the near future the much more statistically significant DUNE experiment, if outfitted with a range of nuclear targets beyond the main Ar of its detectors, can add significantly to this still open question yielding a thorough A-dependent study of nuclear PDFs and better determine the $\nu$-A nuclear correction factor in the DIS region.  Perhaps further in the future a neutrino factory with very intense and well-known neutrino beams will provide a direct comparison between nuclear targets and nucleon (liquid hydrogen and deuterium)  targets.

Beyond this important comparison of nuclear effects depending on the incoming lepton, there are  outstanding questions to be resolved for $\nu/\nub$-A scattering alone. These can be summarized as main questions to ask subsequent neutrino experiments:
\begin{itemize}
\item Does the community have the resources to supplement the decades-old bubble chamber measurements of  $\nu/\nub$-p and $\nu/\nub$-n total and differential cross sections with contemporary high-statistics measurements on free proton and deuteron targets?
\item In experimentally extracting nuclear structure functions from nuclear cross sections, what nuclear biases are being built in through the assumed R (= $\sigma_L / \sigma_T$) and $\Delta (xF_3)$?
%\item At high x, what is the behavior of the valence quarks as x  $\rightarrow$ 1.0?  
\item What is happening in the region with x $\ge$ to 1.0 with $\nu$-A interactions and how is this region to be addressed in global fits to neutrino nPDFs?
\item When will DIS modeling in generators be updated to reflect the recent nuclear parton distributions and $\nu$-A models available.
\item At high-x, at what value of $Q^2$ do the higher-twist contributions become significant after correcting for target mass effects?
\item A study of {\em nuclear} higher-twist effects is necessary to better understand the transition region for $\nu-A$ interactions.
%\item At all x and $Q^2$, what is yet to be learned if we can measure all six $\nu$  and $\nub$ structure functions to yield maximal information on the parton distribution functions?
\item As W decreases and approaches the SIS region, what is the interplay of non-perturbative QCD effects with the approaching resonant/non-resonant region that governs this transition? 
\item Considering the suggested problems with PYTHIA and even KNO at low-W, when will the community re-examine hadronization models in current generators to better describe exclusive hadron production at relevant W values?
\item Considering the importance of $\nu_e$ interactions for current and future experiments, when will our understanding of the impact of radiative corrections and their applicability be improved.
\end{itemize}

 \section{Comparing DIS Theory and DIS Phenomenological Approaches}
\label{comp}
 
 In the previous sections we have presented both  theoretical and phenomenological approaches to describe deep-inelastic scattering.  Here we present a direct comparison of the predictions of these two approaches as well as a comparison of these predictions with past experimental results. We also present expectations for the DIS contributions to on-going and future experiments. 
 
 \subsection{Comparison to Past High-statistics Experimental Results}\label{comparison}
 
The  experimental results of the CCFR, NuTeV and CHORUS experiments that can be compared to these two approaches have been presented in \ref{Subsec-Iron}.

\begin{figure}[tb]
\begin{center}
\includegraphics[width=0.65\textwidth]{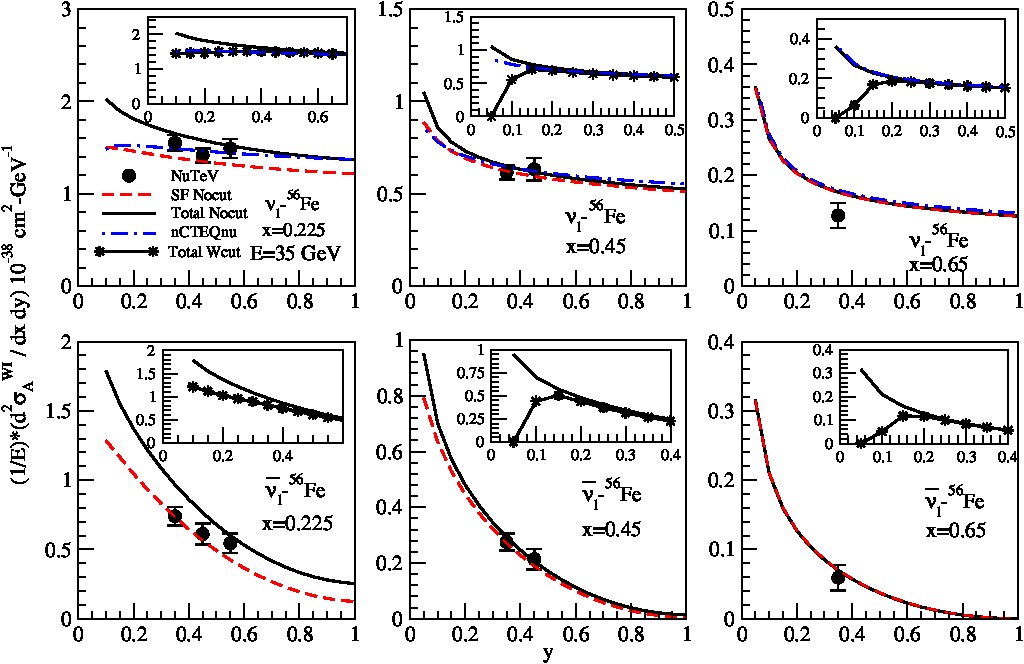}
\caption{Differential cross section vs y for different values of x for the incoming beam of energy E = 35 GeV for $\nu$-Fe DIS (top row) and $\nub$-Fe DIS (bottom row).  Theoretical predictions are shown with the spectral function only (dashed line) and with the full model (solid line) at NNLO. In the inset the effects of an additional kinematical cut of W $\ge$ 2 GeV (solid line with star) for the full theoretical model are shown.  The blue dash-dotted line in the top row is the result from nCTEQnu nPDFs for $\nu$-Fe with $Q^2 \ge 1.0 ~GeV^2$. Solid circles with error bars are the limited experimental data points of NuTeV at this lower energy.}
\label{fig:35GeVFe-nu&nub}
\end{center}
\end{figure}

\begin{figure}[tb]
\begin{center}
\includegraphics[width=0.65\textwidth]{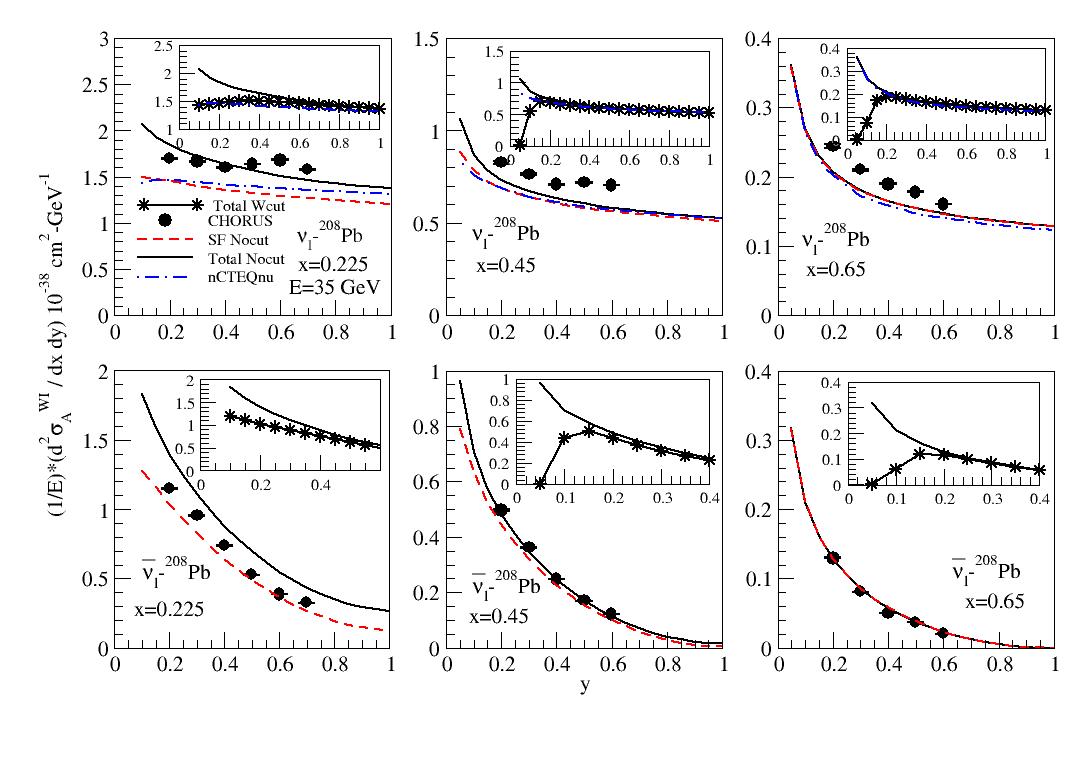}
%\caption{Differential cross section vs y for different values of x for the incoming beam of energy E = 35 GeV. The numerical results for $\nu$-Pb DIS (top panel) and $\nub$-Pb DIS (bottom panel). Predictions are obtained with the spectral function only (dashed line) and with the full model (solid line) at NNLO. In the inset the results for the full model are compared with the corresponding results obtained with a kinematical cut of W $\ge$ 2 GeV (solid line with star). Solid circles are the experimental data points of CHORUS. The blue dash-dotted line in the top panel is the result from nCTEQnu nPDFs for $\nu$-Pb with with $Q^2 \ge 1.0 ~GeV^2$.
%}
\caption{Differential cross section for $\nu$-Pb DIS (top row) and $\nub$-Pb DIS (bottom row) for the incoming beam of energy E = 35 GeV. Lines representing the theoretical and nCTEQ nPDF approaches have the same meaning as in Fig.~\ref{fig:35GeVFe-nu&nub}. Solid circles are the experimental data points of CHORUS.   }
\label{fig:35GeVPb-nu&nub}
\end{center}
\end{figure}

In Figs.~\ref{fig:35GeVFe-nu&nub} for Fe and \ref{fig:35GeVPb-nu&nub} for Pb, the theoretical predictions of the Aligarh-Valencia group for $\nu$ and $\nub$ differential cross sections as well as the phenomenological predictions using the nCTEQnu nuclear PDFs for $\nu$ differential cross sections at E$_\nu$ = 35 GeV are presented.  The results of Aligarh-Valencia group are shown for the spectral function only and using the full model (Eqs.\ref{f1f2_tot} and \ref{f3_tot}) where it can be observed that the mesonic contributions play important role in the region of $x \le 0.5$. In comparing the two approaches for $\nu$, the nCTEQnu-based results are somewhat lower than the theoretical prediction at the lowest-$x$ presented while the results of the two approaches are in reasonable agreement with each other in the region of higher x.

Both approaches are compared with the limited experimental results from NuTeV and CHORUS~\cite{KayisTopaksu:2011mx, Onengut:2005kv} experiments at E$_\nu$ = 35 GeV.  In general, for $\nu$ the results obtained with both the full theoretical model and using the nCTEQnu nuclear PDFs are below the experimental results for $\nu$-Pb at higher x and the full theoretical model predictions are above the experimental results for $\nub$ in the lowest x bin for both nuclei.
% Fig.~\ref{fig:35GeVPb-nu&nub}
%both in reasonable agreement with the experimental data,   
 
\begin{figure}[tb]
\begin{center}
\includegraphics[width=0.65\textwidth]{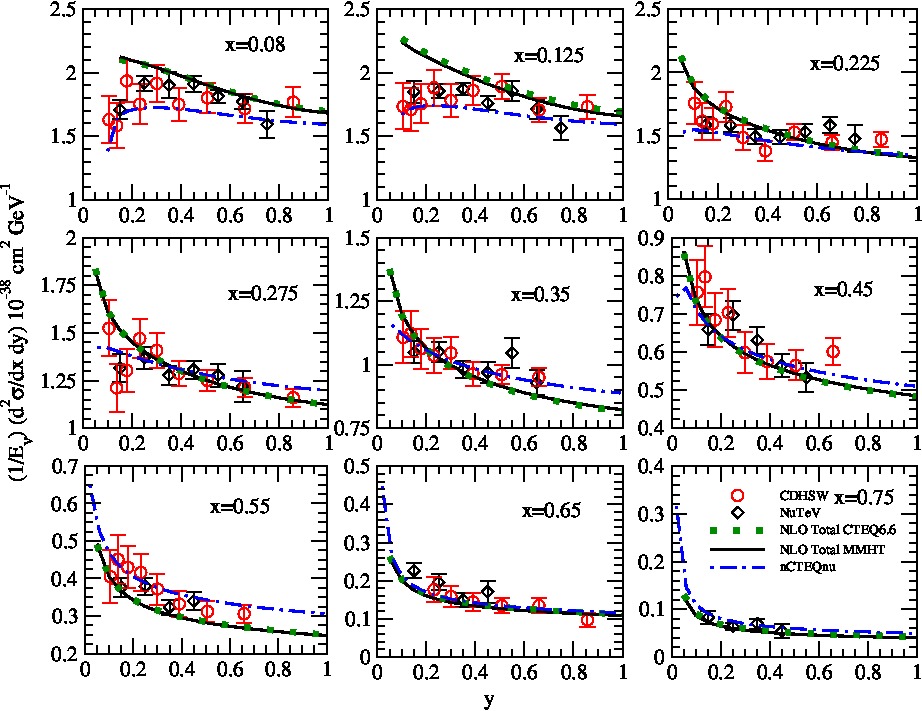}
\caption{Results of the differential scattering cross section vs y, at different x for $\nu$-Fe (treated as an isoscalar target) at $E_\nu$ = 65 GeV.  The theoretical results are obtained for iron  by using (i) CTEQ 6.6 nucleon PDFs at NLO in the MS-bar scheme (dotted line), (ii) MMHT nucleon PDFs at NLO (solid line). The blue dash-dotted line is the result from nCTEQnu nPDFs with $Q^2 \ge 1.0 ~GeV^2$.  The experimental points are the data from CDHSW and NuTeV experiments. }
\label{fig:65GeVFe-nuallx}
\end{center}
\end{figure}

A comparison of the differential cross sections for Fe at E$_\nu$ = 65 GeV with the nPDFs labeled nCTEQnu as well as the theoretical predictions of the Aligarh-Valencia group based on both CTEQ and MMHT nucleon PDFs can be found in Fig.~\ref{fig:65GeVFe-nuallx}.  
Both the approaches are compared with the measured $\nu$-Fe cross sections from the CDSHW and NuTeV experiments. 
%Although the difference between the nCTEQ15 and nCTEQnu predictions are minimal at this energy, 
A first observation is that there is little difference in the full theoretical prediction based on either CTEQ or MMHT nucleon PDFs.  It is also clear that the low-x, low-y (= low-$Q^2$) and medium-x behavior of the NuTeV and CDHSW measurements tend to favor the phenomenological nPDF (nCTEQnu) results rather than the theoretical approach based on applying nuclear effects to nucleon PDF based structure functions. This observations is not surprising since the NuTeV results were used in the fit to determine the nCTEQnu nPDFs.

\begin{figure}[tb]
\begin{center}
\includegraphics[width=0.65\textwidth]{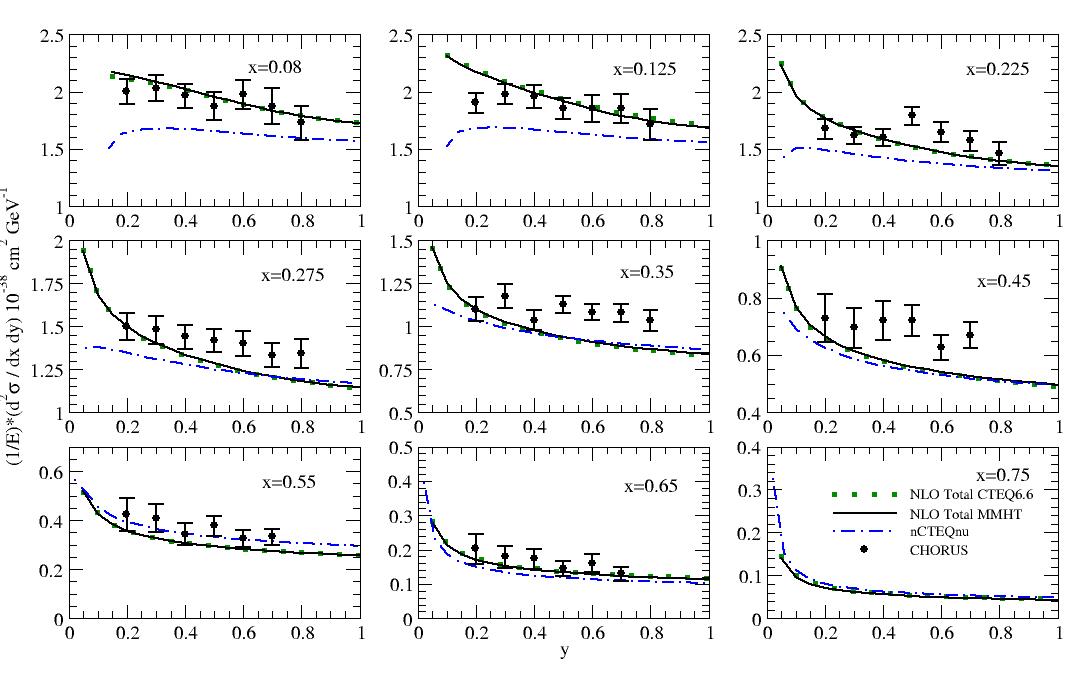}
%\caption{Results of the differential scattering cross section vs y, at different x for $\nu$ induced reaction on Pb at $E_\nu$ = 55 GeV. The results are obtained by using (i) CTEQ6.6 nucleon PDFs at NLO in the MS-bar scheme (dotted line), (ii) MMHT nucleon PDFs at NLO (solid line). The experimental points are the data from the CHORUS experiment. Here lead is treated as isoscalar target. The blue dash-dotted line is the result from nCTEQnu nPDFs with $Q^2 \ge 1.0 ~GeV^2$.
%}
\caption{Differential scattering cross section for $\nu$-Pb (treated as an isoscalar target) at $E_\nu$ = 55 GeV.  The lines representing the theoretical and nPDF approaches have the same meaning as in Fig.~\ref{fig:65GeVFe-nuallx}. Solid circles are the data points from the CHORUS experiment.}
\label{fig:55GeVPb-nuallx}
\end{center}
\end{figure}

%This observation is, except for the very lowest x-bin, similar for 
We draw different conclusions from the comparison of CHORUS $\nu$-Pb results at E$_\nu$ = 55 GeV with theoretical and phenomenological predictions in~Fig.~\ref{fig:55GeVPb-nuallx}. At low-x the data are consistent with the theoretical approach and above the nPDF predictions.  At mid- and high-x both the theoretical and the nPDF approaches agree and the CHORUS data lie above both. The better fit of the nPDF results to Fe compared to Pb is not surprising since the quite small errors on the NuTeV data insured that Fe results would dominate the global fit.

\subsection{Predictions for Future Experimental Measurements}\label{prediction}
%In the previous section we have seen comparisons of both theoretical and phenomenological predictions compared to existing experimental results at neutrino energies greater than 30 GeV. 
Of course, the on-going and future neutrino cross section and oscillation experiments are not using $\nu~and~\nub$ beams with the high energies of past experiments.   In light of this we include our predictions of what on-going cross section experiments and the future DUNE oscillation experiment might expect as DIS contributions to their statistics.  

\begin{figure}[tb]
\begin{center}
\includegraphics[width=0.65\textwidth]{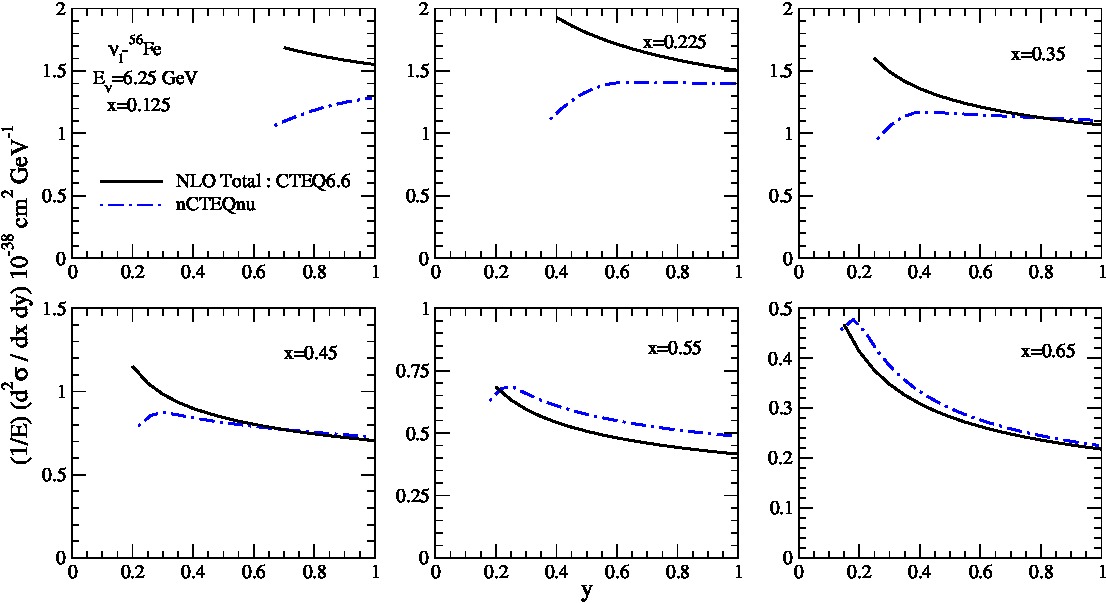}
\caption{Predictions for the differential scattering cross section vs y, at different values of x for 6.25 GeV $\nu$-Fe treated as an isoscalar target. The results are obtained with a $Q^2 \ge 1.0 ~GeV^2$ cut by the Aligarh-Valencia model using CTEQ 6.6 nucleon PDFs at NLO in the MS-bar scheme (solid line).  The nCTEQnu nuclear PDFs based prediction is the blue dash-dotted line. 
%The lower-y downward curve of the prediction at a given x corresponds to the extrapolation of the nCTEQnu global fit below $Q_0^2$ = 1.69 $GeV^2$.
}
\label{fig:625gev-Fe}
\end{center}
\end{figure}

\begin{figure}[tb]
\begin{center}
\includegraphics[width=0.65\textwidth]{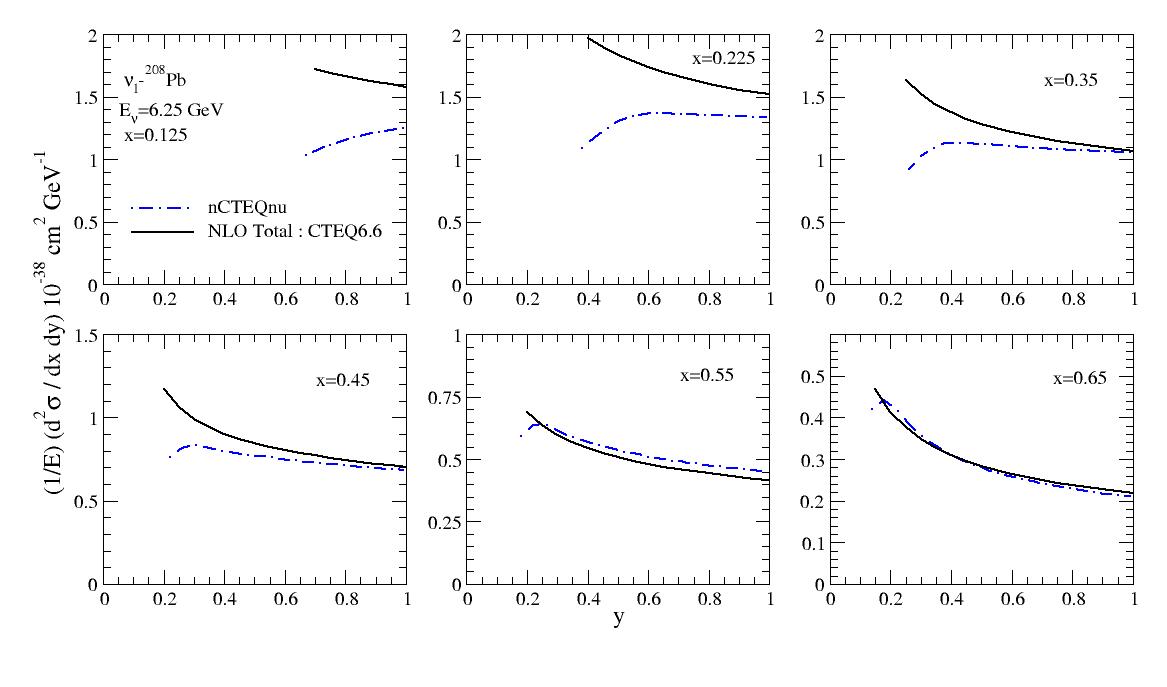}
%\caption{Results of the differential scattering cross section vs y, at different values of x for 6.25 GeV %$\nu$ induced reaction on Pb treated as an isoscalar target. The results are obtained with a $Q^2 \ge 1.0 %~GeV^2$ cut by the Aligarh-Valencia model using CTEQ6.6 nucleon PDFs at NLO in the MS-bar scheme (solid %line).  The nCTEQnu nuclear PDFs based prediction is the blue dash-dotted line. The lower-y downward curve %of the prediction at a given x corresponds to the extrapolation of the nCTEQnu global fit below $Q_0^2$ = %1.69 $GeV^2$.}
\caption{Predictions of the differential scattering cross section vs y at different values of x for 6.25 GeV $\nu$-Pb treated as an isoscalar target. The solid and dash-dotted lines in this figure have the same meaning as in Fig.\ref{fig:625gev-Fe}. }
\label{fig:625gev-Pb}
\end{center}
\end{figure}

Assuming a 6.25 GeV neutrino beam, the average energy of the \minerva\ ME beam, and a $Q^2 \ge 1.0 ~GeV^2$ cut, Fig.~\ref{fig:625gev-Fe} and Fig.~\ref{fig:625gev-Pb} show the expected cross sections from the Aligarh-Valencia theoretical calculations and the CTEQ neutrino-based (nCTEQnu) nuclear PDFs for Fe and Pb respectively. For both nuclei, in the mid-x region from $\approx$ 0.3 to $\approx$ 0.5 the two approaches agree at higher y (= higher $Q^2$). As y decreases, the nPDF approach predicts lower cross sections than the theoretical approach.  For high-x ($\gtrapprox$ 0.5), the nPDF approach and the theoretical approach predicts quite similar cross sections while for low-x ($\lessapprox$ 0.3) the nPDF approach predicts a lower cross section than the theoretical approach.

\begin{figure}[tb]
\begin{center}
\includegraphics[width=0.65\textwidth]{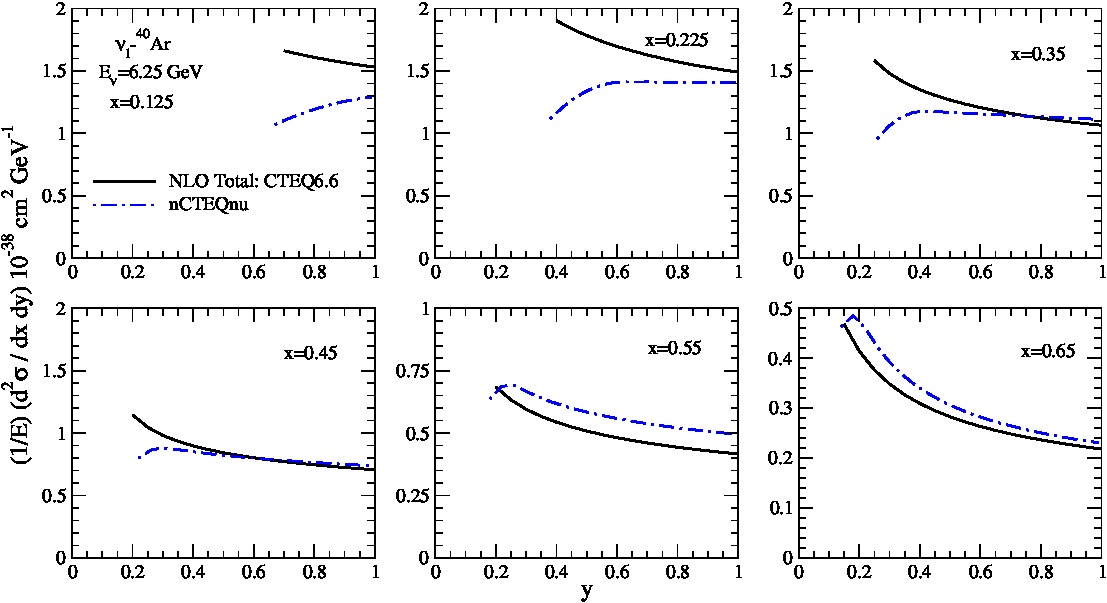}
%\caption{Results of the differential scattering cross section vs y, at different values of x for 6.25 GeV %$\nu$ induced reaction on Ar treated as an isoscalar target. The results are obtained with a $Q^2 \ge 1.0 %~GeV^2$ cut by the Aligarh-Valencia model using CTEQ6.6 nucleon PDFs at NLO in the MS-bar scheme (solid %line).  The nCTEQnu nuclear PDFs based prediction is the blue dash-dotted line. The lower-y downward curve %of the prediction at a given x corresponds to the extrapolation of the nCTEQnu global fit below $Q_0^2$ = %1.69 $GeV^2$.}
\caption{Prediction of the differential scattering cross section vs y at different values of x for 6.25 GeV $\nu$-Ar. The lines in this figure have the same meaning as in Fig.\ref{fig:625gev-Fe}. }
\label{fig:625gev-Ar}
\end{center}
\end{figure}

For the future DUNE neutrino oscillation experiment Fig.~\ref{fig:625gev-Ar} shows predictions of both the full theoretical model and the nCTEQnu nPDFs for the differential cross sections with a 6.25 GeV neutrino beam on Ar. The comparison of the two approaches demonstrated in this figure is quite similar to what has been shown in the Fig.~\ref{fig:625gev-Fe} and Fig.~\ref{fig:625gev-Pb} for Fe and Pb.

In general it should be noted that there are very small differences between the predictions for Ar, Fe and Pb treated as isoscalar targets for the same $E_\nu$.   This also supports the observation that the x-dependent nuclear effects for larger nuclei, such as the three here considered nuclei, have a rather weak A-dependence.  The actual ratios of Fe/Ar and Pb/Ar in this analysis differ by less than 3 \% over the entire allowable x and y kinematic plane.

\begin{figure}[tb]
\begin{center}
\includegraphics[width=0.65\textwidth]{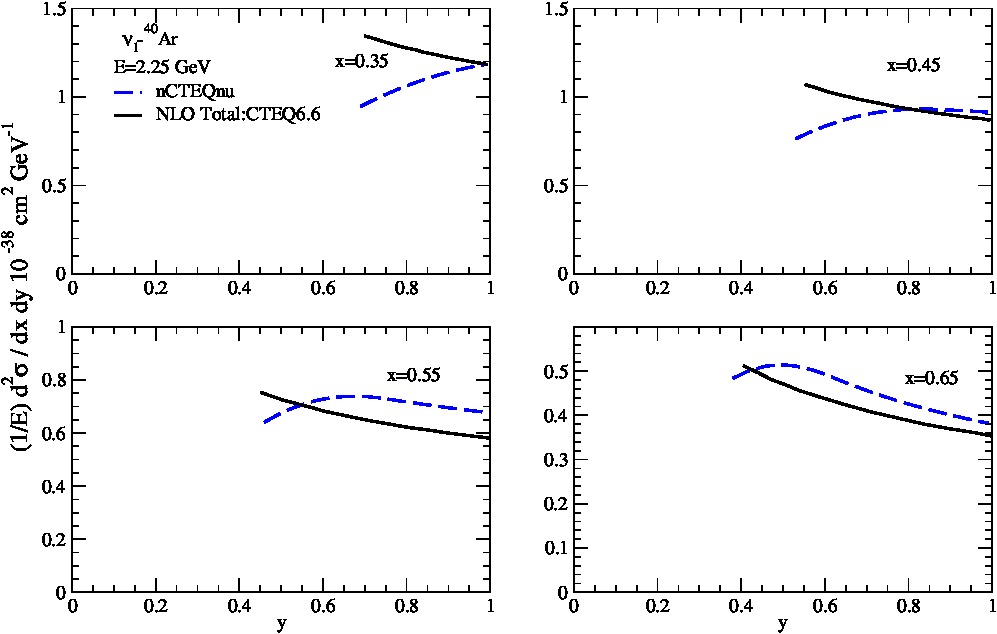}
%\caption{Results of the differential scattering cross section vs y, at different values of x for 2.25 GeV %$\nu$ induced reaction on Ar treated as an isoscalar target. The results are obtained with a $Q^2 \ge 1.0 %~GeV^2$ cut by the Aligarh-Valencia model using CTEQ6.6 nucleon PDFs at NLO in the MS-bar scheme (solid %line).  The nCTEQnu nuclear PDFs based prediction is the blue dash-dotted line.}
\caption{Prediction of the differential scattering cross section vs y at different values of x for 2.25 GeV $\nu$-Ar. The lines in this figure have the same meaning as in Fig.\ref{fig:625gev-Fe}. }
\label{fig:225gev-Ar}
\end{center}
\end{figure}

Fig.~\ref{fig:225gev-Ar} illustrates the much more restricted DIS contribution expected with 2.25 GeV neutrinos. The $Q^2 \geqq~1.0~ GeV^2$ cut restricts lower-x contributions at this energy and further restricts lower-y contributions at a given x.  Over the kinematic regions allowed, there are obvious differences in the predictions of the two approaches that are similar to the observations drawn for the $E_\nu$ = 6.25 GeV Ar example. 

Note that for the predictions of the nCTEQnu nuclear PDFs at 6.25 GeV there is an x-y region corresponding to $Q^2 \ge$ 1 $GeV^2$ however lower than the $Q_0^2$ = 1.69 $GeV^2$ of the nCTEQnu DGLAP expansion. This region  requires an extrapolation that has been performed with the technique provided by the LHAPDF library~\cite{Buckley:2014ana}.  Although the low x - low y NuTeV and CCFR data in Fig.~\ref{fig:65GeVFe-nuallx} support a downward trend of the cross section, the lower-y behavior at a given x is coming mainly from this extrapolation below $Q_0^2$.
%for 1.0 $GeV^2 \le Q^2 \le 1.69 GeV^2$. 

Future global fits of neutrino-nucleus results should consider the well-known lower range of neutrino energies required for current neutrino experiments. The future fits should then take into account the observation of the current theoretical study indicating that, with inclusion of the TMC, any required dynamical higher twist is minimal.  This should allow the introduction of a lower $Q_0$ and lower $Q^2$ cut on the included data than that used in current analyses.
% and the consequent restrictions for using nuclear parton distributions,

 \section{Conclusions}
\label{Sec-Conclusions}
In this review we have examined the higher-W SIS region and the kinematically defined DIS region.  We have found in both the SIS and DIS regions considerable need for further theoretical and experimental efforts to better understand these regions.  We summarize here the main conclusions of our study.  
\subsection{Theoretical Picture of Deep Inelastic $\nu/\nub$ Nucleus Scattering}
%  %We have observed that the difference in the results of free nucleon structure functions $F_{iN}^{WI}(x,Q^2)~(i=1,2)$ with TMC effect, (i) evaluated at NLO with HT effect and the results 
%   (ii) obtained at NNLO without HT, is almost negligible ($<1\%$). This difference is somewhat larger for $F_{3N}^{WI}(x,Q^2)$ at low $x$ and low $Q^2$ which becomes smaller with the increase in $Q^2$.
%  In the case of nucleons bound inside a nucleus, the HT corrections are further suppressed due to the presence of nuclear medium effects. Consequently, the results for 
%  $\nu_l/\bar\nu_l-A$ DIS processes which are evaluated at NNLO have almost negligible difference from the results obtained at NLO with HT effect.  Thus we conclude that as long as TMC effects are applied, the effect of the dynamical higher twist (HT) in nuclei is small in comparison to the free nucleon case and the results obtained at NNLO are very close to the results obtained at NLO with HT(within a percent).
 
  We have studied nuclear medium effects in the structure functions $F_i^A(x,Q^2)$, i=1-3, using Aligarh-Valencia model and obtained the differential scattering cross sections, in $\nu$, $\bar\nu$ scattering from several nuclear targets like C, Ar, Fe and Pb. Starting with free nucleons, using several free nucleon PDF sets, the medium effects were included using many body field theoretical technique to describe the spectral function of the nucleon in the nuclear medium. The local density approximation has been applied to translate results from nuclear matter to nuclei of finite size. The effect of Fermi motion, binding energy, nucleon correlations as well as the effect of mesonic($\pi$ and $\rho$) contributions in $F_i^A(x,Q^2)$, i=1-2 and shadowing have been taken into account leading to a dynamical(nonstatic) treatment of the nucleon and the mesons in the nuclear medium. 
  
  This study has been performed for a wide range of $x$ and $Q^2$.   In general, in comparison to the results obtained for the free nucleon case, we find that the use of the spectral function results in the reduction of the nuclear structure functions (and consequently the differential cross sections) in the intermediate region of $x$ and an enhancement (mainly due to Fermi motion effect) at high $x$,  These results are $Q^2$ dependent with the effect more pronounced at low $Q^2$ and $A$ dependent with the suppression in the intermediate region of $x$ and the enhancement at high $x$ increasing with the the mass number $A$. Furthermore, the inclusion of mesonic contributions results in an enhancement in the nuclear structure functions in the low and intermediate region of $x$ with the enhancement mainly due to pionic rather than rho meson effects.  These mesonic contributions are suppressed with an increase in $x$ and $Q^2$ and are observed to be more pronounced with the increase in mass number $A$ as there are more nucleons and the probability of interactions among nucleons via meson exchange increases. 
  The effect of shadowing is included resulting in a reduction in the nuclear structure functions at low $x$ that increases with increased $A$. 
  
  The nuclear medium effects are found to be significant in the evaluation of nuclear structure functions $F_{1A}^{WI}(x,Q^2)$, $F_{2A}^{WI}(x,Q^2)$ and $F_{3A}^{WI}(x,Q^2)$.  In the free nucleon case  we have shown that the difference in the nucleon structure functions $F_{iN}^{WI}(x,Q^2)~(i=1,2)$ with TMC effect evaluated at NLO with HT effect and evaluated at NNLO without HT are essentially negligible ($<1\%$). This difference is somewhat larger for $F_{3N}^{WI}(x,Q^2)$ at low $x$ and low $Q^2$ which becomes smaller with the increase in $Q^2$.
 In the case of nucleons bound inside a nucleus, the HT corrections are even further suppressed due to the presence of nuclear medium effects. Consequently, the results for $\nu/\bar\nu-A$ DIS processes which are evaluated at NNLO have almost negligible difference from the results obtained at NLO with HT effect.  Thus we conclude that as long as TMC effects are applied, the effect of the dynamical higher twist (HT) in nuclei is small in comparison to the free nucleon case and the results obtained at NNLO are very close to the results obtained at NLO with HT(within a percent).

We find that the nuclear-medium effects are different in $F_{1A}(x, Q^2)$,  $F_{2A}(x, Q^2)$ and $F_{3A}(x, Q^2)$ structure functions and are more pronounced
 in the $\bar\nu-A$ reaction channel than in the case of $\nu-A$ scattering. This can be observed in $F_{3A}(x, Q^2)$, describing the behavior of valence quarks, where the mesonic contributions are absent and in the behavior of the Callan-Gross relation $\frac{F_{2A}(x, Q^2)}{2xF_{1A}(x, Q^2)}$, which is observed to become violated at low $x$. 
 
The correction due to the excess of neutrons over protons (isoscalarity effect) is significantly large for the lead nucleus, for example, $5\%$ at low $x$ and $15\%$ at high $x$, while in argon nucleus it is $\sim 2\%$ at low $x$ and $\sim 4\%$ at high $x$. Significantly, we have found that the nuclear medium effects are different in electromagnetic and weak interaction channels especially for the 
 nonisoscalar nuclear targets. The contribution of strange and charm quarks is found to be different for the electromagnetic and weak interaction induced processes off free
 nucleon target which also gets modified differently for the heavy nuclear targets.
 Furthermore, we have observed that the isoscalarity corrections, significant even at high $Q^2$, and are not the same in $F_{1A}^{WI}(x,Q^2)$ and $F_{2A}^{WI}(x,Q^2)$.
 
 As presented in section~\ref{comp}, the full theoretical model shows reasonable agreement with the experimental data of CCFR, CDHSW, NuTeV and CHORUS data in the mid $x$ and high $Q^2$ regions.  However in the low-x (shadowing) region and the high-x (EMC) region the agreement of the predicted differential scattering cross sections with the NuTeV and CHORUS data is not as good. 
 
%\end{itemize}
It is apparent that in the precision era of neutrino oscillation physics, it is necessary to address differences in predictions compared to the few existing experimental results. Suggesting a need for more measurements of nuclear effects in a wide range of $A$, using neutrino and antineutrino beams in a broad kinematic range of $x$ and $Q^2$.  
%We also need measurements with much-increased statistical and systematic precision of cross-sections (both  total and differential) with neutrino and antineutrino beams on free proton and deuteron targets.

\subsection{Phenomenological Picture of $\nu/\nub$ Nucleus Scattering} 
\paragraph{Shallow Inelastic Scattering}
It should now be quite obvious that the higher-W SIS region in both neutrino nucleon $\nu$-N and neutrino nucleus $\nu$-A scattering is unexplored  experimentally and essentially so theoretically.   Fig.\ref{fig:SIDISwithGens} starkly presents the difference in the simulations of this kinematic region. In increasing W from the $\Delta$ there are only a few $\nu$-N resonance models that treat more than 1-$\pi$ production and it is clear that multi-$\pi$ production can be significant in this high-W region.  As far as non-resonant production is concerned there are several models available for single-$\pi$ non-resonant production including the recent efforts of~\cite{Kabirnezhad:2016nwu} and references therein.  However models of non-resonant two-$\pi$ or more production are not available.  Certainly the careful understanding of how SIS non-resonant $\pi$ production smoothly transforms into DIS pion production is crucial for this transition region and has not been carefully addressed theoretically or experimentally.

Approaching the SIS region from the higher-W DIS region there is no well-defined sharp boundary between the two.  $Q^2 \ge 1 ~GeV^2$ is chosen as the minimum $Q^2$ needed to be interacting with quarks within the nucleon and W $\ge$ 2.0 GeV has been chosen as "safely" out of the resonance region with only very few resonances experimentally defined above this boundary.  This, in principle, allows the so-defined DIS region to be described by perturbative QCD.  At these boundaries and below in $Q^2$ and W is the kinematic region where non-perturbative QCD effects come into serious consideration. A topic very much neglected in $\nu$ nucleon/nucleus physics.  
%What is happening in this transition region between the SIS and  DIS regions?  
Is there a change in the relative strength of SIS and DIS cross sections at this transition?  Is there not a theoretical connection that can be made between increasing W non-resonant pion production and non-perturbative QCD effects?  This is, of course, the goal of the application of duality.

Duality is a concept that supposedly allows phenomena in the DIS region to approximate activity in the SIS region.
%in similar kinematic regions.   
Although duality has been quite thoroughly tested in electroproduction experiments.  It cannot presently be tested in the same manner in $\nu$-N and $\nu$-A scattering due to an obvious lack of experimental data.  However, from the model-dependent studies that have been made of $\nu$-N scattering it appears that duality might be better applied to $\nu$-isoscalar N scattering and not for  $\nu$-p or $\nu$-n scattering individually.  

The many open challenges for this kinematic region can then be summarized as:
\begin{itemize}
    \item A need for much increased experimental investigation of the higher-W kinematic region for single and multi-$\pi$ production.
    \item A need for models of resonant multi-$\pi$ production up to and through the transition into the DIS region.
    \item A need for models of non-resonant multi-$\pi$ production and a better understanding of how non-resonant single and multi-$\pi$ production in the SIS region transitions into DIS single and multi-$\pi$ production.
    \item A much more thorough investigation of non-perturbative QCD effects and how they can be mapped onto non-resonant $\pi$ production in the SIS to DIS transition region is required.
    \item A better understanding of how duality can help address some of these previous listed challenges would be helpful.  The managers of the various simulation programs should check whether their simulations of the SIS and DIS regions for the average nucleon (n+p)/2 are reasonably consistent with the current expectations of duality.
\end{itemize}

\paragraph{Deep Inelastic Scattering}
In contrast to the SIS region, there have been several experimental and many phenomenological studies of the DIS region for both  $\nu$-N and $\nu$-A scattering.  In the DIS region perturbative QCD plus factorization allows a phenomenological approach to the extraction of the parton distribution functions of both the free nucleon (PDFs) and nucleons bound in the nuclear environment (nPDFs) where nuclear medium effects are significant.  While the free nucleon PDFs have been extracted via global fits by many groups, far fewer attempts have been made to extract the nPDFs of nucleons within a nucleus. 

Among the groups concentrating on these nPDFs the nCTEQ group has found a difference in the nPDFs extracted from a global fit using $\ell^\pm$-A scattering and those extracted from a fit using $\nu(\nub)$-A scattering based on the experimental results of CCFR, NuTeV and CHORUS. The difference is most evident in comparing the nuclear correction factors as a function of x for $\nu(\nub)$-A and $\ell^\pm$-A based analyses.  The difference is significant in both location and intensity of the expected nuclear effects of shadowing, antishadowing and the EMC effect.  Other groups fitting nPDFs based on DIS neutrino scattering use different techniques than nCTEQ and are able to find compatible fits including both $\ell^\pm$-A and $\nu$-A.  

It is significant to note that the kinematic regions showing the largest difference between nCTEQ $\nu(\nub)$-A based and $\ell^\pm$-A based analyses are also the regions with the largest differences between the theoretical and nPDF results summarized in this paper.  Particularly the nPDF predicted stronger suppression of the cross section in the low-$Q^2$, low-x shadowing region and the elevated cross section in the EMC region, both directly reflecting the quoted experimental results, emphasize these differences.
When comparing the nCTEQ $\nu(\nub)$-A to $\ell^\pm$-A based analyses, these differences can be attributed to the differences of the weak compared to EM interactions.  However, the theoretical considerations of $\nu(\nub)$-A DIS, summarized in this paper, includes the accepted theoretical considerations of the weak interaction of these two regions in the calculations so the differences here are intriguing.  Since the most recent considerations of shadowing and the EMC effect in $\nu(\nub)$-A DIS interactions presented in the phenomenological section are still speculative, they have not yet been included in the theoretical treatment of these two regions presented here but could indeed provide an explanation of the differences.

The differences in the  $\ell^\pm$-A based and $\nu$-A based results could suggest interesting consequences.  In particular for the low-x region, there are many theoretical and now experimental indications that shadowing is a quite different process in $\ell^\pm$-A and $\nu$-A interactions. The theoretical indications are based on the presence of the axial vector current and the considerably more massive IVB involved in neutrino scattering.  If this fundamental difference does exist, it would follow that there should not be the same universal nPDFs describing this low-x region for $\ell^\pm$-A and $\nu(\nub)$-A analyses unless, for example, a term is incorporated perhaps in the factorization that accounts for IVB-dependent phenomena in the nuclear environment. A resolution of these disagreements is essential for proper simulation of DIS scattering in current and future neutrino oscillation experiments.

%To conclude, the  next generation experiments like DUNE and HyperK, are expected to provide valuable information about  neutrino properties, nevertheless, any precise measurement can only be done by reducing the systematic uncertainties which has presently  about 25$\%$ contributions due to the lack in the understanding of $\nu(\bar\nu_l)-N$ and $\nu(\bar\nu_l)-A$ scattering cross sections in the few GeV energy region, where the major contribution comes from the shallow inelastic and the deep inelastic regions. Therefore, in any forward journey,  it is important to understand theoretically hadron production in the $\nu(\bar\nu_l)$  induced processes as well as the nucleon dynamics in the nuclear medium. In the years to come, greater efforts even experimentally are required, which can better be achieved if the experiments are planned to study  $\nu(\bar\nu_l)$  scattering off proton/deuteron target as well as EMC kind of measurements in the 1-3GeV energy region.

\subsection{Summary} On-going and next generation oscillation experiments like T2K, NOvA, DUNE and HyperK as well as  experiments using atmospheric neutrino such as  IceCube~\cite{Aartsen:2019eht}, JUNO~\cite{Guo:2017mvv} and INO~\cite{Kumar:2017sdq} are expected to provide valuable information about neutrino properties in the roughly 1-10 GeV neutrino energy region.  Significantly, precise measurements of these properties can only be achieved by reducing systematic uncertainties. Currently, considering the target material of these experiments, a large portion of these uncertainties is due to the lack of precise 
%fundamental $\nu(\bar\nu)$–nucleon cross sections and,  the 
cross sections and, most importantly, nuclear effects in $\nu(\bar\nu)$-nucleus scattering. For NOvA and DUNE as well as atmospheric neutrino oscillation studies of SuperK, HyperK, IceCube, JUNO and INO a reasonable or even major fraction of events come from the higher-W shallow inelastic and deep inelastic scattering regions.  This review has highlighted the many current concerns and challenges, both theoretical and experimental, in these regions.  

Therefore, 
%in any forward journey, 
it is important to much improve the nuclear model that covers these two regions, which includes the understanding of nucleon dynamics in the nuclear medium, the resulting hadron production in $\nu(\bar\nu)$-nucleon induced processes as well as the role of final state interactions within the nucleus.  To improve this model in the SIS and DIS regions will take the dedicated efforts of theorists and experimentalists working together with neutrino event simulation experts.  In particular a significant enhancement in the measurement of fundamental $\nu(\bar\nu)$-nucleon scattering as well as precision measurements of $\nu(\bar\nu)$ scattering off a variety of nuclear targets in the SIS and DIS regions would be welcome.  The community and relevant funding agencies should recognize this essential collaborative effort and provide the support necessary for the experiments to reach their stated precision goals.  

\section{Acknowledgements}
We most gratefully acknowledge the invaluable assistance of G. Caceres Vera, H. Haider, I.~Ruiz Simo, A. Kusina and F. Zaidi.  We also appreciate the many informative discussions we had with our NuSTEC colleagues.
M. S. A. is thankful to Department of Science and Technology (DST), Government of India for providing financial assistance under Grant No. EMR/2016/002285.
J. G. M. has been supported by Fermi Research Alliance, LLC under Contract No. DE-AC02-07CH11359 with the U.S. Department of Energy, Office of Science, Office of High Energy Physics.

\section{Appendices}

%\appendix
\begin{appendices}
\section{Neutrino self-energy}\label{app:self}
\begin{figure}[h]
\begin{center}
 \includegraphics[height=2.5 cm, width=12 cm]{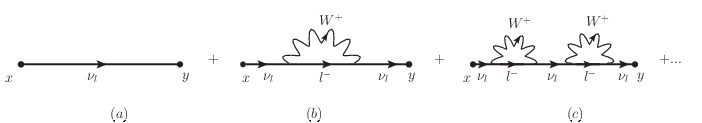}
 \end{center}
 \caption{Representation of neutrino self energy.}
 \label{self_fig}
\end{figure}

When a neutrino interacts with a potential provided by a nucleus (in the present scenario), then the interaction in the language of many body field theory 
can be understood as the modification of the fermion two points function represented by the diagrams shown in Fig.\ref{self_fig}.
\begin{figure}[h]
\begin{center}
 \includegraphics[height=1. cm, width=6 cm]{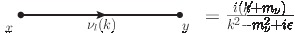}\\
  \includegraphics[height=2. cm, width=8 cm]{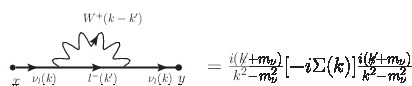}
 \end{center}
 \caption{(Top) Free field fermion propagator, (Bottom) The term that constitutes to neutrino self energy,}
 \label{xyz}
\end{figure}

The first diagram (a) in Fig.\ref{self_fig} is just the free field fermion propagator and the second diagram (b) constitutes to the neutrino self energy which is expressed in Fig.\ref{xyz},
%\begin{figure}[h]
%\begin{center}
% \includegraphics[height=2.2 cm, width=8 cm]{second_part.jpg}
% \end{center}
%\end{figure}
where 
\begin{eqnarray}
 -i\Sigma(k) &=& \int \frac{d^4 k'}{(2\pi)^4}\;\left(-\frac{i g}{2 \sqrt{2}} \gamma^\mu (1-\gamma_5) \right)\;\frac{i(\not k'+m_l)}{k'^2-m_l^2+i\epsilon} \times\nonumber\\
 && \left(-\frac{i g}{2 \sqrt{2}} \gamma^\nu (1-\gamma_5) \right)\;\frac{-ig_{\mu\nu}}{(k-k')^2-M_W^2+i\epsilon}
\end{eqnarray}
Notice that $\Sigma$ has real and imaginary parts. The imaginary part of 
the neutrino self energy accounts for the depletion of the initial 
neutrinos flux out of the 
non-interacting channel, into the quasielastic or the inelastic channels.

By using the Feynman rules the neutrino self-energy corresponding to Fig.\ref{wself_energy} is written as
\begin{eqnarray}
    -i \Sigma(k)&=&\int \frac{d^4 q}{(2 \pi)^4}\Big( \bar u_\nu(k)\frac{-ig}{2\sqrt{2}}\gamma_\mu(1-\gamma_5)\times \frac{i(\not k'+m_l)}{k^{'2}-m_l^2+i\epsilon} \frac{-ig}{2\sqrt{2}}\gamma_\nu(1-\gamma_5)u_\nu(k)\Big)\nonumber\\
    &\times& \Big(-\frac{i g^{\mu\rho}}{q^2-M_W^2} \Big)(-i\Pi_{\rho\sigma(q)})\Big(-\frac{i g^{\sigma\nu}}{q^2-M_W^2} \Big)
\end{eqnarray}
which after simplification modifies to
\begin{eqnarray}
-i \Sigma(k)&=& \frac{g^2}{8 M_W^2}\int \frac{d^4 q}{(2 \pi)^4}Tr\{(\not k+m_\nu)\gamma_\mu(1-\gamma_5)(\not k'+m_l)\gamma_\nu(1-\gamma_5) \}\nonumber\\
&\times&\frac{\Pi^{\mu\nu}(q)}{2m_\nu(k^{'2}-m_l^2+i\epsilon)}\Big(\frac{M_W^2}{q^2-M_W^2}\Big)^2\nonumber
\end{eqnarray}
Now by using the following relations 
\begin{eqnarray}
\frac{g^2}{8 M_W^2} &=& \frac{G_F}{\sqrt{2}};\hspace{4 mm}
d^4 q=d^4 k';\hspace{4 mm}
\sum_r u_r(k) \bar u_r(k)=\frac{\not k+m_\nu}{2 m_\nu}\nonumber
\end{eqnarray}
and the trace properties, neutrino self-energy is further simplified to 
\begin{eqnarray}
\Sigma(k)&=&\frac{- i G_F}{\sqrt{2}} \int \frac{d^4 q}{(2 \pi)^4} \frac{4 L_{\mu\nu}^{WI}}{m_\nu}\frac{1}{(k^{\prime 2}-m_l^2+i\epsilon)}\left(\frac{M_W}{Q^2+M_W^2}\right)^2 \Pi^{\mu\nu}(q)\;,
\end{eqnarray}
To obtain the imaginary part of neutrino self-energy which is required to evaluate the scattering cross section, Cutkosky rules are applied:
\begin{eqnarray}
\Sigma(k) &\rightarrow& 2 i Im \Sigma(k);~~~\textrm{Lepton~ self-energy} \nonumber\\
\Pi^{\mu\nu}(q) &\rightarrow& 2 i \theta(q^0) Im \Pi^{\mu\nu}(q);~~~\textrm{W~boson~ self-energy}\nonumber
\end{eqnarray}
It gives
\begin{eqnarray}
2 i Im \Sigma(k)&=&\frac{-i G_F}{\sqrt{2}} \frac{4}{ m_\nu} \int \frac{d^4 q}{(2 \pi)^4} 2 i Im\Big(\frac{1}{(k^{\prime 2}-m_l^2+i\epsilon)}\Big)2 i  \theta(q^0) \left(\frac{M_W}{Q^2+M_W^2}\right)^2\;\nonumber\\
&\times& Im[L_{\mu\nu}^{WI } \Pi^{\mu\nu}(q)].\nonumber
\end{eqnarray}
\begin{comment}
%In the above expression, the term $Im \left( \frac{1}{k'^{2}-m_l^2 + i \epsilon}\right)$ is solved
%by using the Sokhatsky-Weierstrass theorem which states that
%\begin{equation}\label{cauchy}
%lim_{\epsilon \to 0^+}\left(\frac{1}{x\pm i \epsilon} \right) = {\cal P}\left(\frac{1}{x}\right) \mp i \pi %\delta(x)
%\end{equation}
%\begin{eqnarray} %\label{img}
 %lim_{\epsilon\rightarrow0}\left(\frac{1}{k^{\prime 2}-m_l^2+i\epsilon} \right)&=&{\cal %P}\left(\frac{1}{k^{\prime 2}-m_l^2} \right)~-~i\pi\delta
 %(k^{\prime2}-m_l^2)\nonumber
 %\end{eqnarray}
\end{comment}
Using Sokhotski-Plemelj theorem and equating the imaginary terms on both sides, one may write
 \begin{eqnarray}\label{img}
 %Im \left(\frac{1}{k^{\prime 2}-m_l^2+i\epsilon} \right)&=& -\pi \delta(k^{\prime2}-m_l^2)\nonumber\\
 %&=&-\pi \delta(k^{\prime 2}_0-{\bf  k^{\prime 2}}-m_l^2)=-\pi \delta(k^{\prime 2}_0-E^{\prime 2}({\bf k^\prime}))\nonumber\\
 %&=&-\pi \delta\left[(k^{\prime}_0-E({\bf k^\prime}))(k^{\prime}_0+E({\bf k^\prime}))\right]\nonumber\\
 %&=&-\pi \left[\frac{\delta(k^\prime_0-E({\bf k^\prime}))}{2 E({\bf  k^\prime})}~+~\frac{\delta(k^\prime_0+E({\bf k^\prime}))}{|2 E({\bf k^\prime})|}  \right]\nonumber\\
  Im \left(\frac{1}{k^{\prime 2}-m_l^2 + i\epsilon} \right)&=&\frac{-\pi}{2~E({\bf k^\prime})},\;\; for\;\; E({\bf k}^\prime)\; > \;k_0^\prime
\end{eqnarray}
 where the energy transfer is $q^0=k^0~-~k^{\prime 0}=k^0-E({\bf k} -{\bf q})$. Using the property of the step function, the imaginary part of neutrino self-energy may be written as
%\begin{eqnarray}\label{thita}
%\theta(k_0^\prime)=
%\left\{
%\begin{array}{r}
%1~for~E({\bf k^\prime})>0\\
%0~for~E({\bf k^\prime})<0
%\end{array}\right.
%\end{eqnarray}
\begin{equation}
\Rightarrow Im \Sigma(k)=\frac{ G_F}{\sqrt{2}} \frac{4}{ m_\nu} \int \frac{d^3 q}{(2 \pi)^4} \frac{\pi}{ E({\bf k^\prime})} \theta(q^0) \left(\frac{M_W}{Q^2+M_W^2}\right)^2\;Im[L_{\mu\nu}^{WI } \Pi^{\mu\nu}(q)].
\end{equation}
%new section
\section{Nucleon spectral function}\label{spec_nuc}
The relativistic free nucleon Dirac propagator $G^{0}(p^{0},{{\bf p}})$ is given by
\begin{eqnarray}
 G^{0}(p^{0},{{\bf p}}) = \frac{1}{\not p - M_N+i \epsilon}=\frac{\not p+M_N}{(p^2-M_N^2+i\epsilon)} 
\end{eqnarray}
which may be rewritten in terms of both positive and negative energy states as
\begin{equation} \label{proppm}
G^{0}(p^{0},{{\bf p}}) =\frac{M_N}{E_N({\bf p})}\left\{\frac{\sum_{r}u_{r}({\bf p})\bar u_{r}({\bf p})}{p^{0}-E_N({{\bf p}})+i\epsilon}+\frac{\sum_{r}v_{r}(-{\bf p})
\bar v_{r}(-{\bf p})}{p^{0}+E_N({{\bf p}})-i\epsilon}\right\},
\end{equation}
where $E_N({\bf p})=\sqrt{|{\bf p}|^2+M_N^2}$ is the relativistic energy of an on shell nucleon. As it has been already mentioned in section\ref{spec} that negative energy components are suppressed than the positive energy components, therefore, only first term will contribute. Hence,
\begin{eqnarray}  
G^{0}(p^{0},{{\bf p}})&=&\frac{M_N}{E_N({{\bf p}})}\sum_{r}u_{r}({\bf p})\bar u_{r}({\bf p})
\left[\frac{1-n(\bf{p})}{p^{0}-E_N({{\bf p}})+i\epsilon}+\frac{n(\bf{p})}{p^{0}-E_N({{\bf p}})-i\epsilon}\right] \nonumber
\end{eqnarray}
where $n(\bf{p})$ is the occupation number of the nucleons in the Fermi sea, $n(\bf{p})=1$ for ${\bf p\le p_{F_{N}}}$ while $n(\bf{p})=0$ for ${\bf p > p_{F_{N}}}$. Using the following relation:
\begin{eqnarray}
\sum_r u_r(p) \bar u_r(p)&=&\frac{\not p+M_N}{2 M_N} \nonumber
\end{eqnarray}
the aforementioned expression for nucleon propagator modifies to
\begin{equation}
 G^{0}(p^{0},{{\bf p}})= {\not p +M_N \over p^2-M_N^2 +i\epsilon} + 2\ i \pi \theta(p^0) \delta(p^2 -M_N^2) n({\bf p}) (\not p +M_N)
\end{equation}
In the interacting Fermi sea, the relativistic nucleon propagator is written using Dyson series expansion (shown in Fig.\ref{n_self}) in terms of nucleon self energy $\Sigma^N(p^0, {\bf p})$. This perturbative expansion is summed in a ladder approximation as
\begin{eqnarray}\label{gpseries}
 G(p) &=& G^0(p)~+~G^0(p)\Sigma^N(p)G^0(p)~+~G^0(p)\Sigma^N(p)G^0(p)\Sigma^N(p)G^0(p)~+~.......\nonumber
\end{eqnarray}
One may notice that the aforementioned equation is a geometric progression series and using  Eq.\ref{proppm}, one may write Eq.\ref{gp1} as:
\begin{eqnarray}\label{self_sigma}
%G(p)&=&\frac{M_N}{E_N({\bf p})}\frac{\sum_{r}u_{r}({\bf p})\bar u_{r}({\bf p})}{p^{0}-E_N({\bf p})+ i \epsilon}+\frac{M_N}{E_N({\bf p})}\frac{\sum_{r}u_{r}({\bf p})\bar
%u_{r}({\bf p})}{p^{0}-E_N({\bf p})+i \epsilon}\Sigma^N(p^{0},{\bf p})\nonumber\\
%&\times&\frac{M_N}{E_N({\bf p})} \frac{\sum_{r}u_{r}({\bf p})\bar u_{r}({\bf p})}{p^{0}-E_N({\bf p})+i \epsilon}+..... \nonumber \\
G(p)&=&\frac{M_N}{E_N({\bf p})}\sum_{r}\frac{u_{r}({\bf p})\bar u_{r}({\bf p})}{p^{0}-E_N({\bf p})-\bar u_{r}({\bf p})\Sigma^N(p^{0},{\bf p})u_{r}({\bf p})\frac{M_N}{ E_N({\bf p})}}\nonumber
\end{eqnarray}
This expression contains nucleon self energy in the denominator which is a complex quantity, i.e.
\begin{eqnarray}
 \Sigma^N(p^0,{\bf p}) = Re\{\Sigma^N(p^0,{\bf p})\}~+~i Im\{\Sigma^N(p^0,{\bf p})\}
\end{eqnarray}
Using this definition in Eq.\ref{gp1}, the dressed nucleon propagator may be rewritten as 
 \begin{eqnarray}\label{gpc}
  G(p)&=&\frac{M_N}{E_N({\bf p})}~\sum_{r}u_{r}({\bf p})\bar u_{r}({\bf p})\times\nonumber\\
  &&\left[\frac{\{p^0-E_N({\bf p})-{M_N \over E_N({\bf p})} Re(\Sigma^N)\} + i \{{M_N \over E_N({\bf p})} Im(\Sigma^N)\}}
  {\{p^0-E_N({\bf p})-{M_N \over E_N({\bf p})} Re(\Sigma^N)\}^2 + \{{M_N \over E_N({\bf p})} Im(\Sigma^N) \}^2 }\right]
 \end{eqnarray}
  The use of nucleon Green functions in terms of their
spectral functions offers a precise way to account for Fermi motion and binding
energy. Basically spectral functions are
used to describe the momentum distribution of nucleons in the nucleus. Therefore, to determine the spectral functions of particle and hole let us define
\begin{footnotesize}
\begin{eqnarray}
 \int_{-\infty}^\mu d\omega \frac{S_h(\omega,{\bf p})}{p^0-\omega-i\eta}+ \int^{+\infty}_\mu d\omega \frac{S_p(\omega,{\bf p})}{p^0-\omega+i\eta}&=&{\cal P}\int_{-\infty}^\mu d\omega \frac{S_h(\omega,{\bf p})}{p^0-\omega}\nonumber\\
 +i\pi\int_{-\infty}^\mu d\omega S_h(\omega,{\bf p})\delta(p^0-\omega)+{\cal P}\int^{+\infty}_\mu d\omega \frac{S_p(\omega,{\bf p})}{p^0-\omega}&-&i\pi\int^{\infty}_\mu d\omega S_p(\omega,{\bf p})\delta(p^0-\omega),\nonumber
\end{eqnarray}
\end{footnotesize}
One may write with the help of Eq.\ref{gpc}
\begin{eqnarray}
 {\cal P}\int_{-\infty}^\mu d\omega \frac{S_h(\omega,{\bf p})}{p^0-\omega}&+&{\cal P}\int^{+\infty}_\mu d\omega \frac{S_p(\omega,{\bf p})}{p^0-\omega}+i\pi S_h(p^0,{\bf p})\theta(\mu-p^0)\nonumber\\
-i\pi\times  S_p(p^0,{\bf p})\theta(p^0-\mu)&=&\left[\frac{\{p^0-E_N({\bf p})-{M_N \over E_N({\bf p})} Re(\Sigma^N)\} + i \{{M_N \over E_N({\bf p})} Im(\Sigma^N)\}}
  {\{p^0-E_N({\bf p})-{M_N \over E_N({\bf p})} Re(\Sigma^N)\}^2 + \{{M_N \over E_N({\bf p})} Im(\Sigma^N) \}^2 }\right]\nonumber
 \end{eqnarray}
\begin{comment}
\begin{eqnarray}
 {\cal P}\int_{-\infty}^\mu d\omega \frac{S_h(\omega,{\bf p})}{p^0-\omega}&+&{\cal P}\int^{+\infty}_\mu d\omega \frac{S_p(\omega,{\bf p})}{p^0-\omega}+i\pi (S_h(p^0,{\bf p})\theta(\mu-p^0)-\nonumber\\
S_p(p^0,{\bf p})\theta(p^0-\mu))&=&\frac{p^0-E_N({\bf p})-\frac{M_N}{E_N({\bf p})}Re\Sigma^N}{(p^0-E_N({\bf p})-\frac{M_N}{E_N({\bf p})}Re\Sigma^N)^2+(\frac{M_N}{E_N({\bf p})}Im\Sigma^N)^2}\nonumber\\
&&+\frac{i\frac{M_N}{E_N({\bf p})} Im\Sigma^N}{(p^0-E_N({\bf p})-\frac{M_N}{E_N({\bf p})}Re\Sigma^N)^2+(\frac{M_N}{E_N({\bf p})}Im\Sigma^N)^2}\nonumber
\end{eqnarray}
\end{comment}
On comparing imaginary parts on both sides, we obtain
%\begin{eqnarray}
 %\frac{\frac{M_N}{E_N({\bf p})} Im\Sigma^N}{(p^0-E_N({\bf p})-\frac{M_N}{E_N({\bf %p})}Re\Sigma^N)^2+(\frac{M_N}{E_N({\bf p})}Im\Sigma^N)^2}&=&\pi [S_h(p^0,{\bf %p})\theta(\mu-p^0)\nonumber\\
 %&& -S_p(p^0,{\bf p})\theta(p^0-\mu)]\nonumber
 %\end{eqnarray}
 \begin{eqnarray}
 S_h(p^0,{\bf p})&=&\frac{1}{\pi}\frac{\frac{M_N}{E_N({\bf p})} Im\Sigma^N}{(p^0-E_N({\bf p})-\frac{M_N}{E_N({\bf p})}Re\Sigma^N)^2+(\frac{M_N}{E_N({\bf p})}Im\Sigma^N)^2};~~\mbox{for $p^0 \leq \mu$}\nonumber\\
 %&&~~~~\mbox{for $p^0 \leq \mu$}\nonumber\\
 S_p(p^0,{\bf p})&=&-\frac{1}{\pi}\frac{\frac{M_N}{E_N({\bf p})} Im\Sigma^N}{(p^0-E_N({\bf p})-\frac{M_N}{E_N({\bf p})}Re\Sigma^N)^2+(\frac{M_N}{E_N({\bf p})}Im\Sigma^N)^2};~~\mbox{for $p^0 > \mu$} \nonumber
 %&&~~\mbox{for $p^0 > \mu$} \nonumber
\end{eqnarray}
Using the above two equations in Eq.(\ref{gpc}), the dressed nucleon propagator is obtained in terms of the particle and hole spectral functions as:
\begin{eqnarray}
G(p^0,{\bf p})=\frac{M_N}{E_N({\bf p})}\sum_r u_r({\bf p})\bar u({\bf p})\left[\int_{-\infty}^\mu \frac{S_h(p^0,{\bf p}) d\omega}{(p^0-\omega-i\eta)}+\int^{\infty}_\mu \frac{S_p(p^0,{\bf p}) d\omega}{(p^0-\omega+i\eta)} \right]\nonumber
\end{eqnarray}
%new section
\section{Properties of spectral function}\label{spec_prop}
\begin{figure}
\begin{center} 
 \includegraphics[height=0.28\textheight,width=0.45\textwidth]{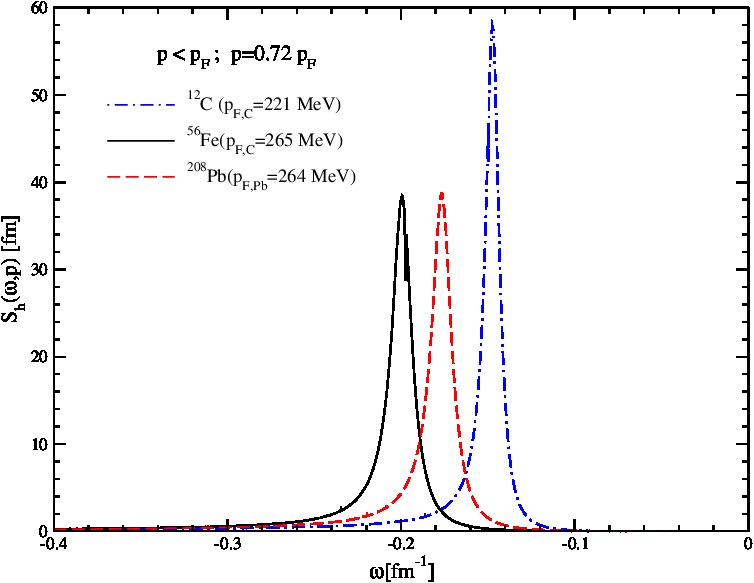}
  \includegraphics[height=0.28\textheight,width=0.45\textwidth]{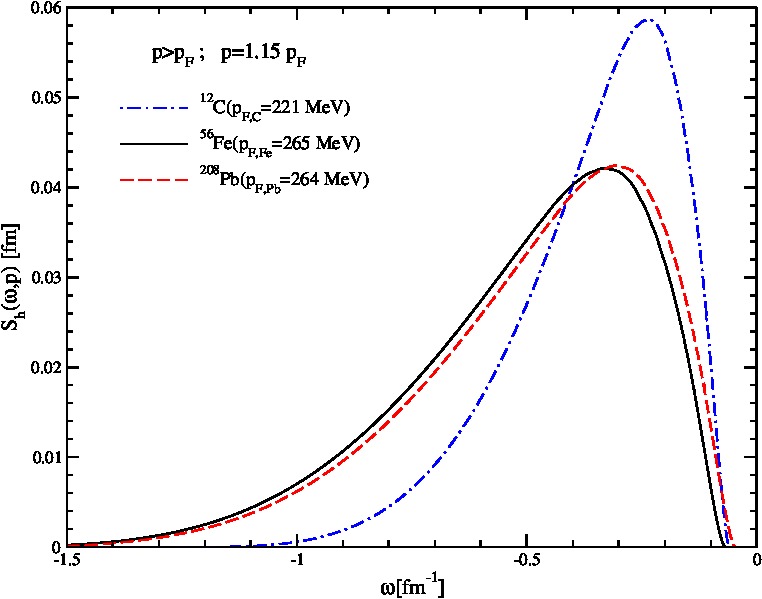}
  \caption{Results for $S_h(\omega,{\bf p})$ vs $\omega$ are shown for {\bf(i)} $p < p_F$(Left panel) and $p > p_F$(Right panel) in various nuclei like $^{12}$C, $^{56}$Fe and $^{208}$Pb.}
 \label{spec_om}
 \end{center}
\end{figure}

 The hole and particle spectral functions fulfill the following relations,
\begin{eqnarray}
 \int_{-\infty}^\mu\,dp^0\;S_h(p^0,\mathbf{p})&=&n(\mathbf{p})\nonumber\\
 \int^{\infty}_\mu\,dp^0\;S_p(p^0,\mathbf{p})&=&1-n(\mathbf{p})\nonumber
\end{eqnarray}
 and thus the spectral functions obey the following sum rule
\begin{equation}
 \int_{-\infty}^\mu\,dp^0\;S_h(p^0,\mathbf{p})+
 \int^{\infty}_\mu\,dp^0\;S_p(p^0,\mathbf{p})=1
\end{equation}
In the absence of interactions (i.e. $\Sigma^N(p)=0$), the nucleon energy $p^0$ is the free relativistic energy $E(\mathbf{p})$ and the dressed propagator $G(p)$ reduces to the free propagator $G^0(p)$ then 
\begin{equation}
 S_h(p^0,\mathbf{p})=S_p(p^0,\mathbf{p})=\delta(p^0-E_N(\mathbf{p}))
\end{equation}
which leads to
\begin{eqnarray*}
 \int_{-\infty}^\mu\,dp^0\;S_h(p^0,\mathbf{p})&=&
 \int_{-\infty}^\mu\,dp^0\;\delta(p^0-E_N(\mathbf{p}))=
 \left\lbrace
\begin{tabular}{ll}
1 & $\textrm{if}\quad \mu>E_N(\mathbf{p})$\\
0 & $\textrm{if}\quad \mu<E_N(\mathbf{p})$
\end{tabular}
\right.\\
 \int^{\infty}_\mu\,dp^0\;S_p(p^0,\mathbf{p})&=&
 \int^{\infty}_\mu\,dp^0\;\delta(p^0-E_N(\mathbf{p}))=
 \left\lbrace
\begin{tabular}{ll}
1 & $\textrm{if}\quad \mu<E_N(\mathbf{p})$\\
0 & $\textrm{if}\quad \mu>E_N(\mathbf{p})$
\end{tabular}
\right.
\end{eqnarray*}
If $E_N(\mathbf{p})$ is the total relativistic energy, then chemical potential $\mu$ must incorporate the 
nucleon mass $M_N$:
\begin{equation}
 \mu=M_N+\epsilon_F
\end{equation}
This definition leads to a constant shift in the integration variable $p^0$ such as:
\begin{equation}
 p^0=\omega+M_N~;\hspace{5 mm} \Rightarrow \omega=p^0-M_N.
\end{equation}
Now the integration of hole and particle spectral functions will be modified to
\begin{eqnarray*}
 \int_{-\infty}^\mu\,d\omega\;S_h(\omega,\mathbf{p})&=&
 \int_{-\infty}^{\mu-M_N}\,d\omega\;\delta(\omega+M_N-E_N(\mathbf{p}))\\
 &=&\left\lbrace
\begin{tabular}{ll}
1 & $\textrm{if}\quad \mu-M_N>E_N(\mathbf{p})-M_N\Rightarrow \epsilon_F>\epsilon(\mathbf{p})$\\
0 & $\textrm{if}\quad \mu-M_N<E_N(\mathbf{p})-M_N\Rightarrow \epsilon_F<\epsilon(\mathbf{p})$
\end{tabular}
\right.\\
 \int^{\infty}_\mu\,d\omega\;S_p(\omega,\mathbf{p})&=&
 \int^{\infty}_{\mu-M_N}\,d\omega\;\delta(\omega+M_N-E_N(\mathbf{p}))\\
 &=&\left\lbrace
\begin{tabular}{ll}
1 & $\textrm{if}\quad \mu-M_N<E_N(\mathbf{p})-M_N\Rightarrow\epsilon_F<\epsilon(\mathbf{p})$\\
0 & $\textrm{if}\quad \mu-M_N>E_N(\mathbf{p})-M_N\Rightarrow\epsilon_F>\epsilon(\mathbf{p})$
\end{tabular}
\right.
\end{eqnarray*}
where $\epsilon(\mathbf{p})=E_N(\mathbf{p})-M_N$ is the nucleon kinetic energy. The behavior of hole spectral function vs removal energy $\omega$ is shown in Fig.~\ref{spec_om} for $^{12}C$, $^{56}Fe$ and $^{208}Pb$. 
From the figure, one may notice that for $p<p_F$, spectral function has a sharp and narrow distribution similar to the delta function
while for $p>p_F$, the distribution has a wide range though very small in magnitude. Furthermore, it may be noticed that the hole spectral function has a smaller magnitude for heavier nuclear targets which is because of the enhancement in the probability of interaction among the nucleons.

%new section
\section{Local Density Approximation}\label{app:lda}
In the local density approximation, Fermi momentum is not fixed but depends upon the 
interaction point ($r$) in the nucleus and is related to the nuclear density as
% In the local density approximation, Fermi momentum is not fixed, but depends upon the  and is given by
\begin{equation}
 p_F(r)=\left(\frac{3 \pi^2 \rho(r)}{2}\right)^{1/3}.
\end{equation}
Thus the Fermi momentum of the nucleon is not a constant number unlike the global Fermi gas model.
In the global Fermi gas model $p_{F}$ is taken to be a constant value like, 221 MeV for $^{12}$C, 251 MeV for $^{40}$Ca, etc. In the local density approximation, the free lepton-nucleon cross section is
folded over the density of the nucleons in the nucleus and integrated over the whole volume of the nucleus. The differential scattering cross section is then given by
\begin{equation}
d\sigma_A = \int \; d^3r\;\rho(r)\;d\sigma_N 
\end{equation}
In a symmetric nuclear matter, each nucleon occupies a volume of $(2\pi \hbar)^3$. However, because of the two possible spin orientations of the nucleon, each unit cell in the configuration space
is occupied by the two nucleons. Therefore, the number of nucleons in a certain volume is given by ($\hbar=1$ in natural units)
\begin{eqnarray}
 N &=& 2 V \;\int_0^{p_{F}}\;\frac{d^3p}{(2 \pi)^3},\\
 \Rightarrow \rho &=& {N \over V} = 2 \int_0^{p_{F}}\;\frac{d^3p}{(2 \pi)^3}\; n({\bf p,r}),
\end{eqnarray}
where $n({\bf p,r})$ is the occupation number of a nucleon lying within the Fermi sea such that
\begin{equation}
n({\bf p,r}) = 
\left\{
 \begin{array}{c}
1 \;\;\;\mbox{for ${\bf p} \le {\bf p_{F}}$}  \\
0 \;\;\;\mbox{for ${\bf p} > {\bf p_{F}}$}
 \end{array}\right.
\end{equation}
In the present model, the spectral
functions of proton and neutron are respectively the function of local Fermi momentum $p_{_{{F}_{p,n}}}(r)=\left[ 3\pi^{2} \rho_{p(n)}({\bf {r}}) \right]^{1/3}$ 
for proton and neutron in the nucleus. The proton and neutron densities $\rho_{p(n)}(r)$
are related to $\rho(r)$ as~\cite{Haider:2015vea, Haider:2016zrk} 
\begin{eqnarray}
 \rho_{p}(r) &=& \frac{Z}{A}\;\rho(r)\nonumber\\
 \rho_{n}(r) &=& \frac{(A-Z)}{A}\;\rho(r)\nonumber
\end{eqnarray}
The equivalent normalization to Eq.(\ref{norm1}) is written as
\begin{eqnarray} \label{norm2}
2\int\frac{d^{3}p}{(2\pi)^{3}}\int_{-\infty}^{\mu}S_{h}(\omega,p,p_{_{{F}_{p,n}}}({\bf {r}})) d\omega= \rho_{p,n}({\bf {r}}),
\end{eqnarray}
These spectral functions are normalized individually for the proton ($Z$) and neutron ($N=A-Z$) 
 numbers in a nuclear target.
\begin{eqnarray}
  2 \int d^3r\;\int \frac{d^3 p}{(2\pi)^3} \;\int_{-\infty}^{\mu_p}\;S_h^p(\omega,{\bf p},\rho_p(r))\;d\omega &=& Z\;, \nonumber\\
    2 \int d^3r\;\int \frac{d^3 p}{(2\pi)^3} \;\int_{-\infty}^{\mu_n}\;S_h^n(\omega,{\bf p},\rho_n(r))\;d\omega &=& N\;, \nonumber
 \end{eqnarray}
where factor 2 is due to the two possible spin projections of nucleon. 
Through the hole spectral function $(S_h(p^0,{\bf p},\rho(r)))$, we incorporate the effects of Fermi motion, Pauli blocking and nucleon correlations.
 The spectral function is properly normalized and checked by
obtaining the correct baryon number and binding energy for a given nucleus such that
 \begin{eqnarray} \label{norm1}
4\int\frac{d^{3}p}{(2\pi)^{3}}\int_{-\infty}^{\mu}S_{h}(\omega,{\bf{p}},p_{_F}(r)) d\omega=\rho(r)
\end{eqnarray}
or equivalently
\begin{eqnarray} \label{norm2}
\int d^3 r\; 4\int\frac{d^{3}p}{(2\pi)^{3}}\int_{-\infty}^{\mu}S_{h}(\omega,{\bf{p}},p_{_F}(r)) d\omega= A.
\end{eqnarray}
Since we are not looking at the final state particles, therefore, we will consider only hole spectral function. The normalization of the hole spectral function is ensured by
obtaining the baryon number $(A)$ of a given nucleus and the
binding energy of the same nucleus.
 Furthermore, for the nonisoscalar nuclear target the spectral function is normalized to the proton and neutron numbers separately. 
\end{appendices}
%\newpage

%Unused bibitems

\end{document}